\documentclass[superscriptaddress,amsmath,amssymb,10pt]{revtex4-2}

\usepackage[margin=0.9in]{geometry}

\usepackage[utf8]{inputenc}
\usepackage[T1]{fontenc}
\usepackage{graphicx}
\usepackage{array}
\newcolumntype{K}[1]{>{\centering\arraybackslash}p{#1}}
\usepackage{braket}
\usepackage{csquotes}
\usepackage{xspace}
\usepackage{algorithmic}

\usepackage{booktabs}
\usepackage{multirow}
\usepackage{color,soul}
\usepackage{mathtools}
\usepackage[hypertexnames=false]{hyperref}

\usepackage{bm}
\usepackage{bbm}
\usepackage{bbold}

\newcommand{\epsdiamond}{\varepsilon_{\diamond}}

\newcommand{\fswap}{f_{\text{swap}}}

\newcommand{\Tdg}{T^{\dagger}}
\newcommand{\TX}{T_{X}}
\newcommand{\TXdg}{T_{X}^{\dagger}}
\newcommand{\Sdg}{S^{\dagger}}
\newcommand{\rX}{\sqrt{X}}

\newcommand{\norm}[1]{\left\lVert#1\right\rVert}

\newcommand{\kmax}{k_{\text{max}}}

\usepackage{xcolor}
\definecolor{riverlane_green}{RGB}{0, 111, 98}
\definecolor{riverlane_light_green}{RGB}{0, 150, 143}
\definecolor{riverlane_orange}{RGB}{255, 117, 0}
\definecolor{riverlane_red}{RGB}{220, 68, 5}
\definecolor{riverlane_pink}{RGB}{207, 111, 127}

\hypersetup{
  colorlinks   = true, 
  urlcolor     = riverlane_green, 
  linkcolor    = riverlane_green, 
  citecolor   = riverlane_green  
}

\usepackage[capitalise]{cleveref}
\usepackage{amsthm}
\theoremstyle{definition}
\newtheorem{definition}{Definition}
\theoremstyle{plain}
\newtheorem{lemma}{Lemma}

\theoremstyle{remark}

\usepackage{siunitx}

\usepackage{enumitem}

\usepackage{tikz}
\usetikzlibrary{quantikz2}

\usepackage{pifont,xcolor}
\definecolor{softgreen}{rgb}{0.1,0.5,0.1}

\usepackage{fancyhdr}
\usepackage{xcolor}

\newcommand{\equalcorr}{These authors contributed equally to this work. Corresponding author: \href{mailto:nick.blunt@riverlane.com}{nick.blunt@riverlane.com}}

\begin{document}

\title{Quantum computing at the MegaQuOp scale: lattice surgery resource estimates for a condensed matter application}

\author{Christopher Erickson}
\thanks{\equalcorr}
\affiliation{Riverlane, Cambridge, CB2 3BZ, United Kingdom}
\author{Annie Ray}
\thanks{\equalcorr}
\affiliation{Riverlane, Cambridge, CB2 3BZ, United Kingdom}
\author{Aleksei V. Ivanov}
\thanks{\equalcorr}
\affiliation{Riverlane, Cambridge, CB2 3BZ, United Kingdom}
\author{Jonathan Kunjummen}
\thanks{\equalcorr}
\affiliation{Riverlane, Cambridge, CB2 3BZ, United Kingdom}
\author{Andreas Juul Bay-Smidt}
\affiliation{NNF Quantum Computing Programme, Niels Bohr Institute, University of Copenhagen, Denmark}
\author{Joonas Majaniemi}
\affiliation{Riverlane, Cambridge, CB2 3BZ, United Kingdom}
\author{Marius Bothe}
\affiliation{Riverlane, Cambridge, CB2 3BZ, United Kingdom}
\affiliation{Astex Pharmaceuticals, 436 Cambridge Science Park, Cambridge, CB4 0QA, United Kingdom}
\author{Christoph Sünderhauf}
\affiliation{Riverlane, Cambridge, CB2 3BZ, United Kingdom}
\author{Marco Radaelli}
\affiliation{Riverlane, Cambridge, CB2 3BZ, United Kingdom}
\author{Modesto Orozco-Ruiz}
\affiliation{Riverlane, Cambridge, CB2 3BZ, United Kingdom}
\author{R\'obert Izs\'ak}
\affiliation{Riverlane, Cambridge, CB2 3BZ, United Kingdom}
\author{Erik G. C. P. van Loon}
\affiliation{NanoLund and Division of Mathematical Physics, Department of Physics, Lund University, Lund, Sweden}
\affiliation{Institute for Molecules and Materials, Radboud University, 6525 AJ Nijmegen, The Netherlands}
\author{Ophelia Crawford}
\affiliation{Riverlane, Cambridge, CB2 3BZ, United Kingdom}
\author{Earl T. Campbell}
\affiliation{Riverlane, Cambridge, CB2 3BZ, United Kingdom}
\author{Nick S. Blunt}
\thanks{\equalcorr}
\affiliation{Riverlane, Cambridge, CB2 3BZ, United Kingdom}

\thispagestyle{plain}
\pagestyle{plain}

\begin{abstract}
As error-corrected quantum computers develop beyond early demonstrations, they are expected to evolve through a series of breakthrough systems at different scales. In this paper, we study error-corrected quantum computing at the MegaQuOp scale, defined informally as the regime where a QEC code can support on the order of a million quantum operations. We develop a quantum simulation to probe magnetic excitations of strongly correlated two-dimensional materials using Trotterised dynamics, taking the transition between Lieb and kagome lattice Hubbard models as a case study, and defining an example of a useful condensed matter simulation for early error-corrected quantum computers. Considering superconducting qubits and the planar surface code, we perform resource estimation by full compilation to lattice-surgery-based computation. We study the range of applicability of the partially fault-tolerant STAR architecture, using logical state tomography with improved non-Clifford simulation techniques. We then perform end-to-end resource estimation using the quasi-probability Clifford+T methodology [arXiv:2605.31544] and magic state cultivation. It is estimated that useful magnetic spectral properties of non-trivial Hubbard models can be obtained by circuits with an expected $4.5$ million quantum operations of space-time volume $d \times d \times d$ each, roughly corresponding to the MegaQuOp regime. For a superconducting noise model with $p=10^{-3}$, we estimate this can be achieved with $d=15$ and 77,000 physical qubits and an expected $0.55$ seconds per shot, or $15.3$ hours per 100,000 shots. This is reduced to $d=11$ and 43,000 physical qubits under more optimistic noise model and Trotter error assumptions. This case study defines a benchmark for assessing emerging QEC architectures in the early fault-tolerant era.
\end{abstract}

\maketitle

\tableofcontents

\clearpage

\section{Introduction}
\label{sec:intro}

Recent years have seen significant advances in digital quantum computing, with a variety of platforms now being routinely used to study a range of problems of scientific importance \cite{kim_2023, quantinuum_trotter, alam2025_1, alam2025_2, granet2026, millar2026}. Despite this, it is expected that physical gate error rates will not be reduced significantly below about $10^{-4}$, which will place limits on the capabilities of non-error-corrected calculations. In parallel, recent years have also seen a series of rapidly developing quantum error correction (QEC) experiments \cite{google2023, google2025, Caune2026, Bluvstein2026, Paetznick2026, Harper2026, Lavasani2026}, taking QEC from the realm of theory to practice, and suggesting that quantum computing may reach significantly lower error rates in the near future. This is necessary, as resource estimation studies for quantum simulation of \emph{ab initio} quantum chemistry typically suggest gate counts of at least $10^8$ and hundreds or thousands of logical qubits \cite{Georges2025, Low2025, low2026denser, ivanov2026}.

Technology will not suddenly leap from current error rates to those supporting large-scale computation, but evolve through a series of breakthrough systems of varying scales. Understanding the capability of error-corrected devices at different scales requires detailed end-to-end resource estimation for relevant algorithms. This in turn informs the better design of such devices, and reveals where research and engineering efforts should be focused.

This paper studies quantum computation at the MegaQuOp scale, which has emerged as an early milestone target for fault-tolerant quantum computing \cite{preskill2025beyond}. The MegaQuOp regime is defined informally as the regime where $10^6$ quantum operations can be performed reliably. More formally, and particularly for lattice-surgery-based computation, MegaQuOp is defined by the QEC code size required to achieve $10^{-6}$ logical error probability per $d$ rounds of a memory experiment \cite{gidney2021fault}. This is significantly lower than gate error rates on a NISQ device. However, fault-tolerant quantum computing (FTQC) introduces a number of overheads which must be properly assessed. For example, certain gates are expected to be more expensive to implement compared to a NISQ device, and idling has a comparable error rate to lattice surgery operations and must be properly accounted for. Therefore, understanding the potential benefit of such a device is non-trivial and requires detailed resource estimation, going beyond simply counting non-Clifford gates, but instead requiring end-to-end compilation \cite{Blunt2024}.

\begin{figure*}
    \centering
    \includegraphics[width=1.0\linewidth]{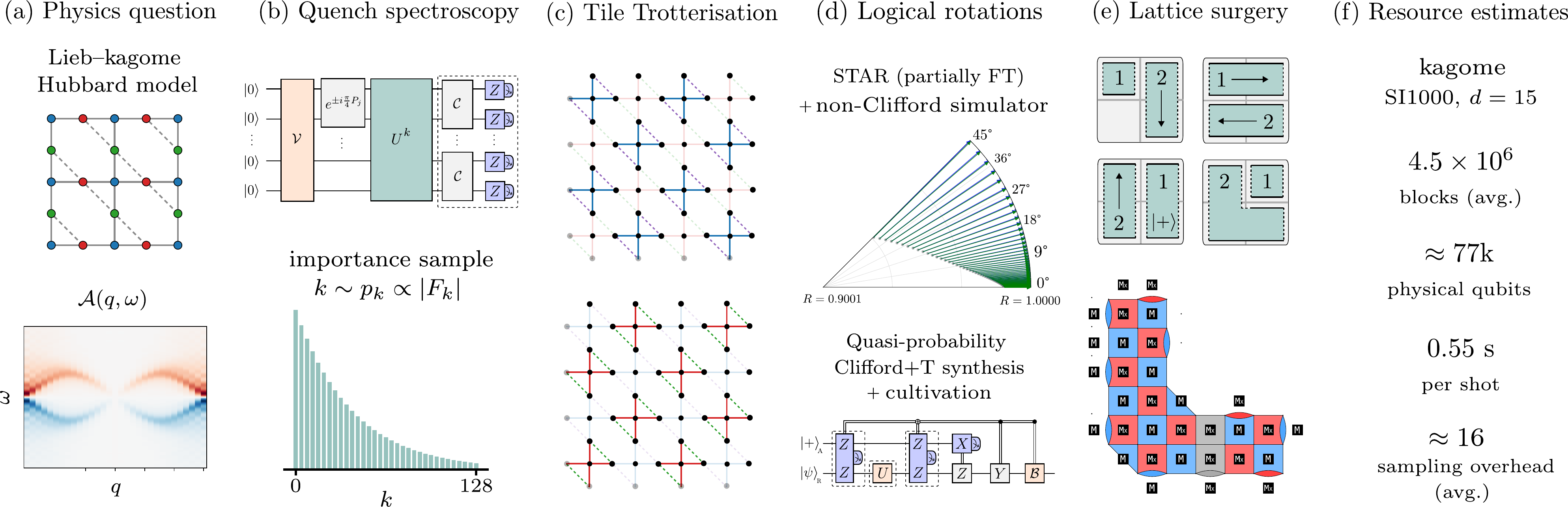}
    \caption{{\bf Computing magnetic spectra at the MegaQuOp scale.} A quantum simulation to probe magnetic excitations for Lieb-kagome Hubbard models requires about 77,000 physical qubits with the surface code (SI1000 noise with $p=10^{-3}$ at $d=15$). (a) The Lieb-kagome Hubbard model and (b) protocol for calculating magnetic spectra are presented in \cref{sec:lieb_kagome_hubbard}. (c) The Tile Trotterisation scheme and QPU mapping are presented in \cref{sec:tile_trotter}. (d) STAR architecture results are presented in \cref{app:star_primary} and \cref{app:star_results}, with the non-Clifford simulation technique used to perform logical state tomography discussed in \cref{app:non_cliff_sim}. Our quasi-probability Clifford+T method is presented in \cref{sec:rotations}, with supporting magic state cultivation simulations in \cref{app:magic_state_cultivation}. (e) Lattice surgery compilation of circuit elements is given in \cref{app:lattice_surgery_constructions}. (f) Final resource estimates are presented and discussed in \cref{sec:resource_estimates}.}
    \label{fig:overview}
\end{figure*}

To investigate a potential application for such a device, we propose an experiment to calculate magnetic excitation spectra for the Hubbard model on Lieb-kagome lattices \cite{LiebKagome2026}. Magnetic excitation spectra are measured directly by inelastic neutron scattering~\cite{boothroyd_principles_2020,scheie_2026_cooper_pair_localization} and resonant inelastic X-ray scattering~\cite{de_groot_resonant_2024,mitrano_2024_explorting_quantum_materials,padma_2025_beyond_hubbard_pairing}, and can reveal properties missed by ground-state properties alone \cite{padma_2025_beyond_hubbard_pairing,scheie_2026_cooper_pair_localization}. Tuning the lattice from Lieb to kagome takes the system from a ferrimagnetic \cite{Lieb89} to a frustrated system with zero magnetic moment \cite{Sun21}, and dramatic changes in magnetic excitations are expected during this transition \cite{LiebKagome2026}. Considering superconducting qubits and the surface code, we perform a detailed compilation of the Trotter circuits for this problem to lattice surgery operations, including full resource estimation for both Clifford and non-Clifford elements of the circuits. Our full approach is outlined in \cref{fig:overview}.

An important question for early fault-tolerant Trotter simulations is how to perform Pauli rotations. We consider both fault-tolerant (Clifford+T) and the partially fault-tolerant STAR \cite{Akahoshi2024, Toshio2025, Akahoshi2025, Chung2026, Kanasugi2026, Toshio2026, Ismail2026_1, Ismail2026_2} methodologies. The STAR architecture is based on performing rotation gates noisily, with error rate $\mathcal{O}(p)$, but with a reduced volume compared to traditional Clifford+T synthesis. We perform detailed simulation of the STAR architecture approach under superconducting noise models and gate sets. This is achieved by logical state tomography on the required resource states, made possible by improvements to a non-Clifford simulation technique. Ultimately, we obtain a negative result: even under optimistic assumptions -- assuming that coherent error can be exactly calibrated away, and choosing rotation angles in the Clifford hierarchy -- the estimated logical errors at physical error rate $p=10^{-3}$ are too large for our case study. Compared to previous works, this is primarily due to longer simulation times required to resolve magnon branches, and a stricter noise model with $p=10^{-3}$. Instead, we focus on Clifford+T synthesis of rotation gates. Clifford+T synthesis is often considered to have a high overhead of tens or even hundreds of T gates per rotation. However, it was recently shown that, under probabilistic compilation, Clifford+T synthesis of Pauli rotations can become substantially cheaper for small $\theta$ \cite{bothe2026}. Here we demonstrate that, even for large rotation angles, and using magic state cultivation~\cite{gidney2024magic} and careful compilation choices, Clifford+T synthesis can be made efficient in the MegaQuOp regime.

Our case study defines a minimal useful condensed matter experiment with which to assess alternative and emerging fault-tolerant architectures. The experiment is useful in the sense that it builds on an important tool in condensed matter physics (calculating magnetic excitation spectra) and reaches simulation parameters sufficient to provide physical insight for a problem of interest to condensed matter theorists, which cannot be calculated trivially, e.g. by exact diagonalisation, and also requiring sufficiently long simulation time to resolve magnon branches. In this paper we consider a monolithic grid architecture for the planar surface code and superconducting qubits; however we hope that this will be valuable as a benchmark against which to compare alternative qubit modalities and emerging architectures \cite{Yoder2025, Geher2026, Tripier2026, Webster2026, Oratomic_Shors, Ismail2026_2}.

\section{MegaQuOp quantum computing}
\label{sec:megaquop_overview}

The phrase TeraQuOp was introduced \cite{gidney2021fault} as referring to the regimes where, informally, a QEC code is sufficiently large it could support approximately $10^{12}$ logical quantum operations. More precisely, it was defined \cite{gidney2021fault} as the code size required to achieve $p_I = 10^{-12}$ logical error probability per $d$ rounds of a memory experiment. Therefore, QuOps are formally a property of a QEC code (and decoder, and device noise), not of a logical circuit. The informal and formal definitions coincide when we perform a trillion logical idling operations (each lasting $d$ syndrome-extraction rounds) and require all operations to succeed with probability exceeding $\exp(-1)\approx 37\%$.  More generally, a $Q=p_{I}^{-1}$ QuOp quantum error correction code can execute $V$ logical idling operations with a success probability $(1-1/Q)^V \approx \exp(-V/Q)$.  How many useful (non-idling) logical operations can be achieved with a $Q$ QuOp code is a subtle question,  leading some researchers to use QuOps as shorthand for different resource metrics, such as the number of Pauli rotations in a logical circuit (e.g. \cite{Chung2026, Ismail2026_2}). Luckily, when using the surface code for lattice surgery computations, we can easily respect the precise definition of QuOps as follows. Note that for any Clifford logical operation $L$, the logical error probability $p_L$ is well approximated by $p_L=V_L p_I$ where $V_L$ is the space-time volume (in units of one $d \times d$ logical qubit over $d$ rounds), for numerical evidence see Refs.~\cite{chamberland2022circuit,geher2024error,gidney2024inplace}, and also \cref{fig:ls_vs_mem}, where we compare the error rate of quantum memory and lattice surgery simulations of equivalent volumes. As such, we can take $V$ as the total space-time volume including both idling logical qubits and the auxiliary workspace used to support logic, and then the algorithm success probability is $\exp(-V/Q)$, requiring $V \ll Q$ for high reliability. Note that we must separately budget for the finite failure probability of distillation or cultivation protocols for non-Clifford gates. This approach is standard across a variety of resource counting frameworks \cite{litinski2019game,litinski2022active,beverland2022assessing,huggins2025fluid}.

In the algorithmic context, therefore, the TeraQuOp regime refers to applications where $V \ll Q = 10^{12}$ to budget for high success probability. Existing resource analyses give a range of QuOp counts; serialised Pauli-based computation \cite{Bravyi2015,litinski2019game} would require GigaQuOp error rates to support Hubbard model simulations of Ref.~\cite{campbell2022early} or PetaQuOp error rates to support RSA2048 circuits of Ref.~\cite{gidney2025factor}. In contrast, the current era of quantum error correction is close to break-even \cite{google2025} and therefore best described as close to KiloQuOp. MegaQuOp quantum computers have been identified as the next significant major milestone \cite{preskill2025beyond}.  While MegaQuOp-scale systems are likely to demonstrate error-corrected circuits that defy classical simulation, it is a significant challenge to design MegaQuOp-scale applications that solve problems of scientific or broader interest.  Pursuing this goal will need advances in algorithm design and compilation, but even then getting close to $V \ll 10^6$ will be challenging and potentially impossible.

We may however significantly extend the space-time budget $V$, without increasing $Q$, by taking inspiration from error mitigation techniques designed for noisy (not-error-corrected) devices \cite{Temme2017, Endo2018}, which can also be applied in various forms at the logical level \cite{shtanko2026, shutman2026, dinca2026, tsubouchi2026}. An established and experimentally successful \cite{Vovrosh2021, Blunt2023, Farrell2024} technique in algorithms that estimate some quantity $M(k)=\mathrm{Tr}[ U^k \rho U^{-k} P ]$ has been to assume that the noise is roughly depolarising so that the noisy signal has the form $\tilde{M}(k) = \gamma^k M(k)$, where $ 0 < \gamma < 1$ is some damping constant, and the noiseless signal can be extracted at the cost of an additional factor of $(1/\gamma)^{2k}$ shots. This scales exponentially in $k$ and therefore cannot be pushed asymptotically, but could extend the computational runway of MegaQuOp devices. In the QEC setting, we let $v$ be the spacetime volume to implement $U$ so that the circuit for $U^k$ has volume $V=vk$.  Considering $v=10^5$, $k=11$ and $Q=10^6$, then $V > Q$ and conventional resource estimation would yield only a $33\%$ success probability where normally over 50\% is demanded.  However, if the depolarising assumption holds, error mitigation allows us to estimate the signal in exchange for a modest overhead in sample complexity. If such a sampling overhead is acceptable, as it has proved in the NISQ era, the scope of MegaQuOp scale applications can be significantly broadened.

In Appendix~\ref{sec:lambda_calc} we simulate quantum memory experiments for the surface code under superconducting qubit assumptions. We primarily focus on the SI1000 noise model \cite{gidney2021fault}, taking CZ as the native two-qubit gate, however, we also consider an ``Optimistic'' noise model where measurement and reset errors are on-par with two-qubit gates. The logical error per round is expressed $\varepsilon_d = p_0 \Lambda^{-(d+1)/2}$, where $p_0$ and $\Lambda = \varepsilon_{d} / \varepsilon_{d+2}$ are fitted constants. From this, the number of QuOps, $Q$, is estimated by $1/Q \approx d \times p_0 \Lambda^{-(d+1)/2}$. For an SI1000 noise model at $p=10^{-3}$, and using the correlated PyMatching decoder \cite{higgott2025sparse}, we obtain $\Lambda = 8.8(4)$, while for the Optimistic noise model at $p=10^{-3}$ we obtain $\Lambda = 12.8(9)$. In the SI1000 noise model the number of QuOps obtained at $d=11$, $13$ and $15$ are $2.5 \times 10^5$, $1.9 \times 10^6$ and $1.4 \times 10^7$, respectively, while under the Optimistic noise model the number of QuOps obtained at distances $d=11$ and $d=13$ are $2.3 \times 10^6$ and $2.5 \times 10^7$, respectively. Therefore, for $p=10^{-3}$ and dependent on the noise assumptions, $\sim 0.25$--$25$ MegaQuOps is achieved with $d=11$--$15$, approximately.

\section{Magnetic spectral functions and the Lieb-kagome Hubbard model}
\label{sec:lieb_kagome_hubbard}

For materials with strong electronic correlations, it is not just the ground state that matters but also the excitations. For sensing applications of correlated materials, for example, the excitations are essential since they determine how the material responds to external stimuli. These excitations can be described by momentum- and energy-resolved spectral functions, which are a powerful tool for characterising the collective motion of electrons in materials. Magnetic spectral functions are directly measurable: both inelastic neutron scattering~\cite{boothroyd_principles_2020,scheie_2026_cooper_pair_localization} and resonant inelastic X-ray scattering~\cite{de_groot_resonant_2024,mitrano_2024_explorting_quantum_materials,padma_2025_beyond_hubbard_pairing} allow one to measure the magnetic excitation spectrum of a material as a function of momentum and energy transfer. These techniques were applied to different cuprate ladder compounds~\cite{padma_2025_beyond_hubbard_pairing,scheie_2026_cooper_pair_localization}, and in both cases, the measured magnetic spectrum ruled out the single-band Hubbard model as a reliable description of such materials. It was identified that a missing ingredient is a large attractive nearest-neighbour interaction -- a conclusion that could not be reached from the ground-state properties alone. Thus, spectral functions allow one to determine the physical mechanisms responsible for certain properties of materials. While classical algorithms such as DMRG used in the works above are very successful for quasi-one-dimensional materials, quantum computation offers a complementary approach for describing materials in 2D and 3D in cases when the entanglement generated during the dynamics becomes large~\cite{alam2025_1,alam2025_2}. In the following, we begin by describing a general set-up to estimate magnetic spectral functions of condensed matter Hamiltonians on a quantum computer. We then specialise this to our case study, the Lieb-kagome Hubbard model.

Mathematically, the spectral function associated with the commutator of operators $A$ and $B$ is defined as~\cite{aarts_nonequilibrium_2001}:
\begin{equation}\label{eq:spectral_function_time}
    C(t, t') = -\mathrm{i} \langle \psi | [ A(t), B(t') ] | \psi \rangle.
\end{equation}
In this work we will be interested in its frequency representation at fixed perturbation time $t' = 0$:
\begin{equation}\label{eq:spectral_function_omega}
    \mathcal{A}(\omega) = \frac{\mathrm{i}}{2\pi} \int_{-\infty}^{\infty} dt \, e^{\mathrm{i} \omega t} C(t), \quad C(t) := C(t, 0).
\end{equation}
The function $\mathcal{A}(\omega)$ is a sum of spectral peaks, whose positions lie at energy differences, $\Delta_{ji} := E_j - E_i$, corresponding to energy transitions (\cref{lemma:spectral_function_approx_init}).
If the initial state $|\psi\rangle$ is a stationary state, and $B = A^{\dagger}$, then $\mathcal{A}(\omega)$ is proportional to the imaginary part of the \emph{dynamical susceptibility}; after taking a spatial Fourier transform (discussed below), and for appropriate spin operators, the dynamical susceptibility measures the linear response of a system in equilibrium to a magnetic field oscillating at wave vector ${\bf q}$ and with frequency $\omega$ \cite{Sachdev_2011}. It therefore allows one to probe for the presence of magnetic excitations with a given ${\bf q}$ and $\omega$, which in turn provides insight into the magnetic phases of a system. In our set-up we will instead assume a non-stationary initial state $|\psi\rangle$. Using an approximate rather than exact eigenstate as the initial state is common practice in both quantum and classical computing~\cite{vilchez-estevez_extracting_2025, millar2026, kemper_efficient_2026}. We note that, even in this case, if \cref{eq:spectral_function_time} is averaged over the perturbation time $t'$ then it corresponds to a strict dynamical susceptibility for a stationary state, specifically a mixed state where each energy eigenstate, $|\Psi_i\rangle$, has weight $p_i = |\langle \psi | \Psi_i \rangle |^2 $ in the mixture. However, for the rest of this paper we continue to analyse the case with $t'=0$, as in \cref{eq:spectral_function_omega}.

In practice, we perform discrete time evolution using Trotterisation and the equation which approximates \cref{eq:spectral_function_omega} is given by:
\begin{equation}\label{eq:spectral_function_omega_approx}
    \mathcal{A}(\omega) \approx \mathrm{i} \sum_{|k| \leq \kmax} F_{k} e^{\mathrm{i}k\omega\delta t} g_k, \quad g_k := C(k \cdot \delta t)
\end{equation}
where $F_k$ are the real Fourier coefficients of the broadening function, $F_k=F_{-k}$, and $\kmax$ is the  maximum number of Trotter steps in each direction. In certain cases, which we consider in this work, positive values of $k$ suffice to recover the full spectral function (see Appendix \ref{sec:appendix_spectral_function} for details).

\begin{figure*}
    \centering
    \includegraphics[width=1.0\linewidth]{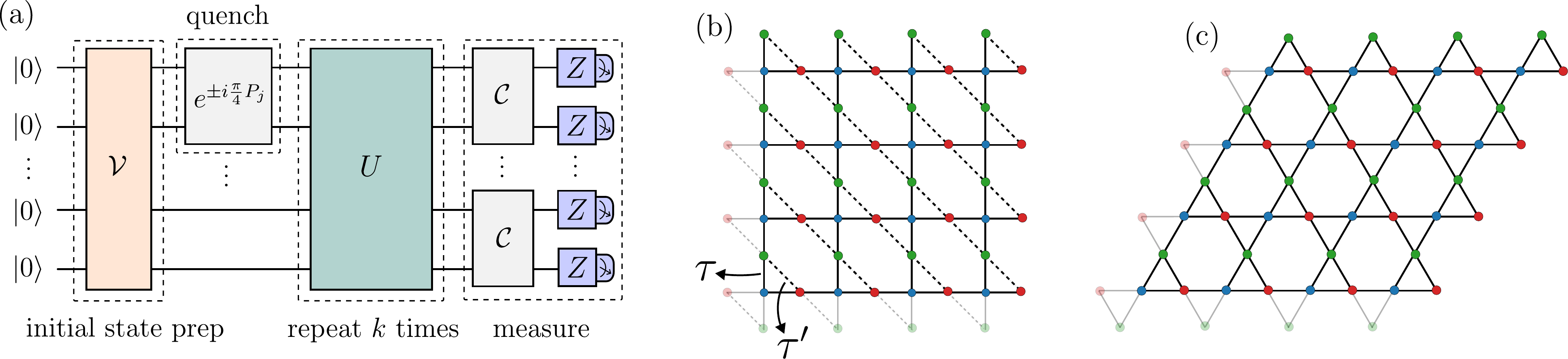}
    \caption{{\bf Quench circuit and Lieb-kagome lattice.} (a) The quench circuit sampled to construct magnetic spectral functions. (i) Apply the unitary $\mathcal{V}$ to prepare the initial state. (ii) Apply a quench $e^{\pm i \frac{\pi}{4} P_j}$ to at most two qubits (corresponding to a single lattice site). (iii) Apply the Trotter step, $U$, a total of $k$ times, where $k$ is importance sampled with probability $p_k$ (\cref{eq:importance_sample_probs}). (iv) Measure in the appropriate basis, which may involve applying Cliffords, $\mathcal{C}$. Note that the fermionic ordering may also be updated before and after applying $U^k$ (not shown). (b) The Lieb-kagome lattice considered in this work, with $48$ sites. The three sites per unit cell are labelled blue, red and green, and periodic boundary connections are shown faded out. Solid lines and dashed lines have hopping strength $\tau$ and $\tau'$, respectively. Varying $\tau'$ from $0$ to $\tau$ interpolates from the Lieb lattice to the kagome lattice (shown in (c)). The Lieb lattice is bipartite and unbalanced, so the ground state is ferrimagnetic by Lieb's theorem \cite{Lieb89}, while magnetic frustration in the kagome lattice favours ground states with zero magnetic moment~\cite{Sun21}. As $\tau'$ is tuned from $0$ to $\tau$, concomitant changes in the magnetic excitation spectrum are expected~\cite{LiebKagome2026}, which our case study is designed to probe.}
    \label{fig:circuit_and_lattice}
\end{figure*}

In order to estimate $C(t)$ on the quantum computer, we use the Pauli decomposition $B = \sum_{j} w_j P_j$,
and as is shown in Ref.~\cite{vilchez-estevez_extracting_2025} (see \cite{Knap2013, Yoshimura2016, Baez2020} for earlier discussion of the same idea, and \cite{andersen2026} for a recent implementation of a related idea), one can write:
\begin{align}\label{eq:quench_formula}
     C(t) = - \sum_{\sigma=0,1}\sum_{j} w_{j,\sigma} \langle \psi_{\sigma, j} | A(t) | \psi_{\sigma, j} \rangle, \quad \psi_{\sigma, j} :=  e^{\mathrm{i}(-1)^{\sigma} \frac{\pi}{4} P_j } | \psi \rangle, \quad w_{j,\sigma} := (-1)^{\sigma} w_j.
\end{align}
Therefore, the calculation of the spectral function, $\mathcal{A}(\omega)$, can be reduced to (i) applying a quench operator $e^{\mathrm{i}(-1)^{\sigma} \frac{\pi}{4} P_j }$ to the initial state $|\psi\rangle$, (ii) performing time evolution for $k$ Trotter steps, (iii) evaluating the expectation value of $A$ with respect to the subsequent state and (iv) classically performing the sums in  \cref{eq:spectral_function_omega_approx,eq:quench_formula}; see \cref{fig:circuit_and_lattice}(a). This is closely related to quench spectroscopy~\cite{villa_unraveling_2019,villa_local_2020,millar2026} where the excitation spectrum is reconstructed from a careful choice of the initial states, observables and quench operators. Since each estimate of $g_k$ contains shot noise, we consider using importance sampling to construct $\mathcal{A}(\omega)$ with reduced variance. Specifically, for each shot we choose to perform a circuit of $k$ Trotter steps with probability $p_k$, defined
\begin{equation}
    p_0 = \frac{|F_0|}{\lambda_F}, \qquad
    p_k = \frac{2|F_k|}{\lambda_F} \quad (1 \le k \le k_{\max}), \qquad
    \lambda_F = \sum_{|k| \le k_{\max}} |F_k|,
    \label{eq:importance_sample_probs}
\end{equation}
This is similar to statistical phase estimation techniques, particularly \cite{Lin2022, Blunt2023}, and we have used similar notation to make the connection clear. The sum in \cref{eq:quench_formula} can also be sampled.

We focus on an experiment for estimating the magnetic spectral function by choosing $A$ and $B$ in \cref{eq:spectral_function_time} as reciprocal-space spin operators:
\begin{align}
    &C^{\alpha\beta}(\mathbf{q}, t) := -\mathrm{i} \langle \psi | [S^{\alpha}(\mathbf{q}, t), S^{\beta}(-\mathbf{q},0) ] | \psi \rangle, \quad \alpha,\beta \in \{x, y, z, +, -\},\\
    &S^{\alpha}(\mathbf{q},t) = \frac{1}{\sqrt{N_s}}\sum_{j}e^{-\mathrm{i} \mathbf{q}\cdot\mathbf{r}_{j}} S^{\alpha}_{j}(t),
\end{align}
where $S^{x}_i, S^{y}_i, S^{z}_i$ are the components of the electron spin operator on site $i$, $S^{\pm}_i = S^{x}_{i} \pm \mathrm{i} S^{y}_{i}$, $N_s$ is the total number of sites, $\mathbf{q}$ is a momentum vector in the first Brillouin zone, and $\mathbf{r}_i$ denotes the position of site $i$. In practice, $\mathbf{q}$ takes discrete values because of the finite size of the lattice.

Next, we explicitly consider a lattice with several sites per unit cell, and simplify equations further. For this, we decompose the site index $i$ into the unit cell index $\mu \in \{1, \dots, N_c\}$ associated with the Bravais lattice vector $\mathbf{R}_{\mu}$ and site index within the unit cell, $a \in \{1, \dots n_s\}$, so $i =(\mu, a)$. Then the physical position of the site in the cell is written as $\mathbf{r}_i = \mathbf{R}_{\mu} + \mathbf{r}_a$, and the total number of sites is $N_s = n_s N_c$. We assume that the initial state used in the  spectral function is an eigenstate of the translational operator. Because the Hamiltonian commutes with the translational operator, we can write:
\begin{align}
    & C^{\alpha\beta}(\mathbf{q}, t) = \frac{1}{n_s}\sum_{\mu=1}^{N_c}\sum_{a,b=1}^{n_s} e^{-\mathrm{i} 
    \mathbf{q}\cdot(\mathbf{R}_{\mu} + \mathbf{r}_a-\mathbf{r}_b)} C^{\alpha\beta}_{(\mu,a), (0, b)}(t),\\
    &C^{\alpha\beta}_{(\mu,a), (0, b)}(t) := -\mathrm{i} \langle \psi | [S^{\alpha}_{(\mu,a)}(t), S^{\beta}_{(0, b)}(0) ] | \psi \rangle, \quad \alpha,\beta \in \{x, y, z, +, -\}.
\end{align}
Then, the final target spectral function is
\begin{equation}\label{eq:spectral_function_final_expression}
    \mathcal{A}^{\alpha\beta}(\mathbf{q},\omega) = \frac{\mathrm{i}}{n_s}\sum_{\mu=1}^{N_c}\sum_{a,b=1}^{n_s} e^{-\mathrm{i} 
    \mathbf{q}\cdot(\mathbf{R}_{\mu} + \mathbf{r}_a-\mathbf{r}_b)}
    \sum_{|k|\leq \kmax} F_{k} e^{\mathrm{i}k\omega\delta t} C^{\alpha\beta}_{(\mu,a), (0,b)}(k \cdot \delta t).
\end{equation}
The time step, $\delta t$, and the largest total evolution time, $\delta t \cdot \kmax$, depend on the Hamiltonian and the desired frequency resolution. Note that the circuit of \cref{fig:circuit_and_lattice}(a) allows operators for all sites $i=(\mu,a)$ to be measured simultaneously. The remaining sums in \cref{eq:spectral_function_final_expression} can be sampled, and again we importance sample according to \cref{eq:importance_sample_probs} to reduce the variance of this estimator; the sampling error on $\mathcal{A}^{\alpha\beta}(\mathbf{q},\omega)$ is analysed in \cref{app:sampling_error_analysis}, and discussed further in the Results.

In this work, we analyse the Hubbard model on the Lieb-kagome lattice \cite{Jiang2019} and choose simulation parameters informed by the recent HF+RPA study of dynamical susceptibilities for this problem~\cite{LiebKagome2026}. The Lieb-kagome lattice, presented in \cref{fig:circuit_and_lattice}, covers both the Lieb and the kagome lattices as limiting cases, and both geometries are realised in real materials. The kagome geometry \cite{kagome_metals, Wang2023} underlies quantum magnets such as ZnCu$_3$(OD)$_6$Cl$_2$, whose magnetic excitation spectrum has been measured using inelastic neutron scattering~\cite{han2012fractionalized}, correlated topological metals such as CsV$_3$Sb$_5$~\cite{ortiz_2020_topological_kagome_metal}, and van der Waals compounds Nb$_3$X$_8$~\cite{aretz_strong_2025}. The Lieb geometry describes the CuO$_2$ plane of the cuprates within the three-band model~\cite{emery_1987_theory_of_high}, and the electronic Lieb lattice has also been experimentally realised~\cite{slot_experimental_2017}. The two-dimensional nature of the lattice leads to enhanced correlation effects compared to three-dimensional systems.

The Lieb-kagome Hubbard Hamiltonian is defined as
\begin{equation}\label{eq:LKH_hamiltonian}
    H = - \tau \sum_{\langle i,j \rangle, \sigma} ( a^{\dagger}_{i\sigma} a_{j\sigma} + \textrm{h.c.}) - \tau' \sum_{\langle \langle i,j \rangle \rangle, \sigma} ( a^{\dagger}_{i\sigma} a_{j\sigma} + \textrm{h.c.}) + U \sum_i \Big( n_{i \uparrow} - \frac{\mathbb{1}}{2} \Big) \Big( n_{i \downarrow} - \frac{\mathbb{1}}{2} \Big),
\end{equation}
where a chemical shift is applied to the interaction term, as is standard in many quantum simulation studies of the Hubbard model, see for example \cite{campbell2022early, BaySmidt2025}. This system is visually demonstrated in Fig.~\ref{fig:circuit_and_lattice}(b) for a $4 \times 4$ lattice, which we study in this paper (see discussion below). We assume half-filling throughout. For $\tau'=0$, this model matches the standard Hubbard model on the Lieb lattice; this is a square Bravais lattice with three sites per unit cell. The parameter $\tau'$ allows one to continuously tune the next-nearest-neighbour hopping amplitude between edge-centred sites; when $\tau' = \tau$, the tight-binding parameters exactly match those of the kagome lattice, Fig.~\ref{fig:circuit_and_lattice}(c). In this sense, tuning $\tau'$ from $0$ to $\tau$ allows us to interpolate between the two lattices.

For intermediate values of $\tau'$, the ground state is not known a priori. As one tunes $\tau'$ from the known ferrimagnetic state with finite magnetisation~\cite{costa_ferromagnetism_2016} (the Lieb lattice) to the antiferromagnetic solution with zero net magnetisation (the kagome lattice), the ground state changes dramatically and concomitant changes in the magnetic excitation spectrum are expected~\cite{LiebKagome2026}. Studying spectral functions of these phases using accurate Trotterised dynamics is an interesting case study for an early fault-tolerant quantum computer.

Simulation parameters for the case study in this paper are chosen as follows. We focus on the periodic $4 \times 4$ lattice, consisting of $48$ sites. This requires $96$ logical qubits, plus routing space to perform lattice surgery. We consider this a realistic target for MegaQuOp-scale applications. While this restricted lattice size introduces limitations, it provides access to several independent momenta after the Fourier transform, sufficient to resolve high-symmetry points (specifically $\Gamma$, $X$ and $M$), which are of particular importance. To choose other simulation parameters in \cref{eq:spectral_function_final_expression}, we are guided by results in \cite{LiebKagome2026}, which presents HF+RPA transverse dynamical susceptibilities for this problem, in particular see \cite[Figs.~6 and 9]{LiebKagome2026}. From this, one can estimate the energy resolution and energy range needed to characterise the magnetic excitations. We estimate that a resolution of $0.5\tau$ is sufficient to distinguish the magnon branches and thereby differentiate between the typical ferrimagnetic and antiferromagnetic excitations. The required maximum evolution time, $T_{\mathrm{max}}$, to achieve this resolution is $T_{\mathrm{max}} = 2\pi/(0.5\tau) \approx 12.6 \, \tau^{-1}$.
The total energy range is determined by the Stoner continuum which extends up to approximately $10\tau$ for the largest interaction strength $U=8\tau$. This sets the upper bound on the time step $\delta t < 2\pi/(20\tau)$ (see Appendix~\ref{app:broadening_function_and_explicit_representation} for details). As discussed in Appendix~\ref{app:broadening_function_and_explicit_representation}, we primarily take a Trotter time step size of $0.1/\tau$ and use a maximum of 128 Trotter steps, but also consider a more optimistic time step of $\delta t=0.2/\tau$ with a maximum of 64 Trotter steps for the full evolution. While these parameters are determined from the transverse susceptibility, we expect that the longitudinal susceptibility is also well captured with this energy range and resolution. In Appendix~\ref{sec:trotter_error_analysis} we perform Trotter error analysis to justify this choice of Trotter step size. This includes full construction of the magnetic susceptibilities for the one-dimensional Hubbard model using time-dependent DMRG simulations, and using the same parameters as in our case study.

Lastly, we discuss the initial state. In our resource estimates, we will disregard the cost of state preparation, as is standard in resource estimation studies. Although this is a limitation, we are helped by the very different physics in the low-energy states of the Lieb and kagome lattices: the ground state of the Lieb lattice is guaranteed to be ferrimagnetic by Lieb's theorem~\cite{Lieb89}, so that a Slater determinant with finite on-site magnetisation (such as the Hartree--Fock approximation) is expected to be representative of the true ground state. The magnetic frustration in the kagome lattice, on the other hand, favours ground states with zero magnetic moment~\cite{Sun21}. Since this leads to radically different magnetic excitations, it is expected that using an approximate initial state with the right magnetic order can already provide physical insight. It may be necessary to take a more accurate initial state near the kagome lattice ($\tau' \approx \tau$), such as a matrix product state of low bond-dimension \cite{millar2026}. In practice, many of these parameters, including the initial state, would be varied to assess for convergence. This experiment should not be viewed as providing a ``complete solution'' to the Lieb-kagome problem, but rather an additional methodology to help probe the physics of magnetic excitations, and to complement existing approaches, such as the HF+RPA and DMFT solutions already considered in \cite{LiebKagome2026}.

\section{Tile Trotterisation}
\label{sec:tile_trotter}

Trotterised simulation of the Hubbard model has become a widely-considered benchmark for early fault-tolerant quantum computing, both in studies that count non-Clifford gates \cite{kivlichan2020improved, Kan2025, BaySmidt2025}, and those performing more detailed resource estimation \cite{Akahoshi2025, Chung2026, Ismail2026_2, Apel2026}. Under the Jordan--Wigner mapping the Lieb-kagome Hubbard Hamiltonian is transformed to
\begin{equation}
H = - \frac{\tau}{2} \sum_{\langle i,j \rangle, \sigma} \Big( X_{i\sigma} \overrightarrow{Z} X_{j\sigma} + Y_{i\sigma} \overrightarrow{Z} Y_{j\sigma} \Big) - \frac{\tau'}{2} \sum_{\langle \langle i,j \rangle \rangle, \sigma} \Big( X_{i\sigma} \overrightarrow{Z} X_{j\sigma} + Y_{i\sigma} \overrightarrow{Z} Y_{j\sigma} \Big) + \frac{U}{4} \sum_{i=1}^{N_s} Z_{i \uparrow} Z_{i \downarrow},
\end{equation}
where $\overrightarrow{Z}$ denotes a string of Pauli $Z$ gates applied to qubits between $i\sigma$ and $j\sigma$ in the fermionic ordering. This non-local Pauli $Z$ string can be eliminated by choosing a fermionic ordering such that $i\sigma$ and $j\sigma$ are adjacent; to achieve this, fermionic swap networks can be applied to update the ordering as needed (see discussion below and Appendix~\ref{sec:fswap_network}).

\begin{figure*}
    \centering
    \includegraphics[width=1.0\linewidth]{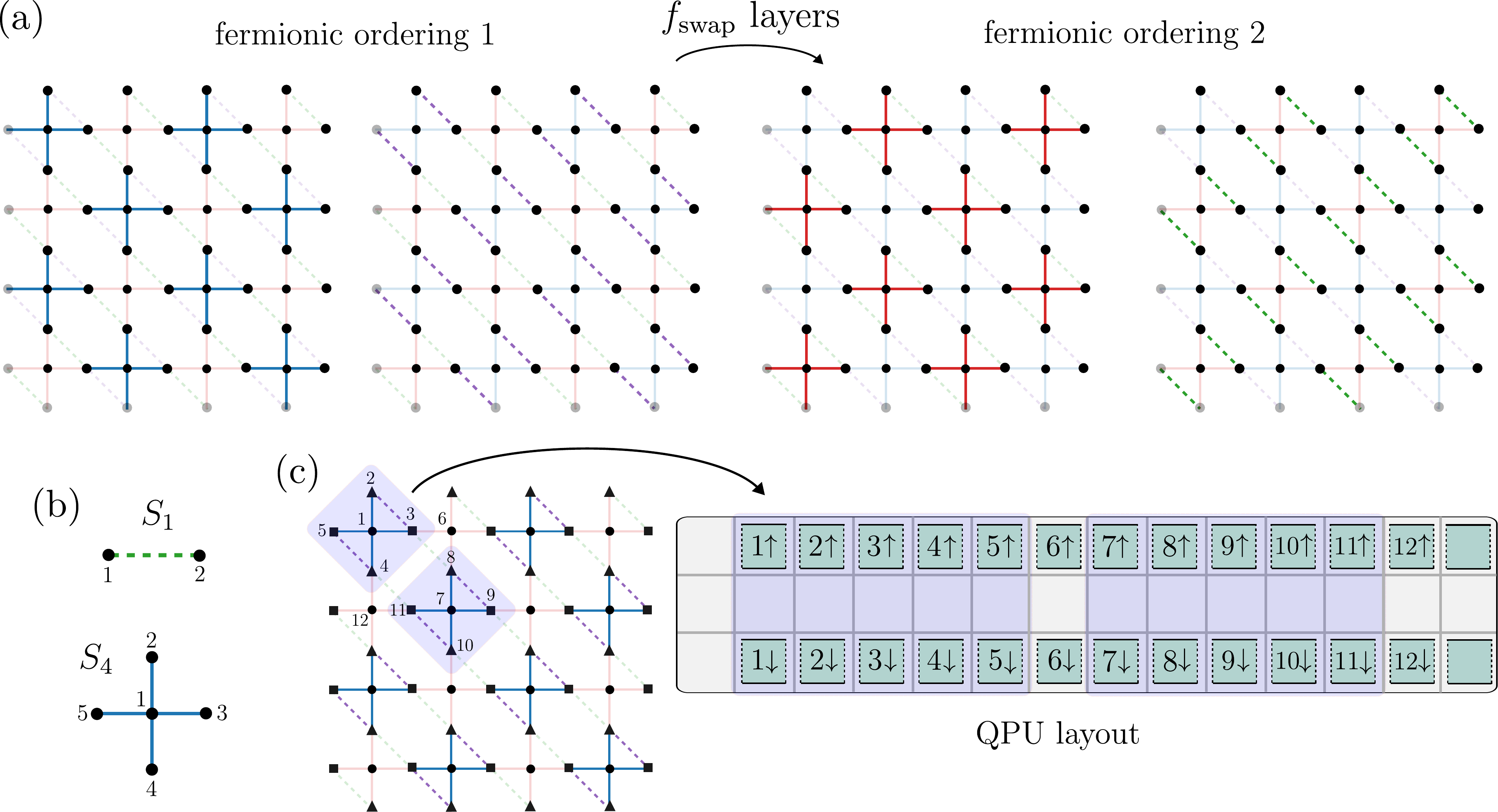}
    \caption{{\bf Trotter hopping partitions and QPU layout.} (a) Partitioning of the hopping Hamiltonian into four sections: blue, purple, red and green. Blue and red sections are comprised of $S_4$ tiles of strength $\tau$, while purple and green sections are comprised of $S_1$ tiles of strength $\tau'$. The $S_4$ tiles are shown as solid lines, while $S_1$ tiles are dashed. A single fermionic ordering is used for blue and purple sections, while a second ordering is used for red and green sections, requiring $25$ $f_{\textrm{swap}}$ layers to swap between orderings (per spin). (b) The $S_1$ and $S_4$ tiles used to implement time evolution for the hopping Hamiltonian, following tile Trotterisation \cite{BaySmidt2025}. (c) The 48 sites and 96 qubits of the $4 \times 4$ lattice are mapped to a 50-by-3 QPU layout, including routing space for lattice surgery. The site ordering shown allows implementing both blue and purple hopping terms, and shows the first two $S_4$ tiles as examples. See Appendix~\ref{sec:fswap_network} for full layouts. The site ordering is chosen to leave a padding site between each $S_4$ tile, allowing additional routing space for lattice surgery, particularly to allow parallel fallback Clifford+T rotations (see Section~\ref{sec:rotations}).}
    \label{fig:sections_and_qpu_layout}
\end{figure*}

Exact time evolution for the interaction term can be implemented trivially by a product of $ZZ$ Pauli rotations,
\begin{equation}
    e^{-i H_I t} = \prod_i e^{-i \frac{U}{4} Z_{i \uparrow} Z_{i \downarrow} t},
    \label{eq:interaction_term}
\end{equation}
since all terms commute. Implementing Trotterisation for hopping terms is less straightforward and, broadly speaking, there are three strategies that are considered: i) using a fermionic Fourier transform (FFT) to diagonalise the hopping Hamiltonian \cite{kivlichan2020improved}; ii) directly time evolving individual hopping terms for each bond on the lattice \cite{Akahoshi2025}; or iii) by implementing a more sophisticated \emph{tiling} of the lattice; see \cite[Figure 1]{BaySmidt2025} for a summary of these strategies. The tiling approach was introduced in Ref.~\cite{campbell2022early} by tiling square lattices with plaquettes, an approach known as \emph{plaquette Trotterisation}. It was shown that time evolution for each plaquette is relatively inexpensive, requiring only two arbitrary-angle Pauli $Z$ rotations per plaquette (per spin). This approach was generalised in \cite{BaySmidt2025} to an extended set of tiles. In particular, it was shown that any tile that is a complete bipartite graph can be implemented with just two arbitrary-angle $Z$ rotations (per spin). This allows a larger set of tiles to be defined, which can be used to efficiently tile other lattices, including the Lieb-kagome lattice. This more general strategy is referred to as \emph{tile Trotterisation}.

\cref{fig:sections_and_qpu_layout}(a) shows the tiling that we use for the periodic Lieb-kagome lattice in this paper, which makes use of two tiles: the (trivial) $S_1$ tile and the $S_4$ tile. The full lattice is covered by four sections of tiles, each labelled by a colour for ease of reference. The four sections are of type $S_4$ (blue), $S_1$ (purple), $S_4$ (red) and $S_1$ (green). For the Lieb lattice where all $S_1$ hopping terms have $\tau'=0$, it is only necessary to implement the two $S_4$ sections, reducing the implementation cost. Given the corresponding tile Hamiltonians (for a single spin), which we denote $H_{\sigma}^{S_1}$ and $H_{\sigma}^{S_4}$, and assuming the fermionic orderings shown in \cref{fig:sections_and_qpu_layout}(b), exact time evolution can be implemented as \cite{BaySmidt2025}
\begin{equation}
    S_1 \textrm{ tile:} \;\;\;\;\; e^{-i H_{\sigma}^{S_1} t} = e^{i \frac{\tau'}{2} (X_1 X_2 + Y_1 Y_2) t},
    \label{eq:s1_tile}
\end{equation}
\begin{equation}
    S_4 \textrm{ tile:} \;\;\;\;\; e^{-i H_{\sigma}^{S_4} t} = F_{32} F_{45} F_{34} \mathrm{CZ}_{12} \, e^{i \tau (X_1 X_3 + Y_1 Y_3) t} \, \mathrm{CZ}_{12} F_{34}^{\dagger} F_{45}^{\dagger} F_{32}^{\dagger}.
    \label{eq:s4_tile}
\end{equation}
Here, $F_{ij}$ denotes the two-qubit fermionic Fourier transform, and $\mathrm{CZ}$ is the controlled-$Z$ gate.

In order to use the above results to implement $S_1$ and $S_4$ tiles, it is necessary to match the fermionic orderings shown in \cref{fig:sections_and_qpu_layout}(b). The blue and purple hopping sections can be implemented with a common ordering that satisfies this criteria, and similarly for the red and green hopping sections. Therefore, only two orderings are needed to implement all terms. The ordering is updated between these sections using a network of fermionic swap gates. Each fermionic swap gate is denoted $f_{\textrm{swap}}$, which can be implemented as a SWAP gate followed by a CZ gate. The two orderings and fermionic swap network for the $4 \times 4$ lattice are given in Appendix~\ref{sec:fswap_network}.

A second-order tile Trotter step for time $\delta t$ is
\begin{equation}
    \mathcal{S}_2(\delta t) := e^{-i H_I \delta t/2} e^{-i H_h^b \delta t/2} e^{-i H_h^p \delta t/2} f_{\mathrm{network}} e^{-i H_h^r \delta t/2} e^{-i H_h^g \delta t} e^{-i H_h^r \delta t/2} f_{\mathrm{network}}^{\dagger} e^{-i H_h^p \delta t/2} e^{-i H_h^b \delta t/2} e^{-i H_I \delta t/2},
\end{equation}
where $H_h^b$, $H_h^p$, $H_h^r$ and $H_h^g$ are hopping Hamiltonians (summed over spin) for blue, purple, red and green sections, respectively, while $f_{\mathrm{network}}$ implements the fermionic swap network to update orderings. While it is straightforward to perform resource estimation for the second-order step, we instead consider a hybrid step. Here, interaction terms are implemented via second-order splitting, while hopping terms are implemented via first-order splitting. In particular, in odd-numbered steps we run through hopping terms forwards,
\begin{equation}
    \mathcal{S}_{\textrm{odd}}(\delta t) := e^{-i H_I \delta t/2} e^{-i H_h^b \delta t} e^{-i H_h^p \delta t} f_{\mathrm{network}} e^{-i H_h^r \delta t} e^{-i H_h^g \delta t} e^{-i H_I \delta t/2},
    \label{eq:trotter_odd}
\end{equation}
while in even-numbered steps we run through hopping terms in reverse order, which we denote $\mathcal{S}_{\textrm{even}}(\delta t)$. This has the benefit of reducing the implementation cost for hopping terms and that only one $f_{\textrm{swap}}$ network is performed per step. This is also justified in terms of Trotter error, because the Trotter error due to splitting $H$ into hopping and interaction terms is much larger than the Trotter error due to further splitting $H_h$ into sections, and this is particularly true at large $U$ where Trotter error is largest (see \cite[Table~II]{BaySmidt2025} and \cite[Figure~4]{BaySmidt2026}).

The QPU layout that we consider is shown in \cref{fig:sections_and_qpu_layout}(c). This is a quasi-linear layout, where all spin-up qubits are placed on the top row, and spin-down qubits are placed on the bottom row. A single row of routing space is included in between. This layout allows significant parallelisation of operations; see \cref{sec:rotations} for a discussion of how rotation gates are implemented with high parallelism in this layout. We note that including two rows of routing space would allow both spin-up and spin-down hopping terms to be implemented in parallel, roughly halving the circuit depth; this approach was taken in Ref.~\cite{Akahoshi2024}. However, we consider the reduction in qubit count to be more important, particularly given the significant engineering challenges in scaling up superconducting devices to this size.

The Trotter steps considered can be implemented using Eqs.~(\ref{eq:interaction_term})--(\ref{eq:s4_tile}). From this, it can be seen that a full implementation of tile Trotterisation for the Hubbard model requires only a small set of gates: a $XX+YY$ Pauli rotation, a $ZZ$ Pauli rotation, the two-qubit fermionic Fourier transform, $\mathrm{CZ}$ and the fermionic swap gate. By providing optimised lattice surgery implementations of these gates, it becomes possible to perform full resource estimation for Trotterisation of the Lieb-kagome Hubbard model. In fact, since all tiles of Ref.~\cite{BaySmidt2025} can be implemented using just these gates and can be used to span any lattice, performing lattice surgery costing of tile Trotterisation for the Hubbard model on \emph{arbitrary} lattices also becomes possible. Lattice surgery implementations of the required gates within the QPU layout of \cref{fig:sections_and_qpu_layout}(c), together with time and volume costs, are given in Appendix~\ref{app:lattice_surgery_constructions}. In Section~\ref{sec:resource_estimates} we will combine these results to perform full time and volume resource estimates for our Lieb-kagome case study circuits.

Let us summarise the parameters of our case study, particularly the required rotation angles, which will aid the discussion in Section~\ref{sec:rotations}. Each Trotter step requires $240$ arbitrary-angle $Z$ rotations: $48$ rotations with $\theta = \delta t \cdot U/4$ (interaction terms), $64$ rotations with $\theta = - \delta t \cdot \tau$ ($S_4$ hopping terms) and $128$ rotations with $\theta = - \delta t \cdot \tau'/2$ ($S_1$ hopping terms). As described in Section~\ref{sec:lieb_kagome_hubbard}, we primarily take $\delta t = 0.1 \tau^{-1}$. As an example, for the kagome lattice ($\tau = \tau' = 1$) with $U = 4$, this gives rotation angles of size $0.1$, $0.1$ and $0.05$, respectively. For a target energy resolution of $0.5 \tau$ we must implement up to $128$ such Trotter steps. In the next section we assess the feasibility of running such a circuit under STAR and Clifford+T methodologies on superconducting qubits and the planar surface code.

\section{Clifford+T synthesis and fault-tolerant quantum computing}
\label{sec:rotations}

In this study, we set out to perform a detailed investigation of both the partially fault-tolerant STAR architecture and Clifford+T methodology for implementing rotation gates in the surface code, targeted at superconducting qubits under realistic noise models and native gates. We primarily consider the superconducting-inspired SI1000 noise model, and taking CZ as the native two-qubit gate \cite{gidney2021fault}.

The STAR architecture approach performs Clifford gates fault-tolerantly, while rotation gates are implemented by teleporting noisy resource states in a repeat-until-success procedure. It was shown \cite{choi2023, Toshio2025} that the error rate of Pauli rotations implemented in this way can be reduced to $\mathcal{O}(p \theta)$; therefore, for small rotation angles, low logical error rates can be achieved. Meanwhile, the total volume associated with this procedure can also be low. In order to perform analysis of the STAR architecture, we developed and optimised a non-Clifford simulation technique based on quasi-probability expansion of non-Clifford gates in the circuit, first considered in \cite{Pashayan2015, Bennink2017}. Ultimately, we find that the error rate of the prepared resource states at $p=10^{-3}$ is too high for our case study. This is not necessarily surprising; directly applying the formulas derived in \cite{Toshio2025} leads to the same result, and here we use a stricter noise model and native gate set. However, we also attempted a number of optimisations: for example, only choosing rotation angles in the Clifford hierarchy by choosing $\delta  t = \pi/32$ instead of $\delta t = 0.1$, and performing perfect calibration of coherent errors. Still, the error rate is found to be too high. This analysis is presented in Appendices~\ref{app:star_primary}, \ref{app:star_results} and \ref{app:non_cliff_sim}. These results do not rule out usefulness of the STAR architecture, which may remain valuable for different qubit modalities, target applications, and under different noise and native gate assumptions. However, they do emphasise that, even in the early fault-tolerant regime, it is necessary to consider and optimise alternative approaches.

In the remainder of this section, we analyse the Clifford+T approach to implementing rotation gates. This approach approximately expresses each $R_z(\theta)$ gate in terms of only Clifford and T gates. This approach has an important practical benefit in terms of calibration, since it requires only a small, discrete set of gates. However, it has been considered costly in practice, particularly for early applications. Two recent results challenge this viewpoint. First, magic state cultivation was introduced \cite{gidney2024magic}, demonstrating that T gates can be performed with significantly lower cost than previously suggested. Magic state cultivation allows magic states to be prepared close to where they are needed, without the need for magic state factories and significant routing, and with a space-time volume comparable to lattice surgery implementations of standard Clifford gates. Second, some of the authors of this study performed analysis of probabilistic and quasi-probabilistic approaches to Clifford+T synthesis, showing that existing formulas can drastically overestimate the cost to implement $R_z(\theta)$, particularly for small $\theta$ \cite{bothe2026}. In the small-angle regime, the average T gate cost to implement $R_z(\theta)$ to diamond norm error $\epsdiamond$ is $\mathcal{\tilde{O}}(\theta^2/\epsdiamond)$. This relies on the simple insight that $R_z(\theta) \approx I$ for small $\theta$, such that including the identity in the probability mixture can reduce the average gate cost significantly. Taking quasi-probability mixtures can also lead to further reductions in gate cost. In this section we analyse our case study using this quasi-probability Clifford+T approach with magic state cultivation.

\begin{figure*}
    \centering
    \includegraphics[width=1.0\linewidth]{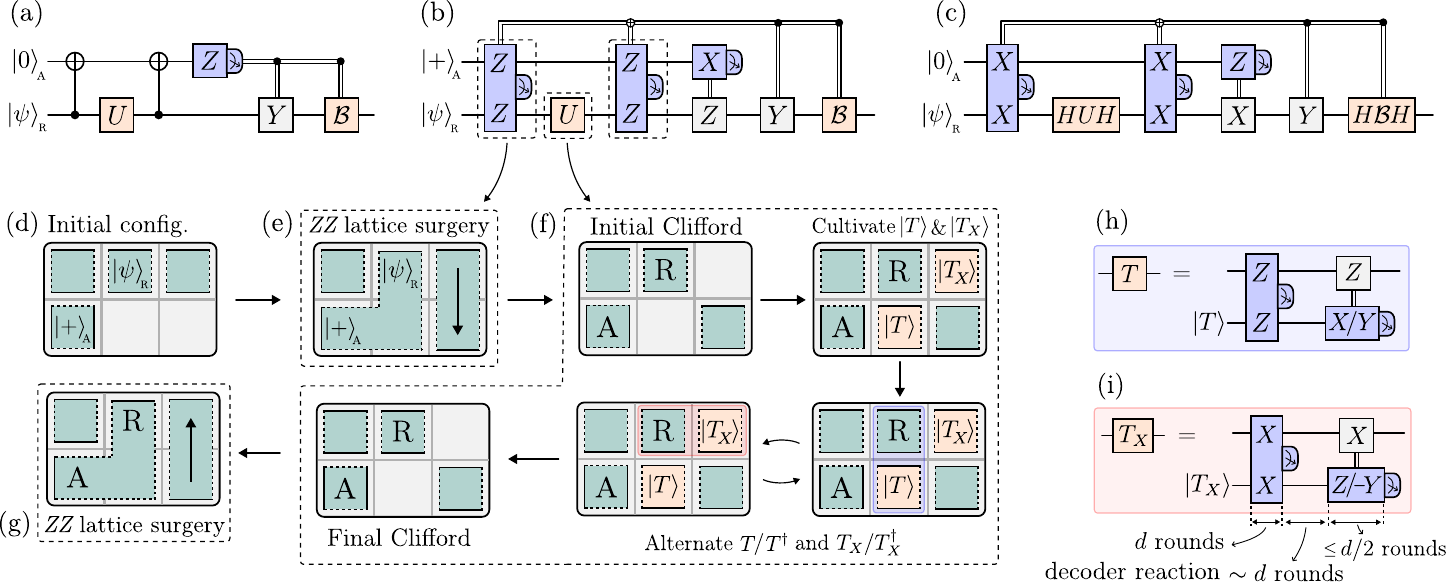}
    \caption{{\bf Fallback Clifford+T rotation implementation.} (a) The fallback circuit to implement a $Z$ rotation. The unitary $U$ is a Clifford+T unitary. If the measurement result is $1$ then a fallback correction is applied. (b) An equivalent circuit in terms of $ZZ$ measurements, with a reduced lattice surgery cost. Corrections are applied if the two $ZZ$ measurements have an XOR equal to $1$. (c) An equivalent circuit for an $X$ rotation. (d)--(g) show a lattice surgery implementation of several steps of circuit (b) for the linear QPU layout considered in this work. Three ancilla tiles are needed per fallback rotation. Labels ``A'' and ``R'' represent the ancilla qubit and the qubit to which the rotation is applied, respectively. (d) Initial configuration, preparing an ancilla patch in $|+\rangle$ ($0$ rounds). (e) $ZZ$ lattice surgery and patch movement ($d$ rounds). (f) Implementation of Clifford+T sequence, $U$. See \cref{tab:clifford_t_solutions} for example sequences. First, apply an initial Clifford, if required (up to $2d$ rounds). Then, cultivate $|T\rangle$ and $|T_X\rangle$ states on the $Z$ and $X$ boundaries of patch R, and apply the alternating sequence of $T/T^{\dagger}$ and $T_X/T_X^{\dagger}$. Apply a final Clifford $S$ if needed ($d$ rounds). The $T$ and $T_X$ gates can be implemented as shown in (h) and (i), respectively ($\sim 2.5 d$ rounds each, excluding magic state cultivation, and assuming $d$ rounds for the decoder reaction time). While teleporting a magic state on one boundary, cultivation can be performed simultaneously on the other boundary. (g) $ZZ$ lattice surgery ($d$ rounds). Afterwards, apply Pauli corrections based on measurement results. The fallback correction, which happens with low probability, is not shown but can be implemented similarly.}
    \label{fig:fallback_rotation}
\end{figure*}

We implement all rotations using quasi-probability mixtures of fallback Clifford+T channels, as described in Ref.~\cite[Section VI]{bothe2026}. The fallback circuit is shown in \cref{fig:fallback_rotation}(a). The Clifford+T unitary, $U$, can be expressed
\begin{equation}
U = 
\begin{pmatrix}
u & -v^* \\
v & u^*
\end{pmatrix}, \; u = r e^{i\varphi}.
\label{eq:su2_clifford_t}
\end{equation}
In the fallback approach, in addition to applying $U$, an ancilla is also used to perform a measurement, leading to a projective rotation. If the measurement result is $0$ (with probability $r^2$) then the measurement is successful, and a $Z$ rotation will be applied with angle $\varphi = \arg(u)$. If $\varphi < \theta$ then this is referred to as an ``under-rotation'', while $\varphi > \theta$ is an ``over-rotation''. If the measurement result is $1$ then a $Z$ rotation is performed with angle $\arg(v)$, which is typically a significant error. In this case, the fallback channel $\mathcal{B}$ is applied to correct the operation. Together, this circuit implements a channel, which we denote $\mathcal{F}(\rho)$.

The desired channel corresponding to $R_z(\theta)$, which we denote $\mathcal{Z}_{\theta}(\rho)$, is then expressed as a linear combination of channels, $\sum_i c_i \mathcal{U}_i$, where $\{ \mathcal{U}_i \}$ contain Clifford+T fallback and Pauli channels. Specifically, we take
\begin{equation}
    \mathcal{Z}_{\theta}(\rho) = c_1 \mathcal{F}_1(\rho) + c_2 \mathcal{F}_2(\rho) + c_I \rho + c_X X \rho X + c_Y Y \rho Y + c_Z Z \rho Z.
    \label{eq:lcc_mixed_fallback}
\end{equation}
Here, the fallback channels $\mathcal{F}_1$ and $\mathcal{F}_2$ contain (twirls of) under- and over-rotation Clifford+T unitaries, respectively. This forms a quasi-probability distribution because $\lambda := \sum_i |c_i| > 1$. The use of quasi-probability mixtures of channels to construct expectation values of observables for quantum circuits is well established \cite{Temme2017}. For a circuit with $N_R$ such rotations, the number of samples required to estimate an expectation value to a given precision in the quasi-probability approach is increased by a factor of $\lambda^{2N_R}$. Therefore, $\lambda^{2}$ is referred to as the sampling overhead for the rotation gate.

\begin{figure*}
    \centering
    \includegraphics[width=1.0\linewidth]{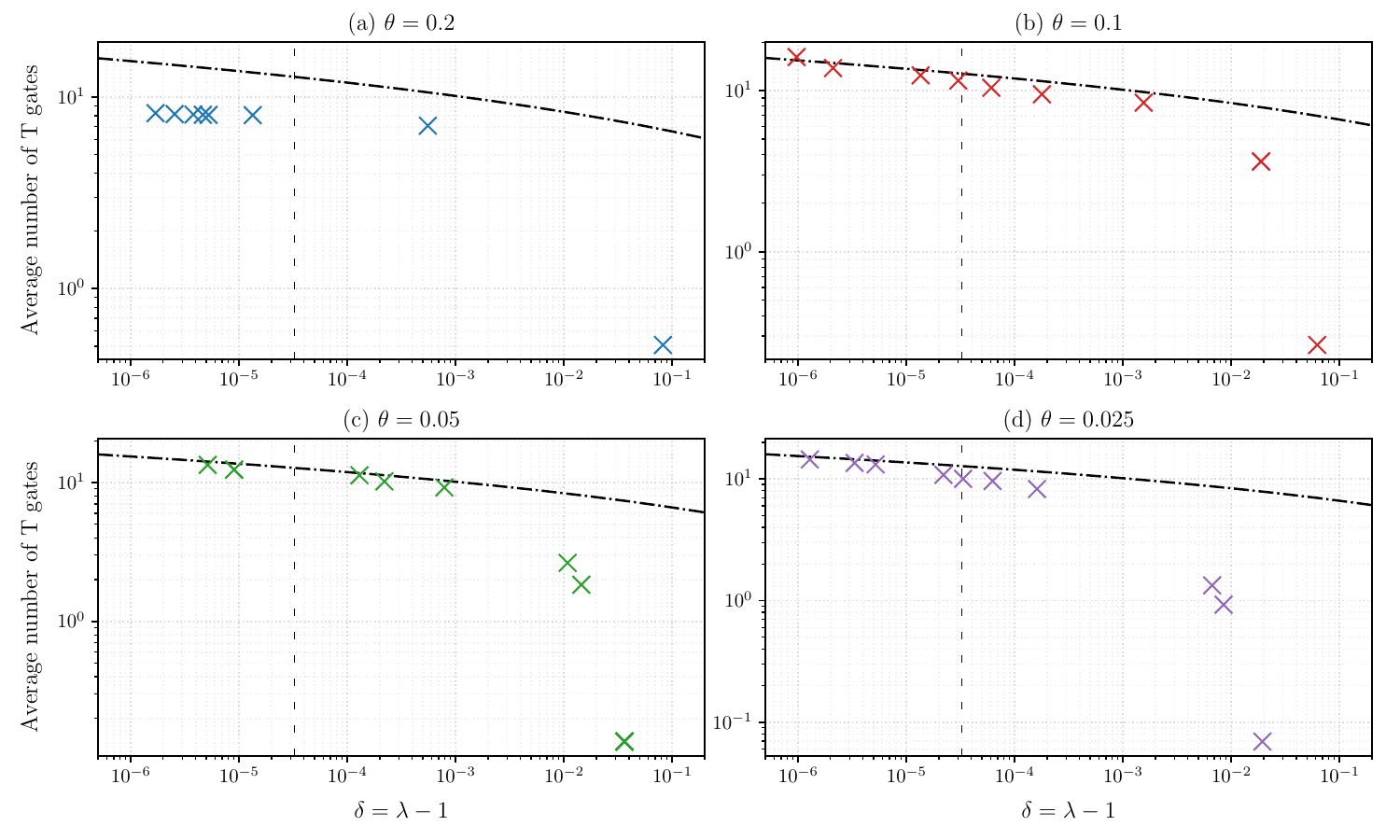}
    \caption{{\bf Results for quasi-probability Clifford+T fallback channels.} The average T gate count versus $\delta = \lambda-1$ using quasi-probability mixtures of Clifford+T fallback channels. The four rotation angles in (a)--(d) are sufficient to implement all Trotter circuits for Lieb and kagome lattices with time step $\delta t=0.1$ and $U=1$, $2$, $4$ and $8$ $\tau$. The vertical dashed line shows the target value of $\delta = 1/N_R$, where $N_R=30{,}720$ is the total number of rotation gates in the largest target circuit. This ensures a small sampling overhead for the full circuit (see discussion in main text). Larger values of $\delta$ may be used in exchange for higher sampling overheads. The dashed-dotted line shows the mixed-fallback formula from Ref.~\cite{Kliunchnikov2023} with $\delta$ in place of $\epsdiamond$. Associated data for the results in this figure is presented in Table~\ref{tab:fallback_results}.}
    \label{fig:av_t_count_fallback}
\end{figure*}

We aim to construct rotations such that the total sampling overhead for the largest circuit, $\prod_i \lambda_i^2$, is within an acceptable value. As a rough guide, to achieve this we require $\delta_i := \lambda_i - 1 \approx 1/N_R$, for $N_R$ rotations total. This ensures a total sampling overhead of $\prod_i \lambda_i^2 = (1+1/N_R)^{2 N_R} \approx e^2 \approx 7.4$. Since our MegaQuOp case study circuit for $128$ Trotter steps contains $N_R = 30720$ rotation gates, this suggests targeting $\delta \approx 3.26 \times 10^{-5}$ or less per rotation. The value of $\delta$ can be reduced arbitrarily by improving the accuracy of $\mathcal{F}_1$ and $\mathcal{F}_2$, by increasing the number of T gates used.

We generate Clifford+T solutions using the procedure described in Ref.~\cite[Section VIB]{bothe2026}, implemented using pygridsynth \cite{pygridsynth}. Results are shown in \cref{fig:av_t_count_fallback} for four rotation angles: $\theta=0.2$, $\theta=0.1$, $\theta=0.05$ and $\theta=0.025$. For a Trotter step size of $\delta t = 0.1$, these are all the angles needed to implement our case study with parameters $\tau=1$, $\tau' = 0$ or $1$ (Lieb or kagome lattices), and $U=1$, $2$, $4$ or $8$. Models with different parameters require other rotation angles, which can be obtained with a similar cost. On each plot we mark the target $\delta$ value with a dashed vertical line. The corresponding solutions to achieve (or almost achieve) this are presented in Table~\ref{tab:fallback_results}. All of the solutions require at most $12$ T gates, and in some cases less. Apart from the example angle $0.2$ that serendipitously admits a particularly cheap and accurate solution, at target $\delta$, our quasi-probability approach is not much better in average T count than the mixed fallback result \cite{Kliunchnikov2023}. This is due to the small-angle method of \cite{bothe2026} giving an advantage for $\theta^2\lesssim\delta$, which our rotations do not satisfy. Yet, using quasi-probabilities rather than a mixed approach is still advantageous as it allows us to implement the rotations with no diamond norm error, and instead with a modest sampling overhead.

\begin{table*}
\centering
{\footnotesize
\renewcommand{\arraystretch}{1.2}
\begin{tabular}{@{\extracolsep{4pt}}lcccccccc}
\hline
\hline
 & & \multicolumn{3}{c}{under-rotation} & \multicolumn{3}{c}{over-rotation} \\
\cline{3-5} \cline{6-8}
angle, $\theta$ & $\delta = \lambda-1$ & $\#$ of T gates & success prob., $r^2$ & selection prob. & $\#$ of T gates & success prob., $r^2$ & selection prob. & $p_{\mathrm{fb}}$ \\
\hline
0.2 & $1.34 \times 10^{-5}$ & 8 & 0.9969 & 0.9960 & 7 & 0.9817 & 0.0040 & 0.05 \\
0.1 & $3.02 \times 10^{-5}$ & 11 & 0.9923 & 0.5026 & 12 & 0.9982 & 0.4973 & 0.07 \\
0.05 & $9.03 \times 10^{-6}$ & 12 & 0.9809 & 0.4867 & 12 & 0.9901 & 0.5132 & 0.21 \\
0.025 & $3.35 \times 10^{-5}$ & 11 & 0.99987 & 0.3817 & 9 & 0.9839 & 0.6183 & 0.15 \\
\hline
\hline
\end{tabular}
}
\caption{{\bf Selected Clifford+T fallback solutions.} For the results in \cref{fig:av_t_count_fallback} we present one solution for each angle that can be used in the target circuits. For each solution, for both the under-rotation $\mathcal{F}_1$ and over-rotation $\mathcal{F}_2$ we present: the number of T gates in $U$ (as in \cref{fig:fallback_rotation}(a) and \cref{eq:su2_clifford_t}); the probability that measurement is successful, $r^2$; and the probability that we select either $\mathcal{F}_1$ or $\mathcal{F}_2$, $|c_1|/\lambda$ and $|c_2|/\lambda$, respectively. Note that the latter two probabilities do not sum to $1$ in general, as it is also possible to select a Pauli gate (\cref{eq:lcc_mixed_fallback}). The value $p_{\textrm{fb}}$ is the probability of implementing at least one fallback step in a parallel layer of $16$ such rotation gates, as in the target circuits.}
\label{tab:fallback_results}
\end{table*}

Let us now consider how the fallback channels may be implemented within our architecture, using lattice surgery and magic state cultivation. First, we consider the efficient implementation of Clifford+T unitaries. We generate Clifford+T unitaries in the Matsumoto--Amano form \cite{matsumoto2008representation, ma_paper}, which is T-optimal. From this, a Clifford+T unitary with T-count $n_T$ can always be written in the form
\begin{equation}
    U = C_1 \, T_{p_1}^{s_1} T_{p_2}^{s_2} \cdots T_{p_{n_T}}^{s_{n_T}} \, C_2 ,
    \qquad s_i = \pm 1, \quad p_i \in \{Z, X\}, \quad p_{i+1} \neq p_i ,
\end{equation}
where $C_1 \in \{ I, S \}$ and $C_2$ is a product of at most two elements of $\{ X, Y, Z, S, S^{\dagger}, \sqrt{X}, \sqrt{X}^{\dagger} \}$. Here, $T_Z$ is the standard $T$ gate and $T_X = HTH$, and the product is applied from right to left. In words, $U$ is an alternating sequence of $T/T^{\dagger}$ and $T_X/T_X^{\dagger}$ gates, with a Clifford on each side in general. This rewriting preserves the T-count. The Clifford $C_1$ is $S$ precisely when the first syllable of the Matsumoto--Amano form is $SHT$. The Clifford $C_2$ can always be reduced to the form stated by using the freedom to exchange the rightmost $T_p^s$ with $T_p^{-s}$ and absorbing the resulting Clifford factor into $C_2$. Since $S$ and $\sqrt{X}$ require $d$ rounds to implement \cite{logical_s_gate}, the Clifford contributions can be implemented in $3d$ rounds in the worst case (and typically less). This can be reduced by certain approaches -- for example, in some scenarios by absorbing a Clifford into state preparation \cite{gidney2025factor} -- but since this is a small cost anyway, we simply assume a worst-case $3d$ Clifford cost.

In Table~\ref{tab:clifford_t_solutions} we present both Matsumoto-Amano and decomposed forms corresponding to the results in Table~\ref{tab:fallback_results}. From these results, we can also see why the T-count for $\theta=0.2$ is so low -- there happens to be a Clifford+T under-rotation with $n_T=8$ which is extremely close in angle, $\varphi$ (and also with large success probability, $r^2$). Particularly for model Hamiltonians, where the number of distinct rotation angles is small, this provides a useful trick -- one can try to choose the Trotter step size, $\delta t$, so that at least one rotation angle can be obtained with high accuracy and low T-count. However, we made no such optimisation here, and the high-quality and cheap solution is a coincidence.

\begin{table*}
  \centering
  \footnotesize
  \setlength{\tabcolsep}{5pt}
  \renewcommand{\arraystretch}{1.4}
  \begin{tabular}{llccll}
    \hline
    \hline
    $\theta$ & type & $n_T$ & $\varphi$ & Matsumoto--Amano form & decomposition \\
    \hline
    \multirow{2}{*}{$0.2$}
      & under & $8$  & $0.19986$ & \texttt{THTSHTSHTSHTSHTHTHTHSH}
                     & $T\,\TXdg\,T\,\TX\,\Tdg\,\TX\,T\,\TXdg\cdot \Sdg$ \\
      & over  & $7$  & $0.2353$ & \texttt{SHTHTHTHTSHTHTHTH}
                     & $S\cdot \TX\,T\,\TX\,\Tdg\,\TXdg\,T\,\TXdg\cdot Z$ \\
    \hline
    \multirow{2}{*}{$0.1$}
      & under & $11$ & $0.0962$ & \texttt{SHTSHTSHTSHTSHTSHTHTSHTHTSHTHTHSH}
                     & $S\cdot \TXdg\,T\,\TX\,\Tdg\,\TXdg\,\Tdg\,\TXdg\,T\,\TXdg\,\Tdg\,\TXdg\cdot X\Sdg$ \\
      & over  & $12$ & $0.1038$ & \texttt{HTHTSHTSHTHTHTHTHTHTSHTHTHTH}
                     & $\TX\,\Tdg\,\TX\,\Tdg\,\TXdg\,\Tdg\,\TXdg\,\Tdg\,\TX\,T\,\TXdg\,\Tdg\cdot \rX S$ \\
    \hline
    \multirow{2}{*}{$0.05$}
      & under & $12$ & $0.0478$ & \texttt{THTHTSHTSHTSHTSHTSHTHTHTHTSHTH}
                     & $T\,\TX\,\Tdg\,\TX\,T\,\TXdg\,\Tdg\,\TXdg\,T\,\TXdg\,\Tdg\,\TX$ \\
      & over  & $12$ & $0.0521$ & \texttt{HTSHTHTSHTSHTHTSHTSHTSHTSHTHTHTSH}
                     & $\TXdg\,\Tdg\,\TXdg\,\Tdg\,\TXdg\,\Tdg\,\TXdg\,T\,\TX\,T\,\TXdg\,T\cdot \rX S$ \\
    \hline
    \multirow{2}{*}{$0.025$}
      & under & $11$ & $0.0198$ & \texttt{SHTHTSHTSHTSHTSHTSHTSHTSHTSHTSHTXH}
                     & $S\cdot \TX\,\Tdg\,\TX\,T\,\TXdg\,\Tdg\,\TX\,T\,\TXdg\,\Tdg\,\TXdg$ \\
      & over  & $9$  & $0.0283$ & \texttt{THTSHTSHTHTSHTHTSHTSHTHSH}
                     & $T\,\TXdg\,T\,\TXdg\,T\,\TX\,T\,\TX\,\Tdg\cdot \Sdg\rX$ \\
    \hline
    \hline
  \end{tabular}
  \caption{{\bf Gate decompositions for selected Clifford+T solutions.}  Clifford+T sequences for results in Table~\ref{tab:fallback_results}. We show the Matsumoto--Amano form in which the solutions were enumerated, and the subsequent decomposition for efficient implementation. The decomposition consists of an arbitrary Clifford, followed by an alternating sequence of $T/T^{\dagger}$ and $T_X/T_X^{\dagger}$, followed by a possible $S$ gate. Gates are applied from right to left. Both under- and over-rotations are shown, together with the number of T gates, $n_T$, and the angle $\varphi$ of the top-left element of $U$ in SU(2) form.}
  \label{tab:clifford_t_solutions}
\end{table*}

\cref{fig:fallback_rotation} shows an example construction. First, note that circuit (a) can be simplified to circuit (b), requiring $ZZ$ measurements instead of full CX gates. We then show an example of how to implement a fallback rotation in our linear QPU layout with a routing corridor. Each rotation requires 3 $d \times d$ patches of routing space -- one for the fallback ancilla, one for cultivating $|T\rangle$ states and one for cultivating $|T_X\rangle := H |T\rangle$ states. Since the routing corridor is of length $48$ qubits, this allows $16$ rotations to be performed in parallel. This is convenient, since $S_4$ layers contain exactly $16$ rotations (per spin), while $S_1$ layers contain $32$ rotations (per spin). Therefore, $S_4$ rotations can be implemented fully in parallel, while $S_1$ rotations are performed in two parallel batches. Our approach for applying $T$ and $T_X$ gates follows that in \cite{gidney2025factor,huggins2025fluid}. By exposing both the $Z$ and $X$ boundaries we can simultaneously cultivate on one boundary while teleporting a magic state on the opposite boundary. This is efficient since the gate sequence alternates between $T/T^{\dagger}$ and $T_X/T_X^{\dagger}$ gates. Combined, the number of logical time steps (each of $d$ syndrome-extraction rounds) to implement a Clifford+T unitary, $t_{\text{clifford+T}}$, is
\begin{equation}
    t_{\text{clifford+T}} = 3 + t_{\textrm{cultivation}} + t_{\text{teleport}} + (n_T-1) \max(t_{\text{cultivation}}, \, t_{\textrm{teleport}}),
    \label{eq:clifford_t_time_steps}
\end{equation}
where $+3$ time steps is the previously discussed Clifford cost. Assuming that the decoder reaction time is roughly $d$ rounds, we take $t_{\textrm{teleport}} = 2.5$ time steps, see \cref{fig:fallback_rotation}(i). We estimate the expected value of $t_{\text{cultivation}}$ following the approach of \cite{huggins2025fluid}. We first calculate the space-time volume of cultivation at the desired logical error rate (LER), and then divide by the number of physical qubits for the patch where cultivation is performed, including a buffer factor of $1.5$ to account for imperfect packing of cultivation attempts and other slack. Using an SI1000 noise model at $p=10^{-3}$ and final distance $d=15$, we estimate that an LER of $2.66 \times 10^{-6}$ is achieved with space time volume $\approx 10^4$ \emph{qubit-rounds}. Therefore, the number of time steps is $t_{\textrm{cultivate}} = 1.5 \times 10^4 /2d (d+1)^2 = 1.95$. As a result, even with the buffer factor, we have $t_{\textrm{cultivate}} < t_{\textrm{teleport}}$. In practice, one must remember that this is an estimate of the average $t_{\textrm{cultivate}}$ value.

The full cost of a rotation gate is then simple to construct by including the additional circuit elements from \cref{fig:fallback_rotation}(b). We note that in our resource estimates we do not include the cost of the fallback stage (apart from the calculation of $\delta$, which uses the exact value). This only happens rarely, less than $2\%$ of the time per rotation gate by construction. The probability of at least one fallback in a layer of $16$ rotation gates is higher, but still between $5\%$ and $21\%$ for our presented solutions, see $p_{\mathrm{fb}}$ in Table~\ref{tab:fallback_results}. Instead, we assume $n_T=12$ for each rotation, taking the worst case of our obtained solutions.

Finally, let us discuss possible caveats for real devices. We assume $p=10^{-3}$ throughout the QPU, which is standard in resource estimation studies. However, large-scale superconducting devices will have defects and varying error rates. It has been noted that the performance of cultivation can vary heavily with physical error rate, for example see \cite[Fig.~29]{gidney2024magic}, where the volume associated with cultivation is significantly higher at $p=2 \times 10^{-3}$ than at $p=1 \times 10^{-3}$. Our resource estimates assume that many rotations can be performed in parallel, using magic states cultivated throughout the device, each with $p=10^{-3}$. In practice, it is possible that the worst-performing cultivation steps in a layer of rotations may act as a bottleneck, particularly in circuit depth. We suggest a few workarounds for this possibility. First, it may be possible to accept higher error rates for a small number of rotations, which suggests either performing cultivation with a higher LER, or using an alternative scheme such as the STAR architecture for these few rotations. Second, these few rotations could be performed in a different region of the device. This would slightly increase our circuit depth and volume estimates due to routing time and some loss of parallelisation, but not significantly. Third, a more drastic change would be to perform all rotations in a small number of rotation units \cite{huggins2025fluid}, with some associated qubit and space-time overhead (due to routing time and loss of parallelisation). For the rest of this paper we continue assuming $p=10^{-3}$, as is standard, but note that such workarounds are possible. In \cref{sec:resource_estimates} we further show the sensitivity of the total circuit volume to the performance of cultivation.

\section{Lattice surgery compilation and MegaQuOp resource estimates}
\label{sec:resource_estimates}

We next perform full lattice surgery compilation of circuit elements to provide resource estimates for our case study. As discussed in Section~\ref{sec:tile_trotter}, the tile Trotter steps for Lieb-kagome lattices require only a small number of circuit elements: $S_1$ hopping terms, $S_4$ hopping terms,  interaction terms, and fermionic swap networks. These together require compilation of only a small set of circuit elements. The full set of lattice surgery constructions, including time and volume estimates, are given in Appendix~\ref{app:lattice_surgery_constructions}. An $S_4$ tile and the two $S_1$ tiles attached can be understood by considering a length-$6$ region of the QPU: $5$ sites from the $S_4$ tile itself, plus $1$ site of padding. This provides $3$ tiles of routing space per rotation, as required for our implementation of fallback rotations in Section~\ref{sec:rotations}. The full $4 \times 4$ lattice can then be spanned by $8$ such regions, and all such regions can be operated in parallel, reducing the total circuit depth and space-time volume compared to a serial circuit implementation.

Our lattice-surgery compilations contain several optimisations. For example, the previous lattice-surgery implementation of $\fswap$ between two adjacent qubits requires $7$ time steps of $d$ syndrome-extraction rounds each~\cite[Fig.~13]{Akahoshi2025}, whereas our implementation requires $3$ time steps. This is made possible by our ZX lattice surgery construction, which preserves the code distance of the patches and combines a background-lattice change with the measurement of the merge-stage or split-stage stabilisers in a single syndrome-extraction round of five two-qubit gate layers (see Appendix~\ref{sec:fswap_lattice_surgery}).

\cref{fig:cost_per_trotter_step} shows the total number of time steps and space-time volume (in units of blocks, i.e. a $d \times d$ patch for $d$ rounds) for each stage of a Trotter step. From this, we obtain the total time and volume cost per Trotter step:
\begin{equation}
    \text{Time: } 1063 \text{ time steps,} \quad \text{Volume: } 1.31 \times 10^5 \text{ blocks}.
\end{equation}
This is the cost for Lieb-kagome lattices with $\tau' > 0$. For $\tau'=0$ (the Lieb lattice) the $S_1$ sections need not be implemented and the cost is reduced accordingly. Note, as described in Section~\ref{sec:rotations}, we have obtained these numbers under the assumption that all rotations require $12$ T gates, as the worst-case of Table~\ref{tab:fallback_results}.

\begin{figure*}
    \centering
    \includegraphics[width=1.0\linewidth]{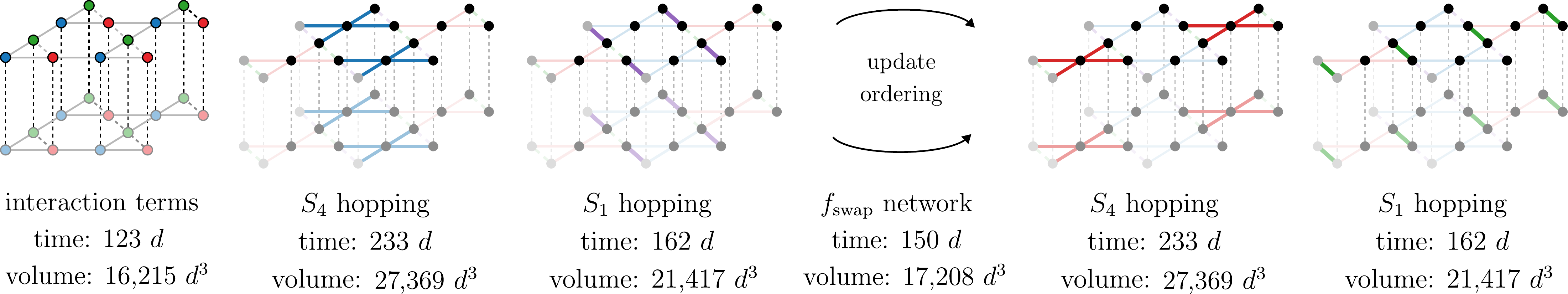}
    \caption{{\bf Time and volume costs per Trotter step for a $4 \times 4$ kagome lattice.} The total number of time steps ($1$ time step is $d$ syndrome extraction rounds) and volume in blocks ($1$ block is a $d \times d \times d$ space-time volume) to perform each term in a single Trotter step for the $4 \times 4$ lattice: the interaction term, two $S_1$ and $S_4$ sections, and the fermionic swap network to update the ordering. Note that the visual representations are shown for a $2 \times 2$ lattice, while the numbers presented are for the $4 \times 4$ case study lattice.}
    \label{fig:cost_per_trotter_step}
\end{figure*}

The magnetic spectral functions considered in this work take the form of $\mathcal{A}^{\alpha\beta}(\mathbf{q},\omega)$ in \cref{eq:spectral_function_final_expression}. This is constructed using importance sampling, whereby a circuit with $k$ Trotter steps is sampled with probability $p_k$ (\cref{eq:importance_sample_probs}). We consider a Lorentzian broadening where $F_k \propto e^{-\eta \delta t |k|}$ such that shorter circuits are sampled more frequently. As such, there are two quantities that we consider: the worst-case cost for the largest circuit ($k=\kmax$), and the expected-case cost over all circuits sampled.

In naive resource estimation, we may try to demand that \emph{all} shots of the circuit succeed with high probability. This would require an extremely high code distance, well beyond those considered for MegaQuOp error rates. Rather, significant experience from equivalent experiments (Trotterised quantum simulation followed by expectation value estimation) on non-error-corrected devices has shown that, even if each shot has a significant error rate, error mitigation techniques may still be applied to obtain accurate results, at the cost of a sampling overhead \cite{Temme2017, Endo2018, Urbanek2021, Vovrosh2021, Blunt2023, Farrell2024}. This is particularly true when performing Pauli twirling, converting the error channel to a Pauli channel \cite{Wallman2016, vandenBerg2023}, which can also be applied at the logical level \cite{Piveteau2021, Jain2023}. We expect that experiments on error-corrected MegaQuOp-scale quantum computers will also rely heavily on error mitigation in a similar way. We therefore seek to provide an estimate of the sampling overhead for our experiment due to error mitigation. As described in \cref{sec:megaquop_overview}, we assume that errors lead to signal damping by some factor $\gamma$ per Trotter step; if this factor can be accurately estimated then we can correct this with a sampling overhead of $(1/\gamma)^{2k}$ for $k$ steps. This error mitigation strategy will inevitably suffer some bias in practice, since the true damping factor cannot be estimated exactly, but is successfully applied on non-error-corrected devices \cite{Urbanek2021, Vovrosh2021, Blunt2023, Farrell2024, alam2025_2}. To estimate a value of $\gamma$ for resource estimation purposes, we first calculate a no-fault (success) probability for both Clifford (memory and lattice surgery) and magic state cultivation components of the case study circuits. We then assume that any logical error leads to a shot contributing no signal on average; this is valid when a measured observable, propagated backwards to the fault, is ``scrambled'', so that a Pauli error flips its sign with probability $50\%$ \cite{Dalzell2024, Nahum2018, Foldager2024}. A fault with probability $p$ then leads to a signal damping factor of approximately $\gamma = 1-p$. We emphasise that these assumptions give an estimate of this overhead, not a rigorous bound. A worst-case bound could be obtained by pessimistically assuming every logical error flips the observable, giving a damping factor $\gamma = 1-2p$, which would also be the cost of probabilistic error cancellation (PEC) \cite{Temme2017, Endo2018, huggins2025fluid}.

As described in Section~\ref{sec:megaquop_overview}, the probability of a logical error per $d \times d \times d$ volume operation (one ``block'') with a $Q$ QuOp QEC code is $1/Q$; note that this accounts for both $Z$ and $X$ errors. Therefore, defining $v$ as the volume per Trotter step (in units of blocks), and $m$ as the number of T gates per Trotter step, the expected sampling overhead per shot for our experiment to construct $\mathcal{A}^{\alpha\beta}(\mathbf{q},\omega)$ is estimated as
\begin{equation}\label{eq:expected_sampling_overhead}
    \sum_{k \ge 0}^{\kmax} p_k \, \lambda_{\text{rs}}^{2k} \, (1 - 1/Q)^{-2vk} \, (1 - p_{\text{cultivation}})^{-2mk},
\end{equation}
where $\lambda_{\text{rs}}^{2}$ is the sampling overhead per Trotter step due to quasi-probabilistic rotation synthesis, and $p_{\text{cultivation}}$ is the failure probability for magic state cultivation. The calculation of $Q$ is described in Section~\ref{sec:megaquop_overview} and Appendix~\ref{sec:lambda_calc}. We perform numerics for magic state cultivation in Appendix~\ref{app:magic_state_cultivation}; as an example, we estimate that $p_{\text{cultivation}} = 2.66 \times 10^{-6}$ can be achieved with space-time volume of $\approx 10^4$ with SI1000 noise and $p=10^{-3}$.

\begin{table*}[t]
\centering
\small
\setlength{\tabcolsep}{5pt}
\renewcommand{\arraystretch}{1.15}
\setlength{\aboverulesep}{0pt}
\setlength{\belowrulesep}{0pt}
\begin{tabular}{@{}l cccc cccc@{}}
\hline\hline
 & \multicolumn{4}{c}{$\delta t = 0.1\,\tau^{-1}$ ($\kmax = 128$)}
 & \multicolumn{4}{c}{$\delta t = 0.2\,\tau^{-1}$ ($\kmax = 64$)} \\
\cmidrule(lr){2-5} \cmidrule(l){6-9}
 & \multicolumn{2}{c}{Lieb} & \multicolumn{2}{c}{kagome}
 & \multicolumn{2}{c}{Lieb} & \multicolumn{2}{c}{kagome} \\
\cmidrule(lr){2-3} \cmidrule(lr){4-5} \cmidrule(lr){6-7} \cmidrule(l){8-9}
 & avg. & max. & avg. & max. & avg. & max. & avg. & max. \\
\hline
T gates                         & $6.0{\times}10^{4}$ & $2.2{\times}10^{5}$ & $1.1{\times}10^{5}$ & $4.2{\times}10^{5}$
                                & $3.0{\times}10^{4}$ & $1.1{\times}10^{5}$ & $5.7{\times}10^{4}$ & $2.1{\times}10^{5}$ \\
Sampling overhead (rot.\ synth.)     & 1.31 & 2.48 & 1.44 & 3.33 & 1.09 & 1.34 & 1.26 & 2.20 \\
\hline
\multicolumn{9}{@{}l}{SI1000: $d = 15$ ($\delta t = 0.1$), $d = 13$ ($\delta t = 0.2$)} \\
\addlinespace[2pt]

\quad Volume ($10^{6}$ blocks)    & \textbf{3.05} & \textbf{11.3} & \textbf{4.53} & \textbf{16.8} & \textbf{1.54} & \textbf{5.69} & \textbf{2.28} & \textbf{8.43} \\
\quad Runtime per shot (s)        & \textbf{0.38} & 1.42 & \textbf{0.55} & 2.04 & \textbf{0.17} & 0.62 & \textbf{0.24} & 0.90 \\
\quad No-fault probability (Clifford)   & 0.82 & 0.46 & 0.75 & 0.31 & 0.53 & 0.049 & 0.43 & 0.011 \\
\quad Sampling overhead (Clifford)     & 1.65 & 4.82 & 2.24 & 10.3 & 19.5 & 423  & 190  & $7.7{\times}10^{3}$ \\
\quad Sampling overhead (cultivation)  & 1.43 & 3.24 & 2.13 & 9.21 & 1.19 & 1.82 & 1.41 & 3.09 \\
\quad Sampling overhead (total)        & \textbf{4.24} & 38.6 & \textbf{15.7} & 317  & \textbf{37.7} & $1.0{\times}10^{3}$ & \textbf{989} & $5.3{\times}10^{4}$ \\
\hline
\multicolumn{9}{@{}l}{Optimistic: $d = 13$ ($\delta t = 0.1$), $d = 11$ ($\delta t = 0.2$)} \\
\addlinespace[2pt]
\quad Volume ($10^{6}$ blocks)    & \textbf{3.00} & \textbf{11.1} & \textbf{4.49} & \textbf{16.6} & \textbf{1.51} & \textbf{5.59} & \textbf{2.26} & \textbf{8.34} \\
\quad Runtime per shot (s)        & \textbf{0.32} & 1.19 & \textbf{0.47} & 1.72 & \textbf{0.14} & 0.51 & \textbf{0.20} & 0.74 \\
\quad No-fault probability (Clifford)   & 0.89 & 0.64 & 0.85 & 0.52 & 0.59 & 0.089 & 0.49 & 0.027 \\
\quad Sampling overhead (Clifford)     & 1.30 & 2.42 & 1.51 & 3.77 & 8.56 & 126 & 46.6 & $1.4{\times}10^{3}$ \\
\quad Sampling overhead (cultivation)  & 1.08 & 1.32 & 1.16 & 1.68 & 1.04 & 1.17 & 1.08 & 1.34 \\
\quad Sampling overhead (total)        & \textbf{2.00} & 7.90 & \textbf{3.10} & 21.1 & \textbf{11.5} & 197  & \textbf{110} & $4.0{\times}10^{3}$ \\

\hline\hline
\end{tabular}
\caption{\textbf{Resource estimates for the Lieb and kagome case studies.} Resource estimates for Lieb and kagome case studies, using Trotter step sizes of $\delta t = 0.1$ and $\delta t = 0.2$. For $\delta t = 0.1$ we consider $d=15$ (SI1000 noise model) and $d=13$ (Optimistic noise model); for $\delta t = 0.2$ we consider $d=13$ (SI1000) and $d=11$ (Optimistic). The constructed spectral functions are averaged over circuits of varying numbers of Trotter steps, $k$, up to a maximum $\kmax$. Columns include both the averaged values and the maximum values. We highlight important values: the averaged and maximum volumes, and the averaged runtimes and total estimated sampling overheads due to error mitigation and quasi-probability rotation synthesis. The estimated number of T gates and sampling overhead due to rotation synthesis are independent of the noise model, so are shown separately. The volume depends slightly on the noise model due to variation in the estimated time to perform cultivation.}
\label{tab:costing_results}
\end{table*}

Results are presented in \cref{tab:costing_results}. The script used to perform costing and produce this table is provided on Zenodo. We consider two scenarios: our default Trotter time-step of $\delta t = 0.1 \tau^{-1}$ and a more optimistic time step of $\delta t = 0.2 \tau^{-1}$. For a kagome lattice (and other Lieb-kagome lattices with $\tau' \ne 0)$, the volume of the largest $\delta t = 0.1 \tau^{-1}$ circuit with $\kmax=128$ Trotter steps is estimated as $16.8 \times 10^6$ blocks, while the average volume is $4.5 \times 10^6$ blocks. For an SI1000 noise model at $d=15$ this gives an average no-fault probability of $0.75$ for the Clifford portion of the circuit, or $0.85$ for the Optimistic noise model at $d=13$. These are sufficiently high that error mitigation can be applied for a reasonable sampling overhead. These distances and noise models correspond to $14$ million QuOps and $25$ million QuOps, respectively. When the step size is increased to $\delta t = 0.2$, an average no-fault probability of $0.49$ is obtained with $d=11$ in the Optimistic noise model, corresponding to $2.3$ million QuOps. Therefore, depending on the Trotter step size, we estimate that this application is within reach with a QEC code supporting a few MegaQuOps to a few tens of MegaQuOps.

The total qubit count is calculated as $150 \times 2(d+1)^2$; this consists of $150$ patches for the assumed $50$-by-$3$ QPU, and taking $(d+1)$-by-$(d+1)$ patches to allow space for lattice surgery. Therefore distances $d=11$ and $d=15$ require 43,200 and 76,800 qubits, respectively. We note that, taking the Optimistic noise model and $p=5 \times 10^{-4}$ allows $8$ MegaQuOps to be reached at $d=9$, as shown in \cref{sec:quantum_memory_simualtions}, reducing the physical qubit count to 30,000, though these latter assumptions are certainly optimistic.

\begin{figure*}
    \centering
    \includegraphics[width=0.7\linewidth]{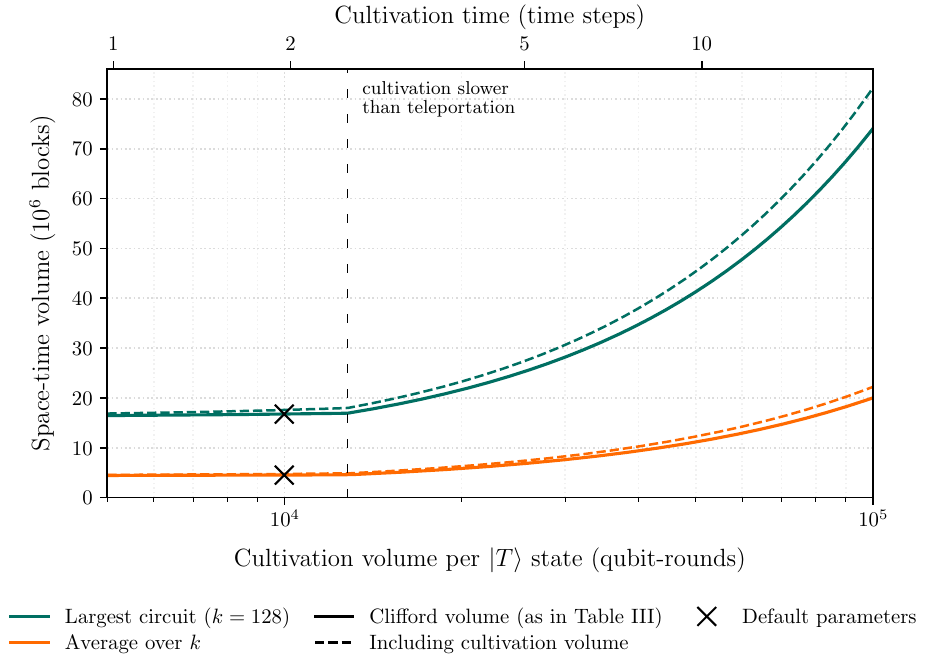}
    \caption{{\bf Sensitivity of circuit volume to cultivation volume.} The space-time volume of the largest circuit (green) and average circuit (orange) when varying the volume to perform magic state cultivation. The ``$\times$'' markers show the estimated volume based on simulations (\cref{app:magic_state_cultivation}) with an SI1000 noise model and final distance $d=15$; the resulting cultivation times are calculated using a factor of $1.5$ for imperfect packing and other slack \cite{huggins2025fluid}. Dashed (solid) lines include (exclude) volume associated with cultivation itself.}
    \label{fig:volume_sensitivity}
\end{figure*}

The average runtime per shot for $d=15$ SI1000 is $0.55$ seconds, assuming a syndrome extraction time of $1$~$\mu$s. The construction of the spectral function $\mathcal{A}^{\alpha\beta}(\mathbf{q},\omega)$ in \cref{eq:spectral_function_final_expression} has associated shot noise. Constructing $\mathcal{A}^{\alpha\beta}(\mathbf{q},\omega)$ also requires summing classically over a number of variables, which are performed by sampling. In addition, as just described, we allow a sampling overhead due to noise and quasi-probability rotation synthesis, ``sampling overhead (total)'' in \cref{tab:costing_results}. Therefore, we must budget a sufficient number of shots. We perform analysis of sampling error in the construction of $\mathcal{A}^{\alpha\beta}(\mathbf{q},\omega)$ in \cref{app:sampling_error_analysis}. Since our goal is to construct spectral functions to probe magnetic excitations, providing a precise number of shots to satisfy this task is challenging. However, for an average sampling overhead of $\overline{\lambda_{\rm sample}^2} = 15.7$ (for kagome circuits under SI1000 noise at $d = 15$, \cref{tab:costing_results}) and $n = 10^5$ shots, we estimate that a spectral line of unit weight is resolved with a relative noise of about $2.5\%$. For the Optimistic noise model at $d = 13$ ($\overline{\lambda_{\rm sample}^2} = 3.10$) this falls to about $1.1\%$. Therefore, we cost $10^5$ and $10^6$ shots as sensible orders of magnitude. As a further reference point, \cite{vilchez-estevez_extracting_2025} calculated magnetic excitation spectra using a quench protocol on a noisy superconducting processor with error mitigation; this experiment used $10^6$ shots, resulting in very well-converged spectra. For $d=15$ SI1000, performing $10^5$ or $10^6$ shots is estimated to take $15.3$ hours or $6.4$ days of runtime, respectively. This demonstrates that constructing $\mathcal{A}^{\alpha\beta}(\mathbf{q},\omega)$ to high precision may require significant runtime. This also highlights the challenges faced by slower qubit types to perform such experiments, and the value of investigating other QEC codes and faster approaches for performing logic when considering these qubits \cite{Ismail2026_2}. We note that, while we allowed a certain amount of sampling overhead in these results, this can be made close to $1$ by allowing higher code distances and higher volume for magic state cultivation. It may be sensible for slower qubit types to avoid this sampling overhead. However, in this work we have optimised for lower code distances, seeking to study what can be achieved with nearer-term devices and a near-MegaQuOp error budget.

Lastly, we investigate the impact of varying cultivation performance on final estimates of circuit volume for our case study. As we discussed in \cref{sec:rotations}, the performance of cultivation is very sensitive to the noise model, and the practical performance of cultivation is hard to predict. In practice, we may need $d_1=5$ cultivation, which comes with a higher associated volume to prepare each magic state. In \cref{fig:volume_sensitivity} we vary the cultivation volume ($V_{\text{cultivate}}$) and show the resulting space-time volume for the final circuits ($V_{\text{avg}}$ and $V_{\text{max}}$) for the kagome lattice and with SI1000 noise at $d=15$. Upon increasing the volume of cultivation by $10$ times, the circuit volume increases by $\sim 4$ times. For $t_{\text{cultivate}} > t_{\text{teleport}}$ the observed relation is
\begin{equation}
V_\text{avg} = 2.31 \times 10^{6} + 177\,V_\text{cultivate} \; \text{blocks}, \qquad
V_\text{max} = 8.56 \times 10^{6} + 654\,V_\text{cultivate} \; \text{blocks},
\end{equation}
again considering the kagome lattice with SI1000 noise at $d=15$. The constant term is from Clifford circuit elements that do not depend on cultivation. Taking $V_{\text{cultivate}} = 10^5$ qubit-rounds, we have $V_{\text{avg}} \approx 2 \times 10^7$. Therefore, even increasing the cultivation volume by a factor of $10$ keeps the average circuit volume within a few tens of millions of blocks, and does not significantly change our conclusions.

\section{Conclusion and perspective}
\label{sec:conclusion}

In this paper, we performed a detailed investigation of a condensed matter application on a MegaQuOp-scale device, i.e.~an error-corrected device with a logical error rate of approximately $10^{-6}$ per logical operation. Our results suggest that, for a device of this scale, it will be possible to reach relevant physical parameters for a non-trivial experiment of interest to condensed matter theorists \cite{LiebKagome2026}. At a physical error rate of $p = 10^{-3}$, and dependent on noise model, we estimate that relevant physical lattice sizes and energy resolutions can be obtained with distances $d=11$--$15$, corresponding to 43,000--77,000 physical qubits for the full device. This is based on a monolithic architecture of superconducting qubits using the planar surface code and lattice surgery computation.

We investigated two major approaches for implementing arbitrary-angle rotation gates on logical qubits, a crucial aspect of Trotterised quantum simulation. The STAR architecture was investigated under SI1000 noise at $p=10^{-3}$ by developing a non-Clifford simulation technique. Ultimately, our results suggest that this partially fault-tolerant approach gives logical error rates too high for this case study. Instead, we performed resource estimation within the framework of Clifford+T synthesis and magic state cultivation, applying the recently-developed quasi-probability Clifford+T approach of Ref.~\cite{bothe2026}. While we focused on a $4 \times 4$ lattice as a MegaQuOp-scale system, this methodology is scalable to larger lattices, or better energy resolutions, by using lower-error magic states (from $d_1 = 5$ cultivation) and higher code distances.

As in any resource estimation study, performing this analysis requires various assumptions. An important assumption of our analysis is that, for a circuit of volume $V$, the error rate of the QEC code needs to be $\sim 1/V$ for unmitigated simulation. However, as shown in Ref.~\cite{granet_2025}, quantum simulations of this type (Trotterised quench dynamics) can see significant dilution of errors, such that the $\sim 1/V$ assumption is pessimistic. This was shown to be particularly true when measuring local observables, as in our case study. Furthermore, the circuits in our case study admit symmetry-verification \cite{Bonet-Monroig2018, Stanisic2022} which allows many faulty shots to be post-selected in a simple manner. As such, our resource estimates may be overestimates of the true cost of such a simulation in some respects. However, some other aspects may be an underestimate: in particular, the cost of initial state preparation is not included. For near-Lieb lattices ($\tau' \approx 0$) this is believed to be well justified, but more sophisticated state preparation circuits may be needed for near-kagome lattices ($\tau' \approx \tau$). Our error mitigation scheme also assumes that signal damping from logical errors can be calibrated; achieving this in practice will be an important research area for early generations of error-corrected devices.

The high degree of parallelism of our circuits is made possible by the low space cost of magic state cultivation. The practical performance of cultivation on real devices is non-trivial, and better understanding and improving the performance of cultivation will be valuable to develop similar resource estimation studies further \cite{Chan2026, Wan2026, Mehta2026}. Finally, it will be interesting to investigate such a case study under higher-rate QEC codes and alternative methods for performing logic \cite{Yoder2025, Geher2026, Tripier2026, Webster2026, Oratomic_Shors, Ismail2026_2}. AMO qubits offer a promising approach to reduce physical qubit counts by allowing easier access to high-rate codes, however, the slower nature of these qubits makes reaching high shot counts for well-resolved spectra a non-trivial challenge to be investigated \cite{Webster2026, Oratomic_Shors, Ismail2026_2}. For superconducting qubits, connectivity requirements are a constraint, for which various approaches are being developed \cite{Yoder2025, Geher2026, Nixon2026, Gu2026, low2026denser}. We hope that the condensed matter case study developed here may act as a valuable benchmark for this ongoing research.

\acknowledgments

We thank Navneeth Ramakrishnan, Luka Skoric, Mark Turner and Adnaan Walayat for reviewing non-Clifford simulator code, Andrew Patterson for helpful discussions on the cost of magic state cultivation, Joan Camps and Setiawan for feedback on this manuscript, and Nina Glaser for providing advice on TD-DMRG simulations. M.B. is a Sustaining Innovation Postdoctoral Research Associate at Astex Pharmaceuticals and thanks Astex Pharmaceuticals for funding, as well as his Astex colleague Patrick Schoepf for his support. This work is supported by the Novo Nordisk Foundation, Grant number NNF22SA0081175, NNF Quantum Computing Programme. 
E.T.C.'s contributions to this work were while employed at Riverlane. The main ideas and work of this study were developed and performed by the authors. LLMs were used to provide feedback on the manuscript, to help with creating certain figures and tables, and to aid with writing code. LLMs also checked mathematical derivations and helped to check and run numerics. Claude (Anthropic) provided a generalised version of Appendix I3, Table II, and helped to write the final costing script with the authors. The authors checked these results and take responsibility for their correctness.

\section*{Author contributions}

The project was led by N.S.B. STAR simulations were performed by C.E. with discussions from N.S.B., A.R. and E.T.C. The non-Clifford simulator was implemented by C.E. with discussions from E.T.C. Magic state cultivation simulations were performed by A.R. Lattice surgery constructions were obtained by A.R. and O.C. The physics problem was proposed by E.G.C.P.V.L, and the quench circuits and case study were developed by A.V.I. with N.S.B. Fermionic swap networks were obtained by J.K. Trotter error analysis was performed by J.K and E.T.C, with DMRG simulations by A.J.B.S and N.S.B. Tile Trotter partitions and circuits were obtained and checked by A.J.B.S. Memory experiments and their analysis were performed by J.M., and memory vs lattice surgery experiments were performed by O.C. Quasi-probability Clifford+T solutions and final resource estimates were obtained by N.S.B. Sampling error analysis was performed by E.T.C and C.S. The ZX lattice surgery construction was provided by M.R. and M.O.R. The manuscript was drafted by C.E., A.R., A.V.I., J.K., J.M., C.S. and N.S.B. Comments and edits were incorporated from all authors.

\section*{Data availability}

Data and circuits used for the memory analysis in \cref{sec:quantum_memory_simualtions} can be found in the \href{https://github.com/riverlane/megaquop_footprints}{megaquop\_footprints} repository. Additional data, including STAR circuits, swap networks, and costing scripts, are included on Zenodo at \url{https://doi.org/10.5281/zenodo.24536356}.

\newpage

\bibliography{main}

@article{preskill2025beyond,
author = {Preskill, John},
title = {{Beyond NISQ: The Megaquop Machine}},
year = {2025},
issue_date = {September 2025},
publisher = {Association for Computing Machinery},
address = {New York, NY, USA},
volume = {6},
number = {3},
url = {https://doi.org/10.1145/3723153},
doi = {10.1145/3723153},
journal = {ACM Transactions on Quantum Computing},
month = apr,
articleno = {18},
numpages = {7}
}

@misc{huggins2025fluid,
      title={{The {FL}uid {A}llocation of {S}urface code {Q}ubits ({FLASQ}) cost model for early fault-tolerant quantum algorithms}}, 
      author={William J. Huggins and Tanuj Khattar and Amanda Xu and Matthew Harrigan and Christopher Kang and Guang Hao Low and Austin Fowler and Nicholas C. Rubin and Ryan Babbush},
      year={2025},
      eprint={2511.08508},
      archivePrefix={arXiv},
      primaryClass={quant-ph},
      url={https://arxiv.org/abs/2511.08508}, 
}

@article{kivlichan2020improved,
  doi = {10.22331/q-2020-07-16-296},
  url = {https://doi.org/10.22331/q-2020-07-16-296},
  title = {{Improved {F}ault-{T}olerant {Q}uantum {S}imulation of {C}ondensed-{P}hase {C}orrelated {E}lectrons via {T}rotterization}},
  author = {Kivlichan, Ian D. and Gidney, Craig and Berry, Dominic W. and Wiebe, Nathan and McClean, Jarrod and Sun, Wei and Jiang, Zhang and Rubin, Nicholas and Fowler, Austin and Aspuru-Guzik, Al{\'{a}}n and Neven, Hartmut and Babbush, Ryan},
  journal = {{Quantum}},
  issn = {2521-327X},
  publisher = {{Verein zur F{\"{o}}rderung des Open Access Publizierens in den Quantenwissenschaften}},
  volume = {4},
  pages = {296},
  month = jul,
  year = {2020}
}

@Article{google_repetition_code,
author={Chen, Zijun
and Satzinger, Kevin J.
and Atalaya, Juan
and Korotkov, Alexander N.
and Dunsworth, Andrew
and Sank, Daniel
and Quintana, Chris
and McEwen, Matt
and Barends, Rami
and Klimov, Paul V.
and Hong, Sabrina
and Jones, Cody
and Petukhov, Andre
and Kafri, Dvir
and Demura, Sean
and Burkett, Brian
and Gidney, Craig
and Fowler, Austin G.
and Paler, Alexandru
and Putterman, Harald
and Aleiner, Igor
and Arute, Frank
and Arya, Kunal
and Babbush, Ryan
and Bardin, Joseph C.
and Bengtsson, Andreas
and Bourassa, Alexandre
and Broughton, Michael
and Buckley, Bob B.
and Buell, David A.
and Bushnell, Nicholas
and Chiaro, Benjamin
and Collins, Roberto
and Courtney, William
and Derk, Alan R.
and Eppens, Daniel
and Erickson, Catherine
and Farhi, Edward
and Foxen, Brooks
and Giustina, Marissa
and Greene, Ami
and Gross, Jonathan A.
and Harrigan, Matthew P.
and Harrington, Sean D.
and Hilton, Jeremy
and Ho, Alan
and Huang, Trent
and Huggins, William J.
and Ioffe, L. B.
and Isakov, Sergei V.
and Jeffrey, Evan
and Jiang, Zhang
and Kechedzhi, Kostyantyn
and Kim, Seon
and Kitaev, Alexei
and Kostritsa, Fedor
and Landhuis, David
and Laptev, Pavel
and Lucero, Erik
and Martin, Orion
and McClean, Jarrod R.
and McCourt, Trevor
and Mi, Xiao
and Miao, Kevin C.
and Mohseni, Masoud
and Montazeri, Shirin
and Mruczkiewicz, Wojciech
and Mutus, Josh
and Naaman, Ofer
and Neeley, Matthew
and Neill, Charles
and Newman, Michael
and Niu, Murphy Yuezhen
and O'Brien, Thomas E.
and Opremcak, Alex
and Ostby, Eric
and Pat{\'o}, B{\'a}lint
and Redd, Nicholas
and Roushan, Pedram
and Rubin, Nicholas C.
and Shvarts, Vladimir
and Strain, Doug
and Szalay, Marco
and Trevithick, Matthew D.
and Villalonga, Benjamin
and White, Theodore
and Yao, Z. Jamie
and Yeh, Ping
and Yoo, Juhwan
and Zalcman, Adam
and Neven, Hartmut
and Boixo, Sergio
and Smelyanskiy, Vadim
and Chen, Yu
and Megrant, Anthony
and Kelly, Julian
and {Google Quantum AI}},
title={{Exponential suppression of bit or phase errors with cyclic error correction}},
journal={Nature},
year={2021},
month={Jul},
day={01},
volume={595},
number={7867},
pages={383-387},
issn={1476-4687},
doi={10.1038/s41586-021-03588-y},
url={https://doi.org/10.1038/s41586-021-03588-y}
}

@Article{google2023,
author={{Google Quantum AI}},
title={{Suppressing quantum errors by scaling a surface code logical qubit}},
journal={Nature},
year={2023},
month={Feb},
day={01},
volume={614},
number={7949},
pages={676-681},
issn={1476-4687},
doi={10.1038/s41586-022-05434-1},
url={https://doi.org/10.1038/s41586-022-05434-1}
}

@Article{google2025,
author={{Google Quantum AI
and Collaborators}},
title={{Quantum error correction below the surface code threshold}},
journal={Nature},
year={2025},
month={Feb},
day={01},
volume={638},
number={8052},
pages={920-926},
issn={1476-4687},
doi={10.1038/s41586-024-08449-y},
url={https://doi.org/10.1038/s41586-024-08449-y}
}

@Article{Caune2026,
author={Caune, Laura
and Skoric, Luka
and Blunt, Nick S.
and Ruban, Archibald
and McDaniel, Jimmy
and Valery, Joseph A.
and Patterson, Andrew D.
and Gramolin, Alexander V.
and Majaniemi, Joonas
and Barnes, Kenton M.
and Bialas, Tomasz
and Bu{\u{g}}dayc{\i}, Okan
and Crawford, Ophelia
and Geh{\'e}r, Gy{\"o}rgy P.
and Krovi, Hari
and Matekole, Elisha
and Topal, Canberk
and Poletto, Stefano
and Bryant, Michael
and Snyder, Kalan
and Gillespie, Neil I.
and Jones, Glenn
and Johar, Kauser
and Campbell, Earl T.
and Hill, Alexander D.},
title={{Demonstrating real-time and low-latency quantum error correction with superconducting qubits}},
journal={Nature Communications},
year={2026},
month={Jun},
day={01},
volume={17},
number={1},
pages={7383},
issn={2041-1723},
doi={10.1038/s41467-026-73331-6},
url={https://doi.org/10.1038/s41467-026-73331-6}
}

@Article{Bluvstein2026,
author={Bluvstein, Dolev
and Geim, Alexandra A.
and Li, Sophie H.
and Evered, Simon J.
and Bonilla Ataides, J. Pablo
and Baranes, Gefen
and Gu, Andi
and Manovitz, Tom
and Xu, Muqing
and Kalinowski, Marcin
and Majidy, Shayan
and Kokail, Christian
and Maskara, Nishad
and Trapp, Elias C.
and Stewart, Luke M.
and Hollerith, Simon
and Zhou, Hengyun
and Gullans, Michael J.
and Yelin, Susanne F.
and Greiner, Markus
and Vuleti{\'{c}}, Vladan
and Cain, Madelyn
and Lukin, Mikhail D.},
title={{A fault-tolerant neutral-atom architecture for universal quantum computation}},
journal={Nature},
year={2026},
month={Jan},
day={01},
volume={649},
number={8095},
pages={39-46},
issn={1476-4687},
doi={10.1038/s41586-025-09848-5},
url={https://doi.org/10.1038/s41586-025-09848-5}
}

@Article{Paetznick2026,
author={Paetznick, A.
and Reichardt, B. W.
and Silva, M. P. da
and Ryan-Anderson, C.
and Aasen, D.
and Bello-Rivas, J. M.
and Campora, J. P.
and Chao, R.
and Chernoguzov, A.
and van Dam, W.
and Dreiling, J. M.
and Foltz, C.
and Frachon, F.
and Gaebler, J. P.
and Gatterman, T. M.
and Grans-Samuelsson, L.
and Gresh, D.
and Hayes, D.
and Hewitt, N.
and Holliman, C.
and Horst, C. V.
and Johansen, J.
and Lucchetti, D.
and Matsuoka, Y.
and Mills, M.
and Moses, S. A.
and Neyenhuis, B.
and Paz, A.
and Pino, J.
and Siegfried, P.
and Sundaram, A.
and Tom, D.
and Wang, Z.
and Wernli, S. J.
and Zanner, M.
and Stutz, R. P.
and Svore, K. M.},
title={{Improved quantum processor logical error rates via correction and detection}},
journal={Nature},
year={2026},
month={Jun},
day={01},
volume={654},
number={8118},
pages={349-355},
issn={1476-4687},
doi={10.1038/s41586-026-10628-y},
url={https://doi.org/10.1038/s41586-026-10628-y}
}

@Article{Harper2026,
author={Harper, Robin
and Lain{\'e}, Constance
and Hockings, Evan T.
and McLauchlan, Campbell
and Nixon, Georgia M.
and Brown, Benjamin J.
and Bartlett, Stephen D.},
title={{Characterising the failure mechanisms of error-corrected quantum logic gates}},
journal={Nature Communications},
year={2026},
month={Apr},
day={10},
volume={17},
number={1},
pages={5039},
issn={2041-1723},
doi={10.1038/s41467-026-71773-6},
url={https://doi.org/10.1038/s41467-026-71773-6}
}

@misc{gidney2025factor,
      title={{How to factor 2048 bit {RSA} integers with less than a million noisy qubits}}, 
      author={Craig Gidney},
      year={2025},
      eprint={2505.15917},
      archivePrefix={arXiv},
      primaryClass={quant-ph},
      url={https://arxiv.org/abs/2505.15917}, 
}

@misc{litinski2022active,
      title={{Active volume: An architecture for efficient fault-tolerant quantum computers with limited non-local connections}}, 
      author={Daniel Litinski and Naomi Nickerson},
      year={2022},
      eprint={2211.15465},
      archivePrefix={arXiv},
      primaryClass={quant-ph},
      url={https://arxiv.org/abs/2211.15465}, 
}

@article{gidney2024inplace,
  title={{Inplace access to the surface code Y basis}},
  author={Gidney, Craig},
  journal={Quantum},
  volume={8},
  pages={1310},
  year={2024},
  publisher={Verein zur F{\"o}rderung des Open Access Publizierens in den Quantenwissenschaften}
}

@misc{beverland2022assessing,
      title={{Assessing requirements to scale to practical quantum advantage}}, 
      author={Michael E. Beverland and Prakash Murali and Matthias Troyer and Krysta M. Svore and Torsten Hoefler and Vadym Kliuchnikov and Guang Hao Low and Mathias Soeken and Aarthi Sundaram and Alexander Vaschillo},
      year={2022},
      eprint={2211.07629},
      archivePrefix={arXiv},
      primaryClass={quant-ph},
      url={https://arxiv.org/abs/2211.07629}, 
}

@article{gidney2021fault,
  title={{A fault-tolerant honeycomb memory}},
  author={Gidney, Craig and Newman, Michael and Fowler, Austin and Broughton, Michael},
  journal={Quantum},
  volume={5},
  pages={605},
  year={2021},
  publisher={Verein zur Förderung des Open-Access-Publizierens in den Quantenwissenschaften}
}

@article{campbell2022early,
doi = {10.1088/2058-9565/ac3110},
url = {https://doi.org/10.1088/2058-9565/ac3110},
year = {2021},
month = {nov},
publisher = {IOP Publishing},
volume = {7},
number = {1},
pages = {015007},
author = {Campbell, Earl T},
title = {{Early fault-tolerant simulations of the Hubbard model}},
journal = {Quantum Science and Technology}
}

@Article{Kan2025,
author={Kan, Angus
and Symons, Benjamin C. B.},
title={{Resource-optimized fault-tolerant simulation of the {Fermi-Hubbard} model and high-temperature superconductor models}},
journal={npj Quantum Information},
year={2025},
month={Aug},
day={11},
volume={11},
number={1},
pages={138},
issn={2056-6387},
doi={10.1038/s41534-025-01091-0},
url={https://doi.org/10.1038/s41534-025-01091-0}
}

@article{litinski2019game,
  title={{A game of surface codes: Large-scale quantum computing with lattice surgery}},
  author={Litinski, Daniel},
  journal={Quantum},
  volume={3},
  pages={128},
  year={2019},
  publisher={Verein zur F{\"o}rderung des Open Access Publizierens in den Quantenwissenschaften}
}

@article{chamberland2022circuit,
  title = {{Circuit-level protocol and analysis for twist-based lattice surgery}},
  author = {Chamberland, Christopher and Campbell, Earl T.},
  journal = {Phys. Rev. Res.},
  volume = {4},
  issue = {2},
  pages = {023090},
  numpages = {11},
  year = {2022},
  month = {May},
  publisher = {American Physical Society},
  doi = {10.1103/PhysRevResearch.4.023090},
  url = {https://link.aps.org/doi/10.1103/PhysRevResearch.4.023090}
}

@article{geher2024error,
  doi = {10.22331/q-2024-07-02-1394},
  url = {https://doi.org/10.22331/q-2024-07-02-1394},
  title = {{Error-corrected {H}adamard gate simulated at the circuit level}},
  author = {Geh{\'{e}}r, Gy{\"{o}}rgy P. and McLauchlan, Campbell and Campbell, Earl T. and Moylett, Alexandra E. and Crawford, Ophelia},
  journal = {{Quantum}},
  issn = {2521-327X},
  publisher = {{Verein zur F{\"{o}}rderung des Open Access Publizierens in den Quantenwissenschaften}},
  volume = {8},
  pages = {1394},
  month = jul,
  year = {2024}
}

@article{LiebKagome2026,
  title = {{Magnons in multiorbital {Hubbard} models: From {Lieb} to kagome lattices}},
  author = {Ying, Teng-Fei and Strand, Hugo U. R. and Zhou, Benjamin T. and van Loon, Erik G. C. P.},
  journal = {Phys. Rev. B},
  volume = {114},
  issue = {17},
  pages = {175109},
  numpages = {16},
  year = {2026},
  month = {Sep},
  publisher = {American Physical Society},
  doi = {10.1103/gcdq-mdgr},
  url = {https://link.aps.org/doi/10.1103/gcdq-mdgr}
}

@article{Jiang2019,
  title = {{Topological band evolution between {Lieb} and kagome lattices}},
  author = {Jiang, Wei and Kang, Meng and Huang, Huaqing and Xu, Hongxing and Low, Tony and Liu, Feng},
  journal = {Phys. Rev. B},
  volume = {99},
  issue = {12},
  pages = {125131},
  numpages = {9},
  year = {2019},
  month = {Mar},
  publisher = {American Physical Society},
  doi = {10.1103/PhysRevB.99.125131},
  url = {https://link.aps.org/doi/10.1103/PhysRevB.99.125131}
}

@article{kim_2023,
author={Kim, Youngseok
and Eddins, Andrew
and Anand, Sajant
and Wei, Ken Xuan
and van den Berg, Ewout
and Rosenblatt, Sami
and Nayfeh, Hasan
and Wu, Yantao
and Zaletel, Michael
and Temme, Kristan
and Kandala, Abhinav},
title={{Evidence for the utility of quantum computing before fault tolerance}},
journal={Nature},
year={2023},
month={Jun},
day={01},
volume={618},
number={7965},
pages={500-505},
doi={10.1038/s41586-023-06096-3},
url={https://doi.org/10.1038/s41586-023-06096-3}
}

@misc{quantinuum_trotter,
      title={{Digital quantum magnetism at the frontier of classical simulations}}, 
      author={Reza Haghshenas and Eli Chertkov and Michael Mills and Wilhelm Kadow and Sheng-Hsuan Lin and Yi-Hsiang Chen and Chris Cade and Ido Niesen and Tomislav Begušić and Manuel S. Rudolph and Cristina Cirstoiu and Kevin Hemery and Conor Mc Keever and Michael Lubasch and Etienne Granet and Charles H. Baldwin and John P. Bartolotta and Matthew Bohn and Julia Cline and Matthew DeCross and Joan M. Dreiling and Cameron Foltz and David Francois and John P. Gaebler and Christopher N. Gilbreth and Johnnie Gray and Dan Gresh and Alex Hall and Aaron Hankin and Azure Hansen and Nathan Hewitt and Ross B. Hutson and Mohsin Iqbal and Nikhil Kotibhaskar and Elliot Lehman and Dominic Lucchetti and Ivaylo S. Madjarov and Karl Mayer and Alistair R. Milne and Steven A. Moses and Brian Neyenhuis and Gunhee Park and Boris Ponsioen and Michael Schecter and Peter E. Siegfried and David T. Stephen and Bruce G. Tiemann and Maxwell D. Urmey and James Walker and Andrew C. Potter and David Hayes and Garnet Kin-Lic Chan and Frank Pollmann and Michael Knap and Henrik Dreyer and Michael Foss-Feig},
      year={2025},
      eprint={2503.20870},
      archivePrefix={arXiv},
      primaryClass={quant-ph},
      url={https://arxiv.org/abs/2503.20870}, 
}

@article{granet_2025,
  title = {{Dilution of Error in Digital {Hamiltonian} Simulation}},
  author = {Granet, Etienne and Dreyer, Henrik},
  journal = {PRX Quantum},
  volume = {6},
  issue = {1},
  pages = {010333},
  numpages = {27},
  year = {2025},
  month = {Feb},
  publisher = {American Physical Society},
  doi = {10.1103/PRXQuantum.6.010333},
  url = {https://link.aps.org/doi/10.1103/PRXQuantum.6.010333}
}

@misc{granet2026,
      title={{Superconducting pairing correlations on a trapped-ion quantum computer}}, 
      author={Etienne Granet and Sheng-Hsuan Lin and Kevin Hémery and Reza Haghshenas and Pablo Andres-Martinez and David T. Stephen and Anthony Ransford and Jake Arkinstall and M. S. Allman and Pete Campora and Samuel F. Cooper and Robert D. Delaney and Joan M. Dreiling and Brian Estey and Caroline Figgatt and Cameron Foltz and John P. Gaebler and Alex Hall and Ali Husain and Akhil Isanaka and Colin J. Kennedy and Nikhil Kotibhaskar and Ivaylo S. Madjarov and Michael Mills and Alistair R. Milne and Annie J. Park and Adam P. Reed and Brian Neyenhuis and Justin G. Bohnet and Michael Foss-Feig and Andrew C. Potter and Ramil Nigmatullin and Mohsin Iqbal and Henrik Dreyer},
      year={2026},
      eprint={2511.02125},
      archivePrefix={arXiv},
      primaryClass={quant-ph},
      url={https://arxiv.org/abs/2511.02125}, 
}

@misc{alam2025_1,
      title={{Fermionic dynamics on a trapped-ion quantum computer beyond exact classical simulation}}, 
      author={Faisal Alam and Jan Lukas Bosse and Ieva Čepaitė and Adrian Chapman and Laura Clinton and Marcos Crichigno and Elizabeth Crosson and Toby Cubitt and Charles Derby and Oliver Dowinton and Norhan Eassa and Paul K. Faehrmann and Steve Flammia and Brian Flynn and Filippo Maria Gambetta and Raúl García-Patrón and Max Hunter-Gordon and Glenn Jones and Abhishek Khedkar and Joel Klassen and Michael Kreshchuk and Edward Harry McMullan and Lana Mineh and Ashley Montanaro and Caterina Mora and John J. L. Morton and Alberto Nocera and Dhrumil Patel and Pete Rolph and Raul A. Santos and James R. Seddon and Evan Sheridan and Wilfrid Somogyi and Marika Svensson and Niam Vaishnav and Sabrina Yue Wang and Gethin Wright and Eli Chertkov and Henrik Dreyer and Michael Foss-Feig},
      year={2025},
      eprint={2510.26300},
      archivePrefix={arXiv},
      primaryClass={quant-ph},
      url={https://arxiv.org/abs/2510.26300}, 
}

@misc{alam2025_2,
      title={{Programmable digital quantum simulation of 2D Fermi-Hubbard dynamics using 72 superconducting qubits}}, 
      author={Faisal Alam and Jan Lukas Bosse and Ieva Čepaitė and Adrian Chapman and Laura Clinton and Marcos Crichigno and Elizabeth Crosson and Toby Cubitt and Charles Derby and Oliver Dowinton and Paul K. Faehrmann and Steve Flammia and Brian Flynn and Filippo Maria Gambetta and Raúl García-Patrón and Max Hunter-Gordon and Glenn Jones and Abhishek Khedkar and Joel Klassen and Michael Kreshchuk and Edward Harry McMullan and Lana Mineh and Ashley Montanaro and Caterina Mora and John J. L. Morton and Dhrumil Patel and Pete Rolph and Raul A. Santos and James R. Seddon and Evan Sheridan and Wilfrid Somogyi and Marika Svensson and Niam Vaishnav and Sabrina Yue Wang and Gethin Wright},
      year={2025},
      eprint={2510.26845},
      archivePrefix={arXiv},
      primaryClass={quant-ph},
      url={https://arxiv.org/abs/2510.26845}, 
}

@misc{millar2026,
      title={{Quench Spectroscopy of Magnetic Excitations on a Superconducting Quantum Processor}}, 
      author={D. A. Millar and G. W. Pennington and N. T. M. Siow and S. Brandhofer and J. Crain and F. H. L. Essler and A. G. Green and S. J. Thomson},
      year={2026},
      eprint={2607.02673},
      archivePrefix={arXiv},
      primaryClass={quant-ph},
      url={https://arxiv.org/abs/2607.02673}, 
}

@misc{andersen2026,
      title={{Precision quantum simulation of magnon spectra and interactions}}, 
      author={{Google Quantum AI
and Collaborators}},
      year={2026},
      eprint={2607.13301},
      archivePrefix={arXiv},
      primaryClass={quant-ph},
      url={https://arxiv.org/abs/2607.13301}, 
}

@misc{bothe2026,
      title={{More efficient {Clifford+T} synthesis for small-angle rotations and application to {T}rotterization}}, 
      author={Marius Bothe and Christoph Sünderhauf and Michael J. Witham and Earl T. Campbell and Nick S. Blunt},
      year={2026},
      eprint={2605.31544},
      archivePrefix={arXiv},
      primaryClass={quant-ph},
      url={https://arxiv.org/abs/2605.31544}, 
}

@misc{choi2023,
      title={{Fault Tolerant Non-{Clifford} State Preparation for Arbitrary Rotations}}, 
      author={Hyeongrak Choi and Frederic T. Chong and Dirk Englund and Yongshan Ding},
      year={2023},
      eprint={2303.17380},
      archivePrefix={arXiv},
      primaryClass={quant-ph},
      url={https://arxiv.org/abs/2303.17380}, 
}

@article{Toshio2025,
  title = {{Practical Quantum Advantage on Partially Fault-Tolerant Quantum Computer}},
  author = {Toshio, Riki and Akahoshi, Yutaro and Fujisaki, Jun and Oshima, Hirotaka and Sato, Shintaro and Fujii, Keisuke},
  journal = {Phys. Rev. X},
  volume = {15},
  issue = {2},
  pages = {021057},
  numpages = {43},
  year = {2025},
  month = {May},
  publisher = {American Physical Society},
  doi = {10.1103/PhysRevX.15.021057},
  url = {https://link.aps.org/doi/10.1103/PhysRevX.15.021057}
}

@article{Akahoshi2024,
  title = {{Partially Fault-Tolerant Quantum Computing Architecture with Error-Corrected {Clifford} Gates and Space-Time Efficient Analog Rotations}},
  author = {Akahoshi, Yutaro and Maruyama, Kazunori and Oshima, Hirotaka and Sato, Shintaro and Fujii, Keisuke},
  journal = {PRX Quantum},
  volume = {5},
  issue = {1},
  pages = {010337},
  numpages = {21},
  year = {2024},
  month = {Mar},
  publisher = {American Physical Society},
  doi = {10.1103/PRXQuantum.5.010337},
  url = {https://link.aps.org/doi/10.1103/PRXQuantum.5.010337}
}

@article{Akahoshi2025,
  title = {{Compilation of {Trotter}-Based Time Evolution for Partially Fault-Tolerant Quantum Computing Architecture}},
  author = {Akahoshi, Yutaro and Toshio, Riki and Fujisaki, Jun and Oshima, Hirotaka and Sato, Shintaro and Fujii, Keisuke},
  journal = {PRX Quantum},
  volume = {6},
  issue = {4},
  pages = {040319},
  numpages = {24},
  year = {2025},
  month = {Oct},
  publisher = {American Physical Society},
  doi = {10.1103/93zr-1ykb},
  url = {https://link.aps.org/doi/10.1103/93zr-1ykb}
}

@misc{Chung2026,
      title={{Partially Fault-Tolerant Quantum Computation for Megaquop Applications}}, 
      author={Ming-Zhi Chung and Ali H. Z. Kavaki and Artur Scherer and Abdullah Khalid and Xiangzhou Kong and Toru Kawakubo and Namit Anand and Gebremedhin A Dagnew and Zachary Webb and Allyson Silva and Gaurav Gyawali and Tennin Yan and Keisuke Fujii and Alan Ho and Masoud Mohseni and Pooya Ronagh and John Martinis},
      year={2026},
      eprint={2603.13093},
      archivePrefix={arXiv},
      primaryClass={quant-ph},
      url={https://arxiv.org/abs/2603.13093}, 
}

@misc{Toshio2026,
      title={{{STAR-Magic Mutation}: Even More Efficient Analog Rotation Gates for Early Fault-Tolerant Quantum Computer}}, 
      author={Riki Toshio and Shota Kanasugi and Jun Fujisaki and Hirotaka Oshima and Shintaro Sato and Keisuke Fujii},
      year={2026},
      eprint={2603.22891},
      archivePrefix={arXiv},
      primaryClass={quant-ph},
      url={https://arxiv.org/abs/2603.22891}, 
}

@misc{Kanasugi2026,
      title={{Enabling Chemically Accurate Quantum Phase Estimation in the Early Fault-Tolerant Regime}}, 
      author={Shota Kanasugi and Riki Toshio and Kazunori Maruyama and Hirotaka Oshima},
      year={2026},
      eprint={2603.22778},
      archivePrefix={arXiv},
      primaryClass={quant-ph},
      url={https://arxiv.org/abs/2603.22778}, 
}

@article{BaySmidt2025,
  title = {{Fault-Tolerant Quantum Simulation of Generalized {Hubbard} Models}},
  author = {Bay-Smidt, Andreas Juul and Klausen, Frederik Ravn and S\"underhauf, Christoph and Izs\'ak, R\'obert and Solomon, Gemma C. and Blunt, Nick S.},
  journal = {PRX Quantum},
  volume = {6},
  issue = {3},
  pages = {030348},
  numpages = {39},
  year = {2025},
  month = {Sep},
  publisher = {American Physical Society},
  doi = {10.1103/gr4t-b1w5},
  url = {https://link.aps.org/doi/10.1103/gr4t-b1w5}
}

@misc{BaySmidt2026,
      title={{Quantum simulation of nanographenes and {Trotter} error cancellation}}, 
      author={Andreas Juul Bay-Smidt and Nina Glaser and Marcel D. Fabian and Earl T. Campbell and Nick S. Blunt and Gemma C. Solomon},
      year={2026},
      eprint={2605.00745},
      archivePrefix={arXiv},
      primaryClass={quant-ph},
      url={https://arxiv.org/abs/2605.00745}, 
}

@misc{logical_s_gate,
      title={{A {$\boldsymbol{2d \times d \times d}$} Spacetime Volume Implementation of a Logical {S} Gate in the Surface Code}}, 
      author={Yuga Hirai and Shota Ikari and Yosuke Ueno and Yasunari Suzuki},
      year={2026},
      eprint={2604.13632},
      archivePrefix={arXiv},
      primaryClass={quant-ph},
      url={https://arxiv.org/abs/2604.13632}, 
}

@misc{low2026denser,
      title={{A Denser Planar Surface Code}}, 
      author={Guang Hao Low and William J. Huggins and Dominic W. Berry and Tanuj Khattar and Alec F. White and Nicholas C. Rubin and Ryan Babbush},
      year={2026},
      eprint={2605.30455},
      archivePrefix={arXiv},
      primaryClass={quant-ph},
      url={https://arxiv.org/abs/2605.30455}, 
}

@misc{gidney2024magic,
      title={{Magic state cultivation: growing {T} states as cheap as {CNOT} gates}}, 
      author={Craig Gidney and Noah Shutty and Cody Jones},
      year={2024},
      eprint={2409.17595},
      archivePrefix={arXiv},
      primaryClass={quant-ph},
      url={https://arxiv.org/abs/2409.17595}, 
}

@article{higgott2025sparse,
  doi = {10.22331/q-2025-01-20-1600},
  url = {https://doi.org/10.22331/q-2025-01-20-1600},
  title = {{Sparse {B}lossom: correcting a million errors per core second with minimum-weight matching}},
  author = {Higgott, Oscar and Gidney, Craig},
  journal = {{Quantum}},
  issn = {2521-327X},
  publisher = {{Verein zur F{\"{o}}rderung des Open Access Publizierens in den Quantenwissenschaften}},
  volume = {9},
  pages = {1600},
  month = jan,
  year = {2025}
}

@article{bausch2024learning,
   title={{Learning high-accuracy error decoding for quantum processors}},
   volume={635},
   ISSN={1476-4687},
   url={http://dx.doi.org/10.1038/s41586-024-08148-8},
   DOI={10.1038/s41586-024-08148-8},
   number={8040},
   journal={Nature},
   publisher={Springer Science and Business Media LLC},
   author={Bausch, Johannes and Senior, Andrew W. and Heras, Francisco J. H. and Edlich, Thomas and Davies, Alex and Newman, Michael and Jones, Cody and Satzinger, Kevin and Niu, Murphy Yuezhen and Blackwell, Sam and Holland, George and Kafri, Dvir and Atalaya, Juan and Gidney, Craig and Hassabis, Demis and Boixo, Sergio and Neven, Hartmut and Kohli, Pushmeet},
   year={2024},
   month=Nov, pages={834–840} }

@misc{senior2026scalable,
      title={{A scalable and real-time neural decoder for topological quantum codes}}, 
      author={Andrew W. Senior and Thomas Edlich and Francisco J. H. Heras and Lei M. Zhang and Oscar Higgott and James S. Spencer and Taylor Applebaum and Sam Blackwell and Justin Ledford and Akvilė Žemgulytė and Augustin Žídek and Noah Shutty and Andrew Cowie and Yin Li and George Holland and Peter Brooks and Charlie Beattie and Michael Newman and Alex Davies and Cody Jones and Sergio Boixo and Hartmut Neven and Pushmeet Kohli and Johannes Bausch},
      year={2026},
      eprint={2512.07737},
      archivePrefix={arXiv},
      primaryClass={quant-ph},
      url={https://arxiv.org/abs/2512.07737}, 
}

@misc{beni2025tesseract,
      title={{Tesseract: A Search-Based Decoder for Quantum Error Correction}}, 
      author={Laleh Aghababaie Beni and Oscar Higgott and Noah Shutty},
      year={2025},
      eprint={2503.10988},
      archivePrefix={arXiv},
      primaryClass={quant-ph},
      url={https://arxiv.org/abs/2503.10988}, 
}

@misc{fowler2013optimal,
      title={{Optimal complexity correction of correlated errors in the surface code}}, 
      author={Austin G. Fowler},
      year={2013},
      eprint={1310.0863},
      archivePrefix={arXiv},
      primaryClass={quant-ph},
      url={https://arxiv.org/abs/1310.0863}, 
}

@article{paler2023pipelined,
   title={{Pipelined correlated minimum weight perfect matching of the surface code}},
   volume={7},
   ISSN={2521-327X},
   url={http://dx.doi.org/10.22331/q-2023-12-12-1205},
   DOI={10.22331/q-2023-12-12-1205},
   journal={Quantum},
   publisher={Verein zur Forderung des Open Access Publizierens in den Quantenwissenschaften},
   author={Paler, Alexandru and Fowler, Austin G.},
   year={2023},
   month=Dec, pages={1205} }

@article{gidney2021stim,
  doi = {10.22331/q-2021-07-06-497},
  url = {https://doi.org/10.22331/q-2021-07-06-497},
  title = {{Stim: a fast stabilizer circuit simulator}},
  author = {Gidney, Craig},
  journal = {{Quantum}},
  issn = {2521-327X},
  publisher = {{Verein zur F{\"{o}}rderung des Open Access Publizierens
                in den Quantenwissenschaften}},
  volume = 5,
  pages = 497,
  month = jul,
  year = 2021
}

@article{swiadek2024enhancing,
   title={{Enhancing Dispersive Readout of Superconducting Qubits through Dynamic Control of the Dispersive Shift: Experiment and Theory}},
   volume={5},
   ISSN={2691-3399},
   url={http://dx.doi.org/10.1103/PRXQuantum.5.040326},
   DOI={10.1103/prxquantum.5.040326},
   number={4},
   journal={PRX Quantum},
   publisher={American Physical Society (APS)},
   author={Swiadek, François and Shillito, Ross and Magnard, Paul and Remm, Ants and Hellings, Christoph and Lacroix, Nathan and Ficheux, Quentin and Zanuz, Dante Colao and Norris, Graham J. and Blais, Alexandre and Krinner, Sebastian and Wallraff, Andreas},
   year={2024},
   month=Nov }

@article{sunada2024photon,
   title={{Photon-Noise-Tolerant Dispersive Readout of a Superconducting Qubit Using a Nonlinear {Purcell} Filter}},
   volume={5},
   ISSN={2691-3399},
   url={http://dx.doi.org/10.1103/PRXQuantum.5.010307},
   DOI={10.1103/prxquantum.5.010307},
   number={1},
   journal={PRX Quantum},
   publisher={American Physical Society (APS)},
   author={Sunada, Yoshiki and Yuki, Kenshi and Wang, Zhiling and Miyamura, Takeaki and Ilves, Jesper and Matsuura, Kohei and Spring, Peter A. and Tamate, Shuhei and Kono, Shingo and Nakamura, Yasunobu},
   year={2024},
   month=Jan }

@misc{beaulieu2026fast,
      title={{Fast, high-fidelity Transmon readout with intrinsic {Purcell} protection via nonperturbative cross-{Kerr} coupling}}, 
      author={Guillaume Beaulieu and Jun-Zhe Chen and Marco Scigliuzzo and Othmane Benhayoune-Khadraoui and Alex A. Chapple and Peter A. Spring and Alexandre Blais and Pasquale Scarlino},
      year={2026},
      eprint={2601.04975},
      archivePrefix={arXiv},
      primaryClass={quant-ph},
      url={https://arxiv.org/abs/2601.04975}, 
}

@article{Kliunchnikov2023,
  doi = {10.22331/q-2023-12-18-1208},
  url = {https://doi.org/10.22331/q-2023-12-18-1208},
  title = {{Shorter quantum circuits via single-qubit gate approximation}},
  author = {Kliuchnikov, Vadym and Lauter, Kristin and Minko, Romy and Paetznick, Adam and Petit, Christophe},
  journal = {{Quantum}},
  issn = {2521-327X},
  publisher = {{Verein zur F{\"{o}}rderung des Open Access Publizierens in den Quantenwissenschaften}},
  volume = {7},
  pages = {1208},
  month = dec,
  year = {2023}
}

@misc{Gidney2024StackOverflow,
    title={{Quantum Computing Stack Exchange}},
    howpublished={\url{https://quantumcomputing.stackexchange.com/questions/37267/uncertainty-of-estimates-computed-by-stim-sinter}},
    note={Accessed: 2026-09-01}
}

@article{aretz_strong_2025,
  title = {{From Strong to Weak Correlations in Breathing-Mode Kagome van der Waals Materials: ${\mathrm{Nb}}_{3}(\mathrm{F},\mathrm{Cl},\mathrm{Br},\mathrm{I}{)}_{8}$ as a Robust and Versatile Platform for Many-Body Engineering}},
  author = {Aretz, Joost and Grytsiuk, Sergii and Liu, Xiaojing and Feraco, Giovanna and Knekna, Chrystalla and Waseem, Muhammad and Dan, Zhiying and Bianchi, Marco and Hofmann, Philip and Ali, Mazhar N. and Katsnelson, Mikhail I. and Grubi{\ifmmode \check{s}\else \v{s}\fi{}}i{\ifmmode \acute{c}\else \'{c}\fi{}}-{\ifmmode \check{C}\else \v{C}\fi{}}abo, Antonija and Strand, Hugo U. R. and van Loon, Erik G. C. P. and R{\"o}sner, Malte},
  journal = {Phys. Rev. X},
  volume = {15},
  issue = {4},
  pages = {041042},
  numpages = {35},
  year = {2025},
  month = {Dec},
  publisher = {American Physical Society},
  doi = {10.1103/wr7w-nfhg},
  url = {https://link.aps.org/doi/10.1103/wr7w-nfhg}
}

@article{costa_ferromagnetism_2016,
  title = {{Ferromagnetism beyond {Lieb}'s theorem}},
  author = {Costa, Natanael C. and Mendes-Santos, Tiago and Paiva, Thereza and Santos, Raimundo R. dos and Scalettar, Richard T.},
  journal = {Phys. Rev. B},
  volume = {94},
  issue = {15},
  pages = {155107},
  numpages = {9},
  year = {2016},
  month = {Oct},
  publisher = {American Physical Society},
  doi = {10.1103/PhysRevB.94.155107},
  url = {https://link.aps.org/doi/10.1103/PhysRevB.94.155107}
}

@article{vilchez-estevez_extracting_2025,
  title = {{Extracting the spin excitation spectrum of a fermionic system using a quantum processor}},
  author = {Vilchez-Estevez, Lucia and Santos, Raul A. and Wang, Sabrina Yue and Gambetta, Filippo Maria},
  journal = {Phys. Rev. B},
  volume = {112},
  issue = {4},
  pages = {045143},
  numpages = {21},
  year = {2025},
  month = {Jul},
  publisher = {American Physical Society},
  doi = {10.1103/ydfw-k83n},
  url = {https://link.aps.org/doi/10.1103/ydfw-k83n}
}

@article{aarts_nonequilibrium_2001,
	title = {{Nonequilibrium time evolution of the spectral function in quantum field theory}},
	volume = {64},
	copyright = {http://link.aps.org/licenses/aps-default-license},
	issn = {0556-2821, 1089-4918},
	url = {https://link.aps.org/doi/10.1103/PhysRevD.64.105010},
	doi = {10.1103/PhysRevD.64.105010},
	number = {10},
	urldate = {2026-09-14},
	journal = {Physical Review D},
	author = {Aarts, Gert and Berges, Jürgen},
	month = oct,
	year = {2001},
	pages = {105010},
}

@article{poole_architecture_2025,
  title = {{Architecture for fast implementation of quantum low-density parity-check codes with optimized {Rydberg} gates}},
  author = {Poole, C. and Graham, T. M. and Perlin, M. A. and Otten, M. and Saffman, M.},
  journal = {Phys. Rev. A},
  volume = {111},
  issue = {2},
  pages = {022433},
  numpages = {11},
  year = {2025},
  month = {Feb},
  publisher = {American Physical Society},
  doi = {10.1103/PhysRevA.111.022433},
  url = {https://link.aps.org/doi/10.1103/PhysRevA.111.022433}
}

@misc{hagge_optimal_2022,
    title = {{Optimal fermionic swap networks for {Hubbard} models}},
    url = {http://arxiv.org/abs/2001.08324},
    doi = {10.48550/arXiv.2001.08324},
    urldate = {2026-04-27},
    publisher = {arXiv},
    author = {Hagge, Tobias},
    month = aug,
    year = {2022},
    note = {arXiv:2001.08324},
}

@misc{Apel2026,
      title={{Compiling the {2D} {Fermi-Hubbard} ground-state energy estimation algorithm for active volume quantum architectures}}, 
      author={Harriet Apel and Athena Caesura and Carys Harvey and Sam Heavey and Angus Kan and Jessica Lemieux and Ryan Levy and Sam Pallister and Joseph Peetz and William Pol and Sukin Sim and William A. Simon and Mark Steudtner and Gideon Uchehara},
      year={2026},
      eprint={2609.05316},
      archivePrefix={arXiv},
      primaryClass={quant-ph},
      url={https://arxiv.org/abs/2609.05316}, 
}

@article{Ismail2026_1,
  title = {Transversal Architecture for Megaquop-Scale Quantum Simulation with Neutral Atoms},
  author = {Ismail, Refaat and Chen, I-Chi and Zhao, Chen and Weiss, Ronen and Liu, Fangli and Zhou, Hengyun and Wang, Sheng-Tao and Sornborger, Andrew and Kornja{\ifmmode \check{c}\else \v{c}\fi{}}a, Milan},
  journal = {PRX Quantum},
  volume = {7},
  issue = {2},
  pages = {020343},
  numpages = {39},
  year = {2026},
  month = {May},
  publisher = {American Physical Society},
  doi = {10.1103/j2fw-ccmy},
  url = {https://link.aps.org/doi/10.1103/j2fw-ccmy}
}

@misc{Ismail2026_2,
      title={{Fast and Parallel High-Rate {STAR} Architecture for Megaquop Quantum Simulation}}, 
      author={Refaat Ismail and Milan Kornjača and Hong-Ye Hu and Nishad Maskara and Sheng-Tao Wang and Hengyun Zhou and Chen Zhao},
      year={2026},
      eprint={2606.25011},
      archivePrefix={arXiv},
      primaryClass={quant-ph},
      url={https://arxiv.org/abs/2606.25011}, 
}

@article{Bennink2017,
  title = {{Unbiased simulation of near-{Clifford} quantum circuits}},
  author = {Bennink, Ryan S. and Ferragut, Erik M. and Humble, Travis S. and Laska, Jason A. and Nutaro, James J. and Pleszkoch, Mark G. and Pooser, Raphael C.},
  journal = {Phys. Rev. A},
  volume = {95},
  issue = {6},
  pages = {062337},
  numpages = {10},
  year = {2017},
  month = {Jun},
  publisher = {American Physical Society},
  doi = {10.1103/PhysRevA.95.062337},
  url = {https://link.aps.org/doi/10.1103/PhysRevA.95.062337}
}

@article{Pashayan2015,
  title = {{Estimating Outcome Probabilities of Quantum Circuits Using Quasiprobabilities}},
  author = {Pashayan, Hakop and Wallman, Joel J. and Bartlett, Stephen D.},
  journal = {Phys. Rev. Lett.},
  volume = {115},
  issue = {7},
  pages = {070501},
  numpages = {5},
  year = {2015},
  month = {Aug},
  publisher = {American Physical Society},
  doi = {10.1103/PhysRevLett.115.070501},
  url = {https://link.aps.org/doi/10.1103/PhysRevLett.115.070501}
}

@misc{ma_paper,
      title={{Remarks on {Matsumoto} and {Amano}'s normal form for single-qubit {Clifford+T} operators}}, 
      author={Brett Giles and Peter Selinger},
      year={2019},
      eprint={1312.6584},
      archivePrefix={arXiv},
      primaryClass={quant-ph},
      url={https://arxiv.org/abs/1312.6584}, 
}

@misc{pygridsynth,
      title={{pygridsynth: A fast numerical tool for ancilla-free {Clifford+T} synthesis}}, 
      author={Shuntaro Yamamoto and Nobuyuki Yoshioka},
      year={2026},
      eprint={2604.21333},
      archivePrefix={arXiv},
      primaryClass={quant-ph},
      url={https://arxiv.org/abs/2604.21333}, 
}

@article{Temme2017,
  title = {{Error Mitigation for Short-Depth Quantum Circuits}},
  author = {Temme, Kristan and Bravyi, Sergey and Gambetta, Jay M.},
  journal = {Phys. Rev. Lett.},
  volume = {119},
  issue = {18},
  pages = {180509},
  numpages = {5},
  year = {2017},
  month = {Nov},
  publisher = {American Physical Society},
  doi = {10.1103/PhysRevLett.119.180509},
  url = {https://link.aps.org/doi/10.1103/PhysRevLett.119.180509}
}

@article{Lin2022,
  title = {{{Heisenberg}-Limited Ground-State Energy Estimation for Early Fault-Tolerant Quantum Computers}},
  author = {Lin, Lin and Tong, Yu},
  journal = {PRX Quantum},
  volume = {3},
  issue = {1},
  pages = {010318},
  numpages = {21},
  year = {2022},
  month = {Feb},
  publisher = {American Physical Society},
  doi = {10.1103/PRXQuantum.3.010318},
  url = {https://link.aps.org/doi/10.1103/PRXQuantum.3.010318}
}

@article{Wan2022,
  title = {{Randomized Quantum Algorithm for Statistical Phase Estimation}},
  author = {Wan, Kianna and Berta, Mario and Campbell, Earl T.},
  journal = {Phys. Rev. Lett.},
  volume = {129},
  issue = {3},
  pages = {030503},
  numpages = {7},
  year = {2022},
  month = {Jul},
  publisher = {American Physical Society},
  doi = {10.1103/PhysRevLett.129.030503},
  url = {https://link.aps.org/doi/10.1103/PhysRevLett.129.030503}
}

@article{Blunt2023,
  title = {{Statistical Phase Estimation and Error Mitigation on a Superconducting Quantum Processor}},
  author = {Blunt, Nick S. and Caune, Laura and Izs\'ak, R\'obert and Campbell, Earl T. and Holzmann, Nicole},
  journal = {PRX Quantum},
  volume = {4},
  issue = {4},
  pages = {040341},
  numpages = {23},
  year = {2023},
  month = {Dec},
  publisher = {American Physical Society},
  doi = {10.1103/PRXQuantum.4.040341},
  url = {https://link.aps.org/doi/10.1103/PRXQuantum.4.040341}
}

@article{childs_theory_2021,
    title = {{Theory of {Trotter} {Error} with {Commutator} {Scaling}}},
    volume = {11},
    issn = {2160-3308},
    url = {https://link.aps.org/doi/10.1103/PhysRevX.11.011020},
    doi = {10.1103/PhysRevX.11.011020},
    number = {1},
    urldate = {2025-10-27},
    journal = {Physical Review X},
    author = {Childs, Andrew M. and Su, Yuan and Tran, Minh C. and Wiebe, Nathan and Zhu, Shuchen},
    month = feb,
    year = {2021},
    pages = {011020},
}

@article{padma_2025_beyond_hubbard_pairing,
  title = {{Beyond-{Hubbard} Pairing in a Cuprate Ladder}},
  author = {Padma, Hari and Thomas, Jinu and TenHuisen, Sophia F. R. and He, Wei and Guan, Ziqiang and Li, Jiemin and Lee, Byungjune and Wang, Yu and Lee, Seng Huat and Mao, Zhiqiang and Jang, Hoyoung and Bisogni, Valentina and Pelliciari, Jonathan and Dean, Mark P. M. and Johnston, Steven and Mitrano, Matteo},
  journal = {Phys. Rev. X},
  volume = {15},
  issue = {2},
  pages = {021049},
  numpages = {10},
  year = {2025},
  month = {May},
  publisher = {American Physical Society},
  doi = {10.1103/PhysRevX.15.021049},
  url = {https://link.aps.org/doi/10.1103/PhysRevX.15.021049}
}

@article{scheie_2026_cooper_pair_localization,
  title = {{{Cooper}-pair localization in the magnetic dynamics of a cuprate ladder}},
  author = {Scheie, A. and Laurell, P. and Thomas, J. and Sharma, V. and Kolesnikov, A. I. and Granroth, G. E. and Zhang, Q. and Lake, B. and Mihalik, M. and Bewley, R. I. and Eccleston, R. S. and Akimitsu, J. and Dagotto, E. and Batista, C. D. and Alvarez, G. and Johnston, S. and Tennant, D. A.},
  journal = {Phys. Rev. B},
  volume = {114},
  issue = {14},
  pages = {144511},
  numpages = {11},
  year = {2026},
  month = {Sep},
  publisher = {American Physical Society},
  doi = {10.1103/1s7m-xkby},
  url = {https://link.aps.org/doi/10.1103/1s7m-xkby}
}

@Article{han2012fractionalized,
author={Han, Tian-Heng
and Helton, Joel S.
and Chu, Shaoyan
and Nocera, Daniel G.
and Rodriguez-Rivera, Jose A.
and Broholm, Collin
and Lee, Young S.},
title={{Fractionalized excitations in the spin-liquid state of a kagome-lattice antiferromagnet}},
journal={Nature},
year={2012},
month={Dec},
day={01},
volume={492},
number={7429},
pages={406-410},
issn={1476-4687},
doi={10.1038/nature11659},
url={https://doi.org/10.1038/nature11659}
}

@article{ortiz_2020_topological_kagome_metal,
  title = {{{$\mathrm{Cs}{\mathrm{V}}_{3}{\mathrm{Sb}}_{5}$}: A {${\mathbb{Z}}_{2}$} Topological Kagome Metal with a Superconducting Ground State}},
  author = {Ortiz, Brenden R. and Teicher, Samuel M. L. and Hu, Yong and Zuo, Julia L. and Sarte, Paul M. and Schueller, Emily C. and Abeykoon, A. M. Milinda and Krogstad, Matthew J. and Rosenkranz, Stephan and Osborn, Raymond and Seshadri, Ram and Balents, Leon and He, Junfeng and Wilson, Stephen D.},
  journal = {Phys. Rev. Lett.},
  volume = {125},
  issue = {24},
  pages = {247002},
  numpages = {6},
  year = {2020},
  month = {Dec},
  publisher = {American Physical Society},
  doi = {10.1103/PhysRevLett.125.247002},
  url = {https://link.aps.org/doi/10.1103/PhysRevLett.125.247002}
}

@article{emery_1987_theory_of_high,
  title = {{Theory of high-{${\mathrm{T}}_{\mathrm{c}}$} superconductivity in oxides}},
  author = {Emery, V. J.},
  journal = {Phys. Rev. Lett.},
  volume = {58},
  issue = {26},
  pages = {2794--2797},
  numpages = {0},
  year = {1987},
  month = {Jun},
  publisher = {American Physical Society},
  doi = {10.1103/PhysRevLett.58.2794},
  url = {https://link.aps.org/doi/10.1103/PhysRevLett.58.2794}
}

@article{slot_experimental_2017,
	title = {{Experimental realization and characterization of an electronic {Lieb} lattice}},
	volume = {13},
	issn = {1745-2481},
	url = {https://doi.org/10.1038/nphys4105},
	doi = {10.1038/nphys4105},
	number = {7},
	journal = {Nature Physics},
	author = {Slot, Marlou R. and Gardenier, Thomas S. and Jacobse, Peter H. and van Miert, Guido C. P. and Kempkes, Sander N. and Zevenhuizen, Stephan J. M. and Smith, Cristiane Morais and Vanmaekelbergh, Daniel and Swart, Ingmar},
	month = jul,
	year = {2017},
	pages = {672--676},
}

@misc{kemper_efficient_2026,
	title = {{Efficient computation of real-time correlators using {Pauli} {Propagation}}},
	url = {http://arxiv.org/abs/2607.24924},
	doi = {10.48550/arXiv.2607.24924},
	urldate = {2026-09-30},
	publisher = {arXiv},
	author = {Kemper, Alexander F. and Jha, Raghav G. and Bachhar, Arnab and Toga, Goksu C. and Perez, Mariano Guerrero and Mayhall, Nicholas J.},
	month = jul,
	year = {2026},
	note = {arXiv:2607.24924 [quant-ph]},
}

@article{villa_unraveling_2019,
  title = {{Unraveling the excitation spectrum of many-body systems from quantum quenches}},
  author = {Villa, Louis and Despres, Julien and Sanchez-Palencia, Laurent},
  journal = {Phys. Rev. A},
  volume = {100},
  issue = {6},
  pages = {063632},
  numpages = {11},
  year = {2019},
  month = {Dec},
  publisher = {American Physical Society},
  doi = {10.1103/PhysRevA.100.063632},
  url = {https://link.aps.org/doi/10.1103/PhysRevA.100.063632}
}

@article{villa_local_2020,
  title = {{Local quench spectroscopy of many-body quantum systems}},
  author = {Villa, L. and Despres, J. and Thomson, S. J. and Sanchez-Palencia, L.},
  journal = {Phys. Rev. A},
  volume = {102},
  issue = {3},
  pages = {033337},
  numpages = {12},
  year = {2020},
  month = {Sep},
  publisher = {American Physical Society},
  doi = {10.1103/PhysRevA.102.033337},
  url = {https://link.aps.org/doi/10.1103/PhysRevA.102.033337}
}

@article{kagome_metals,
  title = {{Kagome metals}},
  author = {Di Sante, Domenico and Neupert, Titus and Sangiovanni, Giorgio and Thomale, Ronny and Comin, Riccardo and Checkelsky, Joseph G. and Zeljkovic, Ilija and Wilson, Stephen D.},
  journal = {Rev. Mod. Phys.},
  volume = {98},
  issue = {1},
  pages = {015002},
  numpages = {53},
  year = {2026},
  month = {Feb},
  publisher = {American Physical Society},
  doi = {10.1103/1g9n-wm38},
  url = {https://link.aps.org/doi/10.1103/1g9n-wm38}
}

@Article{Wang2023,
author={Wang, Yaojia
and Wu, Heng
and McCandless, Gregory T.
and Chan, Julia Y.
and Ali, Mazhar N.},
title={{Quantum states and intertwining phases in kagome materials}},
journal={Nature Reviews Physics},
year={2023},
month={Nov},
day={01},
volume={5},
number={11},
pages={635-658},
issn={2522-5820},
doi={10.1038/s42254-023-00635-7},
url={https://doi.org/10.1038/s42254-023-00635-7}
}

@article{Babbush_loose_trotter_bounds,
  title = {{Chemical basis of {Trotter-Suzuki} errors in quantum chemistry simulation}},
  author = {Babbush, Ryan and McClean, Jarrod and Wecker, Dave and Aspuru-Guzik, Al{\'a}n and Wiebe, Nathan},
  journal = {Phys. Rev. A},
  volume = {91},
  issue = {2},
  pages = {022311},
  numpages = {16},
  year = {2015},
  month = {Feb},
  publisher = {American Physical Society},
  doi = {10.1103/PhysRevA.91.022311},
  url = {https://link.aps.org/doi/10.1103/PhysRevA.91.022311}
}

@misc{matsumoto2008representation,
      title={{Representation of Quantum Circuits with {Clifford} and $\pi/8$ Gates}}, 
      author={Ken Matsumoto and Kazuyuki Amano},
      year={2008},
      eprint={0806.3834},
      archivePrefix={arXiv},
      primaryClass={quant-ph},
      url={https://arxiv.org/abs/0806.3834}, 
}

@article{
Heyl_local_observable_trotter_error,
author = {Markus Heyl  and Philipp Hauke  and Peter Zoller },
title = {{Quantum localization bounds {Trotter} errors in digital quantum simulation}},
journal = {Science Advances},
volume = {5},
number = {4},
pages = {eaau8342},
year = {2019},
doi = {10.1126/sciadv.aau8342}}

@misc{li2025softhighperformancesimulatoruniversal,
      title={{{SOFT}: a high-performance simulator for universal fault-tolerant quantum circuits}}, 
      author={Riling Li and Keli Zheng and Yiming Zhang and Huazhe Lou and Shenggang Ying and Ke Liu and Xiaoming Sun},
      year={2025},
      eprint={2512.23037},
      archivePrefix={arXiv},
      primaryClass={quant-ph},
      url={https://arxiv.org/abs/2512.23037}, 
}

@misc{chase2026clifftfastexactsimulation,
      title={{Clifft: Fast Exact Simulation of Near-{Clifford} Quantum Circuits}}, 
      author={Bradley A. Chase and Farrokh Labib},
      year={2026},
      eprint={2604.27058},
      archivePrefix={arXiv},
      primaryClass={quant-ph},
      url={https://arxiv.org/abs/2604.27058}, 
}

@misc{Chan2026,
      title={{Diagnosing and Restoring the Degraded Fault Distance of Magic State Cultivation}}, 
      author={Tim Chan and Armands Strikis and Zhu Sun and Zhenyu Cai},
      year={2026},
      eprint={2609.17706},
      archivePrefix={arXiv},
      primaryClass={quant-ph},
      url={https://arxiv.org/abs/2609.17706}, 
}

@misc{Wan2026,
      title={{Exact logical error rates for magic state cultivation}}, 
      author={Kwok Ho Wan and Ainhoa Zapirain},
      year={2026},
      eprint={2609.18922},
      archivePrefix={arXiv},
      primaryClass={quant-ph},
      url={https://arxiv.org/abs/2609.18922}, 
}

@article{Lieb89,
  title = {{Two theorems on the {Hubbard} model}},
  author = {Lieb, Elliott H.},
  journal = {Phys. Rev. Lett.},
  volume = {62},
  issue = {10},
  pages = {1201--1204},
  numpages = {0},
  year = {1989},
  month = {Mar},
  publisher = {American Physical Society},
  doi = {10.1103/PhysRevLett.62.1201},
  url = {https://link.aps.org/doi/10.1103/PhysRevLett.62.1201}
}

@article{Sun21,
  title = {{Metal-insulator transition and intermediate phases in the kagome lattice {Hubbard} model}},
  author = {Sun, Rong-Yang and Zhu, Zheng},
  journal = {Phys. Rev. B},
  volume = {104},
  issue = {12},
  pages = {L121118},
  numpages = {6},
  year = {2021},
  month = {Sep},
  publisher = {American Physical Society},
  doi = {10.1103/PhysRevB.104.L121118},
  url = {https://link.aps.org/doi/10.1103/PhysRevB.104.L121118}
}

@article{McEwen2023relaxinghardware,
  doi = {10.22331/q-2023-11-07-1172},
  url = {https://doi.org/10.22331/q-2023-11-07-1172},
  title = {{Relaxing {H}ardware {R}equirements for {S}urface {C}ode {C}ircuits using {T}ime-dynamics}},
  author = {McEwen, Matt and Bacon, Dave and Gidney, Craig},
  journal = {{Quantum}},
  issn = {2521-327X},
  publisher = {{Verein zur F{\"{o}}rderung des Open Access Publizierens in den Quantenwissenschaften}},
  volume = {7},
  pages = {1172},
  month = nov,
  year = {2023}
}

@article{Pinelis1994,
 ISSN = {00911798, 2168894X},
 URL = {http://www.jstor.org/stable/2244912},
 author = {Iosif Pinelis},
 journal = {The Annals of Probability},
 number = {4},
 pages = {1679--1706},
 publisher = {Institute of Mathematical Statistics},
 title = {{Optimum Bounds for the Distributions of Martingales in {Banach} Spaces}},
 urldate = {2026-10-12},
 volume = {22},
 year = {1994}
}

@article{Bonet-Monroig2018,
  title = {{Low-cost error mitigation by symmetry verification}},
  author = {Bonet-Monroig, X. and Sagastizabal, R. and Singh, M. and O'Brien, T. E.},
  journal = {Phys. Rev. A},
  volume = {98},
  issue = {6},
  pages = {062339},
  numpages = {10},
  year = {2018},
  month = {Dec},
  publisher = {American Physical Society},
  doi = {10.1103/PhysRevA.98.062339},
  url = {https://link.aps.org/doi/10.1103/PhysRevA.98.062339}
}

@Article{Stanisic2022,
author={Stanisic, Stasja
and Bosse, Jan Lukas
and Gambetta, Filippo Maria
and Santos, Raul A.
and Mruczkiewicz, Wojciech
and O'Brien, Thomas E.
and Ostby, Eric
and Montanaro, Ashley},
title={{Observing ground-state properties of the {Fermi-Hubbard} model using a scalable algorithm on a quantum computer}},
journal={Nature Communications},
year={2022},
month={Oct},
day={11},
volume={13},
number={1},
pages={5743},
issn={2041-1723},
doi={10.1038/s41467-022-33335-4},
url={https://doi.org/10.1038/s41467-022-33335-4}
}

@misc{Mehta2026,
      title={{The practical cost of magic state cultivation}}, 
      author={Rohan Mehta and Varun Menon and Hengyun Zhou and Mikhail D. Lukin and J. Pablo Bonilla Ataides},
      year={2026},
      eprint={2610.06828},
      archivePrefix={arXiv},
      primaryClass={quant-ph},
      url={https://arxiv.org/abs/2610.06828}, 
}

@misc{Yoder2025,
      title={{Tour de gross: A modular quantum computer based on bivariate bicycle codes}}, 
      author={Theodore J. Yoder and Eddie Schoute and Patrick Rall and Emily Pritchett and Jay M. Gambetta and Andrew W. Cross and Malcolm Carroll and Michael E. Beverland},
      year={2025},
      eprint={2506.03094},
      archivePrefix={arXiv},
      primaryClass={quant-ph},
      url={https://arxiv.org/abs/2506.03094}, 
}

@misc{Tripier2026,
      title={{Fault-Tolerant Quantum Computing with Trapped Ions: The Walking Cat Architecture}}, 
      author={Felix Tripier and Woo Chang Chung and Jacob Young and Safwan Alam and Bryce Bjork and Aharon Brodutch and Finn Lasse Buessen and Nolan J. Coble and Thomas Dellaert and Dmitri Maslov and Martin Roetteler and Edwin Tham and Mark Webster and Min Ye and John Gamble and Andrii Maksymov and J. P. Marceaux and Nicolas Delfosse},
      year={2026},
      eprint={2604.19481},
      archivePrefix={arXiv},
      primaryClass={quant-ph},
      url={https://arxiv.org/abs/2604.19481}, 
}

@misc{Webster2026,
      title={{The Pinnacle Architecture: Reducing the cost of breaking RSA-2048 to 100 000 physical qubits using quantum LDPC codes}}, 
      author={Paul Webster and Lucas Berent and Omprakash Chandra and Evan T. Hockings and Nouédyn Baspin and Felix Thomsen and Samuel C. Smith and Lawrence Z. Cohen},
      year={2026},
      eprint={2602.11457},
      archivePrefix={arXiv},
      primaryClass={quant-ph},
      url={https://arxiv.org/abs/2602.11457}, 
}

@Article{Georges2025,
author={Georges, Timothy N.
and Bothe, Marius
and S{\"u}nderhauf, Christoph
and Berntson, Bjorn K.
and Izs{\'a}k, R{\'o}bert
and Ivanov, Aleksei V.},
title={{Quantum simulations of chemistry in first quantization with any basis set}},
journal={npj Quantum Information},
year={2025},
month={Apr},
day={01},
volume={11},
number={1},
pages={55},
issn={2056-6387},
doi={10.1038/s41534-025-00987-1},
url={https://doi.org/10.1038/s41534-025-00987-1}
}

@article{Low2025,
  title = {{Fast Quantum Simulation of Electronic Structure by Spectral Amplification}},
  author = {Low, Guang Hao and King, Robbie and Berry, Dominic W. and Han, Qiushi and DePrince, A. Eugene and White, Alec F. and Babbush, Ryan and Somma, Rolando D. and Rubin, Nicholas C.},
  journal = {Phys. Rev. X},
  volume = {15},
  issue = {4},
  pages = {041016},
  numpages = {41},
  year = {2025},
  month = {Oct},
  publisher = {American Physical Society},
  doi = {10.1103/pb2g-j9cw},
  url = {https://link.aps.org/doi/10.1103/pb2g-j9cw}
}

@article{Geher2026,
author={Geh{\'e}r, Gy{\"o}rgy P.
and Byfield, David
and Ruban, Archibald},
title={{Directional codes for quantum LDPC error correction on square and hexagonal grid architectures}},
journal={Nature Communications},
year={2026},
month={Nov},
day={25},
issn={2041-1723},
doi={10.1038/s41467-026-79572-9},
url={https://doi.org/10.1038/s41467-026-79572-9}
}

@misc{Nixon2026,
      title={{Vine Codes: Low-Overhead Quantum LDPC Codes on a Planar Square Grid}}, 
      author={Georgia M. Nixon and Campbell K. McLauchlan and Charles C. L. van Rest},
      year={2026},
      eprint={2606.20263},
      archivePrefix={arXiv},
      primaryClass={quant-ph},
      url={https://arxiv.org/abs/2606.20263}, 
}

@misc{Gu2026,
      title={{Nearest-neighbour gates are all you need: High-rate quantum low-density parity-check codes on a planar grid}}, 
      author={Boren Gu and Tamas Noszko and Vincent Steffan and Jens Niklas Eberhardt and Joschka Roffe and Jens Eisert and Stergios Koutsioumpas},
      year={2026},
      eprint={2606.19482},
      archivePrefix={arXiv},
      primaryClass={quant-ph},
      url={https://arxiv.org/abs/2606.19482}, 
}

@article{Vovrosh2021,
  title = {Simple mitigation of global depolarizing errors in quantum simulations},
  author = {Vovrosh, Joseph and Khosla, Kiran E. and Greenaway, Sean and Self, Christopher and Kim, M. S. and Knolle, Johannes},
  journal = {Phys. Rev. E},
  volume = {104},
  issue = {3},
  pages = {035309},
  numpages = {8},
  year = {2021},
  month = {Sep},
  publisher = {American Physical Society},
  doi = {10.1103/PhysRevE.104.035309},
  url = {https://link.aps.org/doi/10.1103/PhysRevE.104.035309}
}

@article{Farrell2024,
  title = {Scalable Circuits for Preparing Ground States on Digital Quantum Computers: The Schwinger Model Vacuum on 100 Qubits},
  author = {Farrell, Roland C. and Illa, Marc and Ciavarella, Anthony N. and Savage, Martin J.},
  journal = {PRX Quantum},
  volume = {5},
  issue = {2},
  pages = {020315},
  numpages = {32},
  year = {2024},
  month = {Apr},
  publisher = {American Physical Society},
  doi = {10.1103/PRXQuantum.5.020315},
  url = {https://link.aps.org/doi/10.1103/PRXQuantum.5.020315}
}

@article{Wallman2016,
  title = {Noise tailoring for scalable quantum computation via randomized compiling},
  author = {Wallman, Joel J. and Emerson, Joseph},
  journal = {Phys. Rev. A},
  volume = {94},
  issue = {5},
  pages = {052325},
  numpages = {9},
  year = {2016},
  month = {Nov},
  publisher = {American Physical Society},
  doi = {10.1103/PhysRevA.94.052325},
  url = {https://link.aps.org/doi/10.1103/PhysRevA.94.052325}
}

@Article{vandenBerg2023,
author={van den Berg, Ewout
and Minev, Zlatko K.
and Kandala, Abhinav
and Temme, Kristan},
title={Probabilistic error cancellation with sparse Pauli--Lindblad models on noisy quantum processors},
journal={Nature Physics},
year={2023},
month={Aug},
day={01},
volume={19},
number={8},
pages={1116-1121},
issn={1745-2481},
doi={10.1038/s41567-023-02042-2},
url={https://doi.org/10.1038/s41567-023-02042-2}
}

@article{Piveteau2021,
  title = {Error Mitigation for Universal Gates on Encoded Qubits},
  author = {Piveteau, Christophe and Sutter, David and Bravyi, Sergey and Gambetta, Jay M. and Temme, Kristan},
  journal = {Phys. Rev. Lett.},
  volume = {127},
  issue = {20},
  pages = {200505},
  numpages = {6},
  year = {2021},
  month = {Nov},
  publisher = {American Physical Society},
  doi = {10.1103/PhysRevLett.127.200505},
  url = {https://link.aps.org/doi/10.1103/PhysRevLett.127.200505}
}

@article{Jain2023,
  title = {Improved quantum error correction with randomized compiling},
  author = {Jain, Aditya and Iyer, Pavithran and Bartlett, Stephen D. and Emerson, Joseph},
  journal = {Phys. Rev. Res.},
  volume = {5},
  issue = {3},
  pages = {033049},
  numpages = {13},
  year = {2023},
  month = {Jul},
  publisher = {American Physical Society},
  doi = {10.1103/PhysRevResearch.5.033049},
  url = {https://link.aps.org/doi/10.1103/PhysRevResearch.5.033049}
}

@article{Kueng2016,
  title = {Comparing Experiments to the Fault-Tolerance Threshold},
  author = {Kueng, Richard and Long, David M. and Doherty, Andrew C. and Flammia, Steven T.},
  journal = {Phys. Rev. Lett.},
  volume = {117},
  issue = {17},
  pages = {170502},
  numpages = {6},
  year = {2016},
  month = {Oct},
  publisher = {American Physical Society},
  doi = {10.1103/PhysRevLett.117.170502},
  url = {https://link.aps.org/doi/10.1103/PhysRevLett.117.170502}
}

@misc{Zeng2025,
    author = "Zeng, Pei and Zheng, Guo and Xu, Qian and Jiang, Liang",
    title = "{Error-structure-tailored early fault-tolerant quantum computing}",
    eprint = "2511.19983",
    archivePrefix = "arXiv",
    primaryClass = "quant-ph",
    month = "11",
    year = "2025"
}

@article{Huang2019,
  title = {Performance of quantum error correction with coherent errors},
  author = {Huang, Eric and Doherty, Andrew C. and Flammia, Steven},
  journal = {Phys. Rev. A},
  volume = {99},
  issue = {2},
  pages = {022313},
  numpages = {17},
  year = {2019},
  month = {Feb},
  publisher = {American Physical Society},
  doi = {10.1103/PhysRevA.99.022313},
  url = {https://link.aps.org/doi/10.1103/PhysRevA.99.022313}
}

@article{Surti2025,
  title = {Efficient Simulation of Logical Magic State Preparation Protocols},
  author = {Surti, Samyak and Daguerre, Lucas and Kim, Isaac H.},
  journal = {PRX Quantum},
  volume = {7},
  issue = {2},
  pages = {020329},
  numpages = {32},
  year = {2026},
  month = {May},
  publisher = {American Physical Society},
  doi = {10.1103/fby6-xjbm},
  url = {https://link.aps.org/doi/10.1103/fby6-xjbm}
}

@misc{Haenel2026,
    author = "Haenel, Rafael and Luo, Xiuzhe and Zhao, Chen",
    title = "{Tsim: Fast Universal Simulator for Quantum Error Correction}",
    eprint = "2604.01059",
    archivePrefix = "arXiv",
    primaryClass = "quant-ph",
    month = "4",
    year = "2026"
}

@misc{Umbrarescu2026,
    author = "Umbrarescu, George and Amaro, David",
    title = "{SyQMA: A memory-efficient, symbolic and exact universal simulator for quantum error correction}",
    eprint = "2604.15043",
    archivePrefix = "arXiv",
    primaryClass = "quant-ph",
    month = "4",
    year = "2026"
}

@misc{Cercelescu2026,
    author = "Cercelescu, Serban and Koch, Mark and Pesah, Arthur",
    title = "{Polynomial-time simulation of non-Clifford quantum error correction}",
    eprint = "2610.06811",
    archivePrefix = "arXiv",
    primaryClass = "quant-ph",
    month = "10",
    year = "2026"
}

@misc{Takahashi2026,
    author = "Takahashi, Shunta and Fujii, Keisuke",
    title = "{Efficient Fidelity Estimation of General Quantum Resource States via Clifford Circuits}",
    eprint = "2610.05222",
    archivePrefix = "arXiv",
    primaryClass = "quant-ph",
    reportNumber = "KUNS-3119",
    month = "10",
    year = "2026"
}

@article{Bravyi2015,
  title = {{Trading Classical and Quantum Computational Resources}},
  author = {Bravyi, Sergey and Smith, Graeme and Smolin, John A.},
  journal = {Phys. Rev. X},
  volume = {6},
  issue = {2},
  pages = {021043},
  numpages = {14},
  year = {2016},
  month = {Jun},
  publisher = {American Physical Society},
  doi = {10.1103/PhysRevX.6.021043},
  url = {https://link.aps.org/doi/10.1103/PhysRevX.6.021043}
}

@article{Bravyi2016,
  title = {{Improved Classical Simulation of Quantum Circuits Dominated by Clifford Gates}},
  author = {Bravyi, Sergey and Gosset, David},
  journal = {Phys. Rev. Lett.},
  volume = {116},
  issue = {25},
  pages = {250501},
  numpages = {5},
  year = {2016},
  month = {Jun},
  publisher = {American Physical Society},
  doi = {10.1103/PhysRevLett.116.250501},
  url = {https://link.aps.org/doi/10.1103/PhysRevLett.116.250501}
}

@article{Bravyi2019,
  doi = {10.22331/q-2019-09-02-181},
  url = {https://doi.org/10.22331/q-2019-09-02-181},
  title = {Simulation of quantum circuits by low-rank stabilizer decompositions},
  author = {Bravyi, Sergey and Browne, Dan and Calpin, Padraic and Campbell, Earl and Gosset, David and Howard, Mark},
  journal = {{Quantum}},
  issn = {2521-327X},
  publisher = {{Verein zur F{\"{o}}rderung des Open Access Publizierens in den Quantenwissenschaften}},
  volume = {3},
  pages = {181},
  month = sep,
  year = {2019}
}

@article{Masot-Llima2024,
  title = {Stabilizer Tensor Networks: Universal Quantum Simulator on a Basis of Stabilizer States},
  author = {Masot-Llima, Sergi and Garcia-Saez, Artur},
  journal = {Phys. Rev. Lett.},
  volume = {133},
  issue = {23},
  pages = {230601},
  numpages = {6},
  year = {2024},
  month = {Dec},
  publisher = {American Physical Society},
  doi = {10.1103/PhysRevLett.133.230601},
  url = {https://link.aps.org/doi/10.1103/PhysRevLett.133.230601}
}

@article{Demkowicz-Dobrzanski2012,
    author = "Demkowicz-Dobrza{\'n}ski, Rafa{\l} and Ko{\l}ody{\'n}ski, Jan and Gu{\c{t}}{\u{a}}, M{\u{a}}d{\u{a}}lin",
    title = "{The elusive Heisenberg limit in quantum-enhanced metrology}",
    eprint = "1201.3940",
    archivePrefix = "arXiv",
    primaryClass = "quant-ph",
    doi = "10.1038/ncomms2067",
    journal = "Nature Commun.",
    volume = "3",
    number = "1",
    pages = "1063",
    year = "2012"
}

@article{Higgins2007,
    author = "Higgins, B. L. and Berry, D. W. and Bartlett, S. D. and Wiseman, H. M. and Pryde, G. J.",
    title = "{Entanglement-free Heisenberg-limited phase estimation}",
    eprint = "0709.2996",
    archivePrefix = "arXiv",
    primaryClass = "quant-ph",
    doi = "10.1038/nature06257",
    journal = "Nature",
    volume = "450",
    pages = "393--396",
    year = "2007"
}

@techreport{Kitaev1995,
  title         = {Quantum measurements and the {A}belian {S}tabilizer {P}roblem},
  author        = {Kitaev, A. Yu.},
  institution   = {Electronic Colloquium on Computational Complexity},
  number        = {TR96-003},
  year          = {1996},
  eprint        = {quant-ph/9511026},
  archivePrefix = {arXiv}
}

@article{Baez2020,
author = {Maria Laura Baez  and Marcel Goihl  and Jonas Haferkamp  and Juani Bermejo-Vega  and Marek Gluza  and Jens Eisert },
title = {{Dynamical structure factors of dynamical quantum simulators}},
journal = {Proceedings of the National Academy of Sciences},
volume = {117},
number = {42},
pages = {26123-26134},
year = {2020},
doi = {10.1073/pnas.2006103117},
URL = {https://www.pnas.org/doi/abs/10.1073/pnas.2006103117},
eprint = {https://www.pnas.org/doi/pdf/10.1073/pnas.2006103117}}

@article{Knap2013,
  title = {Probing Real-Space and Time-Resolved Correlation Functions with Many-Body Ramsey Interferometry},
  author = {Knap, Michael and Kantian, Adrian and Giamarchi, Thierry and Bloch, Immanuel and Lukin, Mikhail D. and Demler, Eugene},
  journal = {Phys. Rev. Lett.},
  volume = {111},
  issue = {14},
  pages = {147205},
  numpages = {6},
  year = {2013},
  month = {Oct},
  publisher = {American Physical Society},
  doi = {10.1103/PhysRevLett.111.147205},
  url = {https://link.aps.org/doi/10.1103/PhysRevLett.111.147205}
}

@article{Yoshimura2016,
  title = {Measuring nonequilibrium retarded spin-spin Green's functions in an ion-trap-based quantum simulator},
  author = {Yoshimura, Bryce T. and Freericks, J. K.},
  journal = {Phys. Rev. A},
  volume = {93},
  issue = {5},
  pages = {052314},
  numpages = {11},
  year = {2016},
  month = {May},
  publisher = {American Physical Society},
  doi = {10.1103/PhysRevA.93.052314},
  url = {https://link.aps.org/doi/10.1103/PhysRevA.93.052314}
}

@misc{Lavasani2026,
      title={Quantum error correction well below the color code threshold}, 
      author={Ali Lavasani and Adam P. Reed and Joan M. Dreiling and Annie J. Park and Michael Mills and Matteo Puviani and Gabriel Matos and M. S. Allman and Noah Berthusen and Natalie C. Brown and Victor E. Colussi and Brian Estey and Cameron Foltz and Daniel Gresh and Reza Haghshenas and Akhil Isanaka and Nikhil Kotibhaskar and Anthony Ransford and Ciarán Ryan-Anderson and Brian Neyenhuis and David Hayes and Tzvetan S. Metodi and Andrew C. Potter},
      year={2026},
      eprint={2610.24077},
      archivePrefix={arXiv},
      primaryClass={quant-ph},
      url={https://arxiv.org/abs/2610.24077}, 
}

@article{mitrano_2024_explorting_quantum_materials,
  title = {Exploring Quantum Materials with Resonant Inelastic X-Ray Scattering},
  author = {Mitrano, M. and Johnston, S. and Kim, Young-June and Dean, M. P. M.},
  journal = {Phys. Rev. X},
  volume = {14},
  issue = {4},
  pages = {040501},
  numpages = {32},
  year = {2024},
  month = {Dec},
  publisher = {American Physical Society},
  doi = {10.1103/PhysRevX.14.040501},
  url = {https://link.aps.org/doi/10.1103/PhysRevX.14.040501}
}

@article{de_groot_resonant_2024,
	title = {Resonant inelastic {X}-ray scattering},
	volume = {4},
	issn = {2662-8449},
	url = {https://doi.org/10.1038/s43586-024-00322-6},
	doi = {10.1038/s43586-024-00322-6},
	number = {1},
	journal = {Nature Reviews Methods Primers},
	author = {de Groot, Frank M. F. and Haverkort, Maurits W. and Elnaggar, Hebatalla and Juhin, Amélie and Zhou, Ke-Jin and Glatzel, Pieter},
	month = jul,
	year = {2024},
	pages = {45},
}

@book{boothroyd_principles_2020,
	title = {Principles of {Neutron} {Scattering} from {Condensed} {Matter}},
	isbn = {978-0-19-886231-4},
	url = {https://doi.org/10.1093/oso/9780198862314.001.0001},
	doi = {10.1093/oso/9780198862314.001.0001},
	publisher = {Oxford University Press},
	author = {Boothroyd, Andrew T},
	month = jul,
	year = {2020},
}

@book{Sachdev_2011, place={Cambridge}, edition={2}, title={{Quantum Phase Transitions}}, publisher={Cambridge University Press}, author={Sachdev, Subir}, year={2011}}

@misc{blunt2025,
      title={{A Monte Carlo approach to bound Trotter error}}, 
      author={Nick S. Blunt and Aleksei V. Ivanov and Andreas Juul Bay-Smidt},
      year={2025},
      eprint={2510.11621},
      archivePrefix={arXiv},
      primaryClass={quant-ph},
      url={https://arxiv.org/abs/2510.11621}, 
}

@article{Zhai2023,
    author = {Zhai, Huanchen and Larsson, Henrik R. and Lee, Seunghoon and Cui, Zhi-Hao and Zhu, Tianyu and Sun, Chong and Peng, Linqing and Peng, Ruojing and Liao, Ke and Tölle, Johannes and Yang, Junjie and Li, Shuoxue and Chan, Garnet Kin-Lic},
    title = {Block2: A comprehensive open source framework to develop and apply state-of-the-art DMRG algorithms in electronic structure and beyond},
    journal = {The Journal of Chemical Physics},
    volume = {159},
    number = {23},
    pages = {234801},
    year = {2023},
    month = {12},
    issn = {0021-9606},
    doi = {10.1063/5.0180424},
    url = {https://doi.org/10.1063/5.0180424},
}

@article{geher2024tangling,
  title={Tangling schedules eases hardware connectivity requirements for quantum error correction},
  author={Geh{\'e}r, Gy{\"o}rgy P and Crawford, Ophelia and Campbell, Earl T},
  journal={PRX Quantum},
  volume={5},
  number={1},
  pages={010348},
  year={2024},
  publisher={APS}
}

@unpublished{Erickson2027,
  title  = {{[Working title: Fast Estimations of Tomography in Noisy Quantum Error Correcting Circuits with Linear Combinations of Clifford Channels]}},
  author = {Erickson, Christopher and S{\"u}nderhauf, Christoph},
  note   = {In preparation},
  year   = {2026}
}

@misc{Oratomic_Shors,
      title={{Shor's algorithm is possible with as few as 10,000 reconfigurable atomic qubits}}, 
      author={Madelyn Cain and Qian Xu and Robbie King and Lewis R. B. Picard and Harry Levine and Manuel Endres and John Preskill and Hsin-Yuan Huang and Dolev Bluvstein},
      year={2026},
      eprint={2603.28627},
      archivePrefix={arXiv},
      primaryClass={quant-ph},
      url={https://arxiv.org/abs/2603.28627}, 
}

@article{Blunt2024,
  title = {{Compilation of a simple chemistry application to quantum error correction primitives}},
  author = {Blunt, Nick S. and Geh{\'e}r, Gy{\"o}rgy P. and Moylett, Alexandra E.},
  journal = {Phys. Rev. Res.},
  volume = {6},
  issue = {1},
  pages = {013325},
  numpages = {20},
  year = {2024},
  month = {Mar},
  publisher = {American Physical Society},
  doi = {10.1103/PhysRevResearch.6.013325},
  url = {https://link.aps.org/doi/10.1103/PhysRevResearch.6.013325}
}

@Article{Dalzell2024,
author={Dalzell, Alexander M.
and Hunter-Jones, Nicholas
and Brand{\~a}o, Fernando G. S. L.},
title={{Random Quantum Circuits Transform Local Noise into Global White Noise}},
journal={Communications in Mathematical Physics},
year={2024},
month={Mar},
day={12},
volume={405},
number={3},
pages={78},
issn={1432-0916},
doi={10.1007/s00220-024-04958-z},
url={https://doi.org/10.1007/s00220-024-04958-z}
}

@article{Foldager2024,
doi = {10.1088/1751-8121/ad0ac7},
url = {https://doi.org/10.1088/1751-8121/ad0ac7},
year = {2023},
month = {dec},
publisher = {IOP Publishing},
volume = {57},
number = {1},
pages = {015306},
author = {Foldager, Jonathan and Koczor, Bálint},
title = {Can shallow quantum circuits scramble local noise into global white noise?},
journal = {Journal of Physics A: Mathematical and Theoretical}
}

@article{Nahum2018,
  title = {{Operator Spreading in Random Unitary Circuits}},
  author = {Nahum, Adam and Vijay, Sagar and Haah, Jeongwan},
  journal = {Phys. Rev. X},
  volume = {8},
  issue = {2},
  pages = {021014},
  numpages = {30},
  year = {2018},
  month = {Apr},
  publisher = {American Physical Society},
  doi = {10.1103/PhysRevX.8.021014},
  url = {https://link.aps.org/doi/10.1103/PhysRevX.8.021014}
}

@article{Endo2018,
  title = {Practical Quantum Error Mitigation for Near-Future Applications},
  author = {Endo, Suguru and Benjamin, Simon C. and Li, Ying},
  journal = {Phys. Rev. X},
  volume = {8},
  issue = {3},
  pages = {031027},
  numpages = {21},
  year = {2018},
  month = {Jul},
  publisher = {American Physical Society},
  doi = {10.1103/PhysRevX.8.031027},
  url = {https://link.aps.org/doi/10.1103/PhysRevX.8.031027}
}

@article{Urbanek2021,
  title = {Mitigating Depolarizing Noise on Quantum Computers with Noise-Estimation Circuits},
  author = {Urbanek, Miroslav and Nachman, Benjamin and Pascuzzi, Vincent R. and He, Andre and Bauer, Christian W. and de Jong, Wibe A.},
  journal = {Phys. Rev. Lett.},
  volume = {127},
  issue = {27},
  pages = {270502},
  numpages = {6},
  year = {2021},
  month = {Dec},
  publisher = {American Physical Society},
  doi = {10.1103/PhysRevLett.127.270502},
  url = {https://link.aps.org/doi/10.1103/PhysRevLett.127.270502}
}

@misc{shtanko2026,
      title={{Measure Now, Mitigate Later: Virtual Error Cancellation for Logical Quantum Circuits}}, 
      author={Oles Shtanko and Saswata Roy and Shashwat Kumar and Zlatko Minev},
      year={2026},
      eprint={2610.12400},
      archivePrefix={arXiv},
      primaryClass={quant-ph},
      url={https://arxiv.org/abs/2610.12400}, 
}

@misc{shutman2026,
      title={Error Mitigation From a Single Unmodified Error-Corrected Circuit}, 
      author={Maor Shutman and Dorit Aharonov and Itai Arad and Yosi Atia and Eyal Bairey and Zvika Brakerski and Barak A. Katzir and Netanel H. Lindner and Omri Golan and Ilya Gurwich},
      year={2026},
      eprint={2610.06792},
      archivePrefix={arXiv},
      primaryClass={quant-ph},
      url={https://arxiv.org/abs/2610.06792}, 
}

@misc{dinca2026,
      title={Error mitigation for logical circuits using decoder confidence}, 
      author={Maria Dincă and Tim Chan and Simon C. Benjamin},
      year={2026},
      eprint={2512.15689},
      archivePrefix={arXiv},
      primaryClass={quant-ph},
      url={https://arxiv.org/abs/2512.15689}, 
}

@misc{tsubouchi2026,
      title={Quantum advantages for syndrome-aware noisy logical observable estimation}, 
      author={Kento Tsubouchi and Hyukgun Kwon and Liang Jiang and Nobuyuki Yoshioka},
      year={2026},
      eprint={2603.05145},
      archivePrefix={arXiv},
      primaryClass={quant-ph},
      url={https://arxiv.org/abs/2603.05145}, 
}

@misc{ivanov2026,
      title={{Quantum Simulation of First-Quantised Chemistry with a Sum-of-Squares Representation}}, 
      author={Aleksei V. Ivanov and Christoph Sünderhauf and Jon Kunjummen and Nick S. Blunt and Róbert Izsák},
      year={2026},
      eprint={2611.24885},
      archivePrefix={arXiv},
      primaryClass={quant-ph},
      url={https://arxiv.org/abs/2611.24885}, 
}

\appendix
\crefalias{section}{appendix}
\crefalias{subsection}{subappendix}

\section{Calculating $\Lambda$ and distances for MegaQuOp error rates}
\label{sec:lambda_calc}

\subsection{Quantum memory simulations}
\label{sec:quantum_memory_simualtions}

We estimate the code distances required to reach MegaQuOp logical error rates through a series of simulated surface code quantum memory experiments under circuit-level noise, sampled in \textsc{Stim} \cite{gidney2021stim}. To ensure alignment of our results with those previously cited in the literature, we base our simulations on two widely used noise models -- the superconducting-inspired SI1000 noise model and the uniform-depolarising SD6 noise model \cite{gidney2021fault}. SI1000 reflects the capabilities of current hardware \cite{google2023, alam2025_2} and allows us to project existing device performance to future low-error rate regimes, while SD6 is a widely used benchmark in the field and allows for comparison of our results against prior works \cite{gidney2024magic, gidney2025factor, low2026denser}. In both models, the physical error rate of each circuit component is scaled by a global parameter $p$, with the scaling coefficients defined in \cref{tab:noise_parametrisations}. The SI1000 model is dominated by measurement and reset noise, while the SD6 model penalises every circuit operation equally. We note that the SI1000 noise model assumes a native $\textrm{CZ}$-gate as its two-qubit entangling gate, while the SD6 model assumes $\textrm{CX}$ as its native two-qubit operation, leading to fewer stages in its syndrome extraction cycle \cite{gidney2021fault}. In addition to the two standard noise models described above, we evaluate qubit footprints under a noise model similar to SI1000, denoted here as Optimistic noise, where we assume that measurement fidelity has reached parity with two-qubit gates. We use this model to showcase how potential hardware improvements such as those seen in recent experimental demonstrations \cite{swiadek2024enhancing, sunada2024photon, beaulieu2026fast} can improve error correction footprints at algorithmically relevant error rates. All the circuits used for this analysis can be found in the \href{https://github.com/riverlane/megaquop_footprints}{megaquop\_footprints} repository.

\begin{table*}[h]
    \centering
\small
    \setlength{\tabcolsep}{5pt}
    \renewcommand{\arraystretch}{1.15}
    \begin{tabular}{@{}l cccccc@{}}
        \hline\hline
        Model & $p_\textrm{2Q}$ & $p_\textrm{1Q}$ & $p_\textrm{idle}$ & $p_\textrm{resonator-idle}$ & $p_\textrm{meas}$ & $p_\textrm{reset}$ \\
        \hline
        SI1000        & $p$ & $p/10$ & $p/10$ & $2p$ & $5p$ & $2p$ \\
        Optimistic    & $p$ & $p/10$ & $p/10$ & $p$  & $p$  & $p$  \\
        Uniform (SD6) & $p$ & $p$    & $p$    & $0$  & $p$  & $p$  \\
        \hline\hline
    \end{tabular}
    \caption{{\bf Parametrisation of circuit component error rates under SI1000, Optimistic and SD6 noise models}. Each circuit component experiences depolarising noise with a specific probability parametrised by the two-qubit gate error rate $p_\textrm{2Q}=p$. The different error contributions in the model affect single-qubit gates, which experience an error probability $p_\textrm{1Q}$, idling errors, which happen with probability $p_\textrm{idle}$, measurements which flip with probability $p_\textrm{meas}$, and resets which have errors with probability $p_\textrm{reset}$. In the resonator-idle channel, applied in SI1000 and Optimistic noise models, additional depolarisation with probability $p_\textrm{resonator-idle}$ occurs on qubits that idle while other qubits are being measured or reset. The SD6 noise model assigns equal probability to all these components and omits resonator-idle noise, making single-qubit gates and idling the primary source of errors, while the SI1000 and the Optimistic models are dominated by measurement, reset and two-qubit gate errors.}
    \label{tab:noise_parametrisations}
\end{table*}

To find the specific code distances needed to reach a target logical error probability for logical operations, we simulate quantum memory experiments for square distance-$d$ rotated planar surface codes under the three noise models described in \cref{tab:noise_parametrisations}. Each circuit initialises a logical state, subjects it to $r=1000$ stabiliser measurement rounds and ends with data qubit measurement and logical read-out. A large number of rounds $r=1000$ is chosen to minimise the effect of state preparation and measurement errors in the overall logical error probability. As our decoder, we use minimum-weight perfect matching (MWPM) as implemented in the PyMatching package \cite{higgott2025sparse}. MWPM operates in a sweet-spot regime of high decoding accuracy while maintaining high decoding throughput on a CPU -- a requirement for our large-scale Monte Carlo simulations needed to resolve low logical error rates. We benchmark both the correlated and uncorrelated variants of MWPM \cite{fowler2013optimal, paler2023pipelined} to demonstrate how a more accurate decoder can reduce the physical qubit footprint required.  We note that more accurate decoders such as machine learning-based models \cite{bausch2024learning, senior2026scalable} and maximum-likelihood decoders exist \cite{beni2025tesseract}, but these are either not publicly available or have prohibitively slow runtimes for real-time applications.

\begin{table*}[h]
    \centering
\small
    \setlength{\tabcolsep}{7pt}
    \renewcommand{\arraystretch}{1.2}
    \begin{tabular}{@{}l c l c c c c@{}}
        \hline\hline
        Noise model & $p$ & Decoder & $\Lambda$ & $p_0$ & $d_\textrm{min}$ for $10^6$ QuOps & $d_\textrm{min}$ for $10^7$ QuOps \\
        \hline
        SI1000        & $10^{-3}$ & MWPM      & $5.6\pm0.2$  & $0.081\pm0.015$ & $15.4\rightarrow d_\textrm{min}=17$ & $18.2\rightarrow d_\textrm{min}=19$ \\
                      & $10^{-3}$ & MWPM-corr & $8.8\pm0.4$  & $0.167\pm0.038$ & $12.4\rightarrow d_\textrm{min}=13$ & $14.7\rightarrow d_\textrm{min}=15$ \\
        \hline
        Uniform (SD6) & $10^{-3}$ & MWPM      & $6.8\pm0.3$  & $0.121\pm0.024$ & $14.0\rightarrow d_\textrm{min}=15$ & $16.6\rightarrow d_\textrm{min}=17$ \\
                      & $10^{-3}$ & MWPM-corr & $10.9\pm0.6$ & $0.266\pm0.067$ & $11.5\rightarrow d_\textrm{min}=13$ & $13.5\rightarrow d_\textrm{min}=15$ \\
        \hline
        Optimistic    & $10^{-3}$ & MWPM      & $7.9\pm0.4$  & $0.082\pm0.019$ & $12.4\rightarrow d_\textrm{min}=13$ & $14.8\rightarrow d_\textrm{min}=15$ \\
                      & $10^{-3}$ & MWPM-corr & $12.8\pm0.9$ & $0.173\pm0.052$ & $10.3\rightarrow d_\textrm{min}=11$ & $12.2\rightarrow d_\textrm{min}=13$ \\
        \hline
        Optimistic    & $5\times10^{-4}$ & MWPM  & $15.9\pm1.0$  & $0.080\pm0.022$ & $8.7\rightarrow d_\textrm{min}=9$ & $10.5\rightarrow d_\textrm{min}=11$ \\
        & $5\times10^{-4}$ & MWPM-corr & $29\pm3$ & $0.29\pm0.10$ & $7.7\rightarrow d_\textrm{min}=9$  & $9.2\rightarrow d_\textrm{min}=11$ \\
        \hline\hline
    \end{tabular}

    \caption{\textbf{Minimum code distances required to reach logical error rate regimes of 1 and 10 MegaQuOps on a rotated planar surface code}. The results are computed for the standard SI1000 and SD6 noise models, as well as a hardware-optimistic noise model with better measurement fidelity, as specified in \cref{tab:noise_parametrisations}. The correlated MWPM decoder (MWPM-corr) outperforms the uncorrelated MWPM decoder in every instance. The $\Lambda$-values observed in our simulations are in line with previous benchmarks such as Ref.~\cite{low2026denser} which reports $\Lambda\sim12$ for $p=10^{-3}$ under a uniform depolarising noise model and Ref.~\cite{gidney2021fault} which shows $\Lambda=6\pm2$ at $p=10^{-3}$ under SI1000 noise.}
    \label{tab:megaquop_distances}
\end{table*}

As a figure of merit for how logical performance scales with code distance, we use the error suppression factor $\Lambda=\varepsilon_d/\varepsilon_{d+2}$~\cite{google_repetition_code}, where $\varepsilon_d$ is the logical error rate per round at distance $d$. Assuming $\Lambda$ is independent of distance at a fixed physical error rate, $\varepsilon_d=p_0\Lambda^{-(d+1)/2}$ for some constant $p_0$. To estimate $\Lambda$ and $p_0$, we follow the protocol of Ref.~\cite{google2025}. For each distance $d$, we simulate $r$-round quantum memory experiments initialised in the $X$ and $Z$ bases, decode $N$ sampled syndromes, and estimate the logical error probability $p^B_\textrm{L}=N_\textrm{fails}/N$ for each basis $B\in\{X,Z\}$, where $N_\textrm{fails}$ is the number of shots in which the decoder's predicted observable disagrees with the true state. Each estimate is converted to a per-round rate $\varepsilon^B_d=\frac{1}{2}\left[1-\left(1-2p^B_\textrm{L}\right)^{1/r}\right]$, and the two bases are combined into the per-round probability of a logical failure in \textit{either} basis,
\begin{equation}\label{eq:pL_combined}
    \varepsilon_d=\varepsilon^X_d+\varepsilon^Z_d-\varepsilon^X_d\varepsilon^Z_d,
\end{equation}
treating $X$ and $Z$ failures as independent.

\begin{figure*}[tbp]
    \centering
    
    \includegraphics[width=\linewidth]{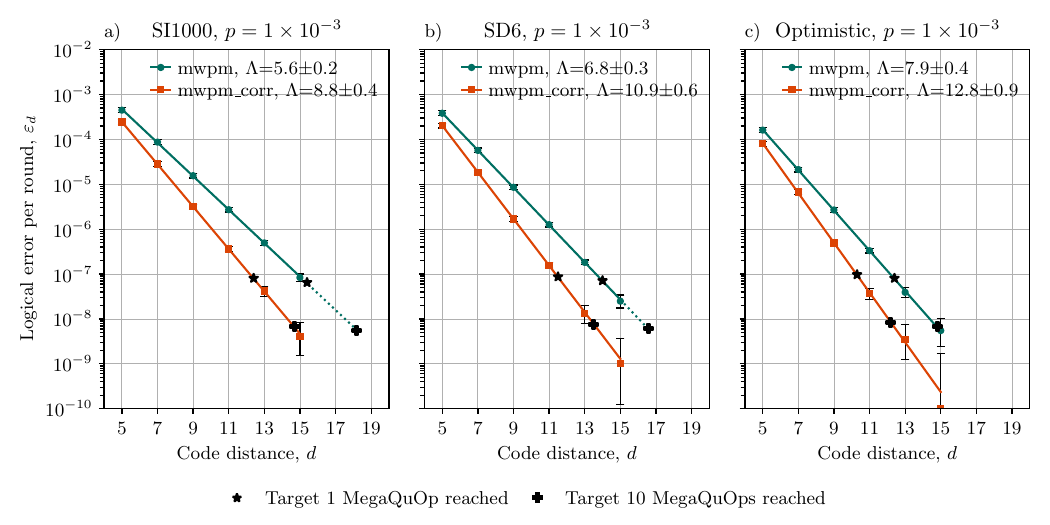}

    \vspace{0.8em}
    
    \includegraphics[width=0.6\linewidth]{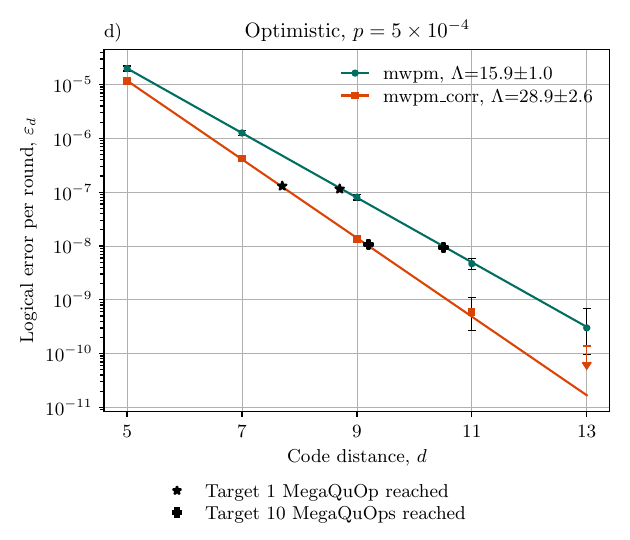}

    \caption{\textbf{Logical error suppression as a function of code distance $d$ for rotated planar surface code quantum memory simulations under physical error rates $p=10^{-3}$~(a-c) and ~$p=5 \times 10^{-4}$~(d).} The results are decoded with the correlated and uncorrelated variants of minimum-weight perfect matching (MWPM), and distances reaching 1-MegaQuOp and 10-MegaQuOp regimes are highlighted with a grey star and a grey cross, respectively. Each data point is estimated from a memory experiment with $r=1000$ rounds, with a maximum sample size of $N=10^7$ (a-c) or $N=10^8$ (d) shots, or $1000$ decoding failures, whichever was reached sooner. Error bars correspond to hypotheses with a likelihood within a factor of 1000 of the maximum likelihood hypothesis, given the sampled errors. For every noise model and decoder variant, the $\Lambda$-fit matches the data within error bars, and the correlated MWPM decoder consistently outperforms its uncorrelated variant.
    (a-c)~The SI1000 noise model is more demanding than the SD6 or the Optimistic noise models, achieving a lower $\Lambda$-factor than the other two models.
    (d)~Halving the error rate of the Optimistic noise model to $p=5 \times 10^{-4}$ results in approximately twice the $\Lambda$-factor and makes $8$ MegaQuOps attainable by distance~$9$. The distance-$13$ experiment for the correlated matching decoder saw no logical failures, so only an upper bound on its logical error probability is shown.}
    \label{fig:lambda_plots}
\end{figure*}

The error bars for each $p^B_\textrm{L}$ are calculated from the likelihood of the observed $N_\textrm{fails}$ as a function of $p^B_\textrm{L}$~\cite{Gidney2024StackOverflow}, computed with the \verb|fit_binomial| function of \textsc{Sinter} (a sub-package of \textsc{Stim}~\cite{gidney2021stim}). The bounds span all values of $p^B_\textrm{L}$ whose likelihood is within a factor of $1000$ of the maximum-likelihood estimate. We propagate these bounds through the per-round conversion and \eqref{eq:pL_combined}, combining lower bounds with lower bounds and upper bounds with upper bounds. We then fit $\ln\varepsilon_d=-\ln(\Lambda)\,x+\ln p_0$, with $x=(d+1)/2$, by weighted least squares, weighting each point by the inverse of its relative error bar. The resulting fits, shown in \cref{fig:lambda_plots} for the three noise models, agree well with the data. The fitted values are listed in \cref{tab:megaquop_distances}.

From the fitted model, the probability of a logical failure in any basis during a $d$-round syndrome extraction block is $p_\textrm{L}(d)=1-\left(1-\varepsilon_d\right)^d$. The crossover distance $d_\textrm{min}$ is the first distance at which $p_\textrm{L}(d)$ is below $10^{-6}$ for the 1-MegaQuOp regime and below $10^{-7}$ for the 10-MegaQuOp regime. \Cref{tab:megaquop_distances} lists these distances for each noise model and decoder.

\subsection{Comparing the performance of memory and lattice surgery}
\label{sec:memory_vs_lattice_surgery}

Our distance estimates used in the costing are based on the quantum memory simulations above. However, in order to perform logical operations, we use lattice surgery, which has different mechanisms for logical errors. In order to justify using the memory performance in our estimates, we present results indicating the performance of a lattice surgery operation is similar to that of quantum memory for the same space-time volume.

\begin{figure*}[h]
   \centering
    \includegraphics[width=0.45\linewidth]{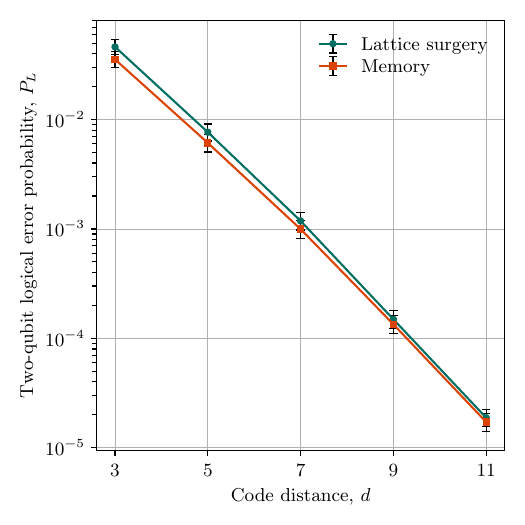}
   \caption{\textbf{Logical error probabilities of lattice surgery and quantum memory for different code distances.} The noise model is SI1000 with $p=10^{-3}$ and the data are decoded using correlated minimum-weight perfect matching. The logical error probability $P_\textrm{L}$ is evaluated for two logical qubits, keeping the lattice surgery error probability as the baseline. This results in $P_L=2P_M-P_M^2$ given a logical failure probability $P_{M}$ for the memory experiment, assuming the two logical qubits do not experience correlated errors. The error bars correspond to hypotheses with a likelihood within a factor of 1000 of the maximum likelihood hypothesis, given the sampled errors.}
   \label{fig:ls_vs_mem}
\end{figure*}

These results are shown in \cref{fig:ls_vs_mem}. For different distances, using the SI1000 noise model with $p=10^{-3}$ and decoding with correlated MWPM, we show the total logical error probability of a lattice surgery operation involving two side-by-side logical qubits and of two idling logical qubits. Each experiment is 3$d$ rounds long -- for lattice surgery, this means we perform $d$ rounds of memory on the two logical qubits, followed by a $d$-round joint logical measurements, followed by $d$ rounds of memory. We obtain the logical probabilities for initialising and measuring in the X and Z bases separately, and combine them using $p_{L} = p_{L}^{X} + p_{L}^{Z} - p_{L}^{X}p_{L}^{Z}$. As described above, we use Monte Carlo sampling, taking batches of samples until we observe at least 400 failures.

\section{Additional details on spectral functions}\label{sec:appendix_spectral_function}

\subsection{Broadening function and explicit representation of the spectral function}\label{app:broadening_function_and_explicit_representation}

\begin{lemma} 
    Let $\{\ket{\Psi_i}\}$ and $E_i$ be eigenstates and eigenvalues of the Hamiltonian $H$, respectively, $\ket{\psi}$ be some quantum state, and let $A$ and $B$ be some operators. Define
    \begin{align}
        C_{ij} & = \braket{\psi|\Psi_i}\braket{\Psi_i|A|\Psi_j}\braket{\Psi_j|B|\psi}, \\
        D_{ij} & = \braket{\psi|B|\Psi_j}\braket{\Psi_j|A|\Psi_i} \braket{\Psi_i|\psi}, \\
        \Delta_{ji} &= E_j - E_i.
    \end{align}
    \cref{eq:spectral_function_omega} can be written as
    \begin{equation}\label{eq:explicit_spectral_function_with_delta}
     \mathcal{A}(\omega) = \sum_{ij}\left(C_{ij}\delta(\omega - \Delta_{ji}) - D_{ij}\delta(\omega + \Delta_{ji}) \right).
    \end{equation}
    \label{lemma:spectral_function_approx_init}
\end{lemma}
\begin{proof}
    Inserting the identity $\sum_{i} \ket{\Psi_i}\!\bra{\Psi_i}=I$ between operators and using $\int_{-\infty}^{\infty} dt e^{i\omega t} =  2\pi \delta(\omega)$, one arrives at \cref{eq:explicit_spectral_function_with_delta}.
\end{proof}

Calculating \cref{eq:explicit_spectral_function_with_delta} requires an infinite time, but in practice the delta functions are replaced by a broadening function of finite width which we denote as $F(\omega)$. In this work we will use a Lorentzian function of width $\eta$
\begin{equation}\label{eq:lorentzian}
F(\omega) = \frac{\eta}{\pi}\frac{1}{\omega^{2} + \eta^2}.
\end{equation}

For our quantum algorithm, we will also take a finite Trotter step size and therefore a finite sampling interval in the time domain. One approach is to then just perform a discrete Fourier transform. However, we can instead take a continuous but periodic broadening function. That is, we replace $F(\omega)$ with a periodic Fourier series approximation, $\tilde{F}(\omega)$, defined
\begin{equation}\label{eq:periodic_broadening}
     \tilde{F}(\omega) = \sum_{|k| \le \kmax} F_k e^{\mathrm{i}k\omega\delta t},\quad F_k = \frac{\delta t}{2\pi}e^{-\eta \delta t |k|},
\end{equation}
where $k$ is an integer, and $\kmax$ is the maximum term in the Fourier series. This is done in analogy with many statistical phase estimation techniques, particularly~\cite{Lin2022, Wan2022, Blunt2023}. With these changes, we obtain
\begin{equation}\label{eq:explicit_spectral_function_with_broadening}
    \mathcal{A}(\omega) = \sum_{ij}\left(C_{ij}\tilde{F}(\omega - \Delta_{ji}) - D_{ij}\tilde{F}(\omega + \Delta_{ji})\right).
\end{equation}
For $\tilde{F}$ to be a good approximation of $F$, the support of the Lorentzian should lie almost entirely within the period of $\tilde{F}$, which is $2\pi/\delta t$, and this leads to the condition 
\begin{equation}\label{eq:Lorentzian_approximation_condition}
    \eta \cdot\delta t \ll 1.
\end{equation}
Additionally, for any $i, j$ the energy differences, $\Delta_{ji}$, should also lie within the period. The maximum energy range for magnetic systems considered in this work is $[-10\tau, 10\tau]$~\cite{LiebKagome2026}. Thus, the time step, $\delta t$, must satisfy $20 \tau <  2\pi / \delta t$ leading to 
\begin{equation}\label{eq:upper_bound_time_step}
\delta t < \pi/(10\tau).
\end{equation}
Because the energy resolution $\epsilon$ that we target is $0.5\tau$, which is sufficient to resolve magnetic excitations~\cite{LiebKagome2026}, we choose $\eta=\epsilon/2$ since the Lorentzian half-max is at $\omega=\pm \eta$ giving the full width at half maximum of $\epsilon$. A simple uncertainty principle gives the simulation time $T_{\rm sim}=2\pi/\epsilon=\pi/\eta=4\pi/\tau$, which leads to the smallest Fourier coefficient $F_{\pm \kmax} = e^{-\pi}F_{0}$. We make the conservative choice $\delta t = 0.1 \tau^{-1}$ (or $\delta t = \pi/(32\tau)$ in STAR simulations) to satisfy the bound \cref{eq:upper_bound_time_step} and to reduce the Trotter error. This leads to $\kmax=T_{\rm sim}/\delta t=128$. We verified numerically that such a choice of $\kmax$ approximates the Lorentzian well. Notice that $\eta \cdot\delta t = \pi/\kmax \approx 0.025$, so the condition \cref{eq:Lorentzian_approximation_condition} is also satisfied.

The next step is to show when positive times suffice to recover the full spectral function. For this, one can revert \cref{eq:explicit_spectral_function_with_broadening} back to the explicit Fourier transformation, to arrive at \cref{eq:spectral_function_omega_approx} which explicitly shows
time evolution in both positive and negative times. However, we can expand \cref{eq:spectral_function_omega_approx} further:
\begin{equation}\label{eq:spectral_function_omega_approx_2}
    \mathcal{A}(\omega) \approx \mathrm{i} F_0g_0 + \mathrm{i}\sum_{k\geq1}F_k(g_k + g_{-k}) \cos(k\omega \delta t) - \sum_{k\geq1}F_k(g_k - g_{-k}) \sin(k\omega \delta t).
\end{equation}
When $g_{-k}=-g_k^{*}$ the spectral function is real and only positive times are required. This holds in particular when (a) the state is an eigenstate of the Hamiltonian and $B=A^{\dagger}$; or (b) both the Hamiltonian and the state are real in a common basis, and $A$, $B$ are either both real or both imaginary in that basis. For example, this is the case when $A$ and $B$ both equal the $z$ component of the spin operator. If $g_{-k}=g_k^{*}$, the spectral function is imaginary and positive times still suffice. This occurs when $A$ and $B$ are of opposite type (imaginary and real), and both the Hamiltonian and the state are real in a common basis. An example of the latter is when $A$ and $B$ are the $x$ and $y$ components of the spin operator, respectively. In other cases, such as when the initial state is complex, one has to perform Trotterisation in negative time to obtain $g_{-k}$.

\subsection{Transverse and longitudinal spectral function}

To evaluate the spectral function \cref{eq:spectral_function_final_expression}, one needs to calculate:
\begin{equation}
    C^{\alpha\beta}_{i, j}(t) =  -\mathrm{i} \langle \psi | [ S^{\alpha}_i(t), S^{\beta}_j(0) ] | \psi \rangle.
\end{equation} 
As discussed in the main text, we use \cref{eq:quench_formula} for different choices of $A$ and $B$ corresponding to the spin operators $S^{\alpha}_i(t)$ and $S^{\beta}_j(0)$. We will first derive results in terms of Majorana operators to make the results independent of the fermion-to-qubit mapping and then specify final expressions in terms of Pauli strings under Jordan--Wigner map for ordering when spin-orbitals of the same site are adjacent with spin-up first. We will use the convention that when the equations concern a single site only, we will omit that index so as not to overload the notations. The same applies to the spin index and to the combined spin-orbital index. We also introduce the combined index $r=(j\sigma,x)$, where $x \in \{0,1\}$ indicates Majorana type.

Majorana operators are defined as
\begin{align}
    \gamma_{0} & = a + a^\dagger,\\
    \gamma_{1} & = -\mathrm{i}(a - a^\dagger),
\end{align}
and equivalently
\begin{align}
    a & = \frac{1}{2}(\gamma_0 + \mathrm{i} \gamma_1)\\
    a^{\dagger} & = \frac{1}{2}(\gamma_0 - \mathrm{i} \gamma_1)
\end{align}
Majorana operators are Hermitian, unitary and obey the anticommutation relations
\begin{equation}
    \{\gamma_{r}, \gamma_{r'}\} = 2\delta_{rr'}.
\end{equation} 
Useful relations which we are going to use in this section are:
\begin{equation}
    a^{\dagger} a = \frac{1}{2}(\mathbb{1} + \mathrm{i} \gamma_{0} \gamma_{1}). 
\end{equation}
\begin{equation}\label{eq:exponential}
    \forall r\neq r',\quad e^{\theta\gamma_{r} \gamma_{r'}} = \cos(\theta) \mathbb{1} + \sin(\theta) \gamma_{r} \gamma_{r'}.
\end{equation}

\begin{align}\label{eq:spin_plus_in_majorana_operators}
    S^{+} &= a^{\dagger}_{\uparrow}a_{\downarrow}
      = \frac{1}{4}\left(\gamma_{\uparrow,0} - \mathrm{i}\gamma_{\uparrow,1}\right)\left(\gamma_{\downarrow,0} + \mathrm{i}\gamma_{\downarrow,1}\right),\\
    S^{-} &= a^{\dagger}_{\downarrow}a_{\uparrow}
      = \frac{1}{4}\left(\gamma_{\downarrow,0} - \mathrm{i}\gamma_{\downarrow,1}\right)\left(\gamma_{\uparrow,0} + \mathrm{i}\gamma_{\uparrow,1}\right).
\end{align}

\begin{equation}\label{eq:spin_pm_aop_majop}
S^{\pm} = \pm\frac{1}{4}(\gamma_{\uparrow,0}\gamma_{\downarrow,0} + \gamma_{\uparrow,1}\gamma_{\downarrow,1}) + \frac{\mathrm{i}}{4}(\gamma_{\uparrow,0}\gamma_{\downarrow,1} - \gamma_{\uparrow,1}\gamma_{\downarrow,0}),
\end{equation}
\begin{equation}\label{eq:spin_y_aop_majop}
    S^{y} = \frac{\mathrm{i}}{2}(a_{\downarrow}^{\dagger} a_{\uparrow} - a^{\dagger}_{\uparrow} a_{\downarrow}) = -\frac{\mathrm{i}}{4}\left(\gamma_{\uparrow,0}\gamma_{\downarrow,0} + \gamma_{\uparrow,1}\gamma_{\downarrow,1}\right),
\end{equation}
\begin{equation}\label{eq:spin_z_aop_majop}
    S^{z} = \frac{1}{2}\left(a^{\dagger}_{\uparrow} a_{\uparrow} - a^{\dagger}_{\downarrow} a_{\downarrow} \right) = \frac{\mathrm{i}}{4}(\gamma_{\uparrow,0}\gamma_{\uparrow,1} - \gamma_{\downarrow,0}\gamma_{\downarrow,1} ),
\end{equation}
\begin{equation}\label{eq:quench_pauli_2}
    P_2 := -\mathrm{i} \gamma_{\uparrow,1}\gamma_{\downarrow,1},
\end{equation}
\begin{equation}
    D := a^{\dagger}_{\uparrow}a^{\dagger}_{\downarrow} - a_{\downarrow}a_{\uparrow} = \frac{1}{2}\left(\gamma_{\uparrow,0}\gamma_{\downarrow,0} - \gamma_{\uparrow,1}\gamma_{\downarrow,1}\right),
\end{equation}
\begin{equation}\label{eq:pauli_2_in_terms_of_D_Spm}
    P_2 = \mathrm{i} D + 2 S^{y} = \mathrm{i}(D - S^{+} + S^{-}).
\end{equation}

The operators $S^{z},S^{+}, S^{-}$ are real in the occupation-number basis as is the Hamiltonian for real hopping and interaction parameters. Provided the initial state is also real, forward time evolution suffices to evaluate the longitudinal and transverse spectral functions (see the discussion following \cref{eq:spectral_function_omega_approx_2}). Additionally, spin operators commute on different sites, $\forall i\neq j,\text{ and } \forall \alpha,\beta \quad [S^{\alpha}_i, S^{\beta}_j]=0$. Thus, we can measure a component $S^{\alpha},\alpha \in\{x,y,z\}$ for all $i$ in the common basis avoiding running several circuits to compute expectation values on each site $i$. Note that the operators $-\mathrm{i}\gamma_{r}\gamma_{r'}$ are unitary; thus, Eqs.~\eqref{eq:spin_pm_aop_majop}, \eqref{eq:spin_z_aop_majop} are linear combinations of unitaries (LCU). Additionally, due to translational symmetry, it is enough to consider $j$ from the unit cell (see \cref{eq:spectral_function_final_expression}) so the total number of quench states from \cref{eq:quench_formula} is proportional to the number of sites in the unit cell, $n_s$.

\subsubsection{Longitudinal spectral function}
For the longitudinal spectral function, we take $A = S^z_i(t)$ and $B = S^z_j(0)$. Using the LCU~\cref{eq:spin_z_aop_majop}, one obtains:
\begin{align}
    P_0 &= -\mathrm{i}\gamma_{\uparrow,0}\gamma_{\uparrow,1} \overset{\mathrm{JW}}{\longmapsto} Z_{\uparrow} , \quad w_{0} = -\frac{1}{4}, \\
    P_1 &= -\mathrm{i}\gamma_{\downarrow,0}\gamma_{\downarrow,1} \overset{\mathrm{JW}}{\longmapsto} Z_{\downarrow}, \quad w_{1} = \frac{1}{4}.
\end{align}
Thus, for two Paulis, we have $4\times n_s$ quantum circuits (due to $\sigma$ dependence of the quench state in \cref{eq:quench_formula} and the number of sites in the unit cell) differing by the choice of the initial quench state and which are used to measure $S^{z}$ on all sites. 
\subsubsection{Transverse spectral function}
For the transverse spectral function $A = S^+_i(t)$ and $B = S^{-}_j(0)$. Using the LCU \cref{eq:spin_pm_aop_majop} for $B = S^{-}_j(0)$, one obtains:
\begin{align}
    P_0 & = -\mathrm{i}\gamma_{\uparrow,1}\gamma_{\downarrow,0} \overset{\mathrm{JW}}{\longmapsto} \phantom{-}X_{\uparrow} X_{\downarrow},\quad w_0 = \frac{1}{4},\\
    P_1 & = -\mathrm{i}\gamma_{\uparrow,0}\gamma_{\downarrow,1} \overset{\mathrm{JW}}{\longmapsto} -Y_{\uparrow} Y_{\downarrow},\quad w_1 = -\frac{1}{4},\\
    P_2 & = -\mathrm{i}\gamma_{\uparrow,1}\gamma_{\downarrow,1} \overset{\mathrm{JW}}{\longmapsto} \phantom{-}X_{\uparrow} Y_{\downarrow},\quad w_2 = -\frac{\mathrm{i}}{4},\\
    P_3 & = -\mathrm{i}\gamma_{\uparrow,0}\gamma_{\downarrow,0} \overset{\mathrm{JW}}{\longmapsto} -Y_{\uparrow} X_{\downarrow},\quad w_3 = -\frac{\mathrm{i}}{4}.
\end{align}
Since $S^{+}=S^{x} + \mathrm{i} S^{y}$, and $S^{x}$, $S^{y}$ do not commute, we have to measure $S^{x}$ and $S^{y}$ separately. Additionally, because $S^-_j$ is a linear combination of four different operators, we will have to prepare $8\times n_s$ different initial quench states (due to $\sigma$-dependence in \cref{eq:quench_formula} and the number of sites in the unit cell). However, this can be reduced to $n_s$ quenches only if the initial state and Hamiltonian satisfy certain symmetries. 
\subsubsection{Reducing the circuit count for the transverse spectral function}
Here, we will show that calculation of the transverse susceptibility can be reduced to a single quench state and the time evolution of the $S^{+}_i$ operator alone (see \cref{lemma:branch-cost}).

\begin{definition}\label{def:charge}
    The operator $O$ carries charge $\nu$ if it satisfies
    \begin{equation}
        [S^z_{\rm tot}, O] = \nu O.
    \end{equation}
\end{definition}

\begin{definition}\label{def:odd_operator}
    The operator $O$ is odd under $G$ if
    \begin{equation}
        G O G = - O.
    \end{equation}
\end{definition}

\begin{definition}\label{def:even_operator}
    The operator $O$ is even under $G$ if
    \begin{equation}
        G O G = O.
    \end{equation}
\end{definition}

\begin{definition}\label{def:spin-up_parity_operator}
Let $N_{\uparrow}$ be the total number of electrons with spin up. The spin-up parity operator is defined as
\begin{equation}
    G = e^{\mathrm{i} \pi N_{\uparrow}} = \left(-\mathrm{i}\right)^{N_s} \prod_{j=0}^{N_s-1} \gamma_{j\uparrow,0}\gamma_{j\uparrow,1}  
\end{equation}
\end{definition}

\begin{lemma}\label{lemma:majorana_parity}
    Let $G$ be the spin-up parity operator of Definition~\ref{def:spin-up_parity_operator}, then for all $j\in [0, N_s)$ and $x\in \{0, 1\}$ $\gamma_{j\downarrow,x}$ is even while   $\gamma_{j\uparrow,x}$ is odd under $G$:
    \begin{align}
        G \gamma_{j\downarrow,x} G & = \gamma_{j\downarrow,x},\\
        G \gamma_{j\uparrow,x} G & = -\gamma_{j\uparrow,x}.
    \end{align}
\end{lemma}
\begin{proof}
    Follows from Definition~\ref{def:spin-up_parity_operator}.
\end{proof}

\begin{lemma}\label{lemma:charge_to_eigenvalues}
    Let $\ket{\psi}$ be an eigenstate of $S^{z}_{\rm tot}$ with eigenvalue $M$:
    \begin{equation}
        S^{z}_{\rm tot} \ket{\psi} = M \ket{\psi},
    \end{equation}
    and let $O$ be an operator carrying charge $\nu$.
    Then $\ket{\phi} = \hat O \ket{\psi}$ is an eigenstate of $S^{z}_{\rm tot}$ with the eigenvalue $M+\nu$.
    \begin{equation}
        S^{z}_{\rm tot} \ket{\phi} = (M + \nu) \ket{\phi}.
    \end{equation}
\end{lemma}
\begin{proof}
    Consider:
    \begin{equation}
        S^{z}_{\rm tot} \ket{\phi} =  S^{z}_{\rm tot} O \ket{\psi} =  O (S^{z}_{\rm tot} + \nu \mathbb{1}) \ket{\psi} = O (M+\nu) \ket{\psi} = (M+\nu) \ket{\phi}, 
    \end{equation}
    where we used Definition~\ref{def:charge}.
\end{proof}

\begin{lemma}\label{lemma:charge_and_parity_of_spin_pm}
    Operators $S^{\pm}_j$ carry charge $\pm1$ and are odd under $G$.
\end{lemma}
\begin{proof}
Consider the commutator:
\begin{align} 
[S^z_{\rm tot}, S^{\pm}_j] =[S^z_{\rm tot}, S^{x}_j] \pm \mathrm{i}[S^z_{\rm tot}, S^{y}_j] = [S^z_j, S^{x}_j] \pm \mathrm{i}[S^z_j, S^{y}_j] = \mathrm{i} S^{y}_j \pm S^{x}_{j} = \pm S^{\pm}_j,
\end{align}
where we used conventional spin-commutation relations. 
Using $G^2=1$ and $G^{\dagger}=G$, \cref{eq:spin_plus_in_majorana_operators}, and Lemma~\ref{lemma:majorana_parity} we obtain:
\begin{equation}
    G S^{+}_j G = \frac{1}{4}G(\gamma_{j\uparrow,0}-\mathrm{i}\gamma_{j\uparrow,1})G G(\gamma_{j\downarrow,0}+\mathrm{i}\gamma_{j\downarrow,1})G = -\frac{1}{4}(\gamma_{j\uparrow,0}-\mathrm{i}\gamma_{j\uparrow,1})(\gamma_{j\downarrow,0}+\mathrm{i}\gamma_{j\downarrow,1}) = -S^{+}_j.
\end{equation}
\begin{equation}
    G S^{-}_j G =(G S^{+}_j G)^{\dagger} = - S^{-}_j.
\end{equation}
\end{proof}
\begin{lemma}\label{lemma:odd_p2_and_charge_d}
    Operator $P_2$ is odd under the spin-up parity operator $G$, and operator $D$ carries charge $0$.
\end{lemma}
\begin{proof}
    Operator $P_2$ contains a single spin-up Majorana operator, so it is odd by Lemma~\ref{lemma:majorana_parity}. Using standard fermionic anticommutation relations, one obtains $[S^{z}_{\rm tot}, D]=0$.
\end{proof}

\begin{lemma}\label{lemma:branch-cost}
Let $S^z_{\rm tot} = \sum_j S^z_j$ be the total $z$-component of the spin operator, $N_{\uparrow}=\sum_j n_{j\uparrow}$ be the total number of spin-up electrons, and let
\begin{equation}
    G = e^{\mathrm{i} \pi N_{\uparrow}}
\end{equation}
be the spin-up parity operator.
Assume:
\begin{enumerate}
    \item $[H, G] = 0$.
    \item $[H, S^z_{\rm tot}] = 0$.
    \item $S^z_{\rm tot}\ket{\psi}=M\ket{\psi}$.
    \item $G\ket{\psi} = g\ket{\psi}$, $g=\pm 1$.
\end{enumerate}
Then, for $\ket{\psi_{0}}=e^{\frac{\pi}{4}\gamma_{j\uparrow,1}\gamma_{j\downarrow,1}}\ket{\psi}$
\begin{equation}
    C^{+-}_{ij}(t) = 2\mathrm{i} \braket{\psi_0|S^{+}_i(t)|\psi_0}.
\end{equation}
\end{lemma}

\begin{proof}
Let $P_2 = -\mathrm{i}\gamma_{j\uparrow,1}\gamma_{j\downarrow,1}$, then
\begin{equation}
    e^{\mathrm{i}\frac{\pi}{4}P_2} = \frac{1}{\sqrt{2}} \left( 1 + \mathrm{i} P_2 \right)
\end{equation}
and 
\begin{equation}
    \braket{\psi_0|S^{+}_i(t)|\psi_0} = \frac{1}{2}
    \left(
        \braket{\psi|S^{+}_i(t)|\psi}
        + \braket{\psi|P_2 S^{+}_i(t)P_2|\psi}
        + \mathrm{i}\braket{\psi|[S^{+}_i(t), P_2]|\psi}
     \right)
\end{equation}
Because $[H, G]=0$ it follows that $e^{\pm \mathrm{i}Ht}$ is even under $G$.
Combining this with the fact that $S^{+}_i$ and $P_2$ are odd under $G$ (Lemma~\ref{lemma:charge_and_parity_of_spin_pm}) and $G^2=1$, one obtains:   
\begin{equation}
    \braket{\psi|S^{+}_i(t)|\psi} = \braket{\psi|G S^{+}_i(t)G|\psi} = \braket{\psi|G e^{\mathrm{i}Ht}G^2S^{+}_iG^2 e^{-\mathrm{i}Ht} G|\psi} = -\braket{\psi|S^{+}_i(t)|\psi} = 0  
\end{equation}
and similarly
\begin{equation}
    \braket{\psi|P_2S^{+}_i(t)P_2|\psi} = 0.
\end{equation}
Using \cref{eq:pauli_2_in_terms_of_D_Spm}:
\begin{equation}
    \braket{\psi|[S^{+}_i(t), P_2]|\psi} = \mathrm{i}\left( \braket{\psi|[S^{+}_i(t), D]|\psi} - \braket{\psi|[S^{+}_i(t),S^{+}_{j}]|\psi} + \braket{\psi|[S^{+}_i(t),S^{-}_j]|\psi}\right).
\end{equation}
Since $[H, S^{z}_{\rm tot}]=0$, the time evolution operator $e^{\mathrm{i}Ht}$ carries the charge $0$. This means that the time-evolved operators $S^{+}_j(t)$ carries the same charge as $S^{+}_j$. Because $D$, $S^{+}_j$, $S^{-}_j$ carry charges $0$, $+1$, $-1$, respectively (Lemmas~\ref{lemma:charge_and_parity_of_spin_pm} and~\ref{lemma:odd_p2_and_charge_d}), the states $\ket{\phi_0} = [S^{+}_i(t), D]\ket{\psi}$, $\ket{\phi_1} = [S^{+}_i(t), S^{+}_j]\ket{\psi}$ will be eigenstates of $S^{z}_{\rm tot}$ with eigenvalues $M+1$ and $M+2$ (Lemma~\ref{lemma:charge_to_eigenvalues}) making them orthogonal to $\ket{\psi}$, while the state $[S^{+}_i(t),S^{-}_j]\ket{\psi}$ has eigenvalue $M$. Thus,
\begin{equation}
\braket{\psi_0|S^{+}_i(t)|\psi_0} = -\frac{1}{2}\braket{\psi|[S^{+}_i(t),S^{-}_j]|\psi} = -\frac{\mathrm{i}}{2}C^{+-}_{ij},
\end{equation}
\end{proof}
In this proof we have not required the initial state to be stationary. In practice, one often chooses the initial state from the $N$-electron sector. If, in addition, the state is an eigenstate of $S^{z}_{\rm tot}$, then it is also an eigenstate of $G$ with eigenvalue $(-1)^{N/2 + M}$ because $2N_{\uparrow}=N + 2M$, and so it satisfies assumptions 3 and 4 of \cref{lemma:branch-cost}. The quench operator maps $N$-electron wave function to a linear superposition of $N+2$, $N-2$, $N$ electron functions (see Eqs.~\eqref{eq:exponential}, \eqref{eq:quench_pauli_2}, \eqref{eq:pauli_2_in_terms_of_D_Spm}). Therefore, if we use symmetry-verification to post-select results based on the number of electrons \cite{Bonet-Monroig2018, Stanisic2022} then we should use $N+2$, $N-2$, $N$ states.

\section{Primary results for the STAR architecture}
\label{app:star_primary}

\begin{figure}
    \centering
    \begin{quantikz}
        \lstick{$\ket{\psi}$} & \ctrl{1} & &
          \gategroup[wires=2,steps=5,style={dashed,rounded corners,inner xsep=-0.3333em},
                     label style={label position=above,anchor=south}]{Based on the decoded outcome}
          & & \ctrl{1} & & &
          \gategroup[wires=2,steps=5,style={dashed,rounded corners,inner xsep=-0.3333em}]{}
          & & \ctrl{1} & & & \\
         \lstick{$R_{Z_L}(\theta)\ket{+}_L$} & \targ{} & \meter{} & \setwiretype{c}
          & \gate[style={draw=none}]{R_{Z_L}(2\theta)\ket{+}_L}\setwiretype{n} & \targ{}\setwiretype{q} & \meter{} & \setwiretype{c}
          & & \gate[style={draw=none}]{R_{Z_L}(4\theta)\ket{+}_L}\setwiretype{n} & \targ{}\setwiretype{q} & \meter{} & \setwiretype{c}
          &
    \end{quantikz}
    \caption{\textbf{Illustration of three rounds of the repeat until success protocol.} Each stage consumes a resource state $\ket{R_Z(2^{k-1}\theta)+}$ and applies an entangling gate and logical measurement. The decoded outcome is maximally random $P(+1)=1/2$. A $+1$ decoded outcome applies $R_Z(2^{k-1}\theta)$ to the data, while a $-1$ outcome applies $R_Z(-2^{k-1}\theta)$. A $-1$ outcome therefore requires a correction by twice the angle, so the next stage (dashed box) is executed. For Clifford-hierarchy angles the ladder terminates once the correction angle reaches a Clifford rotation, which is applied deterministically. Three stages are shown.}
    \label{fig:RUS gadget}
\end{figure}
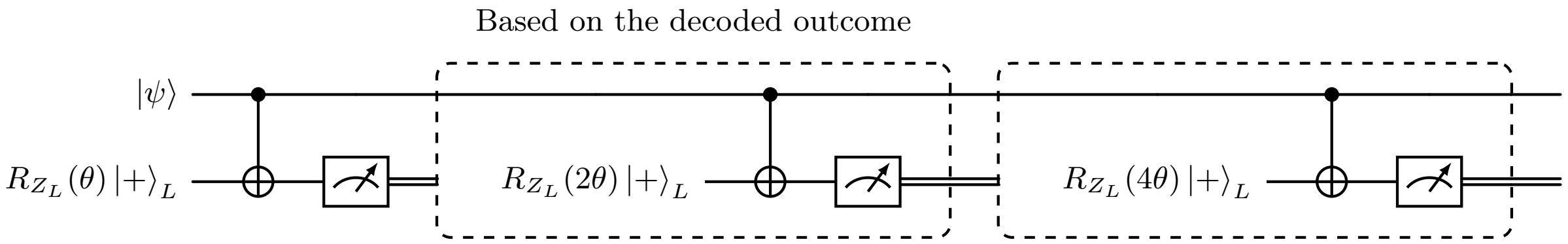

Among the techniques we considered for implementing the case study of this paper was the Space-Time efficient Analogue Rotation protocol (STAR protocol)~\cite{Akahoshi2024,Akahoshi2025,Toshio2025}. STAR denotes a class of partially fault-tolerant architectures in which Clifford operations are performed fault-tolerantly, while analogue rotations are teleported using resource states prepared non-fault-tolerantly (see \cref{fig:RUS gadget}). 

We simulate three constructions, each compiled with CZ as the only native two-qubit gate, and with the following naming convention:

\begin{enumerate}
    \item \emph{Akahoshi}~\cite{Akahoshi2024} is the single-rotation injection protocol, in which one physical rotation is injected into a distance-2 patch, and then discarded if any detector fires while the patch is grown to distance $d$.
    \item \emph{Choi}~\cite{choi2023} decomposes the logical $Z$ rotation into physical $Z$ rotations along the $Z$ logical, applying $k=d$ single-qubit rotations to the qubits supporting $Z_L$.
    \item \emph{Toshio-CZ}~\cite{Toshio2025} is a CZ-native compilation of the transversal multi rotation (TMR) protocol, which generalises the Choi protocol. It pairs the qubits in the $Z$ logical into weight-two rotations, $R_{ZZ}(\phi)$, so that $k=d/2$, i.e. we apply half as many rotation operators as the distance of the protocol. The circuit we use to apply $R_{ZZ}(\phi)$ was suggested in \cite{Akahoshi2024}. It applies the rotation using an ancilla qubit neighbouring the $Z$ logical as a mediator, and to detect some errors that can occur during injection.
\end{enumerate}

\noindent  We include the relevant values of $k$ and $d$ here for reference, and will define $k$, $d$ and explain the relevant choices in detail later. \Cref{app:star_results} gives the circuits in full (\cref{fig:star_circuit}). Using \emph{Akahoshi} for resource preparation versus using TMR for resource preparation is, in our context, the main difference between  STAR version~1~\cite{Akahoshi2024} and STAR version~2~\cite{Toshio2025} (following the terminology of \cite{Toshio2026}).

The TMR state preparation protocol reuses the syndrome extraction already present in the computation to project a product of single- or few-physical qubit analogue rotations onto a single-logical qubit analogue rotation. Within a CSS code patch, such as a surface code patch, it prepares a family of resource states $R_{Z_L}(\theta)\ket{+}_L$ that allow us to apply a logical $R_Z(\theta)$ by gate teleportation (here we take the convention $R_{Z_L}(\theta) = \exp(i Z \theta)$, as we do throughout this paper). The teleportation gadget succeeds on average in two attempts, with a worst-case error of $\mathcal{O}(p \theta)$ for the rotation channel. Here, $p$ is the physical error rate.

Prior applications of TMR injection~\cite{Toshio2025,Akahoshi2025,Chung2026,Ismail2026_1,Ismail2026_2} have primarily focused on the very small-angle regime. We assess TMR in the (relatively) large-angle regime, $\theta > 10^{-2}$, and typically $\theta \sim 0.1$ and under superconducting-inspired noise models. We simulated these circuits by applying several novel enhancements to a known non-Clifford simulation technique (see \cref{app:non_cliff_sim}), and find that the state-preparation error rate (and associated mitigation cost) is too high for STAR to be useful for our case study (\cref{fig:star_trotter}).

We made several favourable choices to verify that our results are due to properties of the TMR state preparation protocol in isolation, as opposed to an unfavourable implementation:

\begin{enumerate}
    \item We restrict to TMR injection within the Clifford hierarchy; thus a bounded number of states is required.
    \item We select the optimal schedule (code distance and angular decomposition) independently at each angle.
    \item We assume that coherent errors during state preparation are stable and may be calibrated away, \cref{fig:harp_calibration}.
    \item We assume that magic state cultivation is also available and that errors occurring during its state preparation or teleportation are subleading to errors occurring during state preparation or teleportation occurring with TMR.
\end{enumerate}

\noindent Even under these choices we find that, at the angles $\theta = \pi/16,\ldots,\pi/128$ ($C^{(4)}$ to $C^{(7)}$ in the Clifford hierarchy, respectively, for our convention of $R_Z(\theta)$), the rotation channel error is $2.53\times10^{-4}$, $2.01\times10^{-4}$, $1.25\times10^{-4}$ and $7.13\times10^{-5}$, respectively (\cref{tab:star_td}). At $p = 10^{-3}$ this limits our case study to fewer than $50$ of the $128$ Trotter steps we require. At $128$ Trotter steps (with PEC) the sampling overhead is $3.6\times10^8$ (\cref{fig:star_trotter}).

To elucidate TMR state preparation, consider a CSS code with a minimum-weight logical operator $Z_L = Z_1 \otimes \ldots \otimes Z_k$. Writing $\phi$ for the physical rotation angle applied to each factor $Z_i$, the central identity of TMR is

\begin{equation}\label{eq:star_operator}
    R_{Z_1}(\phi)\otimes \ldots\otimes R_{Z_k}(\phi)
    = \frac{1}{N_0(\phi)}R_{Z_1\otimes\ldots\otimes Z_k}\!\left(\theta_0(\phi)\right)
    + \frac{1}{N_1(\phi)}R_{Z_1\otimes\ldots\otimes Z_k}\!\left(\theta_1(\phi)\right)\!\left(Z_1 + \ldots + Z_k\right)
    + \mathcal{O}\!\left(Z_i\otimes Z_j\right),
\end{equation}

\noindent where $\theta_a(\phi)$ and $N_a(\phi)$ are the logical rotation angles and (square roots of) normalisations of the operators when collected by the weight ($a$) of $Z_i$ that we factor from the logical operator; their closed forms are given in \cref{app:star_results}. After applying this operator, syndrome measurement is performed and we post-select on the result; in the absence of physical errors, post-selecting on the null syndrome allows one to select the desired logical rotation, $R_{Z_1\otimes\ldots\otimes Z_k}$. The physical angle $\phi$ is chosen so that the corresponding term realises the target logical angle, $\theta_0(\phi) = \theta$. We assume $k$ is odd; the construction generalises to even $k$, but we do not analyse that case here.

In the presence of physical errors, the terms of \cref{eq:star_operator} can mix, such that post-selecting on the null syndrome selects the wrong term, which is a logical error. The $\mathcal{O}(Z_i\otimes Z_j)$ remainder also contains higher weight Pauli Z operators, which do not contribute at first order in the undetected-error probability $p_{UD}$ (defined in \cref{app:star_results}); we will disregard these in the approximate analysis later.

This identity also holds when the $Z_i$ are themselves multi-qubit operators, $Z_i = Z_{i_1}\otimes Z_{i_2}\otimes\ldots\otimes Z_{i_m}$. This was a main insight of Toshio \emph{et al.} \cite{Toshio2025} compared to the original suggestion by Choi \emph{et al.} \cite{choi2023}, and leads to improved post-selection rates and reduced logical error rates.

The TMR state-preparation protocol is therefore:

\begin{enumerate}
    \item prepare a logical encoded $\ket{+}_L$ state by standard CSS transversal state preparation;
    \item apply many single-qubit (or few-qubit) rotations $R_{Z_i}(\phi)$ according to \cref{eq:star_operator};
    \item apply a round of syndrome extraction that projects the operator onto a specific logical rotation; and
    \item accept the state if no detector fires, else, restart the procedure.
\end{enumerate}

These constructions are specified by the weight ($m$) and number of the rotation operators ($k$), which must combine to the distance of the logical ($d$). For instance, we may choose $2 k = d$ for even distances and $2 k = d + 1$ for odd distances. Choi uses $m=1$, $k=d$~\cite{choi2023}. Following Refs.~\cite{Toshio2025, Ismail2026_1, Chung2026}, we also study $m=2$. This has two advantages. First, $p_\textit{UD}$ is higher for the $m=1$ variants because there are single errors that can occur between the syndrome extraction rounds, for instance, resonator noise during measurement or idling during reset, that can mix the terms in \cref{eq:star_operator} and are undetectable; this is not true for $m=2$. Second, reducing $k$ lowers the probability of a fault that mixes the branches of \cref{eq:star_operator} during the rotations, while increasing $d$ suppresses errors in the decoded operations that follow, such as patch growth or lattice surgery. We analyse this trade-off in \cref{app:star_results}. We, for instance, studied $(k=5,\;d=5)$, $(k=5,\;d=9)$, and $(k=5,\;d=10)$, and found that the latter had the best performance for state preparation.

The generated resource states are then consumed to implement a rotation channel. We use the standard state teleportation gadget, which implements either our intended rotation angle or minus our intended rotation angle with $50\%$ probability. When the teleportation gadget's decoded random measurement returns $+1$ we say that gadget ``succeeds'', and we implement our target channel. To analyse this, we consider the case where we have a first time success (FTS), and use this adjectivally to describe the FTS error rate. Contrast this with what happens when the teleportation gadget for $R_{Z_L}(\theta)$ ``fails''. In this eventuality we repeat until success (RUS). That is, we use the same strategy to prepare a resource state for an $R_Z(2 \theta)$ teleportation gadget. See the illustration \cref{fig:RUS gadget} for clarity. The application of multiple TMR resource states with angles $R_Z(2^n\theta)$ at increasing error rates and probabilities $p_n=1/2^n$ defines a channel whose error is given by a formally divergent sum. However, we may use some known digital rotation (such as an $S$ or $T$ gate) to bound the maximum $\theta$ that we must implement. Prior analysis shows that this leads to a channel error rate for the entire channel that scales approximately linearly with the rotation angle, $\varepsilon \approx \alpha\, p\, \theta$, for an implementation-dependent constant $\alpha$ that may depend on $k$ \cite{Toshio2025}.

In our results, we instead tune the Trotter step size so that the rotations lie within the Clifford hierarchy ($\theta = \pi/16,\ldots$) (provided we only consider interaction strengths $U=1,\;2,\;4\;\mathrm{ and }\;8$). The RUS procedure therefore always terminates in a bounded number of steps. In the worst-case scenario we must eventually apply $T$, and we assume that we can prepare a $T$ resource state with lower error than any TMR operator (specifically using magic state cultivation \cite{gidney2024magic}). Therefore, we are unconcerned with the asymptotic behaviour of the injection protocol at large or small angle, and the governing quantity is the error rate from synthesising our discrete angles only. Additionally, we assume that calibration can be performed to remove the coherent error channel.

\begin{figure*}[tbp]
    \centering
    \includegraphics[width=.7\textwidth]{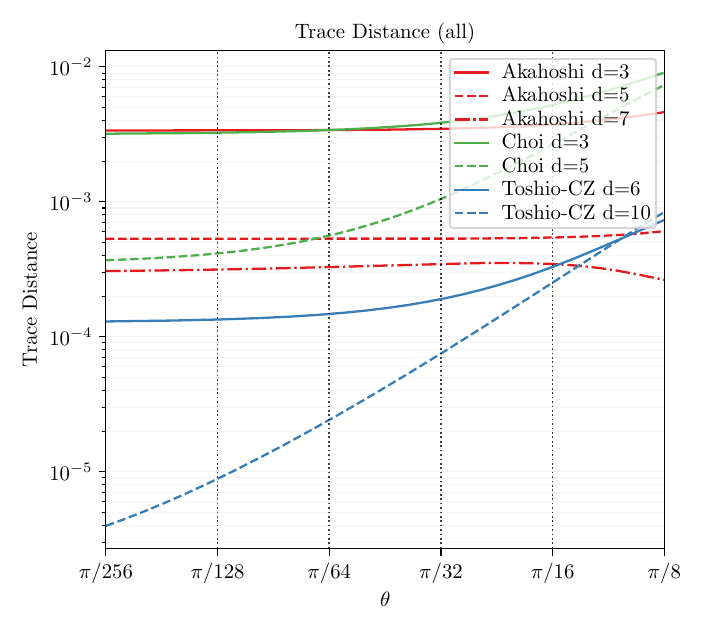}
    \caption{
    \textbf{Trace distance of calibrated resource states across a range of protocols.} We plot the calibrated trace distance between the prepared TMR resource state and the ideal $R_Z(\theta)$ target, as a function of rotation angle $\theta$, under SI1000 noise at physical error rate $p = 10^{-3}$, decoded using correlated PyMatching. We include three rounds of noisy error-correction after the post-selected injection followed by ideal measurement to approximate the noise induced by growth or lattice surgery. Curves show three resource-state constructions (Akahoshi, Choi, and Toshio-CZ) at several code distances $d$; both axes are logarithmic. Two regimes are visible. At small $\theta$, each construction plateaus at a distance-limited floor that falls with increasing $d$ and $k$. At large $\theta$ the trace distance rises with angle as the branch-mixing error dominates, with a slope that steepens towards $2$ as $k$ grows (\cref{app:star_results}). Toshio-CZ with $k=5$ is optimal for all angles in our domain. \Cref{tab:star_td} lists the trace distance for each construction and distance at the four Clifford-hierarchy angles $\theta = \pi/16,\,\pi/32,\,\pi/64,\,\pi/128$ (levels $C^{(4)}$ to $C^{(7)}$), from which we read the optimal schedule per angle used in \cref{fig:star_trotter}.}
    \label{fig:star_td}%
\end{figure*}

\begin{figure*}[tbp]
    \centering
    \includegraphics[width=.7\textwidth]{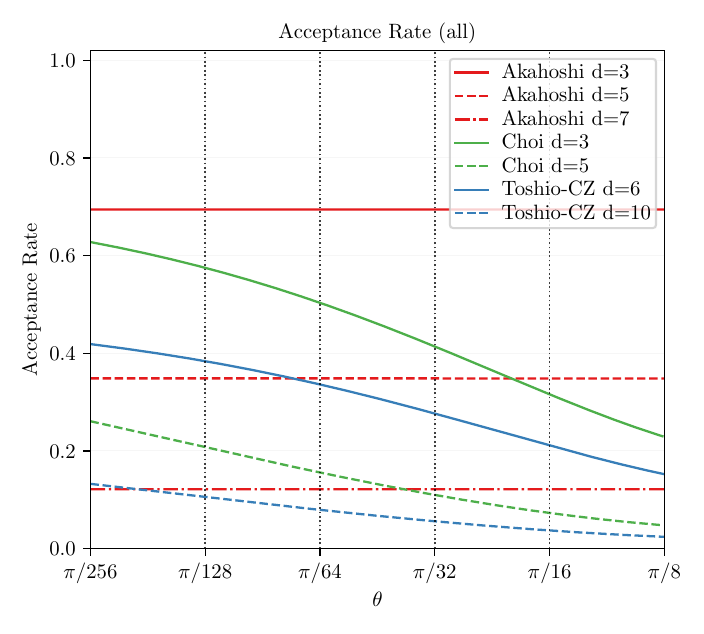}
    \caption{
    \textbf{Acceptance rates for state preparation as a function of angle for every protocol.} Estimated under SI1000 noise at physical error rate $p = 10^{-3}$. These values are independent of decoding or any downstream use of the state. Curves show three resource-state constructions (Akahoshi, Choi, and Toshio-CZ) at several code distances $d$; the $x$ axis is logarithmic and the $y$ axis is linear. The Choi and Toshio-CZ schedules depend on both the injection angle $\theta$ and the distance $d$, while the Akahoshi schedule is angle independent. At small angles the Choi and Toshio-CZ protocols, which only post-select on the next-nearest neighbour stabilisers, have better acceptance rates than the Akahoshi schedule, which post-selects on all stabilisers, even at the same $d$. At large angles, rejection is dominated by the noiseless probability of not selecting the target branch rather than by errors; see our derivation of $N_0$ in \cref{eq:star_coefficients_N_0}}
    \label{fig:star_ac}%
\end{figure*}

\begin{figure*}[tbp]
    \centering
    \includegraphics[width=.6\textwidth]{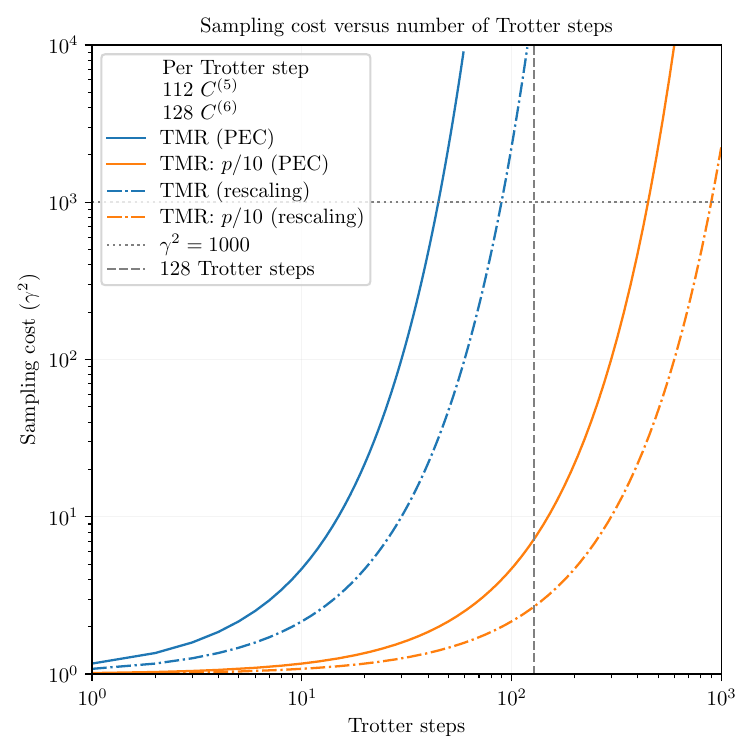}
    \caption{
        \textbf{Cancelling the residual dephasing of STAR rotations is unaffordable at $p = 10^{-3}$ over the Hubbard evolution.} Sampling overhead $\gamma^2$ of cancelling the residual dephasing channel of every STAR rotation, \cref{eq:pec_overhead}, as a function of the number of Trotter steps. Each step contains $112$ rotations by $\pi/32$ and $128$ by $\pi/64$, the Clifford-hierarchy angles ($C^{(5)}$ and $C^{(6)}$) closest to the Trotter angles $0.1$ and $0.05$, each prepared with the optimal schedule of \cref{tab:star_td} and completed by the truncated repeat-until-success ladder including angles only in \cref{tab:star_td} with noiseless $T$ gates. The solid curves indicate using probabilistic error cancellation, with a leading order sampling overhead approximately $(1+2 p_Z)^2$ per rotation. The dashed-dotted curves indicate using rescaling under the same assumptions as the main text, approximately $(1+p_Z)^2$. The blue curve indicates our estimate simulated with SI1000 noise at $p = 10^{-3}$, which exceeds a budget of $\gamma^2 = 10^3$ (dotted) in fewer than $50$ steps when using PEC. The orange line indicates this may be possible with an order of magnitude improvement in the physical error rate. Note these estimates do not include sampling overhead to account for Clifford errors.}
    \label{fig:star_trotter}
\end{figure*}

To estimate these error rates we simulate the TMR circuits, reconstruct the prepared logical state by tomography, calculate the calibrated angle, and compute its trace distance from the target state (\cref{fig:star_td}). We later argue that the residual channel on the resource state after calibration is only a dephasing channel (\cref{fig:harp_calibration}), therefore the trace distance of the resource state equals the diamond distance of the (first-time-success) logical rotation channel, which we prove in \cref{app:star_results}. After removing the coherent error the state infidelity, trace distance, and effective diamond distance of the channel are all related up to a constant. Our simulation also yields the post-selection acceptance rate (\cref{fig:star_ac}).

\begin{table}
    \setlength{\tabcolsep}{5pt}
    \renewcommand{\arraystretch}{1.2}
    \centering
    \begin{tabular}{@{}l c c c c c@{}}
        \hline\hline
        Construction & $d$ & $C^{(4)}$ ($\pi/16$) & $C^{(5)}$ ($\pi/32$) & $C^{(6)}$ ($\pi/64$) & $C^{(7)}$ ($\pi/128$) \\
        \midrule
        \multirow{3}{*}{Akahoshi} & 3 & $3.74\times10^{-3}$ & $3.47\times10^{-3}$ & $3.40\times10^{-3}$ & $3.38\times10^{-3}$ \\
         & 5 & $5.44\times10^{-4}$ & $5.33\times10^{-4}$ & $5.32\times10^{-4}$ & $5.33\times10^{-4}$ \\
         & 7 & $3.46\times10^{-4}$ & $3.47\times10^{-4}$ & $3.29\times10^{-4}$ & $3.15\times10^{-4}$ \\
        \midrule
        \multirow{2}{*}{Choi} & 3 & $5.22\times10^{-3}$ & $3.85\times10^{-3}$ & $3.40\times10^{-3}$ & $3.25\times10^{-3}$ \\
         & 5 & $2.67\times10^{-3}$ & $1.05\times10^{-3}$ & $5.62\times10^{-4}$ & $4.15\times10^{-4}$ \\
        \midrule
        \multirow{2}{*}{Toshio-CZ} & 6 & $3.30\times10^{-4}$ & $1.90\times10^{-4}$ & $1.48\times10^{-4}$ & $1.34\times10^{-4}$ \\
         & 10 & $\mathbf{2.53\times10^{-4}}$ & $\mathbf{7.51\times10^{-5}}$ & $\mathbf{2.42\times10^{-5}}$ & $\mathbf{8.86\times10^{-6}}$ \\
        \midrule
        \multicolumn{2}{c}{Optimal RUS} & $2.53\times10^{-4}$ & $2.01\times10^{-4}$ & $1.25\times10^{-4}$ & $7.13\times10^{-5}$ \\
        \hline\hline
    \end{tabular}
    \caption{Calibrated numeric trace distance at the four Clifford-hierarchy angles under SI1000 noise at $p=10^{-3}$. Bold entries are the per-angle minima. The Optimal RUS row uses $R_4=m_4$ and $R_j=m_j+2^{-(j-j_{\rm prev})}R_{j_{\rm prev}}$, where $j_{\rm prev}$ is the preceding hierarchy level and $m_j$ is the per-angle minimum.}
    \label{tab:star_td}
\end{table}

Error mitigation can be used to remove this remaining dephasing error. This requires an exponential sampling overhead; that is, exponential in the number of Trotter steps we apply, or in the total evolution time. From this we can calculate the total multiplicative sampling overhead as a function of the number of Trotter steps (\cref{fig:star_trotter}). At $p = 10^{-3}$ this overhead exceeds $\gamma^2 = 10^3$ in fewer than 50 Trotter steps (solid lines). These sampling overhead estimates assume the use of probabilistic error cancellation (PEC), which follows the convention in the STAR literature. However, in the main text we consider an alternative error mitigation technique based on rescaling the obtained signal. This latter technique could also likely be applied to STAR, and would require only the square root of these overheads (under the ``scrambled observable'' assumption used in the main text). We plot these more optimistic error mitigation overheads as dashed-dotted lines. However, this does not change our conclusion: $128$ Trotter steps would then require a sampling overhead of $2 \times 10^4$ at $p=10^{-3}$, before accounting for Clifford errors, and assuming perfect calibration of coherent errors. Therefore we dismiss using the state preparation techniques used in STAR versions 1 and 2 at this error rate for our simulation; a tenfold reduction in every trace distance, a conservative proxy for the results that would be achieved at $p = 10^{-4}$, would keep all $128$ Trotter steps within this additional sampling budget.

This conclusion is specific to our noise model, our angle range, and the enhancements that we modelled. Restating our results narrowly: our simulations show that, under SI1000 noise, and using TMR to apply rotations higher in the Clifford hierarchy than $C^{(3)}$, a sampling overhead greater than $1000$ is required before we can evolve for the total number of Trotter steps. Therefore we cannot simulate our system of interest without further modifications at the logical level.

Finally, we note results that could modify this conclusion. Ref.~\cite{Zeng2025} argues that, in certain setups, the leading order of the error rate should be $\mathcal{O}(p^2\theta)$ for the full RUS channel, based on constructions where a two-qubit rotation $R_{Z_{i_1}\otimes Z_{i_2}}(\phi)$ does not admit the stochastic error $Z_{i_1}\otimes Z_{i_2}$ at leading order in $p$, so at least two weight-one errors are needed to mix the terms of \cref{eq:star_operator}.
Constructions that post-select on fewer stabilisers, in particular the distance-2 data-qubit $R_{ZZ}$ approach of \cite{Zeng2025} (distinct from the $\mathrm{CX} \,R_Z(\phi)\, \mathrm{CX}$ construction we simulate), may lower the per-rotation error at large angles; we have not modelled these. At the level of simulation this is equivalent to defining a bespoke noise model for these two-qubit non-Clifford gates. Overall, we therefore emphasise that our results and conclusion are dependent on the noise model and gate set assumptions made.

\section{Further details of STAR architecture}\label{app:star_results}

This appendix supports \cref{app:star_primary}. In this appendix, we recapitulate the TMR variants we consider; derive the three error mechanisms that set the achievable state-preparation error at each angle; state the exact expressions of $\theta_a(\phi)$ and $N_a(\phi)$; establish the trace-distance and diamond-distance results that let us report a single figure of merit; distinguish the two mitigation strategies that we use; describe how we model growth; recall the RUS rates and their protocol-switching regimes; and finally compare with values of $\alpha_{\mathrm{RUS}}$ from the literature. Throughout, $p$ is the physical error rate, $d$ the code distance, and $k$ the number of terms in the logical decomposition of the TMR operator below.

As discussed above, for our final Trotter-step costing we considered \textit{Akahoshi}, \textit{Choi}, and \textit{Toshio-CZ}. The error-rate of \textit{Akahoshi} has only a weak dependence on $\theta$, while the latter two protocols have strong $\theta$ dependence, as discussed in \cref{app:star_primary}. Several other constructions for the TMR protocol require SWAP between data and ancilla qubits, treat $R_{ZZ}$ as a native gate, or apply two-qubit gates between data qubits. We did not consider these for our final Trotter costing, though we did reproduce one of them, \textit{Toshio-SWAP}, to compare with the existing literature \cite{Toshio2025}. We compiled our three candidate protocols using a standard superconducting gate set and performed state tomography under SI1000 noise. In particular, we focused on circuits where CZ is the only native two-qubit gate used, as is the case in many current QEC experiments on real devices \cite{google2023, google2025, Caune2026}. This is a standard assumption, and we believe that it would be optimistic to assume that two separate two-qubit gates (e.g. CX and SWAP) are both sufficiently well calibrated on a large-scale superconducting device. However, note that this differs from other studies of the STAR architecture, which have, for example, taken SWAP to be a native gate \cite{Toshio2025, Chung2026}.

Each of these constructions defines a unique set of noise operators that can mix the branches. The collective probability of one of these errors occurring during any shot on each individual $R_Z$ rotation is related to the physical error rate $p$ by a noise-model dependent constant. We use $p_\textit{UD}$ to denote the probability of such an undetectable error. This sets the prefactor for the error rate, but not the exponent, thus the error rate of the protocol shifts linearly in changes to $p_\textit{UD}$, at least in the small-angle limit. The value of $p_\textit{UD}$ depends on which version of STAR, whether we are using single-qubit or multi-qubit rotations, what noise model and what native gates we use. Specifically, $p_\textit{UD}$ is significantly higher if we have any single-qubit rotation, such as originally defined for the \emph{Choi} operator. This is because, in addition to errors on the one- and two-qubit gates that implement the rotation, idling errors and measurement resonator errors also mix the relevant branches, and these effects are significantly worse, for SI1000 noise, when an error on any single data qubit can mix the terms of \cref{eq:star_operator}. Therefore, considering only $m=2$ or $k=d/2$, is optimal.

\begin{figure*}[tbp]
    \centering
    \newsavebox{\starbox}
    \tikzset{
        xstab/.style={fill=red!35},
        zstab/.style={fill=blue!30},
        dq/.style={circle, draw=black, fill=white, inner sep=0pt, minimum size=4pt},
        zq/.style={circle, draw=black, fill=green!55!black, inner sep=0pt, minimum size=4pt},
        rot/.style={draw=violet, very thick, rounded corners=2pt},
        anc/.style={rectangle, draw=black, fill=orange, inner sep=0pt, minimum size=3.5pt},
        growbox/.style={draw=black, dashed},
        growarrow/.style={->, thick},
    }
    \begin{minipage}[t]{0.58\linewidth}%
        \centering
        (a)\\[0.2em]
        \begin{lrbox}{\starbox}%
        \begin{quantikz}[row sep=0.18cm, column sep=0.14cm]
\lstick{$a_1\;\ket{+}$} & \ctrl{1} & \qw & \qw & \qw & \qw & \qw & \qw & \ctrl{2} & \gate{H} & \ctrl{2} & \gate{H} & \ctrl{2} & \qw & \qw & \qw & \gate[7]{\text{grow to }d} & \qw \\
\lstick{$q_1\;\ket{+}$} & \control{} & \qw & \ctrl{2} & \gate{H} & \gate{R_Z(\phi)} & \gate{H} & \ctrl{2} & \qw & \qw & \qw & \qw & \qw & \qw & \qw & \qw & \qw & \qw \\
\lstick{$q_2\;\ket{+}$} & \qw & \qw & \qw & \qw & \qw & \qw & \qw & \control{} & \gate{H} & \control{} & \gate{H} & \control{} & \gate{H} & \qw & \qw & \qw & \qw \\
\lstick{$a_2\;\ket{+}$} & \ctrl{2} & \gate{H} & \control{} & \qw & \qw & \qw & \control{} & \qw & \qw & \ctrl{1} & \gate{H} & \ctrl{1} & \gate{H} & \ctrl{1} & \qw & \qw & \qw \\
\lstick{$q_3\;\ket{0}$} & \qw & \qw & \qw & \qw & \qw & \qw & \qw & \qw & \qw & \control{} & \gate{H} & \control{} & \gate{H} & \control{} & \gate{H} & \qw & \qw \\
\lstick{$q_4\;\ket{+}$} & \control{} & \gate{H} & \qw & \qw & \qw & \qw & \qw & \qw & \qw & \qw & \qw & \qw & \qw & \qw & \qw & \qw & \qw \\
\lstick{$a_3\;\ket{+}$} & \qw & \qw & \qw & \qw & \qw & \qw & \qw & \qw & \qw & \qw & \qw & \qw & \qw & \qw & \qw & \qw & \qw
        \end{quantikz}%
        \end{lrbox}%
        \resizebox{\linewidth}{!}{\usebox{\starbox}}%
    \end{minipage}\hfill%
    \begin{minipage}[t]{0.38\linewidth}%
        \centering
        (b)\\[0.2em]
        \begin{lrbox}{\starbox}%
        \begin{quantikz}[row sep=0.25cm, column sep=0.2cm]
\lstick{$q_1$} & \ctrl{1} & \qw & \qw & \qw & \qw & \qw & \ctrl{1} & \qw & \qw \\
\lstick{$a\;\ket{+}$} & \control{} & \ctrl{1} & \gate{H} & \gate{R_Z(\phi)} & \gate{H} & \ctrl{1} & \control{} & \gate{M_X} & \qw \\
\lstick{$q_2$} & \qw & \control{} & \qw & \qw & \qw & \control{} & \qw & \qw & \qw
        \end{quantikz}%
        \end{lrbox}%
        \resizebox{\linewidth}{!}{\usebox{\starbox}}\\[0.8em]
        (c)\\[0.2em]
        \begin{lrbox}{\starbox}%
        \begin{quantikz}[row sep={0.6cm,between origins}, column sep=0.2cm]
\lstick{$q_1$} & \swap{1} & \qw &  \swap{1} & \qw \\
\lstick{$a$} & \targX{} & \gate[2]{R_{Z_1 X_2}(\phi)} &  \targX{} & \qw \\
\lstick{$q_2$} & \gate{H} & \qw & \gate{H}  & \qw
        \end{quantikz}%
        \end{lrbox}%
        \resizebox{0.55\linewidth}{!}{\usebox{\starbox}}%
    \end{minipage}

    \vspace{1em}
    \begin{minipage}[t]{0.24\linewidth}%
        \centering
        (d)\\[0.2em]
\begin{tikzpicture}[scale=0.3, every node/.style={transform shape=false}]
\fill[xstab, opacity=0.25] (1,3) arc[start angle=90,end angle=270,radius=1]--cycle;
\fill[xstab, opacity=0.25] (1,7) arc[start angle=90,end angle=270,radius=1]--cycle;
\fill[xstab, opacity=0.25] (1,3)--(3,3)--(3,5)--(1,5)--cycle;
\fill[xstab, opacity=0.25] (1,7)--(3,7)--(3,9)--(1,9)--cycle;
\fill[xstab, opacity=0.25] (3,1)--(5,1)--(5,3)--(3,3)--cycle;
\fill[xstab, opacity=0.25] (3,5)--(5,5)--(5,7)--(3,7)--cycle;
\fill[xstab, opacity=0.25] (5,3)--(7,3)--(7,5)--(5,5)--cycle;
\fill[xstab, opacity=0.25] (5,7)--(7,7)--(7,9)--(5,9)--cycle;
\fill[xstab, opacity=0.25] (7,1)--(9,1)--(9,3)--(7,3)--cycle;
\fill[xstab, opacity=0.25] (7,5)--(9,5)--(9,7)--(7,7)--cycle;
\fill[xstab, opacity=0.25] (9,3) arc[start angle=-90,end angle=90,radius=1]--cycle;
\fill[xstab, opacity=0.25] (9,7) arc[start angle=-90,end angle=90,radius=1]--cycle;
\fill[zstab, opacity=0.25] (3,1) arc[start angle=180,end angle=360,radius=1]--cycle;
\fill[zstab, opacity=0.25] (7,1) arc[start angle=180,end angle=360,radius=1]--cycle;
\fill[zstab, opacity=0.25] (1,1)--(3,1)--(3,3)--(1,3)--cycle;
\fill[zstab, opacity=0.25] (5,1)--(7,1)--(7,3)--(5,3)--cycle;
\fill[zstab, opacity=0.25] (3,3)--(5,3)--(5,5)--(3,5)--cycle;
\fill[zstab, opacity=0.25] (7,3)--(9,3)--(9,5)--(7,5)--cycle;
\fill[zstab, opacity=0.25] (1,5)--(3,5)--(3,7)--(1,7)--cycle;
\fill[zstab, opacity=0.25] (5,5)--(7,5)--(7,7)--(5,7)--cycle;
\fill[zstab, opacity=0.25] (3,7)--(5,7)--(5,9)--(3,9)--cycle;
\fill[zstab, opacity=0.25] (7,7)--(9,7)--(9,9)--(7,9)--cycle;
\fill[zstab, opacity=0.25] (3,9) arc[start angle=0,end angle=180,radius=1]--cycle;
\fill[zstab, opacity=0.25] (7,9) arc[start angle=0,end angle=180,radius=1]--cycle;
\fill[zstab] (1,1)--(3,1)--(3,3)--(1,3)--cycle;
\fill[xstab] (1,3) arc[start angle=90,end angle=270,radius=1]--cycle;
\fill[xstab] (3,1) arc[start angle=-90,end angle=90,radius=1]--cycle;
\node[zq] at (1,1) {};
\node[zq] at (1,3) {};
\node[dq, opacity=0.25] at (1,5) {};
\node[dq, opacity=0.25] at (1,7) {};
\node[dq, opacity=0.25] at (1,9) {};
\node[dq] at (3,1) {};
\node[dq] at (3,3) {};
\node[dq, opacity=0.25] at (3,5) {};
\node[dq, opacity=0.25] at (3,7) {};
\node[dq, opacity=0.25] at (3,9) {};
\node[dq, opacity=0.25] at (5,1) {};
\node[dq, opacity=0.25] at (5,3) {};
\node[dq, opacity=0.25] at (5,5) {};
\node[dq, opacity=0.25] at (5,7) {};
\node[dq, opacity=0.25] at (5,9) {};
\node[dq, opacity=0.25] at (7,1) {};
\node[dq, opacity=0.25] at (7,3) {};
\node[dq, opacity=0.25] at (7,5) {};
\node[dq, opacity=0.25] at (7,7) {};
\node[dq, opacity=0.25] at (7,9) {};
\node[dq, opacity=0.25] at (9,1) {};
\node[dq, opacity=0.25] at (9,3) {};
\node[dq, opacity=0.25] at (9,5) {};
\node[dq, opacity=0.25] at (9,7) {};
\node[dq, opacity=0.25] at (9,9) {};
\draw[rot] (0.45,0.45) rectangle (1.55,3.55);
\draw[growbox] (-0.9,-0.9) rectangle (4.3,4.3);
\draw[growarrow] (4.4,4.4) -- (7.6,7.6);
\end{tikzpicture}

    \end{minipage}\hfill%
    \begin{minipage}[t]{0.24\linewidth}%
        \centering
        (e)\\[0.2em]
\begin{tikzpicture}[scale=0.3, every node/.style={transform shape=false}]
\fill[xstab] (1,3) arc[start angle=90,end angle=270,radius=1]--cycle;
\fill[xstab] (1,7) arc[start angle=90,end angle=270,radius=1]--cycle;
\fill[xstab] (1,3)--(3,3)--(3,5)--(1,5)--cycle;
\fill[xstab] (1,7)--(3,7)--(3,9)--(1,9)--cycle;
\fill[xstab] (3,1)--(5,1)--(5,3)--(3,3)--cycle;
\fill[xstab] (3,5)--(5,5)--(5,7)--(3,7)--cycle;
\fill[xstab] (5,3)--(7,3)--(7,5)--(5,5)--cycle;
\fill[xstab] (5,7)--(7,7)--(7,9)--(5,9)--cycle;
\fill[xstab] (7,1)--(9,1)--(9,3)--(7,3)--cycle;
\fill[xstab] (7,5)--(9,5)--(9,7)--(7,7)--cycle;
\fill[xstab] (9,3) arc[start angle=-90,end angle=90,radius=1]--cycle;
\fill[xstab] (9,7) arc[start angle=-90,end angle=90,radius=1]--cycle;
\fill[zstab] (3,1) arc[start angle=180,end angle=360,radius=1]--cycle;
\fill[zstab] (7,1) arc[start angle=180,end angle=360,radius=1]--cycle;
\fill[zstab] (1,1)--(3,1)--(3,3)--(1,3)--cycle;
\fill[zstab] (5,1)--(7,1)--(7,3)--(5,3)--cycle;
\fill[zstab] (3,3)--(5,3)--(5,5)--(3,5)--cycle;
\fill[zstab] (7,3)--(9,3)--(9,5)--(7,5)--cycle;
\fill[zstab] (1,5)--(3,5)--(3,7)--(1,7)--cycle;
\fill[zstab] (5,5)--(7,5)--(7,7)--(5,7)--cycle;
\fill[zstab] (3,7)--(5,7)--(5,9)--(3,9)--cycle;
\fill[zstab] (7,7)--(9,7)--(9,9)--(7,9)--cycle;
\fill[zstab] (3,9) arc[start angle=0,end angle=180,radius=1]--cycle;
\fill[zstab] (7,9) arc[start angle=0,end angle=180,radius=1]--cycle;
\node[zq] at (1,1) {};
\node[zq] at (1,3) {};
\node[zq] at (1,5) {};
\node[zq] at (1,7) {};
\node[zq] at (1,9) {};
\node[dq] at (3,1) {};
\node[dq] at (3,3) {};
\node[dq] at (3,5) {};
\node[dq] at (3,7) {};
\node[dq] at (3,9) {};
\node[dq] at (5,1) {};
\node[dq] at (5,3) {};
\node[dq] at (5,5) {};
\node[dq] at (5,7) {};
\node[dq] at (5,9) {};
\node[dq] at (7,1) {};
\node[dq] at (7,3) {};
\node[dq] at (7,5) {};
\node[dq] at (7,7) {};
\node[dq] at (7,9) {};
\node[dq] at (9,1) {};
\node[dq] at (9,3) {};
\node[dq] at (9,5) {};
\node[dq] at (9,7) {};
\node[dq] at (9,9) {};
\draw[rot] (0.45,0.45) rectangle (1.55,1.55);
\draw[rot] (0.45,2.45) rectangle (1.55,3.55);
\draw[rot] (0.45,4.45) rectangle (1.55,5.55);
\draw[rot] (0.45,6.45) rectangle (1.55,7.55);
\draw[rot] (0.45,8.45) rectangle (1.55,9.55);
\end{tikzpicture}

    \end{minipage}\hfill%
    \begin{minipage}[t]{0.24\linewidth}%
        \centering
        (f)\\[0.2em]
\begin{tikzpicture}[scale=0.3, every node/.style={transform shape=false}]
\fill[xstab] (1,3) arc[start angle=90,end angle=270,radius=1]--cycle;
\fill[xstab] (1,7) arc[start angle=90,end angle=270,radius=1]--cycle;
\fill[xstab] (1,3)--(3,3)--(3,5)--(1,5)--cycle;
\fill[xstab] (1,7)--(3,7)--(3,9)--(1,9)--cycle;
\fill[xstab] (3,1)--(5,1)--(5,3)--(3,3)--cycle;
\fill[xstab] (3,5)--(5,5)--(5,7)--(3,7)--cycle;
\fill[xstab] (5,3)--(7,3)--(7,5)--(5,5)--cycle;
\fill[xstab] (5,7)--(7,7)--(7,9)--(5,9)--cycle;
\fill[xstab] (7,1)--(9,1)--(9,3)--(7,3)--cycle;
\fill[xstab] (7,5)--(9,5)--(9,7)--(7,7)--cycle;
\fill[xstab] (9,3) arc[start angle=-90,end angle=90,radius=1]--cycle;
\fill[xstab] (9,7) arc[start angle=-90,end angle=90,radius=1]--cycle;
\fill[zstab] (3,1) arc[start angle=180,end angle=360,radius=1]--cycle;
\fill[zstab] (7,1) arc[start angle=180,end angle=360,radius=1]--cycle;
\fill[zstab] (1,1)--(3,1)--(3,3)--(1,3)--cycle;
\fill[zstab] (5,1)--(7,1)--(7,3)--(5,3)--cycle;
\fill[zstab] (3,3)--(5,3)--(5,5)--(3,5)--cycle;
\fill[zstab] (7,3)--(9,3)--(9,5)--(7,5)--cycle;
\fill[zstab] (1,5)--(3,5)--(3,7)--(1,7)--cycle;
\fill[zstab] (5,5)--(7,5)--(7,7)--(5,7)--cycle;
\fill[zstab] (3,7)--(5,7)--(5,9)--(3,9)--cycle;
\fill[zstab] (7,7)--(9,7)--(9,9)--(7,9)--cycle;
\fill[zstab] (3,9) arc[start angle=0,end angle=180,radius=1]--cycle;
\fill[zstab] (7,9) arc[start angle=0,end angle=180,radius=1]--cycle;
\node[zq] at (1,1) {};
\node[zq] at (1,3) {};
\node[zq] at (1,5) {};
\node[zq] at (1,7) {};
\node[zq] at (1,9) {};
\node[dq] at (3,1) {};
\node[dq] at (3,3) {};
\node[dq] at (3,5) {};
\node[dq] at (3,7) {};
\node[dq] at (3,9) {};
\node[dq] at (5,1) {};
\node[dq] at (5,3) {};
\node[dq] at (5,5) {};
\node[dq] at (5,7) {};
\node[dq] at (5,9) {};
\node[dq] at (7,1) {};
\node[dq] at (7,3) {};
\node[dq] at (7,5) {};
\node[dq] at (7,7) {};
\node[dq] at (7,9) {};
\node[dq] at (9,1) {};
\node[dq] at (9,3) {};
\node[dq] at (9,5) {};
\node[dq] at (9,7) {};
\node[dq] at (9,9) {};
\draw[rot] (0.45,0.45) rectangle (1.55,3.55);
\draw[rot] (0.45,4.45) rectangle (1.55,7.55);
\draw[rot] (0.45,8.45) rectangle (1.55,9.55);
\node[anc] at (0,2) {};
\node[anc] at (0,6) {};
\end{tikzpicture}

    \end{minipage}\hfill%
    \begin{minipage}[t]{0.24\linewidth}%
        \centering
        (g)\\[0.2em]
\begin{tikzpicture}[scale=0.3, every node/.style={transform shape=false}]
\fill[xstab] (1,3) arc[start angle=90,end angle=270,radius=1]--cycle;
\fill[xstab] (1,7) arc[start angle=90,end angle=270,radius=1]--cycle;
\fill[xstab] (1,11) arc[start angle=90,end angle=270,radius=1]--cycle;
\fill[xstab] (1,3)--(3,3)--(3,5)--(1,5)--cycle;
\fill[xstab] (1,7)--(3,7)--(3,9)--(1,9)--cycle;
\fill[xstab] (3,1)--(5,1)--(5,3)--(3,3)--cycle;
\fill[xstab] (3,5)--(5,5)--(5,7)--(3,7)--cycle;
\fill[xstab] (3,9)--(5,9)--(5,11)--(3,11)--cycle;
\fill[xstab] (5,3)--(7,3)--(7,5)--(5,5)--cycle;
\fill[xstab] (5,7)--(7,7)--(7,9)--(5,9)--cycle;
\fill[xstab] (7,1)--(9,1)--(9,3)--(7,3)--cycle;
\fill[xstab] (7,5)--(9,5)--(9,7)--(7,7)--cycle;
\fill[xstab] (7,9)--(9,9)--(9,11)--(7,11)--cycle;
\fill[xstab] (9,3)--(11,3)--(11,5)--(9,5)--cycle;
\fill[xstab] (9,7)--(11,7)--(11,9)--(9,9)--cycle;
\fill[xstab] (11,1) arc[start angle=-90,end angle=90,radius=1]--cycle;
\fill[xstab] (11,5) arc[start angle=-90,end angle=90,radius=1]--cycle;
\fill[xstab] (11,9) arc[start angle=-90,end angle=90,radius=1]--cycle;
\fill[zstab] (1,1)--(3,1)--(3,3)--(1,3)--cycle;
\fill[zstab] (1,5)--(3,5)--(3,7)--(1,7)--cycle;
\fill[zstab] (1,9)--(3,9)--(3,11)--(1,11)--cycle;
\fill[zstab] (3,1) arc[start angle=180,end angle=360,radius=1]--cycle;
\fill[zstab] (3,3)--(5,3)--(5,5)--(3,5)--cycle;
\fill[zstab] (3,7)--(5,7)--(5,9)--(3,9)--cycle;
\fill[zstab] (5,11) arc[start angle=0,end angle=180,radius=1]--cycle;
\fill[zstab] (5,1)--(7,1)--(7,3)--(5,3)--cycle;
\fill[zstab] (5,5)--(7,5)--(7,7)--(5,7)--cycle;
\fill[zstab] (5,9)--(7,9)--(7,11)--(5,11)--cycle;
\fill[zstab] (7,1) arc[start angle=180,end angle=360,radius=1]--cycle;
\fill[zstab] (7,3)--(9,3)--(9,5)--(7,5)--cycle;
\fill[zstab] (7,7)--(9,7)--(9,9)--(7,9)--cycle;
\fill[zstab] (9,11) arc[start angle=0,end angle=180,radius=1]--cycle;
\fill[zstab] (9,1)--(11,1)--(11,3)--(9,3)--cycle;
\fill[zstab] (9,5)--(11,5)--(11,7)--(9,7)--cycle;
\fill[zstab] (9,9)--(11,9)--(11,11)--(9,11)--cycle;
\node[zq] at (1,1) {};
\node[zq] at (1,3) {};
\node[zq] at (1,5) {};
\node[zq] at (1,7) {};
\node[zq] at (1,9) {};
\node[zq] at (1,11) {};
\node[dq] at (3,1) {};
\node[dq] at (3,3) {};
\node[dq] at (3,5) {};
\node[dq] at (3,7) {};
\node[dq] at (3,9) {};
\node[dq] at (3,11) {};
\node[dq] at (5,1) {};
\node[dq] at (5,3) {};
\node[dq] at (5,5) {};
\node[dq] at (5,7) {};
\node[dq] at (5,9) {};
\node[dq] at (5,11) {};
\node[dq] at (7,1) {};
\node[dq] at (7,3) {};
\node[dq] at (7,5) {};
\node[dq] at (7,7) {};
\node[dq] at (7,9) {};
\node[dq] at (7,11) {};
\node[dq] at (9,1) {};
\node[dq] at (9,3) {};
\node[dq] at (9,5) {};
\node[dq] at (9,7) {};
\node[dq] at (9,9) {};
\node[dq] at (9,11) {};
\node[dq] at (11,1) {};
\node[dq] at (11,3) {};
\node[dq] at (11,5) {};
\node[dq] at (11,7) {};
\node[dq] at (11,9) {};
\node[dq] at (11,11) {};
\draw[rot] (0.45,0.45) rectangle (1.55,3.55);
\draw[rot] (0.45,4.45) rectangle (1.55,7.55);
\draw[rot] (0.45,8.45) rectangle (1.55,11.55);
\node[anc] at (0,2) {};
\node[anc] at (0,6) {};
\node[anc] at (0,10) {};
\end{tikzpicture}

    \end{minipage}
    \caption{\textbf{Resource-state preparation circuits and rotation layouts.} (a)~CZ-native compilation of the single-rotation injection of Ref.~\cite{Akahoshi2024} (Akahoshi), shown on the seven qubits of the distance-2 corner of the patch: data qubits $q_1,\ldots,q_4$ and the measure qubits $a_1,a_2,a_3$ of the $d=2$ stabilisers. A single physical rotation $R_Z(\phi)$ on $q_1$ is entangled into the $d=2$ code by an $H$/CZ network, after which the patch is grown to distance $d$ and post-selected. (b)~Toshio-CZ weight-two rotation $R_{Z_{q_1}Z_{q_2}}(\phi)$ through an ancilla qubit $a$, with CZ as the only two-qubit gate: $a$ is prepared in $\ket{+}$, entangled with both data qubits, rotated, disentangled, and measured in the $X$ basis, with post-selection on $+1$. A single operator on $a$ after $R_Z(\phi)$ propagates to $Z_{q_1}Z_{q_2}$. (c)~Toshio-SWAP/RZX, the construction of Ref.~\cite[Fig.~5(c)]{Toshio2025}, which assumes native SWAP and $R_{ZX}$ gates. (d)--(g)~Rotation layouts on rotated surface-code patches of distance $d=5$ (d--f) and $d=6$ (g), after Ref.~\cite[Fig.~4]{Toshio2025}. Red (blue) faces are $X$ ($Z$) stabilisers, green qubits support $Z_L$, and purple boxes group the qubits of each physical rotation. (d)~$(m,k,d)=(2,1,2)$: a single weight-two rotation on a $d=2$ patch, which is then grown (arrow); this is essentially equivalent to (a). (e)~Choi: $k=d=5$ single-qubit rotations. (f)~Toshio-CZ at odd $d=5$: $k=3$, two weight-two rotations implemented as in (b) through ancilla qubits (orange) and one single-qubit rotation. (g)~Toshio-CZ at even $d=6$: $k=3$ weight-two rotations, with no single-qubit rotation left over.}
    \label{fig:star_circuit}
\end{figure*}
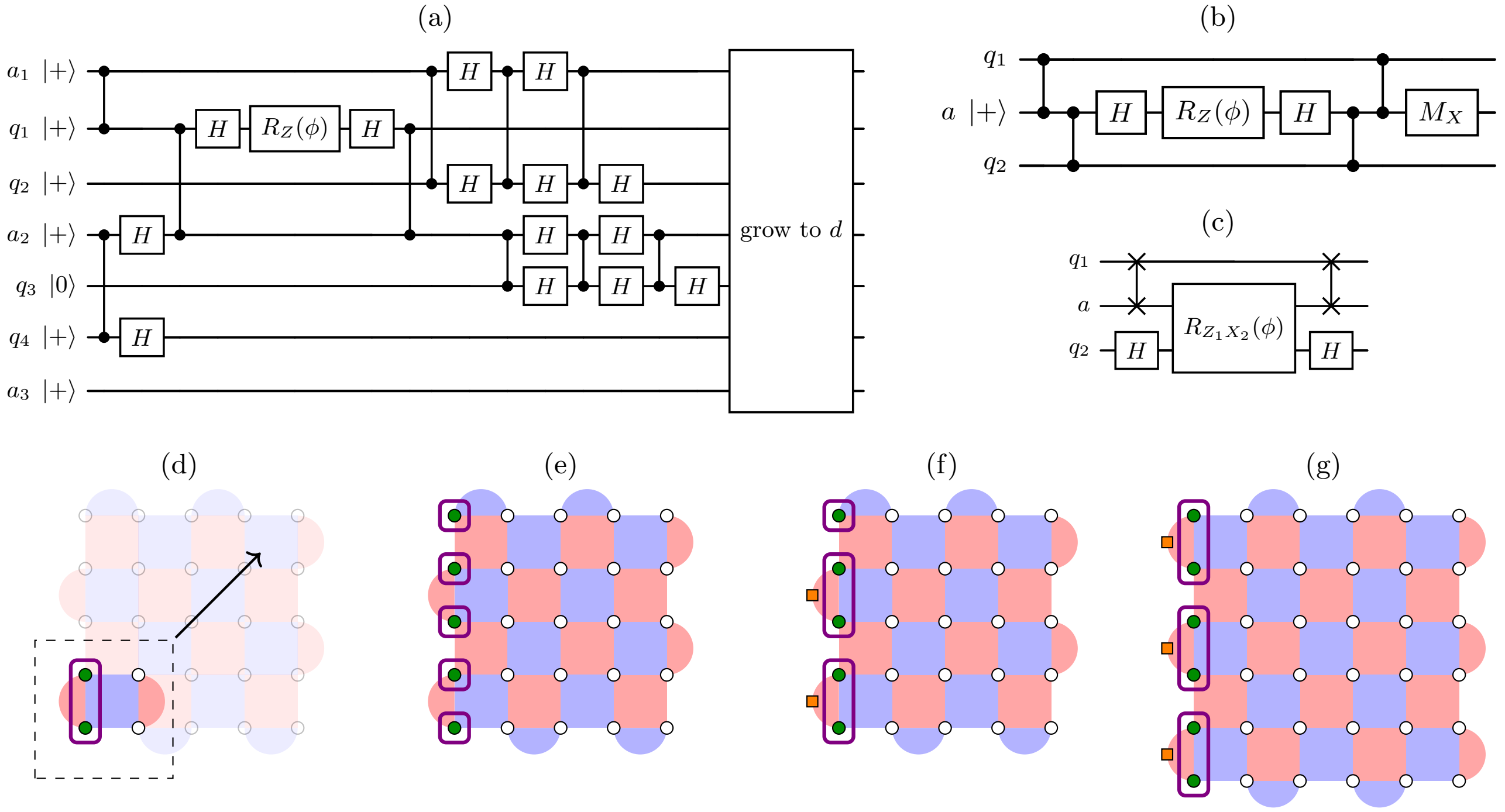

Next, we review generic facts about error channels that act entirely in the equator of the Bloch sphere. We study several error metrics of a channel such as the diamond norm error or channel infidelity. However, for a teleportation gadget where the Clifford operators are noiseless, it is easier to study metrics on the space of states, such as the trace distance between two states or the state infidelity. In the general case the trace distance bounds the diamond norm. However, given a protocol for removing or handling the coherent error, we can show these are equal. Further, we can convert these to the sampling cost of performing subsequent error mitigation.

The trace distance between two states $\sigma, \sigma'$ is $D(\sigma,\sigma') = \tfrac12\lVert \sigma - \sigma'\rVert_1 = \tfrac12\sum_i |\lambda_i|$, with $\{\lambda_i\}$ the eigenvalues of $\sigma - \sigma'$. The diamond distance of an error channel $\mathcal{E}$ from the identity is $\varepsilon_\diamond(\mathcal{E}) = \tfrac12\lVert \mathcal{E} - \mathcal{I}\rVert_\diamond$. For a channel combining a coherent $Z$-rotation by $\delta\theta$ with dephasing of probability $p$,

\begin{equation}\label{eq:rot_plus_dephasing}
    \mathcal{E}(\rho) = e^{-i\delta\theta Z}\big[(1-p)\rho + p Z\rho Z\big]e^{i\delta\theta Z}
    = (1-x)\rho + x Z\rho Z - iy\,(Z\rho - \rho Z),
\end{equation}

\noindent the diamond distance is known exactly \cite{Kueng2016}, and in the form used here \cite{Huang2019} and \cite{Toshio2025},

\begin{equation}\label{eq:diamond_general}
    \varepsilon_\diamond(\mathcal{E}) = \tfrac12\big|1 - (1-2p)e^{-2i\delta\theta}\big| = \sqrt{x^2 + y^2} \approx \sqrt{p^2 + \delta\theta^2}.
\end{equation}

The prepared resource state, $\sigma$, is always equatorial on the Bloch sphere. Any such state is an $R_Z(\theta)$ rotation of $\ket{+}$ dephased with probability $p_Z$. Writing $\ket{\theta} = R_Z(\theta)\ket{+}$,

\begin{equation}\label{eq:resource_state}
    \sigma = (1-p_Z)\ket{\theta}\!\bra{\theta} + p_Z\, Z\ket{\theta}\!\bra{\theta}Z = (1-p_Z)\ket{\theta}\!\bra{\theta} + p_Z\ket{\theta+\pi/2}\!\bra{\theta+\pi/2}.
\end{equation}

\noindent Let us now denote the desired resource state as $\sigma'$. When the target angle matches, we have $\sigma' = \ket{\theta}\!\bra{\theta}$, so $\sigma - \sigma' = -p_Z\ket{\theta}\!\bra{\theta} + p_Z\ket{\theta+\pi/2}\!\bra{\theta+\pi/2}$. This has eigenvalues $\pm p_Z$, and so the trace distance is $D(\sigma,\sigma') = p_Z$. Teleporting $\sigma$, on a successful first attempt, implements

\begin{equation}\label{eq:rotation_channel_with_dephasing}
    \mathcal{N}_\theta(\rho) = (1-p_Z)\,\mathcal{R}_\theta(\rho) + p_Z\,\mathcal{Z}\circ\mathcal{R}_\theta(\rho),
    \qquad \mathcal{R}_\theta(\rho) = R_Z(\theta)\rho R_Z^\dagger(\theta),\quad \mathcal{Z}(\rho) = Z\rho Z,
\end{equation}

\noindent that is, the target rotation composed with the residual error channel $\mathcal{E} = (1-p_Z)\mathcal{I} + p_Z\mathcal{Z}$. This is pure dephasing, so by \cref{eq:diamond_general} with $\delta\theta = 0$,

\begin{equation}\label{eq:calibrated_floor}
    D(\sigma,\sigma') = \varepsilon_\diamond(\mathcal{E}) = p_Z.
\end{equation}

The calibrated resource-state trace distance and the teleported-channel diamond distance coincide, so we may report the trace distance in \cref{fig:star_td} as the operationally relevant quantity. When the prepared angle is instead offset from the target by $\delta\theta$, the residual channel acquires a coherent correction and \cref{eq:diamond_general} gives $\varepsilon_\diamond(\mathcal{E}) \approx \sqrt{p_Z^2 + \delta\theta^2}$.

Previous studies of STAR \cite{Toshio2025} introduce probabilistic coherent error cancellation (PCEC). They consider the error to be a small probability $q_1$ of some coherent error $R_{Z_L}(\Delta)$, where $\Delta=\theta_1-\theta_0$. After a successful teleportation, with probability $q_1$ they apply an additional rotation by $-\Delta$. This converts the coherent offset into Pauli-$Z$ dephasing of probability $2 q_1\sin^2\Delta$. The dephasing error, both from the TMR state preparation and induced by PCEC, can then be removed by probabilistic error cancellation (PEC) which we also use and is explained below after we define calibration. This mitigates all error, at some sampling cost, which is related to the size of the dephasing error that PEC must remove.

\begin{figure*}
    \centering
    \includegraphics[width=0.65\textwidth]{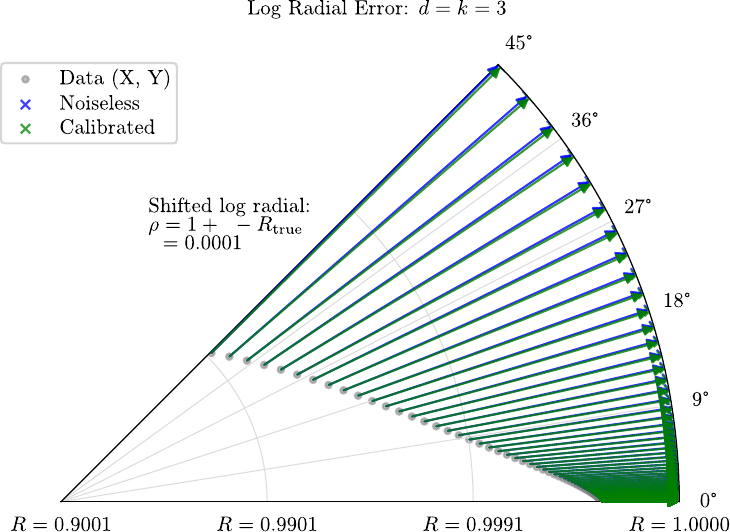}
      \caption{\textbf{Calibration removes the coherent angle error but not the radial dephasing which sets the error floor.}
          Each grey point is the measured equatorial Bloch vector $(\langle X\rangle, \langle Y\rangle)$ of a prepared resource state, shown in polar coordinates with azimuth equal to the rotation angle ($0^\circ$ to $45^\circ$) and a shifted logarithmic radial axis that plots $1 + \varepsilon - r$ with $\varepsilon = 10^{-4}$. The edge points ($r = 1$) correspond to pure states and radial distance inwards shows the loss of Bloch-vector length, $1 - r$ due to an effective dephasing channel. Each point has two arrows. The blue arrow points to the noiseless target at angle $\theta$, and the green arrow points to the calibrated target at angle $\theta'$, chosen so that $\theta' + \delta\theta(\theta') = \theta$.}
      \label{fig:harp_calibration}
\end{figure*}

In our work, we instead consider the error to be a definite shift in the target angle, together with a dephasing error of size $p_Z$. That is, we use our tomographic results to define the ``calibrated'' target angle, $\theta_0'(\phi)$ which is different from the ``noiseless'' angle $\theta_0(\phi)$, as illustrated in \cref{fig:harp_calibration}. This requires that we know $\delta\theta_0(\phi)$ in advance and that it is stable across the computation, which is also a requirement of PCEC. To restate, our protocol would be to experimentally find the change in angle from a tomography experiment on the quantum hardware used for the experiment, and then to invert the function of $\theta_0'(\phi)$ based on data to eliminate the coherent error. This returns the channel to the pure-dephasing form of \cref{eq:rotation_channel_with_dephasing}. We then eliminate the dephasing error by error mitigation. To apply error mitigation, we write the desired rotation channel as a quasi-probability mixture,

\begin{equation}\label{eq:pec_dephasing}
    \mathcal{R}_\theta(\rho) = \frac{1-p_Z}{1-2p_Z}\,\mathcal{N}_\theta(\rho) - \frac{p_Z}{1-2p_Z}\,\mathcal{Z}\circ\mathcal{N}_\theta(\rho).
\end{equation}

\noindent The $1$-norm of the coefficients of this decomposition gives the per-rotation sampling overhead, $\lambda^2$, with

\begin{equation}\label{eq:pec_overhead}
    \lambda = \frac{1}{1-2p_Z} \approx 1 + 2p_Z,
\end{equation}

\noindent so that $\lambda - 1 \approx 2p_Z$ is approximately twice the diamond distance.

We note that the coherent part of the uncalibrated channel $\mathcal{N}'_\theta = \mathcal{E}\circ\mathcal{R}_\theta$, with $\mathcal{E}$ of \cref{eq:rot_plus_dephasing}, could be cancelled in the same way at the cost of Clifford post-rotations. We state this here for completeness, but do not use this in our subsequent analysis, as we always assume calibration. On the equatorial block of the Pauli transfer matrix, $\mathcal{I}$, $\mathcal{Z}$, $\mathcal{S}$ and $\mathcal{S}^\dagger$ span every possible rotation. Therefore for $\delta \theta > 0$ we can remove the channel with a different choice of quasi-probability distribution, 

\begin{equation}\label{eq:pec_coherent}
    \mathcal{R}_\theta = w_I\,\mathcal{N}'_\theta + w_Z\,\mathcal{Z}\circ\mathcal{N}'_\theta + w_{S^\dagger}\,\mathcal{S}^\dagger\circ\mathcal{N}'_\theta,
    \qquad w_{S^\dagger} = \frac{\sin 2\delta\theta}{1-2p_Z},\quad
    w_{I,Z} = \frac12\left(1 - w_{S^\dagger} \pm \frac{\cos 2\delta\theta}{1-2p_Z}\right),
\end{equation}

\noindent with $1$-norm

\begin{equation}\label{eq:pec_coherent_overhead}
    \gamma = \frac{\cos 2\delta\theta + \sin 2\delta\theta}{1-2p_Z} \approx 1 + 2p_Z + 2\delta\theta,
\end{equation}

\noindent which reduces to \cref{eq:pec_overhead} at $\delta\theta = 0$. 

We now move on to further discuss types of errors that can occur in the TMR approach. We separate the errors into three distinct mechanisms:

\begin{enumerate}
    \item hook errors during TMR rotation, whose product is a logical operator (the ``post-selection floor'');
    \item errors that mix the terms of \cref{eq:star_operator} (the ``Choi error''); and
    \item errors during the un-post-selected portion of the circuit, e.g. patch growth (the ``decoded floor'').
\end{enumerate}

The true error for simulation and on hardware will be approximately equal to the sum of these three error mechanisms. The post-selection floor depends primarily on the number of terms, $k$, in the decomposition of \cref{eq:star_operator}. The Choi error rate depends on both $k$ and on the angle for the resource state. Finally, the decoded floor depends primarily on the original distance during growth. In simulation we did not simulate growth for every schedule independently, but instead approximated the effect of this error mechanism by adding a representative number of un-post-selected rounds at the original distance (\cref{fig:bulk_versus_growth_clifford}).

Post-selection makes the preparation an error-detecting experiment, so its floor is set by the smallest number of faults that produce an undetected logical error. During injection this number is $k$, not $d$. Each rotation gadget can suffer a hook error, in which a single fault after the $R_Z(\theta)$ propagates to the whole factor $Z_i$ on the data qubits (\cref{fig:star_circuit}). A single hook error is detected, but the product of all $k$ factors is $Z_L$, so coincident hook errors on every gadget go undetected, giving

\begin{equation}\label{eq:detfloor_scaling}
    p_{\mathrm{post}} \approx A_k\, p^{\,k},
\end{equation}

\noindent with $A_k$ an entropic prefactor counting the number of distinct combinations of hook errors that fulfil this criterion. This $k$ is the same as the number of decomposition terms, because the shortest logical reduces to $k$ during injection. This tends to be extremely small, but is competitive with the Choi error schematically when $\theta \sim p^{k/2}$, or $\theta \sim p^{3/2}$ for $k = 3$.

The Choi error, by contrast, grows with the number of factors $k$ asymptotically. More factors increase both the branch mixing under weight-one errors and the post-selection discard rate. To see this, recall \cref{eq:star_operator}; grouping the terms of the product by syndrome, the branch with $a<k/2$ excitations collects the $Z_S$ with $|S| = a$ and $|S| = k-a$, and its subnormalisation and effective angle are (including the $a=0$ case)

\begin{align}
    N_0(\phi) &= \sqrt{\cos^{2k}\left(\phi\right)+\sin^{2k}\left(\phi\right)}^{-1}\;;\label{eq:star_coefficients_N_0}\\
    \theta_0(\phi) &= (-1)^{\frac{k-1}{2}}\arcsin\left(N_0 \sin^k\left(\phi\right)\right)\;;\label{eq:star_coefficients_theta_0}\\
    N_a(\phi) &= \sec^a(\phi)\csc^a(\phi)\sqrt{\cos^{2k-4a}\left(\phi\right)+\sin^{2k-4a}\left(\phi\right)}^{-1}\;;\label{eq:star_coefficients_N_a}\\
    \theta_a(\phi) &= (-1)^{\frac{k-1}{2}}\arcsin\left((-1)^aN_a \cos^a(\phi)\sin^{k-a}(\phi)\right).\label{eq:star_coefficients_theta_a}
\end{align}

A toy noise model that we consider to help understand the error structure is a sum of weight-one dephasing channels at each rotation operator, $\mathcal{D}(\rho) = (1-kp)\,\rho + p\sum_{j=1}^{k} Z_j\rho Z_j$, with $p$ the probability of a weight-one fault on each term, acting between the rotations and syndrome extraction. Note, we do not use this model in our simulations, only here to derive the effective error channel approximately. A fault $Z_j$ maps the $j$'th single-excitation term of \cref{eq:star_operator} onto the null syndrome, while a fault on the null branch moves it to a single-excitation syndrome and is rejected. Higher errors, such as those at $\mathcal{O}(p^2)$ in a more realistic error model, can cause mixing of the two-excitation branches and the null branch, but we disregard those here. Post-selecting on the null syndrome therefore yields a classical mixture of the two possible angles, up to a global phase. Writing $\ket{\theta} = R_Z(\theta)\ket{+}$, the accepted, unnormalised state is

\begin{equation}\label{eq:mixed_null_branch}
    \tilde{\sigma} = \frac{1-kp}{N_0^2}\ket{\theta_0}\!\bra{\theta_0} + \frac{kp}{N_1^2}\ket{\theta_1}\!\bra{\theta_1}.
\end{equation}

\noindent We define the noisy probability of accepting the state as $1/{N_0'}^2$ and the effective angle as $\theta_0'$. After normalisation we may rewrite the state as a partially dephased $R_Z(\theta_0')$ rotation of $\ket{+}$. The errorful constants are

\begin{align}\label{eq:mixed_null_parameters}
    \frac{1}{N_0'^2} &= \frac{1-kp}{N_0^2} + \frac{kp}{N_1^2},\\
    \theta_0' &= \theta_0 + \tfrac12\operatorname{atan}\!\Big(\frac{q\sin\big(2(\theta_1-\theta_0)\big)}{1 - q + q\cos\big(2(\theta_1-\theta_0)\big)}\Big).
\end{align}

\noindent Here, we have defined the useful constant $q = \frac{kp\,N_0'^2}{N_1^2}\;$. The magnitude of the relevant dephasing channel ($p_Z$) is

\begin{equation}\label{eq:non-perturbative_error}
    p_Z = \tfrac12\left(1 - \sqrt{1 - 4q(1-q)\sin^2(\theta_1-\theta_0)}\right).
\end{equation}

To leading order in $p$ these are

\begin{align}\label{eq:mixed_null_leading}
    N_0' &\approx N_0\left(1 + \frac{kp}{2}\left(1 - \frac{N_0^2}{N_1^2}\right)\right),\\
    \theta_0' - \theta_0 &\approx q\,\sin(\theta_1-\theta_0)\cos(\theta_1-\theta_0),\\
    p_Z &\approx q\,\sin^2(\theta_1-\theta_0).
\end{align}

\noindent In the small-angle limit, \cref{eq:star_coefficients_theta_0} gives $\tan\theta_0 = \tan^k\phi$ and $\tan\theta_1 = -\tan^{k-2}\phi$, so for odd $k \geq 3$ we have $\phi \approx \theta_0^{1/k}$ and $\theta_1 - \theta_0 \approx -\theta_0^{1-2/k}$. Since $N_0^{-2} \approx 1$ and $N_1^{-2} \approx \phi^2$, the mixing probability is $q \approx kp\,N_0^2/N_1^2 \approx kp\,\theta_0^{2/k}$. The wrong branch is therefore offset by $\theta_0^{1-2/k}$ but weighted by $\theta_0^{2/k}$, giving

\begin{equation}\label{eq:choi_error}
    \theta_0' - \theta_0 \approx -kp\,\theta_0,\qquad p_Z \approx kp\,\theta_0^{2(1-1/k)}.
\end{equation}

\noindent With this analysis we can see that the size of the dephasing channel has an optimum value of $k$ for every angle independently. In the small-angle limit we may minimise \cref{eq:choi_error} to find a $p$-independent optimum, $k_\mathrm{opt}$:

\begin{equation}\label{eq:kopt}
    k_\mathrm{opt} \approx -2\log(\theta_0),\qquad\frac{p_Z}{p}\bigg|_{k_\mathrm{opt}} \approx 2e\,\theta_0^2\ln(1/\theta_0) = k_\mathrm{opt}\exp(1-k_\mathrm{opt}).
\end{equation}

\begin{figure}[tbp]
    \centering
    \includegraphics[width=1.0\textwidth]{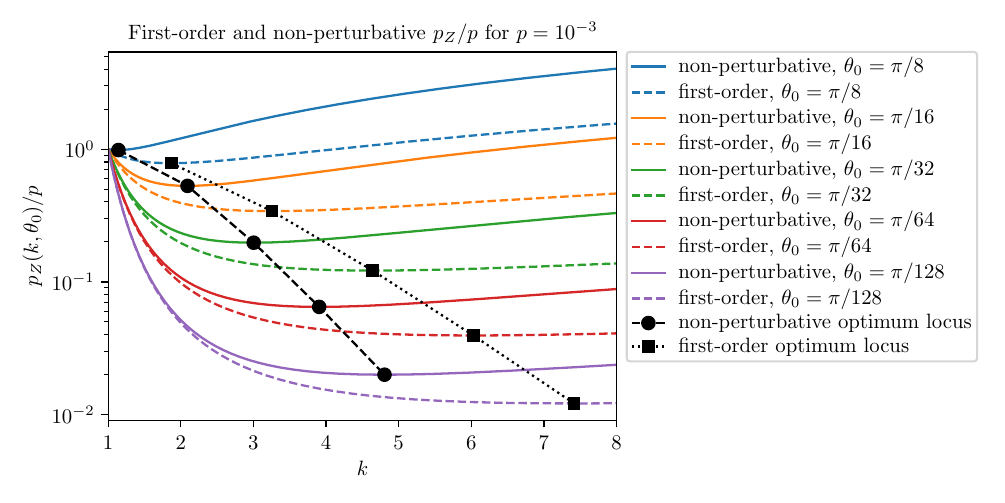}
    \caption{{\bf Optimal decomposition size for the Choi error.} The dephasing prefactor $p_Z/p$ given at first order $k\,\theta_0^{2(1-1/k)}$ and non-perturbatively, as a function of the decomposition size $k$, for $\theta_0 = \pi/8$ to $\pi/128$. The minimum of each curve is shown for the first-order noise (squares), and non-perturbative noise (circles). All curves meet at $k = 1$, where the prefactor is independent of angle.}
    \label{fig:star_kopt}
\end{figure}

\begin{figure*}
    \centering
    \includegraphics[width=0.7\textwidth]{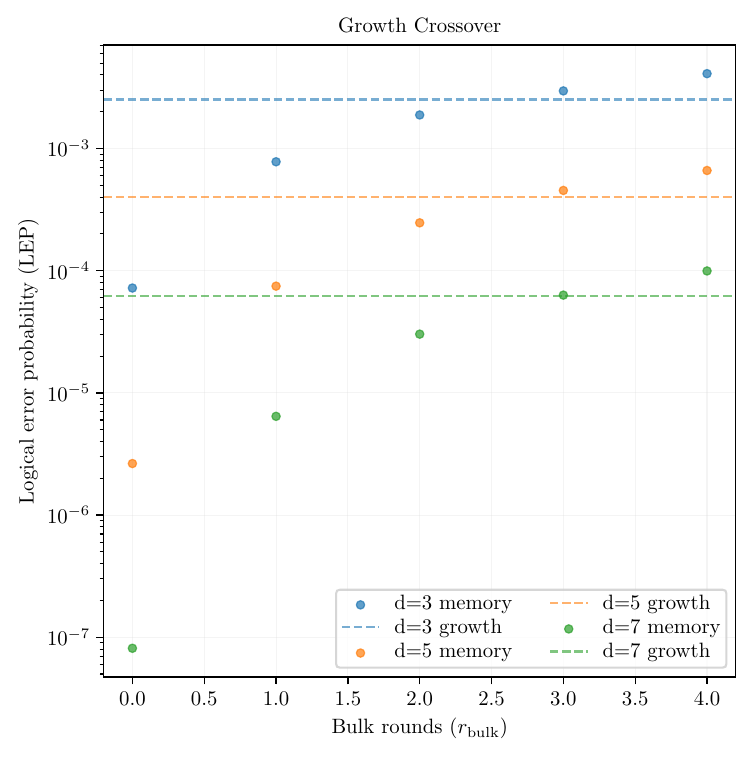}
    \caption{{\bf Bulk rounds reproduce the error of patch growth.} Logical error probability (LEP) under SI1000 noise at $p=10^{-3}$, for a state held at distance $d=3$, $5$ and $7$ through $r_{\mathrm{bulk}}$ noisy syndrome-extraction rounds followed by an ideal measurement (circles). Dashed lines give the LEP of growing a patch from distance $d$ to the final distance $d=15$. At each distance the dashed line is closest to $r_\mathrm{bulk}=3$ steps, therefore we use three bulk rounds to model the decoded floor in \cref{fig:star_td}.}
    \label{fig:bulk_versus_growth_clifford}
\end{figure*}

\noindent We may also consider solving \cref{eq:non-perturbative_error} at beyond first-order in $\theta_0$. This increases the effective error channel from errors of weight one, and \textit{decreases} the $k_\mathrm{opt}$ to use for injection. Importantly for our case at $\theta_0=\pi/128$ we find that $k_\mathrm{opt} \approx 5$. Therefore, considering larger patch-sizes for injecting this operator will exacerbate the leading source of error. We show the value of $p_Z$ as a function of $k$ both at first-order and the non-perturbative result, and illustrate the optimal value of $k$ using either expression in \cref{fig:star_kopt}.

The final error mechanism, the decoded floor, is due to errors that occur outside the post-selection portion of the circuit, such as during possible patch growth or the subsequent lattice surgery. There are different options for consuming this resource state once we have created it. We may either immediately grow the surviving state from its injection distance to the computational distance before performing lattice surgery, or we may directly perform lattice surgery between patches of different distances. Ref.~\cite{Chung2026} considered the case where patch growth is performed first. However, performing lattice surgery without growth may be preferable at large rotation angles, since in this regime the error rate of injection will dominate the error rate of the subsequent lattice surgery anyway, and the rejection rate will be reduced by avoiding the growth step. In the following, we consider the decoded floor given by growth for our final analysis, since this gives a lower error rate, and our results are ultimately bottlenecked by the error rate rather than by the post-selection success rate.

We did not simulate growth directly in our non-Clifford simulator for each protocol. Instead, we reproduced this error channel by holding the injected state through some number of un-post-selected noisy syndrome-extraction rounds, which we call bulk rounds, before a perfect measurement (which also permits $Y$-basis readout). We determined how many bulk rounds to inject through experimentation (\cref{fig:bulk_versus_growth_clifford}). To determine the correct number of bulk rounds we simulated memory with perfect measurement for multiple initial distances $d=3,\;5,\;7,\;\ldots$ and with differing numbers of bulk rounds $r_\textrm{bulk}=0,\;1,\;2,\;3,\;\ldots$ and then simulated growth to our suggested final distance $d=15$ and found the number of bulk rounds that reproduced the LEP of growth most accurately (cf. \cref{fig:bulk_versus_growth_clifford}). Note that this choice does not affect the final results for the sampling overhead of our case study, since all resource states selected in \cref{tab:star_td} are dominated by the Choi error, sufficiently above the decoded error floor.

One additional simplification that we consider is a slight generalisation of protocol switching. Ref.~\cite{Toshio2025} reports a total channel error $\varepsilon = \alpha_{\mathrm{RUS}}\,p\,\theta$, where $\alpha_{\mathrm{RUS}}$ absorbs the error accumulated over the RUS trials, and evaluates $\alpha_{\mathrm{RUS}}$ both with and without protocol switching. The literature offers three strategies:

\begin{itemize}
    \item \emph{No switching.} Every RUS step uses TMR with fixed $k$, so $\alpha_{\mathrm{RUS}}$ grows linearly with $k$, because they are dominated by the errors in the largest angle injections.

    \item \emph{Switching to Akahoshi.} Every RUS step uses TMR with fixed $k$ until the angle grows large enough, roughly $0.1$. Here the FTS error of TMR exceeds that of the $[[4,1,1,2]]$ protocol of \cite{Akahoshi2024}, so we use the latter protocol to prepare states above the threshold. This makes $\alpha_{\mathrm{RUS}}$ nearly independent of $k$, $\alpha_{\mathrm{RUS}} \approx 1.5$ \cite{Toshio2025}.

    \item \emph{Switching to fault-tolerant synthesis.} STAR Magic Mutation~\cite{Toshio2026} synthesises rotations above a threshold angle $\theta_{\mathrm{th}}$ from cultivated $T$ states, which yields $\mathcal{O}\big(\theta^{2(1-\Theta(1/d))}p\big)$.
\end{itemize}

We consider a modified version similar to~\cite{Toshio2026}. Since we only require states within the Clifford hierarchy, we may choose a unique protocol at every angle until $\pi/8$, where we apply a $T$ gate. Interestingly, this approach applies Toshio-CZ with $k=d/2=5$ at all angles we require. However, if we were to extend to smaller angles, such as $\pi/256$ or beyond, then we would likely swap between Toshio-CZ with $k=d/2=5$ and $k=d/2=7$.

\begin{figure*}[tbp]
    \centering
    \includegraphics[width=1\textwidth]{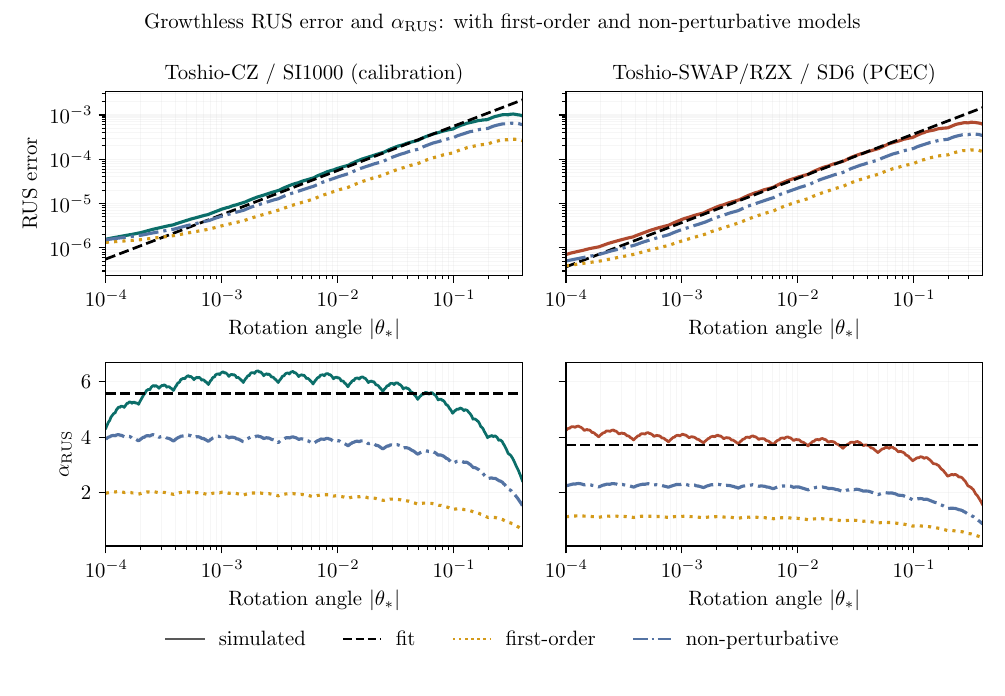}
    \caption{\textbf{Growthless RUS error and $\alpha_{\mathrm{RUS}}$ for our schedule and for the schedule of Ref.~\cite{Toshio2025}.} Left: Toshio-CZ under SI1000 noise, with the coherent error removed by calibration. Right: Toshio-SWAP/RZX under SD6 noise, which assumes the native SWAP and $R_{ZX}$ gates of Ref.~\cite{Toshio2025}, with the coherent error removed by PCEC. Both use $p=10^{-3}$ and omit patch growth. Top: total error of the repeat-until-success channel as a function of the target angle $|\theta_*|$. Bottom: the prefactor $\alpha_{\mathrm{RUS}} = \varepsilon/(p\,|\theta_*|)$. Solid lines are simulations and dashed lines a constant-$\alpha_{\mathrm{RUS}}$ fit. Dotted lines propagate the first-order FTS dephasing $p_Z = k\, p_{\mathit{UD}}\,\theta^{2(1-1/k)}$ (\cref{eq:choi_error}) through the RUS ladder; dash-dotted lines propagate the non-perturbative dephasing of \cref{eq:non-perturbative_error}. The non-perturbative model accounts for roughly half of the gap between the first-order model and simulation; we attribute the remainder to higher-weight, $\mathcal{O}(p^2)$, branch mixing.}
    \label{fig:star_alpha}
\end{figure*}

As a further check we also simulated Toshio-SWAP/RZX with SD6 noise, to gauge how strongly the conclusion depends on the noise model. Results are shown in \cref{fig:star_alpha}. We recalculated the estimate using the circuits described in~\cite{Toshio2025}, and found that we are able to roughly match their values of $\alpha_{\mathrm{RUS}}$ with the FTS derived with $p_{UD}=(k/15)p$ and using the \textit{non-perturbative} expression of~\cref{eq:non-perturbative_error}, although there is a small increase. We speculate that this increase is due to higher-weight faults, at $\mathcal{O}(p^2)$. A general property of these values of $\alpha_{\mathrm{RUS}}$ is that they are dominated by the \textit{large} angle contribution of the FTS error. That is why, even though the non-perturbative and first-order expressions agree at small angle, the $\alpha_\textrm{RUS}$ we derive does not agree at small angle. Thus, while disregarding higher-order errors at small angles will give an accurate FTS value, it will not give an accurate RUS value.

However, this analysis need not give an overly pessimistic view of TMR or STAR. A small increase in the value of $\alpha_\textrm{RUS}$ found during an exact simulation does imply a small decrease in the maximum evolution time or total angle of a simulation. However, our exact simulations indicate that a more precise study of this architecture, with a more realistic noise model only requires a factor of $\sim3$ from the analysis currently found in literature, a factor that can be attributed to both noise model and to higher-weight mechanisms revealed by simulation. 

\section{Non-Clifford simulation of STAR circuits}\label{app:non_cliff_sim}

Over roughly the last year, there has been a flurry of works introducing new non-Clifford simulation tools in precisely our regime. That is, for circuits with a relatively small number of non-Clifford gates, many qubits, many Clifford gates, and many stochastic Pauli noise mechanisms \cite{li2025softhighperformancesimulatoruniversal,Surti2025,Haenel2026,Umbrarescu2026,chase2026clifftfastexactsimulation,Cercelescu2026,Takahashi2026}. These works have focused on the important task of adapting scientifically known simulation techniques, or very close generalisations thereof, and making them more useful and accessible from an engineering standpoint. We make a point to note the exceptions of \cite{Cercelescu2026} and \cite{Takahashi2026}, which use alternative methods for achieving similar estimates, and find similar bounds for simulation time.

The prior state of the art that these works drew from for simulating quantum circuits dominated by Clifford gates, whether by stabiliser-rank decompositions \cite{Bravyi2015,Bravyi2016,Bravyi2019}, or by stabiliser tensor networks \cite{Masot-Llima2024}, however, was aimed at sampling the outputs or calculating the probability of some specific output of generic quantum circuits rather than quantifying the impact of noise on quantum error correcting circuits. This left untreated several problems within the context of sampling  non-Clifford operators in quantum error correcting codes specifically, several of which the new non-Clifford tools have addressed through a combination of circuit compilation (as in Tsim \cite{Haenel2026}) and clever simulation-state caching (as in Clifft \cite{chase2026clifftfastexactsimulation}). These tools nonetheless remain focused on sampling the underlying circuit.

The task of sampling from a non-Clifford circuit is not our current scientific question. We want to find the precise nature of a logical quantum error channel. For Clifford QEC circuits this is summarised as finding the logical error rate or logical error probability (LER or LEP). For a memory experiment under stochastic Pauli noise, the LEP fully describes the error on the measured observable. For a general quantum channel, including non-Clifford state preparation, this is an underdetermined description. Full state or process tomography, or more precisely the density matrix or Pauli transfer matrix, are maximal descriptions. However, they are significantly more expensive to estimate in the non-Clifford QEC context than in the Clifford QEC context. Even if some tool gives us the same time-to-sample a non-Clifford circuit as a Clifford circuit, it would still be significantly more expensive to estimate the outcome of a tomography experiment than a standard memory experiment. This is because of a relative property of samples from Clifford circuits and non-Clifford circuits, which is best explained for Clifford circuits first.

In the noiseless limit, every measurement in a Clifford circuit is either deterministic or maximally random.

A maximally random variable remains maximally random under unbiased noise, so a random Clifford observable carries no information except through its correlation with observables, such as the two qubit-basis measurements of a Bell state. We will return to this when we discuss post-selection. However, for a single observable, we can foresee the distribution of any maximally random variable in the presence of noise. Therefore, we only need to consider deterministic observables.

Given a deterministic logical observable, which without loss of generality reads $+1$ in the noiseless limit, we add noise and sample the logical measurement together with the syndrome. The decoder then solves for (or approximates) whether the syndrome is more likely to belong to the class in which the measurement was flipped or not, and we reconstruct the corrected outcome as the product of the measurement and the decoder's predicted flip. There are only two possibilities: either the decoder predicted the flip correctly, or it did not. In the latter case we say we have a decoding ``failure''.  Call this noisy corrected observable $O$ and the samples $O_i \in \{+1, -1\}$ for the corrected outcome on shot $i$, so that a failure is the event $O_i = -1$, which occurs with probability $q \equiv \mathrm{LEP}$.

This probability is small. Every error mechanism in our circuits fires independently and rarely, and an ideal decoder fails only when at least $\lceil d/2 \rceil$ faults align on a logical operator of the distance-$d$ code (or approximately this many for an approximate decoder), so that $q = \mathcal{O}\!\left((p/p_{\mathrm{th}})^{\lceil d/2 \rceil}\right)$ below threshold. The outcome, $O_i$, has mean $\mu = 1 - 2q$, further $O^2=1$ regardless of the outcome, therefore the variance is

\begin{equation}\label{eq:det_variance}
    \sigma^2 = \mathbb{E}(O^2)-\mathbb{E}(O)^2=1-\mu^2=4 \,q\,(1-q).
\end{equation}

\noindent That is, the variance of a deterministic observable is of the same order as the shift in its mean, and so proportional to the LEP.
 
The sampling complexity of such a low-variance, small-shift, distribution is lower than one would naively expect. Let $\bar{O} = \tfrac{1}{N}\sum_{i=1}^{N} O_i$ be the sample mean over $N$ independent shots, from which we estimate $\hat q = (1-\bar{O})/2$, let $M$ be the uniform bound, or the maximum absolute difference between the sample and mean ($M=2$), and let $t > 0$ be a deviation of $\bar{O}$ from its mean. Bernstein's inequality bounds the probability that our sample mean deviates from the true mean by

\begin{equation}\label{eq:bernstein}
    \Pr\!\left[\,|\bar{O} - \mu| \ge t\,\right] \le 2\exp\!\left(-\frac{N t^2}{2\sigma^2 + \tfrac{2}{3} M\,t}\right).
\end{equation}

\noindent We ask for relative precision $\varepsilon \in (0,1]$ on the shift, that is, $t = 2\varepsilon q$, at confidence $1-\delta$, so we require the right-hand side of \cref{eq:bernstein} to be at most $\delta$. Substituting $t$ and solving for $N$ gives

\begin{equation}\label{eq:det_samples}
    \Pr\!\left[\,|\bar O - \mu| \ge 2\varepsilon q\,\right] \le 2\exp\!\left(-\frac{4\,\varepsilon^2 q^2 N}{8\,q\,(1-q) + \tfrac{8}{3}\,\varepsilon\,q}\right) \le \delta
    \quad\Longrightarrow\quad
    N \geq \frac{2\left(1-q+\frac{1}{3}\varepsilon\right)}{\varepsilon^2 q\,}\log\left(\frac{2}{\delta}\right) = \mathcal{O}\!\left(\mathrm{LEP}^{-1}\right).
\end{equation}

If we relax the requirement that the observable is deterministic in the $p\rightarrow0$ limit, then we have $0 < |\mu_0| < 1$. If we consider the same \textit{relative} shift $\mu=\mu_0(1-2q)$, then the variance of the distribution has

\begin{equation}\label{eq:nondet_variance}
    \sigma^2 = \mathbb{E}(O^2)-\mathbb{E}(O)^2=1-\mu^2=1-\mu_0^2(1-4\,q\,(1-q)),
\end{equation}

\noindent which is no longer small: it is dominated by the noiseless variance $1-\mu_0^2$. We now estimate $\hat q = \tfrac12(1 - \bar{O}/\mu_0)$, so relative precision $\varepsilon$ on $q$ corresponds to a deviation $t = 2\varepsilon q |\mu_0|$ of the sample mean, with $M = 2$ as before. Substituting $t$, $\sigma^2$ and $M$ into \cref{eq:bernstein} gives

\begin{equation}\label{eq:nondet_samples}
    \Pr\!\left[\,|\bar O - \mu| \ge 2\varepsilon q |\mu_0|\,\right] \le 2\exp\!\left(-\frac{2\varepsilon^2 q^2 \mu_0^2 N}{1-\mu_0^2 + \tfrac{16}{3}q}\right) \le \delta
    \quad\Longrightarrow\quad
    N \ge \frac{1-\mu_0^2 + \tfrac{16}{3}q}{2\varepsilon^2 \mu_0^2\, q^2}\log\frac{2}{\delta},
\end{equation}

\noindent where we have used $\mu_0^2 \le 1$, $1-q \le 1$ and $\varepsilon|\mu_0| \le 1$ to bound the denominator. This is $\mathcal{O}(\mathrm{LEP}^{-2})$ whenever the noiseless variance dominates, $q \ll 1 - \mu_0^2$. Since $\sigma^2 \ge 1-\mu_0^2$ is bounded below, the central limit theorem shows that no estimator from independent shots can do better. When instead $1 - \mu_0^2 \lesssim q$, the noise terms dominate the denominator and we recover the linear scaling of \cref{eq:det_samples}.

We may summarise this finding with the maxim: ``Estimating the shift in a deterministic observable under noise requires linear time in the inverse LEP, estimating the shift in a non-deterministic observable under noise requires quadratic time in the inverse LEP''.

There is a simple circumvention of this problem on real hardware. Rather than measuring a single application of the channel, we apply it $m$ times in succession before measuring, amplifying the shift $\mu-\mu_0 = \mathcal{O}(q)$ approximately linearly to $\mathcal{O}(mq)$. Resolving this against the same $\mathcal{O}(1)$ variance requires $N = \mathcal{O}((mq)^{-2})$ shots, or $mN = \mathcal{O}((mq^2)^{-1})$ channel uses, which falls with $m$ until the accumulated stochastic error decoheres the signal at $m = \mathcal{O}(q^{-1})$ \cite{Demkowicz-Dobrzanski2012}. At this bound the total cost is $\mathcal{O}(\mathrm{LEP}^{-1})$, recovering the linear scaling of the deterministic case. This is the amplification underlying Heisenberg-limited phase estimation \cite{Higgins2007}, \cite{Kitaev1995}.

This route is closed to a classical simulator (when we consider non-Clifford channels). Applying the channel $m$ times means simulating a circuit with $m$ times as many non-Clifford operations, and the per-sample cost of the near-Clifford simulators of which we are aware grows with the stabiliser rank, which is exponential in the non-Clifford count \cite{Bravyi2016}. Amplifying the signal by a factor $m$ therefore multiplies the simulation cost by $\exp(\mathcal{O}(m))$, and this exponential blow-up overwhelms the polynomial saving the amplification was meant to provide.

The solution to this conundrum is to directly use the soft-information our simulator generates. This access can be exploited in several ways. Existing tools such as Tsim and Clifft could be augmented to, from one stochastic-Pauli noise sample, return both possible outcomes $+1$ and $-1$ together with the probability of each. We then estimate the probability that the decoder will correctly estimate the correction given the outcome, rather than estimating the mean of the outcome given the decoder. In this work we chose a different option, to use the linearity of quantum channels and to apply the soft-information \textit{prior} to sampling.

We rewrite the non-deterministic channel as a distribution, or quasi-distribution, of deterministic and maximally random channels. Each such piece is exactly the kind of observable we showed can be estimated at only $\mathcal{O}(\mathrm{LEP}^{-1})$, so a decomposition into them trades the quadratic sampling cost for a linear one, at the price of a branch complexity or quasi-distribution negativity complexity we quantify below.

We pause before detailing our construction to track its provenance, since attribution requires subtlety. We arrived at the linear-combination-of-Clifford-channels representation independently, while using the stabiliser rank to teleport the injected state. After reinventing this technique for non-Clifford simulation we found that the essential idea is already established. The role of negativity, specifically the $1$-norm of a quasi-distribution representation, as the quantity controlling the cost of Monte Carlo estimation was set out by Pashayan, Wallman, and Bartlett \cite{Pashayan2015}. The exact representation of a quantum channel as a real linear combination of Clifford channels (together with Pauli reset channels), at the level of channels rather than states, is due to Bennink \textit{et al.} \cite{Bennink2017}; their single-qubit rotation decomposition coincides with the one we use, and they already identify the $1$-norm as the per-gate variance factor. The construction at the level of unitaries, via the stabiliser extent and a lifting lemma, was given by Bravyi \textit{et al.} \cite{Bravyi2019}. Our contribution is therefore not the decomposition itself, which is known, but its adaptation to partially post-selected QEC circuits: the per-branch, post-selected channel estimator of \cref{eq:lccc_postselect,eq:branch_accepted_state}, the observation that the deterministic branches are estimable at $O(\mathrm{LEP}^{-1})$ rather than $\mathcal{O}(\mathrm{LEP}^{-2})$, and culling, which we discuss after introducing the basic technique.

Concretely, we represent our non-Clifford channel as a linear combination of Clifford channels, using the identity

\begin{equation}
    R_Z\left(\theta/2\right) = \cos(\theta) I + \frac{1}{2}\left(1 - \cos(\theta) + \sin(\theta)\right) S^\dagger + \frac{1}{2}\left(1 - \cos(\theta) - \sin(\theta)\right) S.
\end{equation}

\noindent This holds at the level of the unitary (up to a convention-dependent global phase), and at the level of the channel. That is, for any density matrix, $\rho$, observable $O$, and product of $R_Z(\theta/2)^{\otimes t} =\sum_{\mathbf{b}}w_\mathbf{b} C_\mathbf{b}$, we have

\begin{equation}
    tr\Big\{O R_Z(\theta/2)^{\otimes t}\rho {R_Z(\theta/2)^{\otimes t\dagger}}\Big\} = \sum_\mathbf{b} w_\mathbf{b} tr\Big\{O C_\mathbf{b}\rho C_\mathbf{b}^\dagger\Big\}.
\end{equation}

\noindent The channel weights ($w_\mathbf{b}$) sum to one, but at least one value is negative when $R_Z(\theta/2)$ defines a non-Clifford operator. We can sample this distribution using a deterministic sampler, whose complexity scales as $3^t$ (where $t$ is the number of non-Clifford gates), or with a Monte Carlo sampler, whose complexity scales as the $\|w_\mathbf{b}\|^2$ norm of $\{w_\mathbf{b}\}$.

Considering this not as a simulation technique of arbitrary quantum circuits, but as a simulation technique for partially post-selected QEC circuits, we have a slight subtlety to handle. Sometimes $C_\mathbf{b}$ will create an entangled state that is no longer an eigenstate of some stabiliser of the code. These states, even in the noiseless case, will have some stabiliser which is no longer deterministic; alternatively they will have some gauge detector. Further, sometimes this stabiliser and the final logical observable will obey the same statistics as a qubit-basis measurement of a Bell state. In the case of TMR, post-selecting on these detectors projects us back into a deterministic (Clifford) codespace. In the language of QEC we might say that some of the Clifford branches of our non-Clifford operator cause some deterministic detectors to become gauge detectors. Post-selecting on these gauge detectors is equivalent to a partial trace over the gauge register, so the final observable obeys

\begin{equation}\label{eq:lccc_postselect}
    \operatorname{tr}\!\Big\{O\, R_Z(\theta/2)^{\otimes t}\rho\, R_Z(\theta/2)^{\otimes t\,\dagger}\Big\}_{\mathrm{ps}}
    = \frac{1}{p_{\mathrm{sel}}}\sum_{\mathbf{b}} w_{\mathbf{b}}\, p_{\mathbf{b}}\,
      \operatorname{tr}\!\big\{O\, \rho^{\mathrm{acc}}_{\mathbf{b}}\big\},
\end{equation}

\noindent where each branch contributes its accepted (normalised) logical state

\begin{equation}\label{eq:branch_accepted_state}
    \rho^{\mathrm{acc}}_{\mathbf{b}}
    = \frac{1}{p_{\mathbf{b}}}\,
      \operatorname{tr}_{d}\!\Big\{\Pi\, C_{\mathbf{b}}\,\rho\, C_{\mathbf{b}}^{\dagger}\, \Big\},
    \qquad
    p_{\mathbf{b}} = \operatorname{tr}\!\Big\{\Pi\, C_{\mathbf{b}}\,\rho\, C_{\mathbf{b}}^{\dagger}\Big\},
\end{equation}

\noindent $p_{\mathbf{b}}$ is the post-selection rate of branch $\mathbf{b}$, $\Pi$ projects onto its accepted detector (including gauge detector) outcomes, and $\operatorname{tr}_{d}$ traces out the corresponding post-selected registers. The normalisation is the post-selection rate over all channels,

\begin{equation}\label{eq:total_postselect_rate}
    p_{\mathrm{sel}} = \sum_{\mathbf{b}} w_{\mathbf{b}}\, p_{\mathbf{b}}.
\end{equation}

In the presence of noise the sampling complexity of estimating $\operatorname{tr}\!\big\{O\, \rho^{\mathrm{acc}}_{\mathbf{b}}\big\}\;$ for a deterministic branch, scales as the inverse of the post-selection of that branch, $p_\mathbf{b}$, times the LEP within that branch, $\mathrm{LEP}_\mathbf{b}$. That is the sample complexity of the finding the LEPs for just the deterministic branches is

\begin{equation}
    N = \mathcal{O}\!\left(\sum_\mathbf{b}\left(p_\mathbf{b}\;\mathrm{LEP}_\mathbf{b}\right)^{-1}\right).
\end{equation}

There are two more simplifications to apply. First, if we apply only unbiased noise we need not sample the maximally random branches to find the relevant trace, though we still need to determine $p_\mathbf{b}$. However, calculating $p_\mathbf{b}$ in the presence of stochastic Pauli noise can be done analytically, and does not require Monte Carlo sampling as does the LEP. In our implementation we do sample these branches, however a future engineering enhancement is to remove this overhead. Second, if two noiseless branches have the same number of distinct possible syndromes, number of gauge detectors, and same final logical given we post-select on those gauge detectors, then they have approximately the same LEP and noisy post-selection rate up to minor corrections. In this work we verified this simplification for several sample circuits; however, a more rigorous analysis of this approximation and further improvements to the sampling technique must wait for future work.

We refer to the set of branches we do not need to sample as ``culled'' as they do not contribute to the overall time to estimate. The two simplifications are ``culling of maximally random branches'' and ``culling of degenerate branches'', respectively. And the process of determining which branches require sampling is ``baking''. There is one further wrinkle, which is that we must estimate our observable in each basis to reconstruct the density matrix or Pauli transfer matrix. In the case of state tomography this requires an extra factor of $3$; in the case of process tomography the extra factor increases to $12$. Therefore without any culling our total sample complexity for estimating the full density matrix goes as

\begin{equation}
    N = \mathcal{O}\!\left(3^{t+1}\;{p_\mathrm{sel}}^{-2}\mathrm{LEP}^{-2}\right).
\end{equation}

\noindent Culling the maximally random branches drops this from $\mathrm{LEP_b}^{-2}$ to $\mathrm{LEP_b}^{-1}$. This is competitive with any tomography experiment from a sampler as long as $3^t<\mathrm{LEP}^{-1}$, even if the time to sample the full non-Clifford circuit is the same as the time to sample the Clifford branch. Culling the statistically redundant branches reduces the $3^{t+1}$ to $3 \kappa$ where we call $\kappa$ the branch complexity. Calculating the branch complexity requires specific detail about the circuit and operator; in the case of TMR it is bounded by the number of possible gauge detectors. That is, it is at most $\mathcal{O}(d^2)$, and we will present a thorough study of this simplification in a future work \cite{Erickson2027}.

In conclusion, our final time to estimate is

\begin{equation}
    N = \mathcal{O}\!\left(3\sum_{\mathbf{b}\,\in\,\mathrm{kept}}\left(p_\mathbf{b}\;\mathrm{LEP}_\mathbf{b}\right)^{-1}\right) \le \mathcal{O}\!\left(3\,\kappa\,\max_\mathbf{b}\left(p_\mathbf{b}\;\mathrm{LEP}_\mathbf{b}\right)^{-1}\right).
\end{equation}

\section{Resource estimates for magic state cultivation}
\label{app:magic_state_cultivation}

\begin{figure*}
    \centering
    \includegraphics[width=1.0\linewidth]{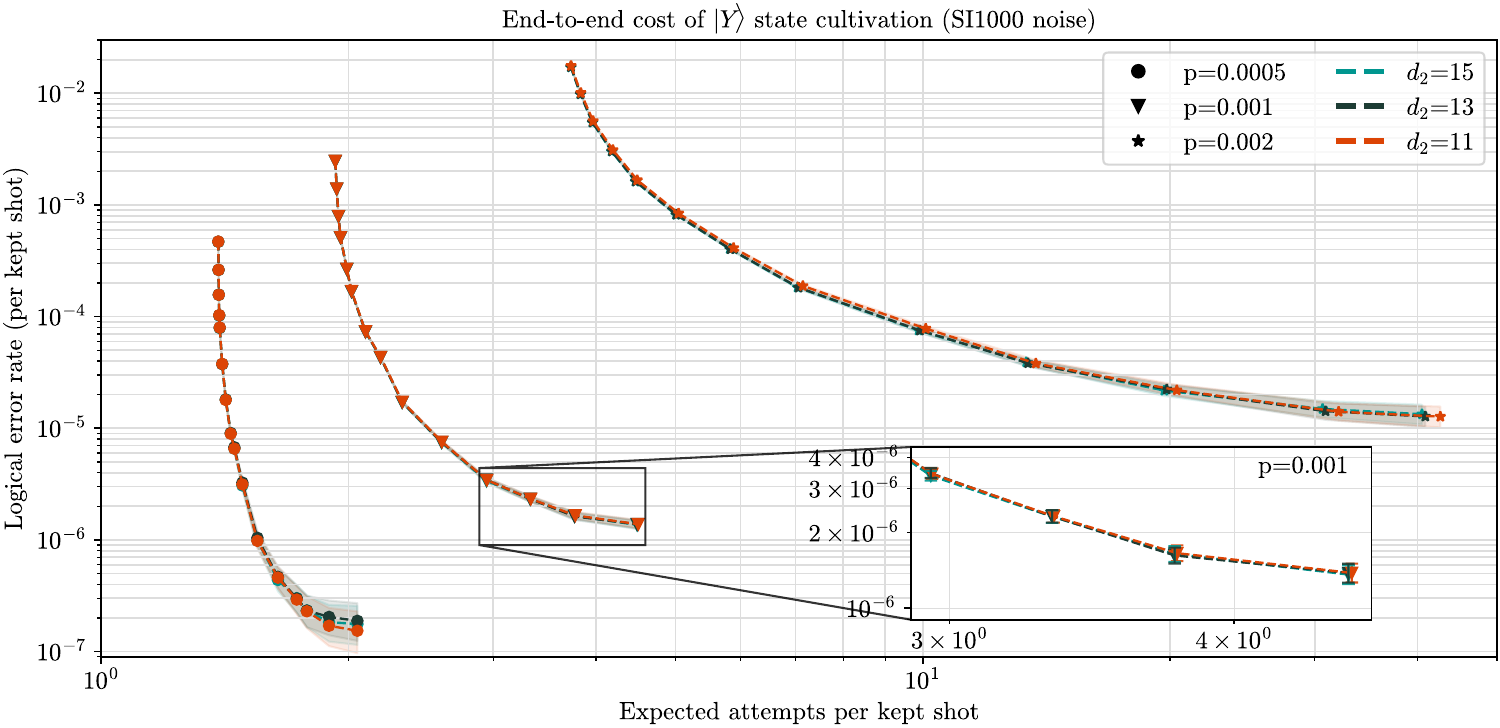}
    \caption{{\bf Numerical simulation of the cultivation protocol for the $\ket{Y}$ state.} 
    We report logical error rates as a function of the expected number of reattempts under SI1000 noise, for the fault-distance $f=3$ protocol with escape to grafted codes of distance 11 (orange), 13 (dark green), and 15 (teal).
    For each value of physical error rate, the plots for the three escape distances coincide, indicating that $d_2 = 11$ is sufficient for protecting the magic state cultivated at fault-distance $3$.
    Assuming that the $\ket{T}$ state cultivated using this protocol will have at most twice the LER of the $\ket{Y}$ state obtained similarly, we conclude that
    at $p_{\text{phys}}=$\SI{1e-3}, it is possible to obtain a cultivated $\ket{T}$ state with a logical error rate of \SI{2.7e-6} after roughly 5 attempts on average.
    }
    \label{fig:msc_si1000}

    \vspace{2.5em}
    \includegraphics[width=1.0\linewidth]{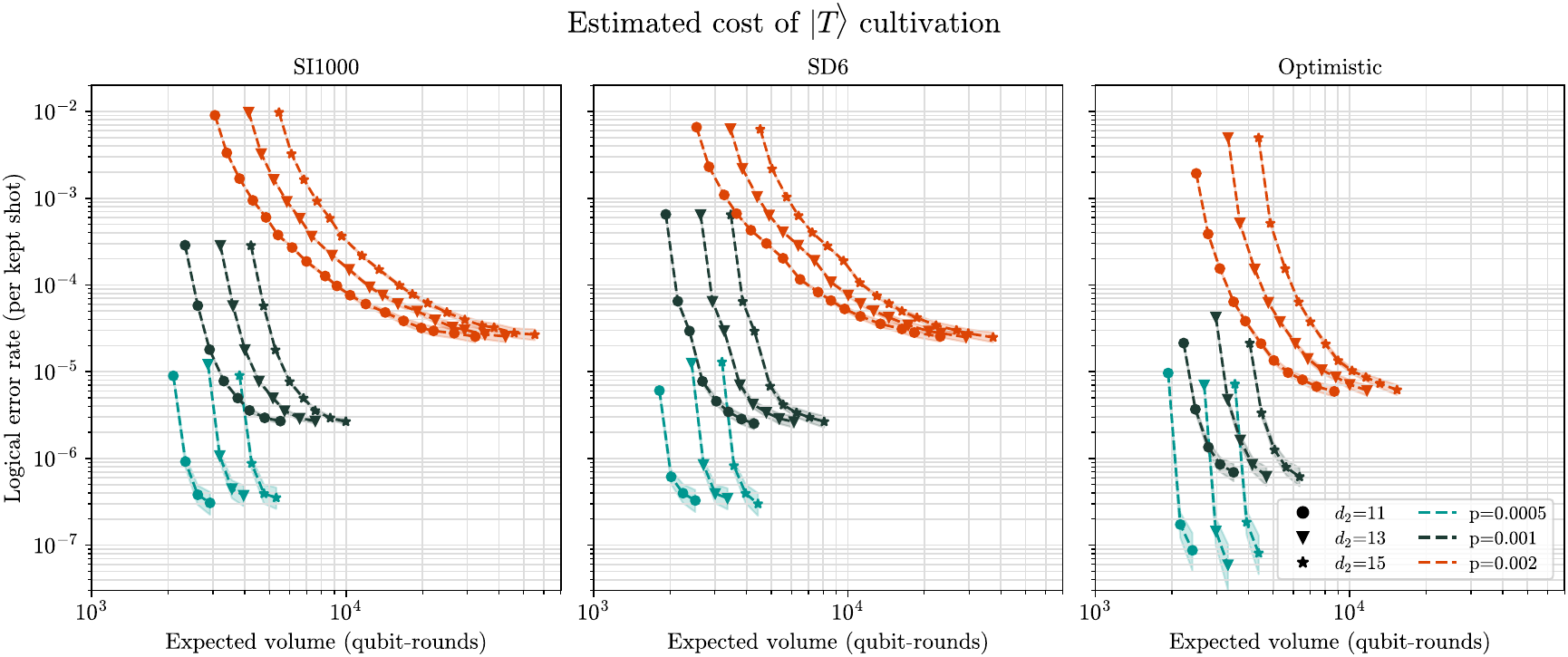}
    \caption{{\bf Expected space-time volume cost of magic state cultivation.}
    Plots of logical error rate (LER) of the $\ket{T}$ state as a function of the expected volume of the end-to-end cultivation protocol. 
    We performed Clifford simulations of $\ket{Y}$ state cultivation and report the $\ket{T}$ state LER as double the LER of the $\ket{Y}$ state cultivated using the same protocol under different noise models (see the text for discussion of this assumption in this parameter range).
    We observe that at $p_{\text{phys}}=$\SI{1e-3}, we can cultivate $\ket{T}$ states of LER$=$\SI{2.66e-6} at an expected volume cost of \SI{9.99e3} qubit-rounds (dark green star plot on leftmost panel).
    }
    \label{fig:msc_volume}
\end{figure*}

Magic state cultivation~\cite{gidney2024magic} is the leading candidate for preparing magic states in the early fault-tolerant regime.
The protocol works by injecting a magic state (typically an eigenstate of the Hadamard operator) into a quantum code that has a transversal logical Hadamard operator. 
This logical Hadamard operator is then measured fault-tolerantly and behaves as a logical check for the encoded magic state.
The protocol involves post-selecting on states that satisfy the stabiliser checks as well as the logical Hadamard check.
The surviving encoded state is then grown to the target code distance while discarding shots for which the decoder is insufficiently confident about the correctness of the logical state.
Due to the post-selection involved in various stages of this protocol, it can take several attempts to produce a usable magic state with the target logical error rate (see \cref{fig:msc_si1000}).
In this work, we follow the original magic state cultivation protocol which first injects the magic state into a colour code of distance 3, and after applying the logical Hadamard checks on the distance 3 colour code, grows into a grafted surface code patch of final distance $d_2$.
We simulated the $\ket{Y}$ state prepared via this protocol, under three different noise models -- SI1000, SD6, and optimistic (see Table~\ref{tab:noise_parametrisations}) -- while varying the distance of the final grafted surface code patch.
\cref{fig:msc_si1000} is a plot of the $\ket{Y}$ state cultivation protocol (a Clifford-proxy for the $\ket{T}$ state protocol) under SI1000 noise with the grafted code distances of $d_2 = 11, 13, 15$.
We find that the protocol at a given physical error rate can achieve a given target logical error rate while requiring the same number of expected attempts for all three values of $d_2$. 
This is to be expected since the fault-distance of this protocol is 3, i.e., it can be protected via error-correction in any code whose logical memory error-rate is lower than the requisite logical error rate of the cultivated state.
More specifically, we only require that the final grafted surface code distance is $d_2 \gtrsim 9$ (see discussion in Ref.~\cite{gidney2024magic}).
We observed similar trends in the expected number of attempts for a given target logical error rate when we simulate the protocol for the SD6 and optimistic noise models (which we do not report here).

From the numerical simulations of $\ket{Y}$ state cultivation, we estimate the expected space-time volume cost of magic state cultivation in \cref{fig:msc_volume} by assuming that the logical error rate (LER) of $\ket{T}$ state cultivation is roughly 2 times higher than that of $\ket{Y}$ state cultivation~\cite{gidney2024magic}.
We remark that although recent works have identified fault-distance reducing mechanisms~\cite{Chan2026, Wan2026} in the original cultivation protocol, exact classical simulations~\cite{chase2026clifftfastexactsimulation,li2025softhighperformancesimulatoruniversal} of $\ket{T}$ state cultivation via the original method at fault-distance $f=3$ suggest that for the physical error rate and target LER that we consider, the assumption that the $\ket{T}$ state LER is at most double the $\ket{Y}$ state LER remains valid. 
We conclude from the numerics (\cref{fig:msc_volume}) that at $p=$\SI{1e-3}, we can cultivate $\ket{T}$ states of LER$=$\SI{2.66e-6} at an expected volume cost of \SI{9.99e3} qubit-rounds, of LER$=$\SI{2.65e-6} at an expected volume cost of \SI{6.14e3} qubit-rounds, and of LER$=$\SI{6.92e-7} at an expected volume cost of \SI{3.5e3} qubit-rounds under SI1000 (with $d_2=15$), SD6 (with $d_2=13$), and optimistic noise (with $d_2=11$), respectively. 
The numerical simulations of cultivation reported in this work were obtained from slightly modified versions of the circuits made available by the authors of Ref.~\cite{gidney2024magic}.
All data points were obtained by sampling \SI{1e9} shots, with the exception of the $p=$\SI{1e-3} data for SI1000 noise which was obtained by sampling \SI{6e9} shots.

\section{Lattice surgery construction of circuit elements}
\label{app:lattice_surgery_constructions}

\begin{figure*}
    \centering
    \includegraphics[width=1.0\linewidth]{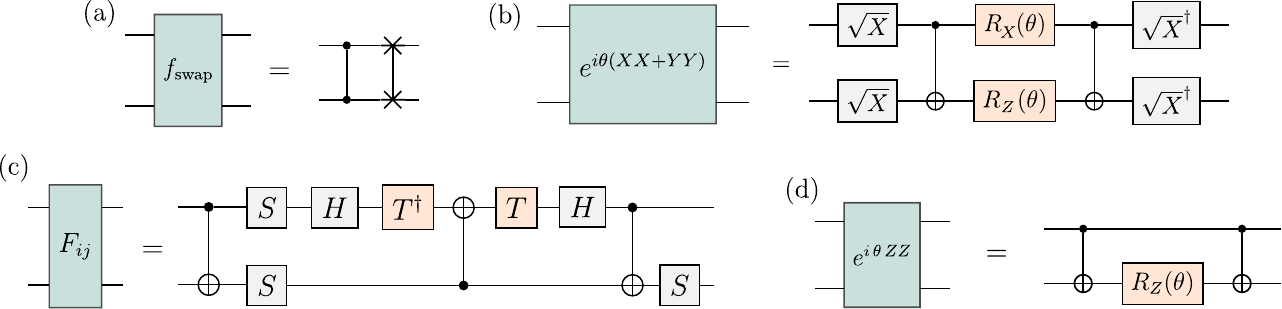}
    \caption{{\bf Circuit components for Hubbard model simulation.} Basic circuit components used to implement plaquette or tile Trotterisation for the Hubbard model. These are compiled to lattice surgery operations to perform resource estimation. We use the conventions $R_Z(\theta) = e^{i \theta Z}$ and $R_X(\theta) = e^{i \theta X}$, as we do throughout this paper. (a) A fermionic swap gate, which is equal to the controlled-Z gate followed by a swap gate. (b) The $XX+YY$ rotation gate and (c) two-qubit fermionic Fourier transform are used to implement time evolution of hopping terms. (d) A $ZZ$ rotation, used to implement time evolution of interaction terms. Note that these implementations of $\fswap$ and $F_{ij}$ are only valid when the two qubits are adjacent in the Jordan-Wigner ordering.}
    \label{fig:hubbard_basic_circuits}
\end{figure*}

In this appendix, we provide lattice surgery constructions for all circuit elements to implement tile Trotter circuits, for the linear QPU layout used in our case study (\cref{fig:sections_and_qpu_layout}(c)). As described in \cref{sec:tile_trotter}, the implementation of time evolution using the tile Trotterisation algorithm \cite{BaySmidt2025} requires just a small number of circuit elements: a $XX+YY$ Pauli rotation, a $ZZ$ Pauli rotation, the two-qubit fermionic Fourier transform ($F_{ij}$), controlled-$Z$, and the fermionic swap gate ($\fswap$). In \cref{fig:hubbard_basic_circuits} we show each of these elements, together with an implementation in terms of standard gates. In the following subsections, we provide lattice surgery constructions based on these implementations.

For each circuit we provide both a time and space-time volume cost. Time is measured in multiples of $d$ rounds (one ``time step''). For example, a Hadamard gate (including restoration of the background lattice), which takes $3d+3$ rounds \cite{geher2024error} would be estimated as $3d$ rounds, or $3$ time steps. Meanwhile, a transversal $Z$ or $X$ gate would be estimated as $0$ time steps. This approach is standard in many resource estimation studies \cite{litinski2019game,Akahoshi2024, huggins2025fluid}. Meanwhile, volume is calculated in multiples of $d \times d \times d$, i.e. a $d \times d$ surface code patch for $d$ rounds (one ``block''). This is obtained by summing the volumes from each stage of a given lattice surgery construction. As discussed in \cref{sec:megaquop_overview}, and evidenced in \cref{sec:memory_vs_lattice_surgery}, quantum memory and lattice surgery experiments of equal volumes have similar error rates. Therefore, we can estimate the error rate of a full logical circuit (excluding contributions from magic state cultivation) by the error rate of a quantum memory experiment of equal volume. This is the costing approach used in \cref{sec:resource_estimates}. Again, these assumptions are standard in many resource estimation studies \cite{litinski2019game,Akahoshi2024, huggins2025fluid}. We note that the original cultivation proposal \cite{gidney2024magic} gives a final magic state in a grafted code, which has worse memory performance than the surface code; therefore, this assumption might be slightly optimistic for subsequent teleportation steps. However, we speculate that cultivation will be improved with further development, escaping into a code with better performance. The number of such teleportations is also small compared to the total Clifford volume.

In the following subsections, $\kappa$ denotes the number of syndrome-extraction rounds required to implement a $T$, $T_X$, $T^{\dagger}$ or $T_X^{\dagger}$ gate; this includes both cultivation of a $\ket{T}$ state and subsequent teleportation, see \cref{fig:fallback_rotation}(h) and (i) for teleportation circuits for $T$ and $T_X$ gates, respectively. Meanwhile, $\gamma$ denotes the syndrome-extraction rounds required to implement a Clifford+T sequence for a rotation gate; see \cref{eq:clifford_t_time_steps} (which gives the number of time steps) and the surrounding discussion.

The quoted volume estimates in this appendix only include the patches shown in the figures for the constructions, unless stated otherwise; in particular, they only include either spin-up or only spin-down qubits. To obtain the full volume, one must also add the volume for the opposite-spin qubits, which are idling, and sum over all regions of the QPU. The Python script used to obtain the final resource estimates is included on Zenodo.

\subsection{Fermionic SWAP gate}
\label{sec:fswap_lattice_surgery}

\begin{figure*}
    \centering
    \includegraphics[width=1.0\linewidth]{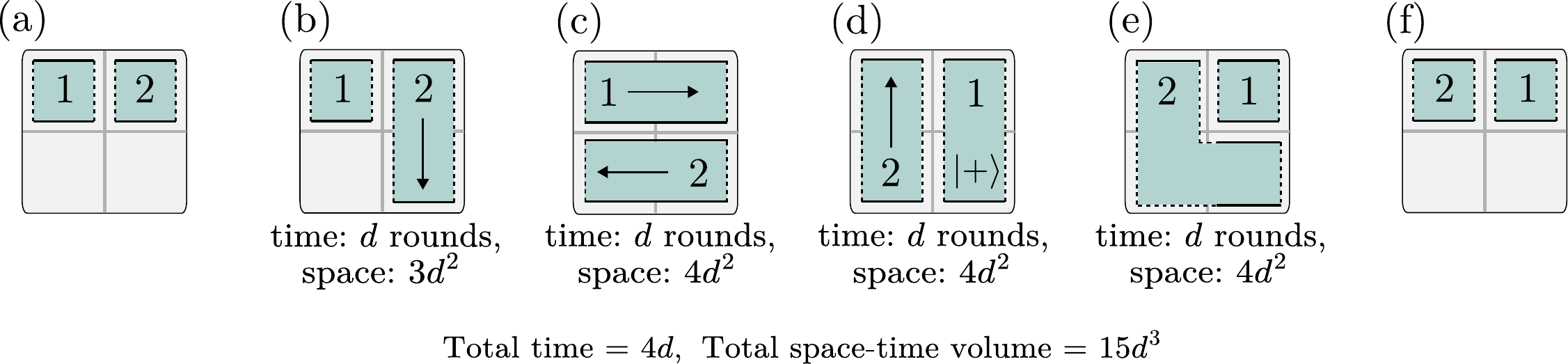}
    \caption{{\bf Lattice surgery implementation of the fermionic SWAP gate in $4d$ rounds.} (a) Initial position of patches 1 and 2 on which the $\fswap$ gate is to be applied. (b)--(c) Patches 1 and 2 are swapped via code growth assisted repositioning. (d) ZZ lattice surgery between patch 1 and an ancilla patch initialised in state $\ket{+}$. Patch 2 is repositioned in parallel via code growth and measurement. (e) ZX lattice surgery between patch 2 and the ancilla patch, concluding the CZ gate between 1 and 2. (f) The resulting patches after $\fswap$ implementation.}
    \label{fig:fswap_ls}

    \vspace{1.5em}
    \includegraphics[width=1.0\linewidth]{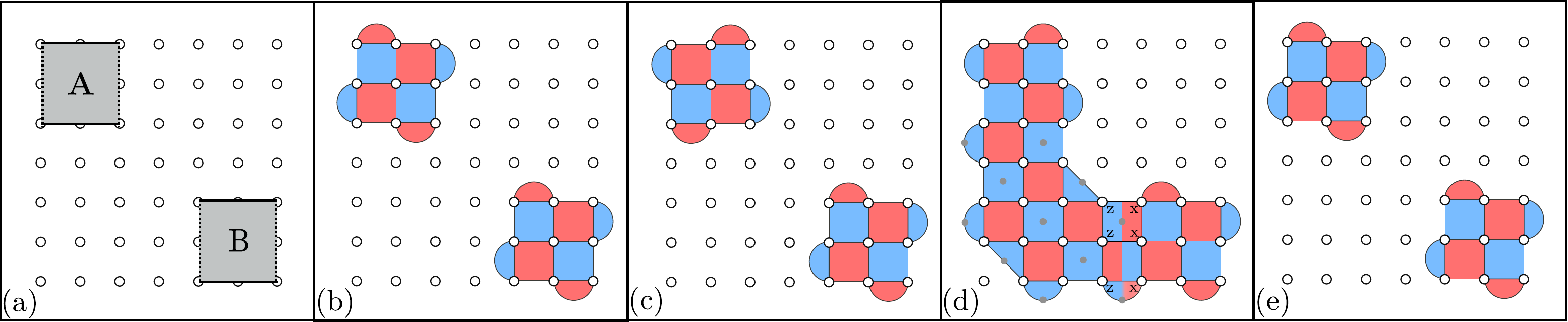}
    \caption{{\bf Schematic of ZX lattice surgery between $d=3$ surface code patches.} (a) Layout of two surface code patches, A and B, in preparation for ZX lattice surgery. The solid (dashed) boundaries indicate the Z (X) logical operators of the patches. (b) The underlying X (red) and Z (blue) stabilisers of the patches. The two patches initially lie on the same checker board pattern or ``background lattice''. (c) Patch A with a background lattice change (BLC). Note that although patch A is no longer on the same checker board pattern as patch B, its new weight-2 stabilisers ensure that the Z (X) logical operators of the patch remain horizontal (vertical). This BLC of patch A enables the measurement of the requisite stabilisers in the merge stage. (d) Merge stage of ZX lattice surgery between patches A and B. The product of the stabiliser measurements indicated by grey solid dots gives the logical ZX measurement value. The blue-red mixed cells represent the mixed-type stabilisers (ZZXX or ZX) which constitute a domain wall. (e) Patch A undergoes another BLC to return to the same checkerboard pattern as patch B.}
    \label{fig:zx_ls_schematic}
\end{figure*}

\begin{figure*}
    \centering
    \includegraphics[width=1.0\linewidth]{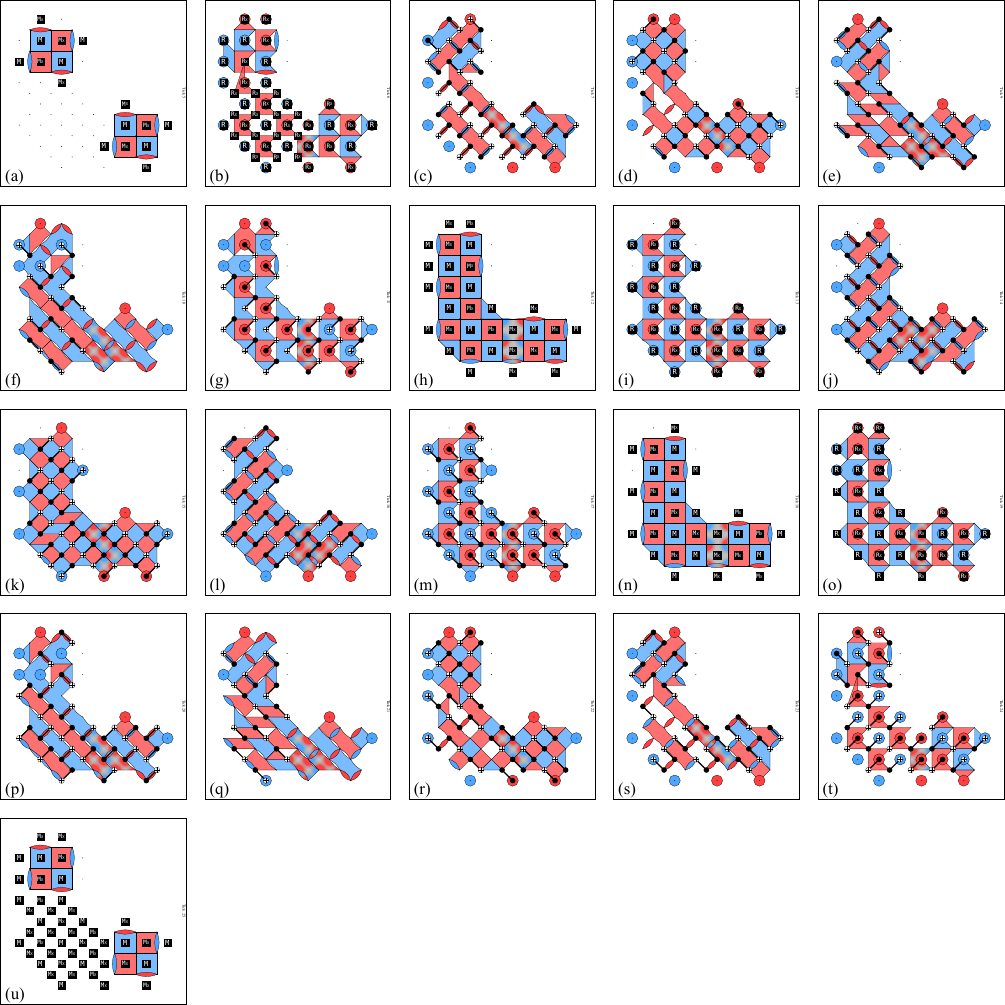}
    \caption{{\bf Illustrative detector-slice diagram of ZX lattice surgery between $d=3$ surface code patches.} (a) Stabilisers of two $d=3$ surface code patches are measured. (b)--(h) First round of syndrome extraction of merge stage stabilisers which also simultaneously implements the background lattice change (BLC) on the top left patch as described in \cref{fig:zx_ls_schematic}(c). Unlike the typical 4-layer syndrome extraction round (excluding measurements and resets) of the surface code, this round requires 5 layers. (i)--(n) One round of normal 4-layer syndrome extraction of the merge stage. For lattice surgery between patches of distance $d$, this would be replaced by $(d-2)$ rounds of 4-layer syndrome extraction. (o)--(u) Final round of 5-layer syndrome extraction of merge stage stabilisers which also simultaneously implements the BLC on the top left patch and returns it to the same background lattice as the lower patch, see \cref{fig:zx_ls_schematic}(e).}
    \label{fig:zx_ls_detslices}
\end{figure*}

The fermionic swap gate can be implemented as a CZ gate followed by a SWAP gate (\cref{fig:hubbard_basic_circuits}(a)), assuming that the two qubits are adjacent in the Jordan-Wigner ordering. 
The lattice surgery compilation of the $\fswap$ gate was previously estimated to require $7d$ rounds, see \cite[Fig.~13]{Akahoshi2025}, in which it was implemented by decomposing the CZ gate into a combination of CX (implemented via ZZ and XX lattice surgery) and Hadamard gates.
Here, we reduce the cost of implementing the $\fswap$ by avoiding the decomposition of CZ to CX.
Instead, we directly implement the CZ gate between two patches via ZZ lattice surgery between the first patch and an ancilla patch initialised in state $\ket{+}$, followed by ZX lattice surgery between the second patch and the ancilla patch, which is then measured out in the Z basis.
This is similar to the way that the CX gate between two patches would be implemented, but with the XX lattice surgery operation required for the CX gate transformed into the ZX lattice surgery operation for the CZ gate.
In order to perform a ZX lattice surgery operation, the logical qubits must be on different background lattices~\cite{geher2024tangling}. Therefore, to perform the operation, we must perform a background lattice change (BLC) on one of the patches before the operation, and again afterwards to return the patch to the original background lattice, as shown in \cref{fig:zx_ls_schematic}.
We develop a 5-layer syndrome extraction circuit that preserves the code distance $d$ of the patches and combines the required BLC into a single round of measurements of the merge-stage stabilisers of the ZX lattice surgery operation, effectively implementing the BLC instantaneously. An example, used in the $4d$-round $\fswap$ construction is shown in~\cref{fig:zx_ls_detslices}; ZX lattice surgery operations with qubits in different layouts can be performed analogously.
The combination of the direct implementation of the CZ gate along with the instantaneous BLC required for the ZX lattice surgery involved allows us to reduce the time cost of the $\fswap$ gate. We present two versions of the operation.

The first, shown in~\cref{fig:fswap_ls}, is uniform across all qubit pairs and all time steps and takes $4d$ rounds with a space-time volume of $15d^3$ (15 blocks). This is a self-contained implementation, in that it returns both qubits to their starting positions, apart from being swapped.

\begin{figure*}
    \centering
    \includegraphics[width=0.9\linewidth]{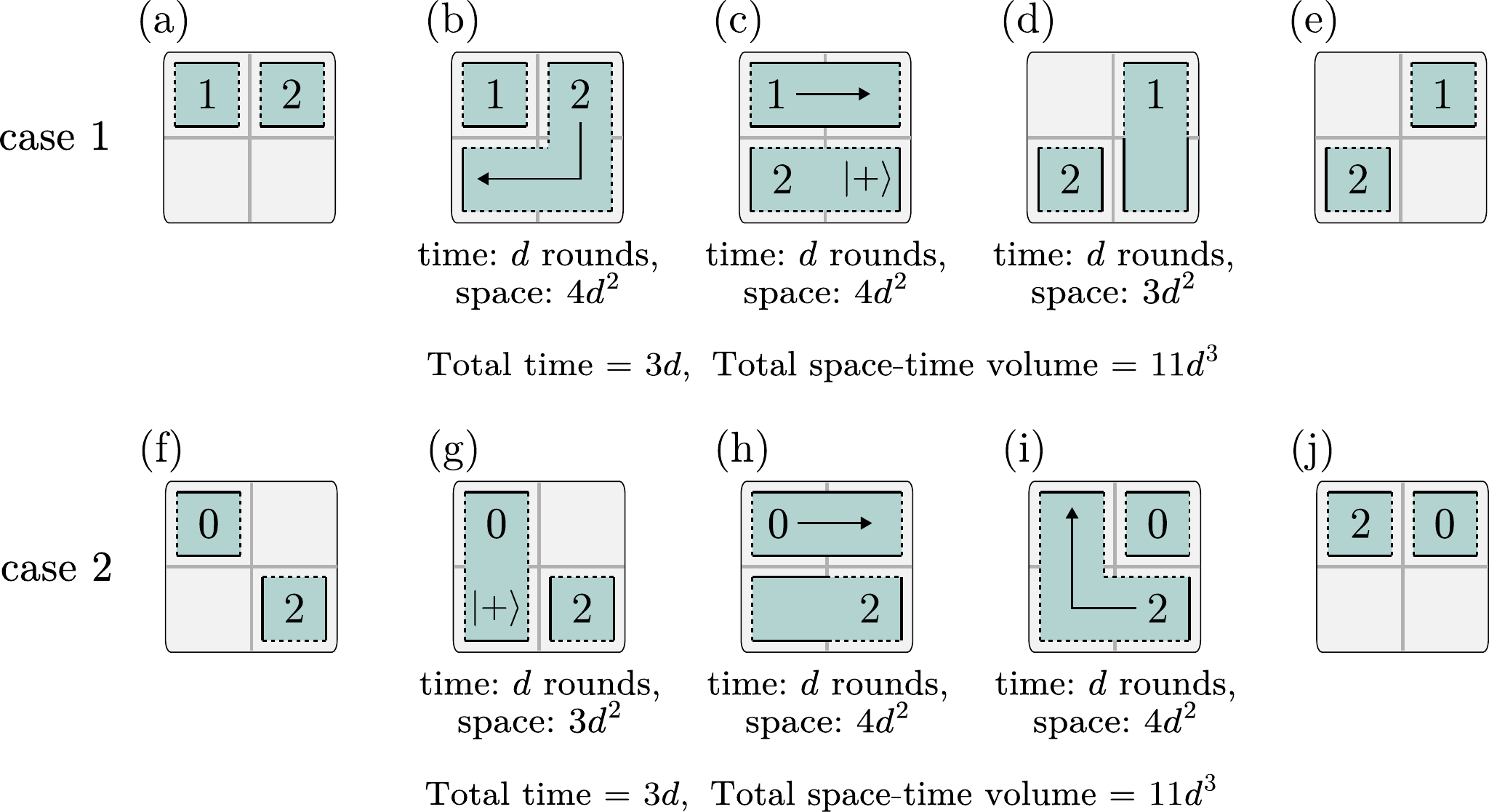}
    \caption{{\bf Lattice surgery implementation of the fermionic SWAP gate in $3d$ rounds.} There are two possible cases, shown in panels (a)--(e) and (f)--(j), respectively. (a) Possible initial position of patches 1 and 2 on which the $\fswap$ gate is to be applied. (b) Patch 2 is moved (and rotated) via code growth assisted reposition. (c) ZZ lattice surgery between patch 2 and an ancilla patch initialised in state $\ket{+}$. Patch 1 is repositioned in parallel via code growth and measurement. (d) ZX lattice surgery between patch 1 and the ancilla patch, followed by measuring the ancilla patch in the Z basis, concluding the CZ gate between 1 and 2. (e) The resulting patches after $\fswap$ implementation. (f) The other possible initial position of patches 1 and 2 on which the $\fswap$ gate is to be applied. (g) Patch 2 is moved (and rotated) via code growth assisted reposition. (h) ZZ lattice surgery between patch 2 and an ancilla patch initialised in state $\ket{+}$. Patch 1 is repositioned in parallel via code growth and measurement. (i) ZX lattice surgery between patch 1 and the ancilla patch, followed by measuring the ancilla patch in the Z basis, concluding the CZ gate between 1 and 2. (j) The resulting patches after $\fswap$ implementation. For both cases, it is also possible to perform the operation on qubits in reflected (about a vertical axis between the two qubits) starting and ending configurations by applying a similar reflection to the steps shown.}
    \label{fig:fswap_ls_3d}
\end{figure*}

The second takes $3d$ rounds and has a space-time volume of $11d^3$ (11 blocks). This approach relies on considering the subsequent $\fswap$ operations to understand why it is applicable. There are two possible cases, shown in \cref{fig:fswap_ls_3d}, plus their reflections about a vertical axis, depending on which operations have happened in previous $\fswap$ layers and the operation that is to happen in the next layer. We see that each qubit has two possible configurations at the start of an $\fswap$ round: on the top row in its original orientation or in the routing corridor in a rotated orientation. The $\fswap$ can be performed on any combination of the two configurations, except where both are in the routing corridor. To avoid this arising, we decide the operation on qubits $i$ and $i+1$ based on the final location of qubit $i-1$ after that round. There are four possibilities for qubits $i$ and $i+1$, taking all combinations of needing to perform an $\fswap$ and not needing to perform an $\fswap$, and starting in case 1 locations and starting in case 2 locations. If we do not need to perform an $\fswap$ between the qubits and they start in case 1 locations, or if we do need to perform an $\fswap$ between the qubits and they start in case 2 locations, both qubits $i$ and $i+1$ end in the top row configuration and so we will not have a problem in the next round. If instead we do not need to perform an $\fswap$ but the qubits are in case 2 locations with qubit $i$ on the bottom row then, if qubit $i-1$ is in the routing corridor and will be involved in an $\fswap$ in the next round, we perform a move-and-rotate operation, in parallel with the $\fswap$ of that round, to restore qubit $i$ to the top row in the correct orientation. This ``restore'' operation takes two time steps and has a volume of four blocks. In the final case, where we need to perform an $\fswap$ and the qubits begin in the case 1 configuration then, if qubit $i-1$ is in the routing corridor, we choose to perform the reflection of the case 1 operation, so that the left qubit after the operation is on the top row. If qubit $i-1$ is not in the routing corridor, we perform case 1 as shown in the figure. The total space-time volume of a fermionic swap network will depend on the number of restorations of qubits to the top row we have to perform; these are given in \cref{app:4x4_fswap} for the specific fermionic swap network used in our case study.

Note, the above descriptions consider applying fermionic swaps to top-row (spin-up) qubits, but trivially also extends to bottom-row (spin-down) qubits. Given that we only have one row of routing space, we first swap all spin-up qubits, followed by all spin-down qubits, sequentially.

\subsection{XX+YY rotation gate}
\begin{figure*}
    \centering
    \includegraphics[width=1.0\linewidth]{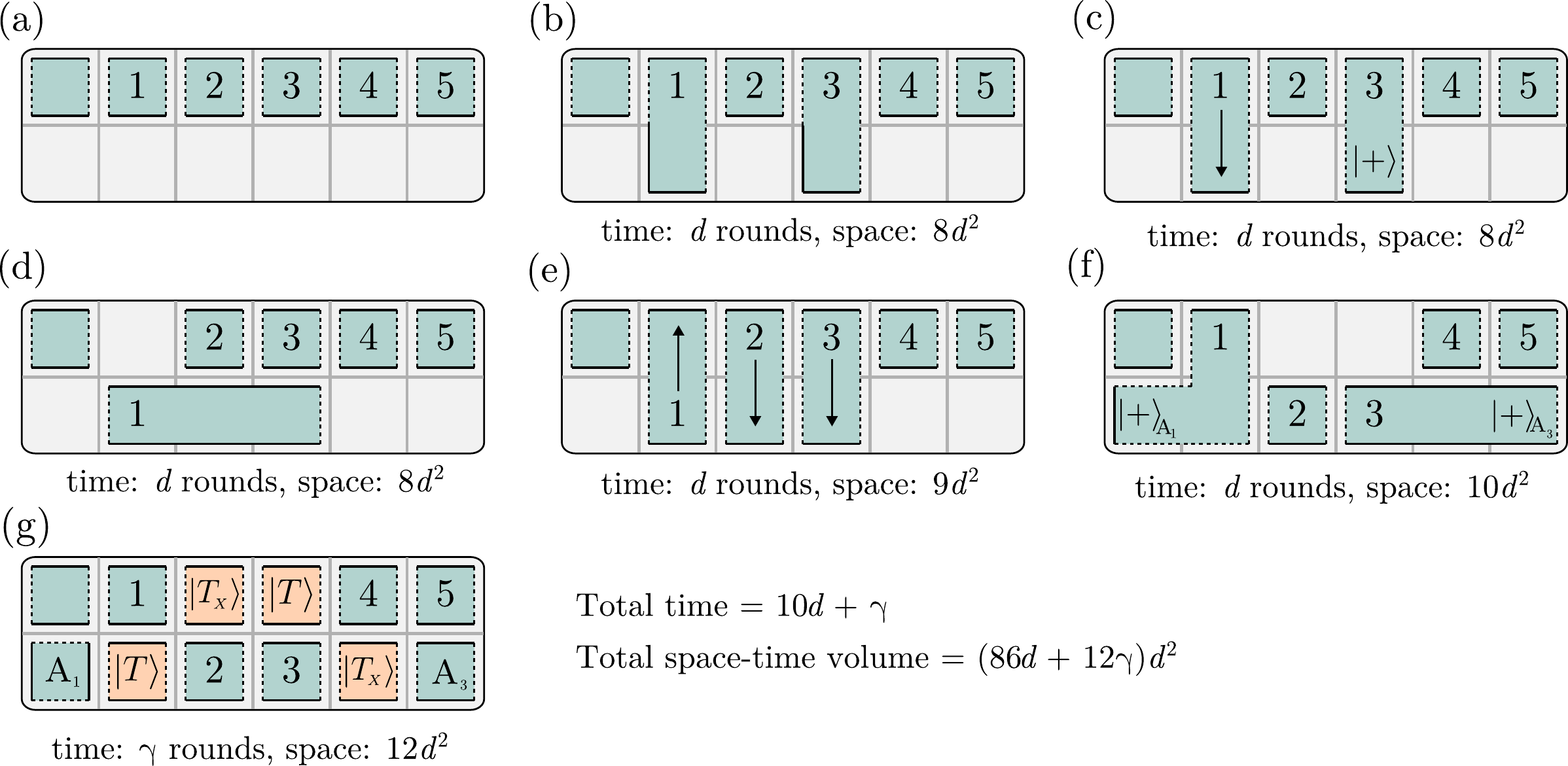}
    \caption{{\bf Lattice surgery implementation of $X_1 X_3 + Y_1 Y_3$ rotation.} See \cref{fig:hubbard_basic_circuits}(b) for the circuit diagram. (a) An example showing six qubits, including those labelled $1$--$5$, which refer to the five qubits in an $S_4$ tile. (b) Perform $\sqrt{X}$ on qubits 1 and 3 (using the method of \cite{logical_s_gate}) to reach the configuration shown, and then return the patches to their original shape. (c) Move qubit $1$ down; prepare a $|+\rangle$ state as an ancilla for the CNOT, and then perform $ZZ$ lattice surgery with qubit $3$. (d) Perform $XX$ lattice surgery between the ancilla and qubit $1$. (e) Return qubit $1$ to its original shape and move it up; move qubits $2$ and $3$ down. (f) Prepare two ancilla qubits in state $|+\rangle$ for use in fallback rotation circuits; then perform $ZZ$ lattice surgery with qubits $1$ and $3$. (g) Perform the Clifford+T unitaries on qubits $1$ and $3$, cultivating $|T\rangle$ and $|T_{\text{x}}\rangle$ on $Z$ and $X$ boundaries, respectively. After step (g), undo steps (a)--(f), doubling the cost of these steps.}
    \label{fig:xx_yy_rotation}
\end{figure*}

The $XX+YY$ rotation gates are the central component of both the $S_1$ and $S_4$ tiles (see \cref{eq:s1_tile,eq:s4_tile}). 
Crucially, as indicated in \cref{fig:hubbard_basic_circuits}(b), the two-qubit rotation gate is compiled into a sequence of single-qubit gates, CNOT gates and two single-qubit rotations $R_X(\theta)$, $R_Z(\theta)$. \cref{fig:xx_yy_rotation} shows a step-by-step lattice surgery implementation of the $X_1X_3+Y_1Y_3$ rotation gate used in the $S_4$ tile, with a total time cost of $10d +\gamma$ rounds and a total space-time cost of $(86d+12\gamma)d^2$.
The single-qubit rotations $R_X(\theta)$, $R_Z(\theta)$ are applied via the fallback Clifford$+$T implementation described in \cref{fig:fallback_rotation} of Section~\ref{sec:rotations}.
\cref{fig:s1_tile} similarly implements two sets of $XX+YY$ rotations for the $S_1$ tile.

\subsection{$F_{ij}$ gate}
\begin{figure*}
    \centering
    \includegraphics[width=0.95\linewidth]{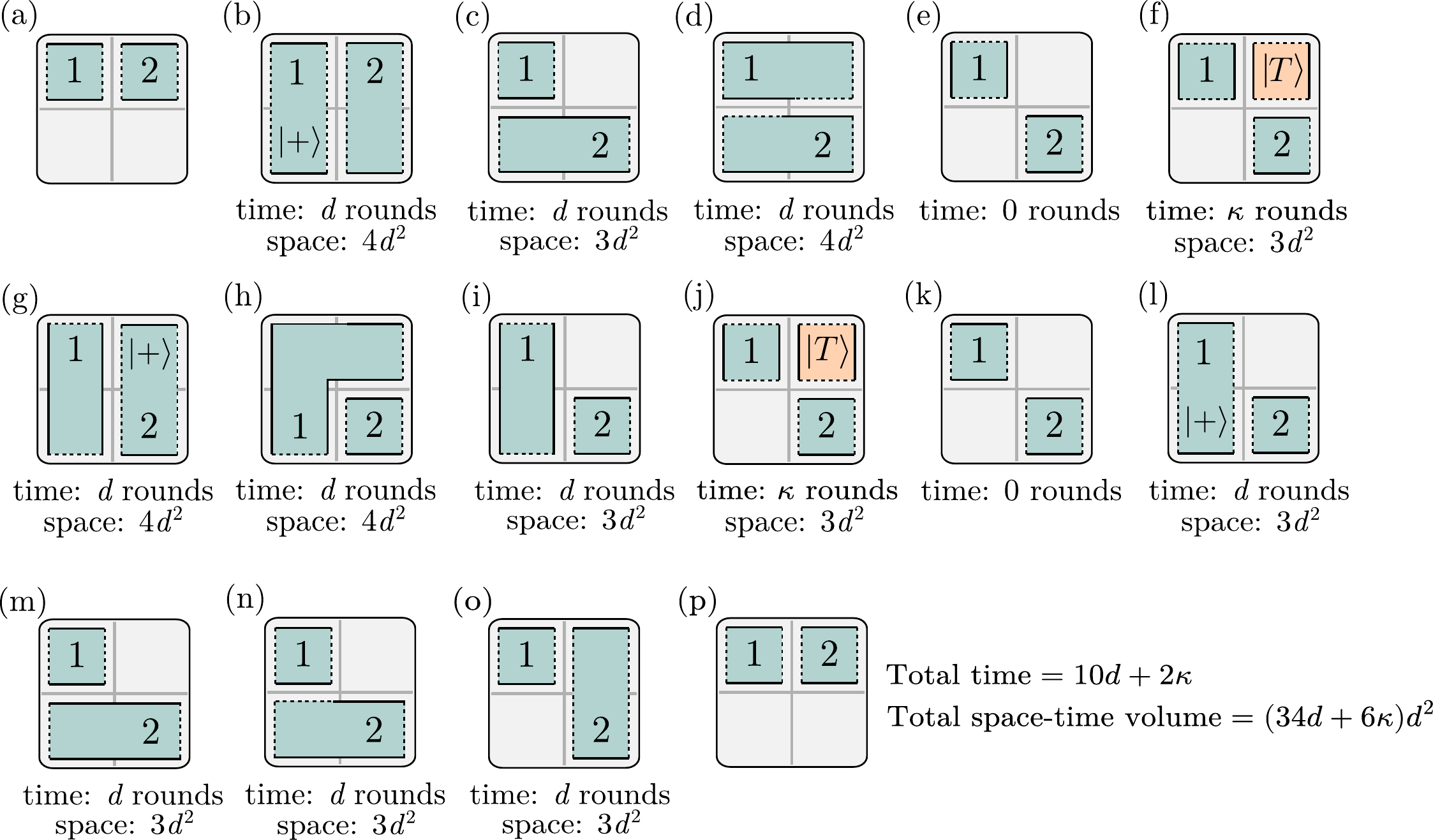}
    \caption{{\bf Lattice surgery implementation of $F_{ij}$ gate.} See \cref{fig:hubbard_basic_circuits}(c) for the circuit diagram. 
    Beginning in the initial configuration shown in (a), we apply the gates
    (b)--(c) $\text{CNOT}_{12}$ via ZZ and XX lattice surgery with the aid of an ancillary patch (patch 2 is also repositioned to allow the XX lattice surgery operation), (d) $S_1$, $S_2$, i.e., S gates on patches 1 and 2, (e) (instantaneous) transversal Hadamard on patch 1, (f) $T^{\dagger}$ gate via teleportation, (g)--(i) $\text{CNOT}_{21}$ followed by repositioning of patch 1, (j) $T$ gate via teleportation, (k) (instantaneous) transversal Hadamard on patch 1, (l)--(m) $\text{CNOT}_{12}$, (n) $S_2$, (o) repositioning patch 2 to return to the initial configuration in (p). 
    Here $\kappa$ denotes the number of rounds required to cultivate a $\ket{T}$ state and then teleport a $T$ (or $T^{\dagger}$) gate with it.}
    \label{fig:fij_gate}
\end{figure*}

The $F_{ij}$ gate is a two-qubit fermionic Fourier transform that is used in the $S_4$ tile. It can be implemented by \cref{fig:hubbard_basic_circuits}(c), assuming that the two qubits are adjacent in the Jordan-Wigner ordering.
\cref{fig:fij_gate} shows a lattice surgery implementation of the gate which takes $(10d + 2\kappa)$ rounds and has a space-time volume cost of $(34d + 6\kappa)d^2$.

\subsection{$S_1$ tile hopping term}

\begin{figure*}
    \centering
    \includegraphics[width=1.0\linewidth]{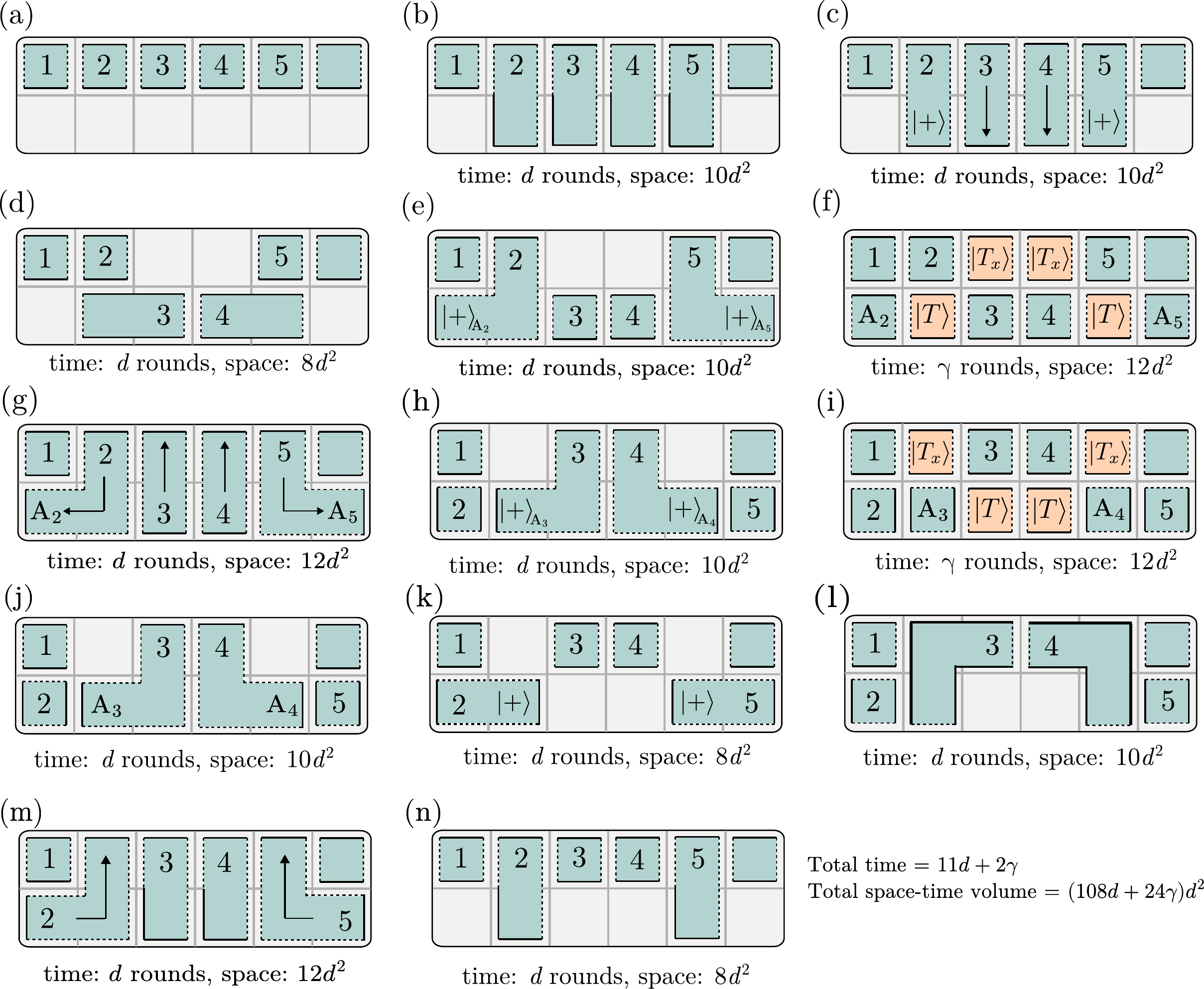}
    \caption{{\bf Lattice surgery implementation of the $S_1$ tile hopping terms.} This tile involves applying rotations $X_2 X_3 + Y_2 Y_3$ as well as $X_4 X_5 + Y_4 Y_5$. As per the circuit in \cref{fig:hubbard_basic_circuits}(b), this can be decomposed into a layer of 4 rotation gates on qubits 2, 3, 4 and 5. However due to the compact arrangement of patches in our layout, the 4 rotations are split into two sequential rotation layers, with rotations on qubits 2 and 5 implemented in parallel in (f) while rotations on qubits 3 and 4 are implemented in (i). The remainder of the compilation of the tile follows that of \cref{fig:xx_yy_rotation} with some additional steps for repositioning patches appropriately.}
    \label{fig:s1_tile}
\end{figure*}

The $S_1$ tile is an $XX+YY$ rotation, which can be implemented as shown in \cref{fig:hubbard_basic_circuits}(b). The fermionic orderings used allow us to implement both $S_1$ tiles attached to a given $S_4$ tile without updating this ordering. As such, we can look at $6$-site regions of the QPU and perform lattice surgery compilation for both of the $S_1$ tiles within this region. The full lattice is then spanned by $8$ such regions, which can operate in parallel (see \cref{fig:sections_and_qpu_layout}(c) and \cref{fig:fswap_network}(c) for visualisations of this partitioning). The two $S_1$ tiles are implemented by combining $X_2 X_3 + Y_2 Y_3$ and $X_4 X_5 + Y_4 Y_5$ rotations (following the labelling of the $S_4$ tile in \cref{fig:sections_and_qpu_layout}(b)).
The lattice surgery compilation of the two $S_1$ tiles costs $11d + 2\gamma$ rounds and a volume of $(108d + 24\gamma)d^2$, as illustrated in \cref{fig:s1_tile}.

\subsection{$S_4$ tile hopping term}
The $S_4$ tile comprises the circuit elements $F_{ij}$, $X_1X_3 + Y_1Y_3$ rotation, as well as the $CZ_{12}$ gate as per \cref{eq:s4_tile}.
\begin{figure*}
    \centering
    \includegraphics[width=0.5\linewidth]{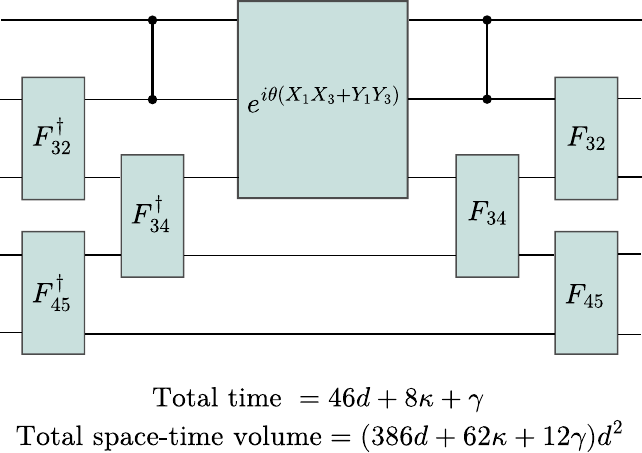}
    \caption{{\bf Circuit implementing the $S_4$ tile hopping term on patches 1--5.} The cost of this tile is obtained by combining the cost of the circuit components described in \cref{fig:hubbard_basic_circuits} as per \cref{eq:s4_tile}. Since the CZ gates can be applied in parallel to the $F_{34}$ gate, they do not contribute to the time cost of the tile. Additionally, two S gates on qubit 4 from adjacent $F_{ij}$ gates can be combined into a transversal $Z$. This allows us to simplify circuits, reducing $4d$ total from the combined time cost of the tile. The quoted volume is for a six-site region as in \cref{fig:xx_yy_rotation}, including one padding site.}
    \label{fig:s4_tile}
\end{figure*}
Combining the circuit elements, each tile can be implemented in time $46d + 8\kappa + \gamma$ rounds and requiring a total space-time volume of $(386d + 62\kappa + 12\gamma)d^2$, as illustrated in \cref{fig:s4_tile}. This volume includes that of idling qubits (for example while applying $F_{ij}$, which is not included in the numbers quoted in \cref{fig:fij_gate}) and the padding site, six sites total; the full lattice is spanned by $8$ such regions.

\subsection{Interaction term}
\begin{figure*}
    \centering
    \includegraphics[width=\linewidth]{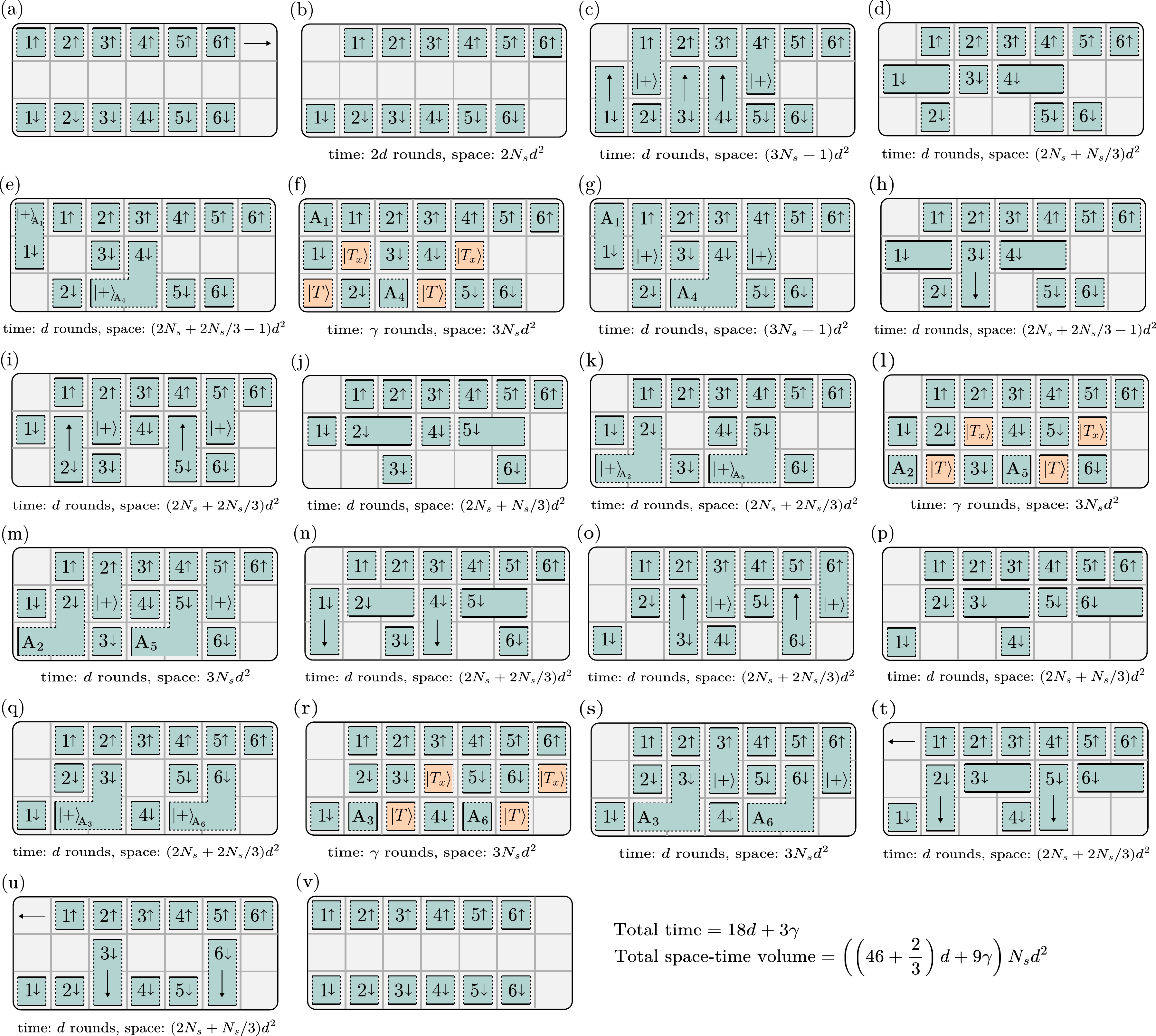}
    \caption{{\bf Representative lattice surgery implementation of the interaction term.} This example with $N_s = 6$ lattice sites illustrates how the $ZZ$ rotation between the up and down spins of each site needs to be implemented in 3 batches of parallel rotations. For the lattice of interest, $N_s = 48$.}
    \label{fig:zz_interac}
\end{figure*}

The interaction term entangles the logical qubits corresponding to the up and down spins of each lattice site.
To enable the appropriate entangling operations via lattice surgery, we use the walking surface code~\cite{huggins2025fluid, McEwen2023relaxinghardware} to shift the spin-up patches one patch to the right of the corresponding spin-down patches. Note, we assume that the QPU has padding on either side (see \cref{fig:sections_and_qpu_layout}(c)), and so this is possible.
The rotations in the interaction term are then applied according to the circuit in \cref{fig:hubbard_basic_circuits}(d), in a manner similar to the rotations involved in the $XX + YY$ rotation terms. 
Since the up- and down-spins of each lattice site will interact, they could all be implemented in parallel.
However, because our fallback implementation (\cref{sec:rotations}) requires three ancilla patches per rotation, and due to the space constraints of the compact layout, the $ZZ$ rotations are implemented in three sequential batches, with all rotations in each batch occurring in parallel.
The lattice surgery operations involved in this implementation are illustrated in \cref{fig:zz_interac} with a toy-example comprising 6 lattice sites. 
For our lattice of 48 sites, the $ZZ$ interaction term can be implemented in $18d + 3\gamma$ rounds with a space-time volume cost of $\left(\left(46 + \frac{2}{3}\right)d + 9\gamma\right)48d^2$.

Note that the interaction term has an identical form for the Hubbard model on any lattice. Therefore, this compilation and the corresponding resource estimates could be trivially applied to future studies of the Hubbard model on arbitrary lattices, not just the Lieb-kagome study of this work.

\section{Fermionic SWAP networks for the Lieb-kagome lattice}
\label{sec:fswap_network}

\begin{figure*}
    \centering
    \includegraphics[width=1.0\linewidth]{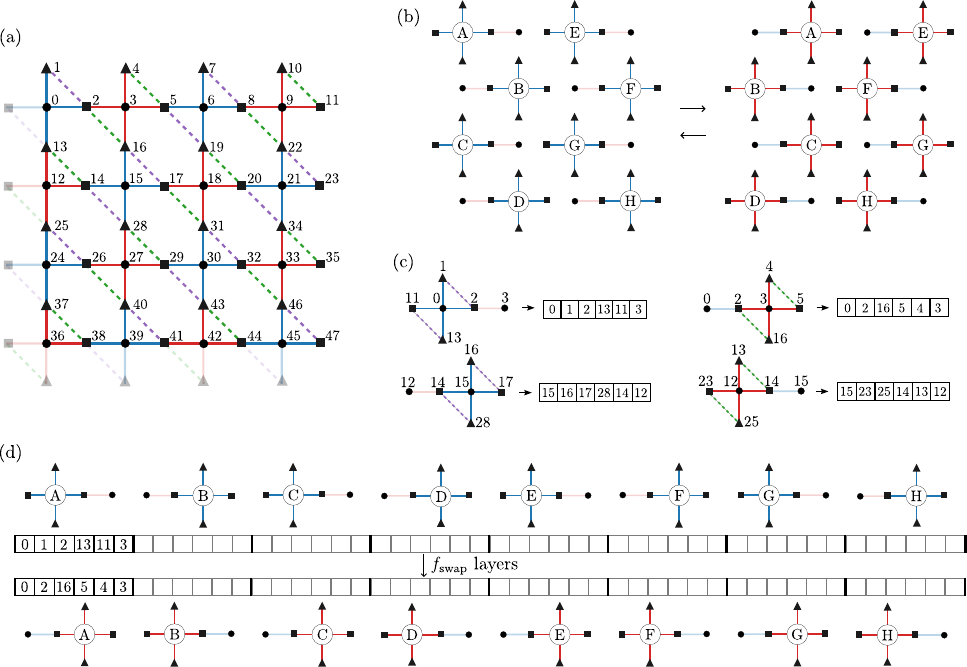}
    \caption{(a) A $4 \times 4$ kagome lattice with $48$ sites. The three sites per unit cell are labelled with a circle, triangle, and square. Periodic boundary connections are shown faded out. (b) The two clustering schemes to implement the hopping term of the Hamiltonian. In the first Jordan--Wigner ordering used, all sites in a single blue cluster live in a contiguous block. Each cluster has a dangling site attached which does not participate in any hopping terms for that ordering. (c) How each possible cluster type (blue and red, of which there are left- and right-facing types) gets mapped to a 1D segment in the QPU. (d) The ordering of the clusters in the QPU. By switching between these two Jordan--Wigner orderings we implement all hopping terms.}  
    \label{fig:fswap_network}
\end{figure*}

\subsection{Optimising fermionic swaps for the $4 \times 4$ Lieb-kagome lattice}\label{app:4x4_fswap}

In this appendix we describe the fermionic swap networks used to implement Trotter steps $\mathcal{S}_{\mathrm{even}}$ and $\mathcal{S}_{\mathrm{odd}}$, see \cref{eq:trotter_odd}. Recall that we implement the hopping term using a combination of $S_1$ and $S_4$ tiles: to implement all terms without using long non-local Pauli strings, we switch between two different Jordan--Wigner orderings, as shown in \cref{fig:sections_and_qpu_layout}(a). The two orderings used are given in \cref{fig:fswap_network}. First, we decompose the lattice into disjoint clusters in two different ways as shown in \cref{fig:fswap_network}(b). The ordering of sites within a given cluster for the four possible types is shown in \cref{fig:fswap_network}(c), and the overall sequence of clusters that we use is shown in \cref{fig:fswap_network}(d).
In the first ordering, the clusters are structured so that both the blue $S_4$ tiles and purple $S_1$ tiles can be implemented with no additional fermionic swaps, and likewise for red and green tiles in the second ordering.

Given the two orderings, we cost the depth of the fermionic swap network required to go from one to the other using a parallelised version of bubble sort (see Ref.~\cite[Fig.~2]{Akahoshi2025} for a visual example). To explain the algorithm, recall that in our proposed QPU layout each lattice site is mapped to a pair of logical qubits living in a quasi-1D architecture as shown in \cref{fig:sections_and_qpu_layout}(c). Numbering the cells on the QPU from left to right, all nearest-neighbour fermionic swaps whose left-hand cells have the same parity (call this the parity of the fermionic swap) can be executed in parallel. The sorting algorithm then takes the current ordering to some final target ordering by alternating between layers of even and odd nearest-neighbour fermionic swaps. Within a given layer, to decide whether to implement a fermionic swap of allowed parity between cells $i$ and $i+1$, compare the final target cell for the \textit{lattice site} currently in cell $i$ to that of cell $i+1$, and swap them if they are in the wrong order. Implement all such fermionic swaps in parallel, then do the same for fermionic swaps of the opposite parity, etc. After the first non-trivial layer, this algorithm will only fail to find an implementable fermionic swap for subsequent layers when all lattice sites are in the target order. Fix the starting parity by choosing the one that minimises circuit depth. We fix the order in which clusters are arranged in \cref{fig:fswap_network}(d) so as to minimise the depth of the fermionic swap network. The search was carried out by sampling uniformly from the $(8!)^2 \approx 2 \times 10^9$ possible cluster orderings by taking roughly $10^6$ samples. Note that our orderings may not be optimal.

The explicit sequence of orderings (and therefore the fermionic swap network) used in this work is given in the ancillary file \texttt{fswap\_network.json} (available on Zenodo) as a list of orderings, beginning with the ordering shown on the top of \cref{fig:fswap_network}(d), and ending with the bottom ordering. To switch back to the first ordering the sequence is run in reverse. A complete reordering requires 25 layers of fermionic swaps.

Next, let us discuss the depth and volume of this network, as implemented using $\fswap$ gates that take $3$ time steps each, described in \cref{sec:fswap_lattice_surgery}. In this approach, each $\fswap$ has a volume of $11$ blocks. In addition, we need to insert move-and-rotate operations, or ``restore operations'', into the fermionic swap network at certain points. Each restore operation requires $2$ time steps and has a volume of $4$ blocks. These restore operations can always be performed in parallel with $\fswap$ gates, so they contribute no additional time cost, but their volume must be accounted for. We find that we also need to perform some restore operations at the end of the network, to return the remaining routing-space qubits to the top row; however, these too can be performed in parallel with the last round of $\fswap$ gates. The script to calculate the required number of restore operations is included on Zenodo as \texttt{fswap\_restores.py}. We find that running through the network forwards requires $12$ restore operations, while the backwards network can be achieved at the same cost by running the lattice surgery schedule in reverse. The total time cost is therefore exactly $2 \times 25 \times 3 = 150$ time steps, where the $2$ accounts for updating both spin-up and spin-down orderings. This is the time cost quoted in \cref{fig:cost_per_trotter_step}. Calculating the volume cost requires adding the volume of all fermionic swaps, move-and-rotate restore operations, and idle operations (including both spin-up and spin-down rows of qubits). This is calculated in the costing script provided on Zenodo. As quoted in \cref{fig:cost_per_trotter_step}, this is calculated as 17,208 blocks for the full network.

\subsection{General results for $L \times L$ lattices}
\label{app:fswaps_LxL}
\newcommand{\fold}{ \textsf{fold}}

We found the orderings above using the cluster heuristic and optimising by a random search which is not scalable as the system size increases. In this section, for the simpler case of Trotterisation with only $S_1$ tiles, we give explicit strategies for implementing the hopping Hamiltonian for two-dimensional square and kagome lattices with $L$ unit cells along each dimension. These results are not relevant for the case study in the main text; however, we provide them in case they may be useful for other studies of the square and kagome lattices with $S_1$ tilings. Note that these results are for a full second-order step, including both forwards and reverse swap networks, while the results in the previous section are for only one direction (consistent with the Trotter steps defined in the main text).

\begin{table}[h]
    \centering
    \setlength{\tabcolsep}{8pt}
    \begin{tabular}{ccc}
    \hline
    \hline
    Lattice & $D(L)$ & $L=4$, exact \\
    \hline
    square, open & $2L-2$ & 6\\
    square, periodic & $12 L + 12$ & 40 \\
    kagome, open & $6L + 8$ &  32 \\
    kagome, periodic & $12L + 10$ & 58  \\
    \hline
    \hline

    \end{tabular}
    \caption{Fermionic swap depth to implement one full second-order Trotter step for different lattices of size $L \times L$. We quote both the asymptotic scaling with linear size $L$, and the explicit costing for $L=4$, though they only disagree for the periodic square lattice.}
    \label{tab:fswaps_for_LxL}
\end{table}

For concreteness, take an ordering $f$ to be an invertible function from the sites $\mathcal{S}$ of the lattice  to $\left[\, |\mathcal{S}| \,\right]$, zero-indexed. We will use a sequence of orderings $\{f_i\}_{i=1}^M$, where for every pair of sites $s_1,s_2$ in the lattice that share an edge, we can implement this hopping using the $S_1$ tiling strategy if we have some $f_i$ with $|f_i(s_1)- f_i(s_2)|=1$. Given a costing function $D_{\rm layer}(f,g)$ which gives the fermionic swap depth to go from ordering $f$ to $g$ (we use the costing algorithm from App.~\ref{app:4x4_fswap}), the total fermionic swap depth $D$ is
\begin{align}
    D = 2 \sum_{i=1}^{M-1} D_{\rm layer}(f_i, f_{i+1}).
\end{align}
Note that if we were working with first-order Trotterisation, we would need to include the cost of $f_M \to f_1$. Working at second order runs the reorders in reverse for the second half of the Trotter step. The specific orderings used for the four lattices considered are described below, and the numerically observed scaling of the fermionic swap depth with $L$ is reported in Table~\ref{tab:fswaps_for_LxL}.

\subsubsection{Open square lattice}

There is a known and provably optimal ordering sequence for the square lattice with open boundary conditions \cite{Akahoshi2025, hagge_optimal_2022}. The strategy uses two orderings and moving between them has a minimum fermionic swap depth of $L-1$. We are able to reproduce this scaling in numerics with our greedy compilation. Remembering that we use second-order Trotterisation, the cost for one full step is
\begin{equation}
    D^{\rm sq, obc}(L) = 2L - 2.
\end{equation}

\begin{figure}
    \centering
    \includegraphics[width=0.65\linewidth]{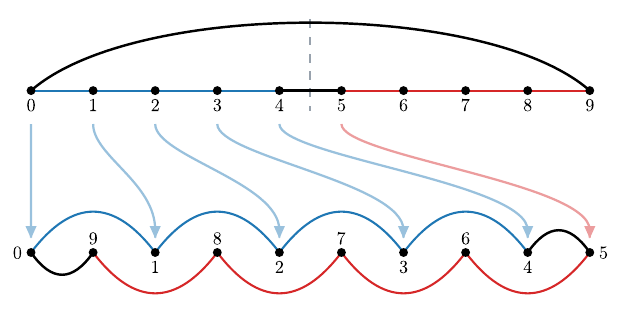}
    \caption{Folding a 1D lattice with periodic boundary conditions. This ensures that all connected sites are a constant distance apart rather than some being $\Omega(L)$ apart. Folding each dimension independently extends this trick to a periodic square lattice of arbitrary dimension.}
    \label{fig:folding_trick}
\end{figure}

\subsubsection{Periodic square lattice}

Beginning from the first ordering used by Akahoshi et al. \cite{Akahoshi2025} will clearly give suboptimal -- i.e. $\Omega(L^2)$ -- fermionic swap depth for a square lattice with periodic boundary conditions, because it begins at one corner of the lattice and places the opposite corner near the end of the ordering, but in the periodic lattice these are next-nearest neighbours. To achieve $O(L)$ scaling in this case, we use orderings inspired by a folded version of the lattice \cite{poole_architecture_2025}. This fold makes sure that all connected pairs in the lattice are a constant (Manhattan) distance apart from each other without adding long-range connectivity, though at the cost of inflating the typical distance between nearest-neighbours from 1 to 2, as shown in \cref{fig:folding_trick}.

The ordering strategy is simple. We use four different orderings, where in each one we implement only horizontal or only vertical hoppings. Concretely, label a single lattice site with the tuple $(i,j)$ to respectively index the row and column. Note $i,j$ are zero-indexed, and $L$ is always taken to be even. We can most conveniently specify our orderings relative to the row-major ordering of the folded lattice,
\begin{align}\label{eq:inital_periodic_square_ordering}
    (i,j) &\mapsto L \,\fold_L(i) + \fold_L(j) ,\\
    \fold_L(n) &\equiv \begin{cases}
        2 n & n = 0,1,\dots, \frac{L}{2}-1\\
        2 \left(\frac{L}{2} - (n \mod \frac{L}{2})\right)-1 & n = \frac{L}{2}, \dots, L-1
    \end{cases}. \label{eq:fold}
\end{align}
This defined, our sequence of four actual orderings is as follows:
\begin{enumerate}
    \item Starting from the baseline ordering of \cref{eq:inital_periodic_square_ordering}, fix the relative order of all sites $(i,j)$ with $j$ even. Then require that each such $(i,j)$ is followed in the ordering by $(i,j+1)$.
    \item From the Step 1 ordering, fix the location of all sites with even $j$ and now require each to be followed by $(i,j-1 \mod L)$.
    \item Fix the relative ordering of all even $i$ sites in the Step 2 ordering, then require each to be followed by $(i+1, j)$.
    \item From Step 3 ordering, fix the location of all even $i$ sites and now require each to be followed by $(i-1 \mod L, j)$.
\end{enumerate}

\subsubsection{Periodic kagome lattice}
\begin{figure*}
    \centering
    \includegraphics[width=0.35\linewidth]{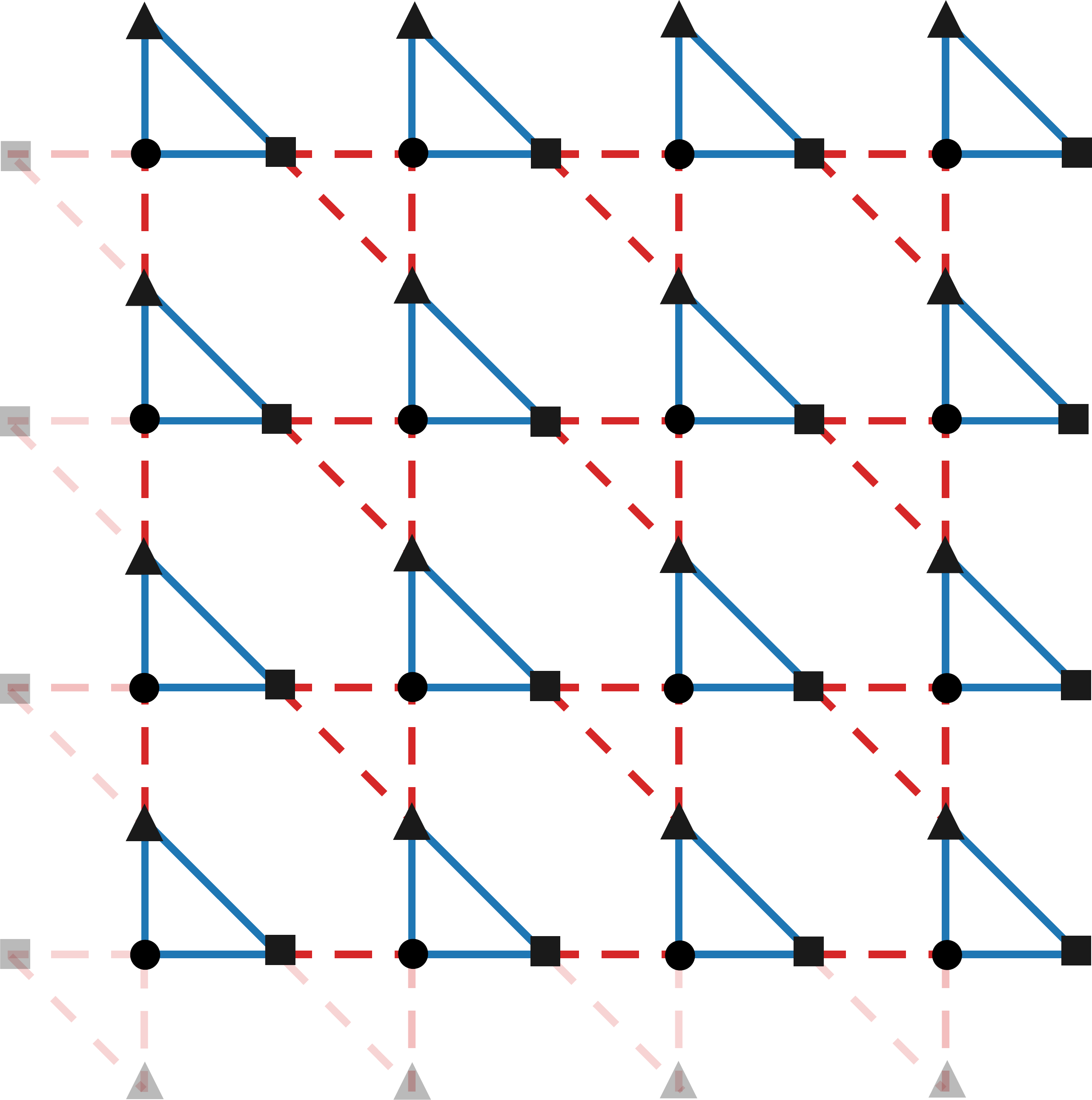}
    \caption{Kagome lattice decomposed into two sets of S1 tiles. The lattice decomposes into disjoint blue clusters of size 3, then into red clusters of size 3. }
    \label{fig:s1_kagome}
\end{figure*}

We now discuss the kagome lattices, starting with the periodic case as it is simpler. Label a single lattice site with the tuple $(i,j,k)$ to respectively index the row, column, and in-cluster index $k \in \{0,1,2\}$ which agrees with the sites $0,1,2$ of \cref{fig:fswap_network}. Our ordering strategy again comes from breaking up the edges into sets of disjoint clusters, but in this case the clusters are built out of $S_1$ tiles. For instance, we can begin by implementing all the edges within each unit cell, the blue tiles in \cref{fig:s1_kagome}. Then the red edges which remain separate the graph into disconnected fragments with again 3 sites each. Our initial ordering then comes from taking the original square lattice of 3-clusters, applying the folding trick, and then ordering the folded lattice of clusters naively starting with the entire first row, then the second, etc. Explicitly, our initial ordering is given by 
\begin{equation}\label{eq:inital_periodic_kagome_ordering}
    (i,j,k) \mapsto 3L\, \fold_L(i)   + 3\,\fold_L(j) + k.
\end{equation}
where $\fold_L$ is defined in \cref{eq:fold}. For the explicit way we assign the index $k$ within a cluster, see sites 0--2 in \cref{fig:fswap_network}(a). We then cover all hopping terms with the following sequence of orderings:
\begin{enumerate}
    \item Order as in \cref{eq:inital_periodic_kagome_ordering}
    \item Take the ordering from Step 1, but for every cluster swap $(i,j,0) $ and $(i,j,1)$, which can be done with a single fermionic swap layer.
    \item From the ordering of Step 2, fix the location of all $k=0$ sites: this completely fixes the order of the red clusters. Require the $k = 1,2$ sites to be next to the $k=0$ site of their red clusters. Empirically we find the in-cluster ordering $(1,0,2)$ minimises the constant term in the final depth.
    \item As in Step 1 $\to$ Step 2, swap $(i,j,0)$ and $(i+1\! \mod L,j,1)$, which can be done with a single fermionic swap layer.
\end{enumerate}
Note that some orderings require multiple layers of rotation gates, as for instance in ordering 1 the kinetic term on $(i,j,0),(i,j,1)$ must be implemented in a different layer from the term on $(i,j,1),(i,j,2)$. Using two layers of rotation gates in orderings 1 and 3, we can implement every kinetic term once using a total of 6 rotation gate layers.

\subsubsection{Open kagome lattice}

Compared to the previous example, the only additional subtlety for the open kagome case is how to deal with those red clusters that have fewer than 3 sites because of the open boundary conditions (\cref{fig:s1_kagome}, and note that with open boundary conditions the topmost sites, rightmost sites, and bottom left corner only participate in blue edges). We begin with the ordering
\begin{equation}\label{eq:initial_open_kagome_ordering}
    (i,j,k) \mapsto 3L i + 3j + k
\end{equation}
from which the total sequence of orderings is as follows:
\begin{enumerate}
    \item Order as in \cref{eq:initial_open_kagome_ordering}
    \item Take the ordering from Step 1, but for every cluster swap $(i,j,0) $ and $(i,j,1)$.
    \item From the ordering of Step 2, fix the relative order of all $k=0$ sites, which fixes the relative order of the red clusters. For $k = 1,2$ sites which participate in a red cluster, require the $k=1$ ($k=2$) sites to come before (after) the $k=0$ site in their red cluster as for the periodic case. Fix the relative order of the leftover $k=1,2$ sites. The final ordering is chosen so that red clusters are never interrupted and leftover $k=1,2$ sites have the smallest position greater than or equal to their position in Step 2, subject to the constraints on relative order.
    \item For every red cluster with three sites, swap the $k=0$ and $k=1$ to implement the last hopping term in each.
\end{enumerate}

\begin{figure}
    \centering
    \includegraphics[width=0.6\linewidth]{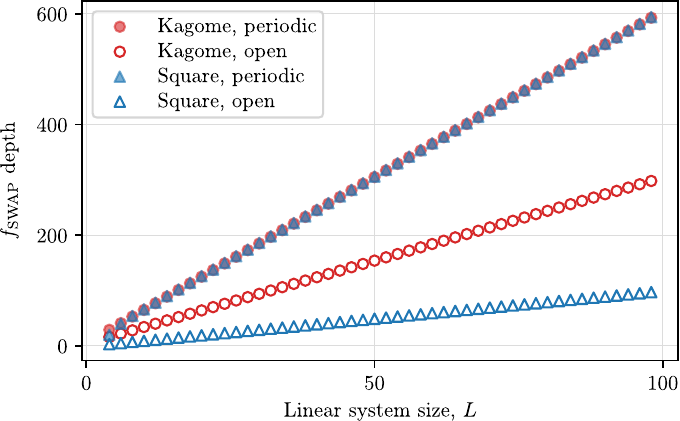}
    \caption{Fermionic swap depths of the ordering strategies of App.~\ref{app:fswaps_LxL} to implement a full second-order Trotter step on various lattices with $L\times L$ unit cells. The asymptotic costs for square and kagome periodic lattices are nearly identical.}
    \label{fig:fswap_depth}
\end{figure}
\subsection{Summary}
The exact numerical results for the orderings described, costed using our greedy compilation, are shown in \cref{fig:fswap_depth}. For $L>6$, the costing agrees exactly with the asymptotic scaling given for each lattice type. These asymptotic scalings, as well as the exact cost for $L=4$, are given in Table~\ref{tab:fswaps_for_LxL}.

\section{Trotter and sampling error analysis of spectral functions}
\label{sec:trotter_error_analysis}

In this appendix, we perform error analysis of spectral functions, beginning with Trotter error analysis and then investigating finite sampling error.

Our Trotter error analysis aims to support our choice of Trotter step size in our case study. In particular, for the Lieb-kagome case study we primarily consider a step size of $\delta t = 0.1 \tau^{-1}$ (and also perform resource estimation for a more optimistic $\delta t = 0.2 \tau^{-1}$). In our STAR analysis we take a slightly smaller step size of $\pi/32 \approx 0.098$, as it allows rotations to remain in the Clifford hierarchy (for appropriate $\tau$, $\tau'$ and $U$). Relying on worst-case Trotter error bounds \cite{childs_theory_2021, kivlichan2020improved, campbell2022early} results in a much smaller Trotter step size which, while rigorous, would result in extremely pessimistic resource estimates that put our case study well beyond the MegaQuOp regime and overestimate the true cost. This is particularly true if these worst-case bounds are evaluated by further applying the triangle inequality to evaluate the required commutator spectral norms, which loosens worst-case bounds even further, see \cite{blunt2025}. Instead, we rely on the following heuristic to motivate our choice. First, we perform time-dependent DMRG simulations (using the time-dependent variational principle, or ``TDVP'') to construct the dynamical susceptibility for a one-dimensional Hubbard model, performed both with and without Trotterisation, and using a step size of $0.1 \tau^{-1}$. This experiment uses the same parameters and quench protocol as described in the main text, and we will see that Trotter error is negligible, even with $\delta t = 0.1 \tau^{-1}$. Then, we perform worst-case analysis of various small ``fragment'' lattices in one- and two-dimensions, showing that the worst-case error for kagome fragments is only a small factor higher than for one-dimensional lattices. This argument is ultimately a heuristic to motivate our choice of step size for the Lieb-kagome case study, not a rigorous proof. We note that non-error-corrected quantum experiments calculating similar spectral functions for the Hubbard model typically use even larger step sizes \cite{vilchez-estevez_extracting_2025}. Similarly, other dynamics experiments for the Hubbard model performed on non-error-corrected devices have used much larger step sizes \cite{alam2025_1, alam2025_2}.

\subsection{TDVP simulation of one-dimensional Hubbard models}
\label{app:tdvp_trotter}

In this appendix, we construct dynamical magnetic susceptibilities for a one-dimensional Hubbard model with $N=30$ sites and periodic boundaries, and using the same quench protocol as considered in our Lieb-kagome case study in the main text. This is performed using both non-Trotterised time evolution, and Trotterised evolution with a time step of $\delta t = 0.1\tau^{-1}$, a Lorentzian broadening with $\eta=0.25 \tau$ and $128$ Trotter steps, using the same parameters as in our case study. Similarly to the case study, we define ``odd'' and ``even'' Trotter steps, in which we run through hopping terms forwards and backwards, respectively. Hopping terms are partitioned into two sections of $S_1$ tiles. In both Trotterised and non-Trotterised cases, time evolution is performed using the time-dependent DMRG method, specifically the time-dependent variational principle (TDVP) approach. The non-Trotterised simulations may contain other DMRG and TDVP errors, though these are well-converged for our example. In particular, TDVP still contains a splitting approximation, though it is typically much smaller than the Trotter splitting error that we study here. All simulations were performed using the Block2 code \cite{Zhai2023}.

\begin{figure*}
    \centering
    \includegraphics[width=0.98\linewidth]{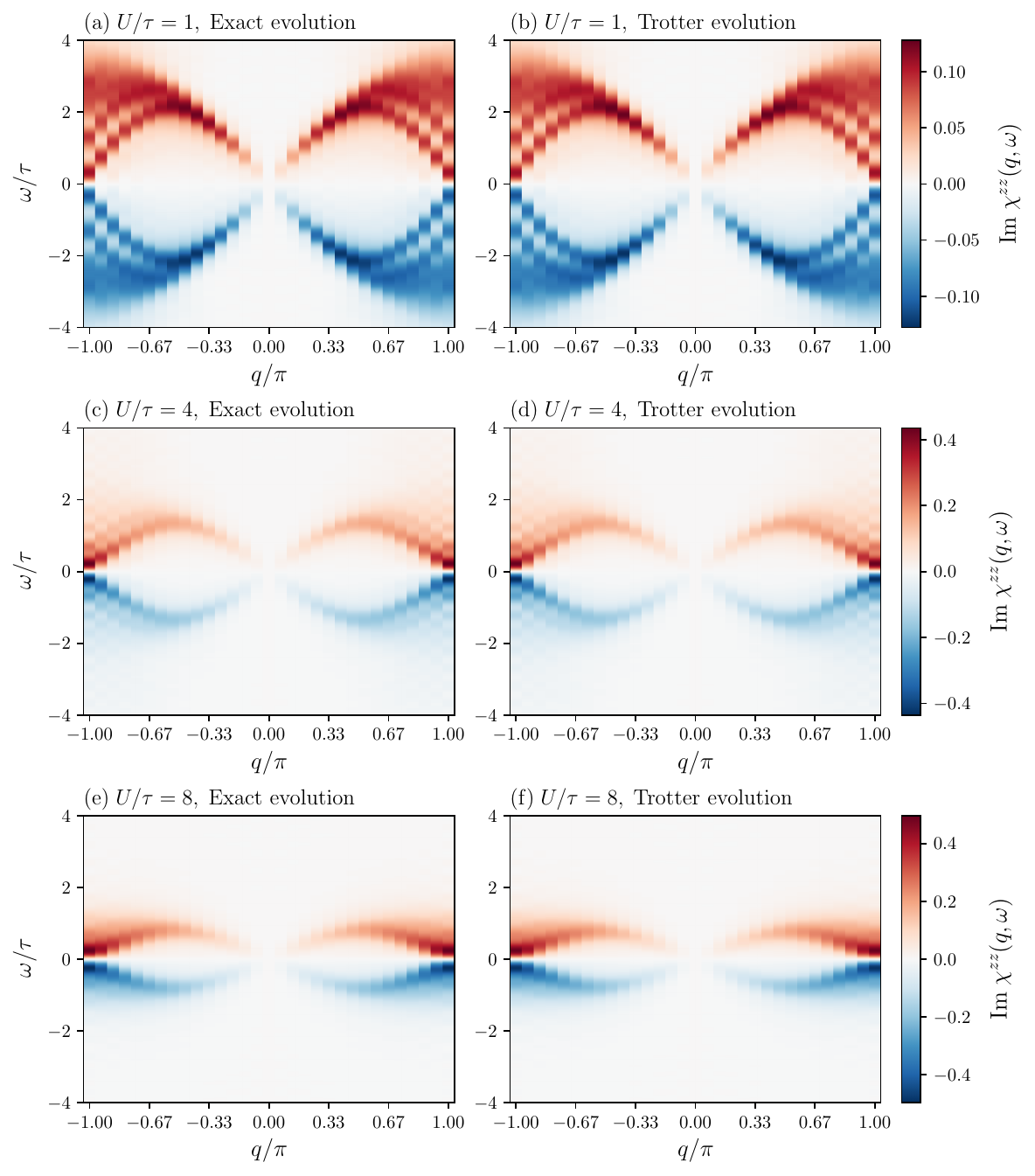}
    \caption{{\bf Dynamical susceptibilities with exact and Trotterised evolution.} Dynamical susceptibilities, $\mathrm{Im} [\chi^{zz}(q,\omega)]$, for a one-dimensional $30$-site Hubbard model ($N_s = 30$) with periodic boundary conditions with (a)--(b) $U/\tau = 1$, (c)--(d) $U/\tau = 4$, and (e)--(f) $U/\tau = 8$. All results use time-dependent DMRG using the TDVP approach with bond dimension $M=600$. In (a), (c) and (e) we perform ``exact'' evolution with the full Hamiltonian, introducing no approximation beyond the use of TDVP and restricted bond dimension. In (b), (d) and (f) we Trotterise the evolution with time step $\delta t = 0.1 \tau^{-1}$. All parameters match those from the case study in the main text: the total evolution time, $T = 12.8 \tau^{-1}$, the Trotter time step size $\delta t = 0.1 \tau^{-1}$, and broadening parameter $\eta = 0.25 \tau$. It is seen that the exact and Trotterised evolution are essentially identical, and all qualitative features are reproduced; comparisons for specific $q$ values are given in \cref{fig:exact_vs_trotter_zoomed}.}
    \label{fig:exact_vs_trotter_spectral}
\end{figure*}

\begin{figure*}
    \centering
    \includegraphics[width=0.95\linewidth]{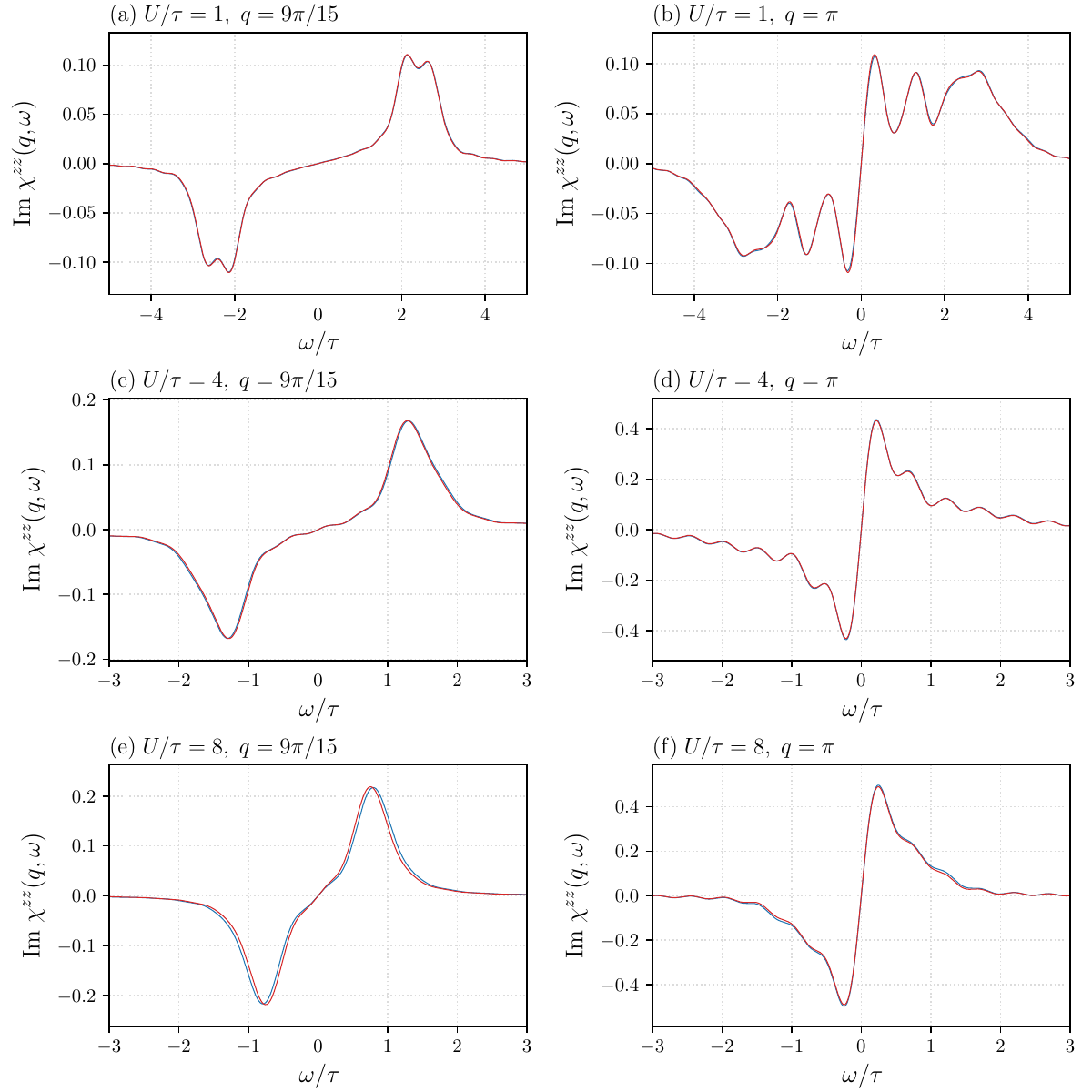}
    \caption{{\bf Trotter vs exact evolution at specific $q$ values.} Comparison of ground-state dynamical susceptibilities for a $30$-site one-dimensional Hubbard model at two wave vectors, $q=9 \pi/15$ and $q=\pi$, comparing Trotterised (plotted in red) and non-Trotterised (plotted in blue) evolution. Results are shown for interaction strengths $U/\tau = 1$, $4$ and $8$. Note that both sets of results contain error from the TDVP approximation and finite bond dimension, $M=600$. Results are seen to be almost identical in all cases. The largest discrepancy is for $U/\tau=8$, $q = 9\pi/15$, where the peak location is slightly shifted. Still, this is negligible compared to the energy resolution, set by the broadening $\eta=0.25\tau$. All simulation parameters match the case study in the main text. In particular, these results are the same as those plotted in \cref{fig:exact_vs_trotter_spectral}.}
    \label{fig:exact_vs_trotter_zoomed}
\end{figure*}

Results are shown in \cref{fig:exact_vs_trotter_spectral}, performed at interaction strengths $U/\tau$ of $1$, $4$ and $8$, with $U/\tau = 8$ expected to have the largest Trotter error. Non-Trotterised or ``exact'' results are shown on the left, while Trotterised results are shown on the right. Note that different colour scales are used for different values of $U/\tau$; in particular, the response is much weaker at $U/\tau=1$. We see that the Trotterised and non-Trotterised spectral functions are essentially identical by eye. To investigate more closely, we plot the corresponding spectral functions at two values of $q$ in \cref{fig:exact_vs_trotter_zoomed}, specifically for $q = 9\pi/15$ and $q = \pi$. Here we can see more clearly that the Trotter error is essentially negligible at this Trotter step size. The largest error is observed at $U/\tau = 8$, $q = 9\pi/15$, where the peak is slightly shifted. Since the peak's position is determined by an energy difference, this demonstrates that the Trotter error causes a very small shift in this energy difference.

Lastly, we provide speculation on why the observed Trotter error may be so small. The spectral function is a sum of peaks whose positions lie at energy differences, $\Delta_{ji} = E_j - E_i$, see \cref{eq:explicit_spectral_function_with_broadening}. The effect of Trotterisation is to work with an effective Hamiltonian, $H_{\rm eff}$, instead of the original Hamiltonian, $H$. We denote the eigenvalues of the effective Hamiltonian $\{ \tilde{E}_i \}$. Thus, each $\Delta_{ji}$ will be replaced by
\begin{equation}
\tilde{\Delta}_{ji} : = \tilde{E}_j - \tilde{E}_i.
\end{equation}
There will also be a shift in the peak amplitudes due to the eigenvectors of $H_{\rm eff}$ being slightly different to those of $H$. The main contributions to the spectral function are from low-energy states (assuming that the initial state is sufficiently accurate). Ref.~\cite{BaySmidt2026} provides numerical evidence that the Trotter error on low-lying energy differences is significantly lower than Trotter error on individual energy eigenvalues; that is, there is significant \emph{Trotter error cancellation} between low-lying energies. Given that the accuracy of these spectral functions depends on such energy differences, it seems likely that this error cancellation mechanism will allow relatively large Trotter step sizes to be used when simulating such excitation spectra.

\subsection{Scaling of worst-case Trotter error bounds}
\label{app:trotter_bounds}

\begin{figure}
    \centering
    \includegraphics[width=0.7\linewidth]{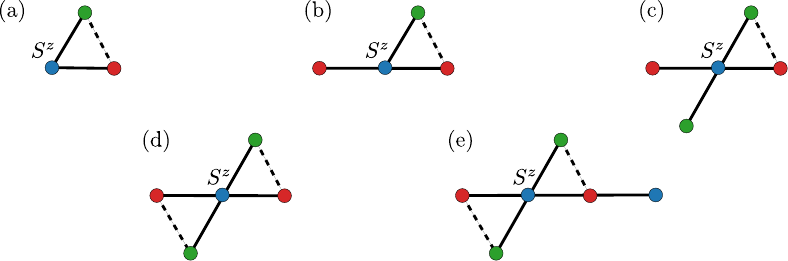}
    \caption{Fragments of the kagome lattice used to evaluate numerically exact Trotter error. We always look at the operator norm error of a Heisenberg-evolved spin $S_z$ on the labelled site.}
    \label{fig:kagome_fragments}
\end{figure}

Next, we present empirical evidence that Trotter error for the kagome lattice is not significantly worse than the 1D lattice, further justifying our use of a $\delta t = 0.1 \tau^{-1}$ time step. Our arguments are heuristic and driven by small-system numerics, as using rigorous error bounds often leads to overly pessimistic estimates of the error in practice \cite{Babbush_loose_trotter_bounds,BaySmidt2026}. We calculate the Trotter error with exact diagonalisation for small systems of differing sizes and connectivity. We examine a variety of systems to explore how Trotter error for time-evolved single-site observable scales with system size and connectivity. We consider three types of systems: chains with open boundary conditions, chains with periodic boundary conditions, and fragments of the kagome lattice. For all types we consider systems from 3 to 6 sites in size to allow exact diagonalisation. The kagome fragments that we consider are shown in \cref{fig:kagome_fragments}. Note that the 3-site fragment is the same as the 3-site periodic chain.

\begin{figure*}
    \centering
    \includegraphics[width=\linewidth]{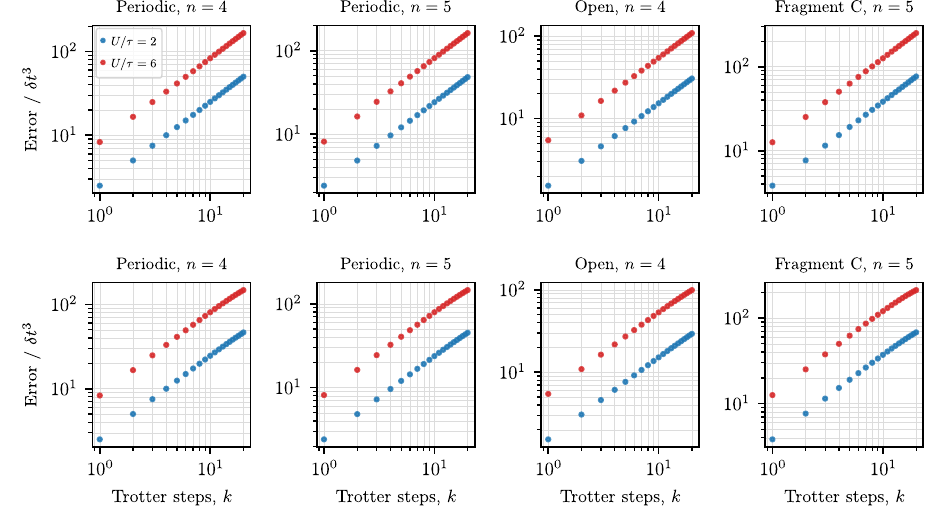}
    \caption{Operator norm error in $S_z(t)$ as a function of the number of Trotter steps $k$ for several different 1D chains and kagome fragment \cref{fig:kagome_fragments}(c). The top row shows results for $\delta t = 0.001\tau^{-1}$, the bottom row for $\delta t = 0.01\tau^{-1}$. Slight deviation from linear scaling is visible for $\delta t= 0.01\tau^{-1}$. }
    \label{fig:k_scaling}
\end{figure*}

We track the error in the time-evolved observable $S^z_i$, for some site $i$. We take $S^z$ on a boundary point for the open chain, a bulk point for the periodic chain, and the highest-connectivity site for the kagome fragments as shown in \cref{fig:kagome_fragments}. We then compare the time evolution of this observable under the ideal and Trotterised time evolutions, which give rise to the exact and Trotterised Heisenberg observables $S_i^\alpha(t), \tilde S_i^\alpha(t)$ respectively. The operator-norm error of this observable is directly related to the spectral function that we ultimately wish to extract. Recall that the spectral function~\cref{eq:spectral_function_final_expression} is calculated from the quantities
\begin{equation}
    C^{\alpha\beta}_{i, j}(t) =  -\mathrm{i} \langle \psi | [ S^{\alpha}_i(t), S^{\beta}_j(0) ] | \psi \rangle.
\end{equation} 
\newcommand{\Strott}{\mathcal{S}_{\mathrm{Trotter}}}
\newcommand{\Sideal}{\mathcal{S}_{\mathrm{ideal}}}
We can bound the error of a single term in the commutator in terms of the operator norm error in the time-evolved observable as
\begin{equation}
     \left |\langle \psi |  S_i^\alpha(t) S^\beta_j| \psi \rangle - \langle \psi |  \tilde S_i^\alpha(t)  S^\beta_j| \psi \rangle \right | \leq \left| \left|S_i^\alpha(t) -\tilde S_i^\alpha(t) \right| \right|_{\rm op} \left| \left|S_j^\beta \right| \right|_{\rm op} \leq \left| \left|S_i^\alpha(t) -\tilde S_i^\alpha(t) \right| \right|_{\rm op},
\end{equation}
where $|| \cdot||_{\rm op}$ denotes the operator norm. For a second-order Trotter formula this can be bounded by
\begin{equation}
    \left| \left|S_i^\alpha(t) -\tilde S_i^\alpha(t) \right| \right|_{\rm op} \leq W k \delta t^3,
\end{equation}
where we Trotterise using $k$ steps of length $\delta t$ and $W$ may depend on the system size and Hamiltonian coefficients \cite{childs_theory_2021}. We examine the error introduced by approximating the ideal unitary $\Sideal$ with a simple second-order Trotter unitary $\Strott$
\begin{align}
\Sideal &= e^{-\mathrm{i} (H_h+H_I )\delta t},\\
\Strott &= e^{-\mathrm{i} H_I\delta t/2} e^{-\mathrm{i} H_h\delta t} e^{-\mathrm{i} H_I\delta t/2}, 
\end{align}
where $H_h$ and $H_I$ are the hopping and interaction terms respectively. Note that in this analysis we do not Trotterise the hopping term as we expect the leading error contribution to be from the potential-kinetic split, particularly for large interaction parameter $U$, as demonstrated numerically in \cite{BaySmidt2025, BaySmidt2026}. We keep the step size $\delta t$ constant, but vary the total evolution time with the number $k$ of applications of $\Strott$ and $\Sideal$.
For each system we calculate the error
\begin{equation}
    \varepsilon(k) = \left|\left|\Strott^{k\dag} \,S^z_i\, \Strott^k  -  \Sideal^{k\dag} \, S^z_i\, \Sideal^k \right|\right|_{\rm op}.
\end{equation}
The error is fitted to the ansatz 
\begin{equation}
    \varepsilon_{\rm fit}(k) =  W k^\gamma \delta t^3
\end{equation}
with $W,\,\gamma$ as fit parameters from linear least squares in log-log space. We expect $\gamma \approx 1$, validated in \cref{fig:k_scaling}.

\begin{figure*}
    \centering
    \includegraphics[width=0.95\linewidth]{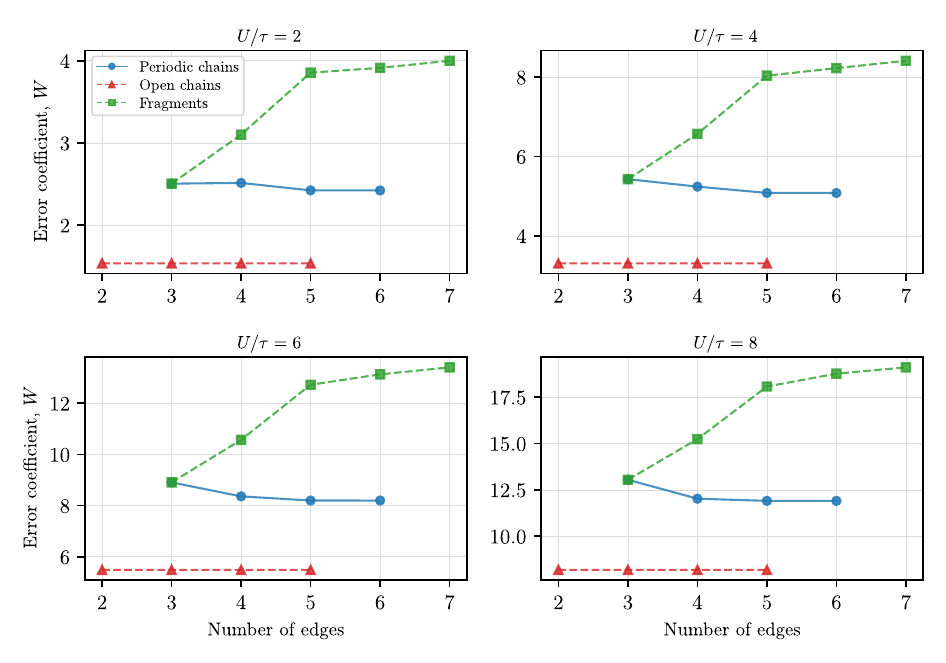}
    \caption{Trotter error coefficient $W$ versus number of edges in the fragment for several different values of $U$. The high connectivity examples (periodic chain and especially kagome fragments) show increased Trotter error, but the effect tapers as system size increases. The Trotter step size is $\delta t = 0.01 \tau^{-1}$.}
    \label{fig:fragment_trotter_error}

    \vspace{2.0em}

    \includegraphics[width=0.95\linewidth]{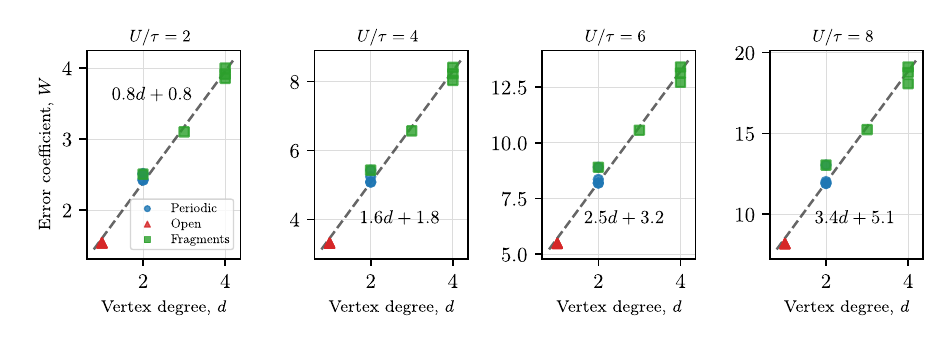}
    \caption{Trotter error coefficient $W$ as a function of vertex degree. Linear fits are shown next to dashed trend lines. Different fragment types are well described by the same linear trend in vertex degree. The Trotter step size is $\delta t = 0.01 \tau^{-1}$.}
    \label{fig:vertex_degree_trotter_error}
\end{figure*}

\cref{fig:fragment_trotter_error} plots the error amplitude parameter $W$ for different systems, where the systems are organised by the total number of edges in the connectivity graph. For 1D systems the Trotter error constant $W$ very quickly converges to a constant value as the number of edges or system size increase. The large increase in the error for the periodic chain compared to the open chain also showcases the importance of connectivity. Extrapolating the trends in the fragments is more difficult, but we can again see evidence of convergence to a constant $W$, though the data has not quite converged for the small systems examined here. Nevertheless, we only need a rough order of magnitude estimate for $W$ in the large system limit and these small systems are sufficient to show that $W$ is not growing with the number of edges and is on the order of 1 to 10 for $U$ values between 2 and 8, and furthermore within a factor of 2 of the periodic chain results. Note that a rigorous bound on $W$ using standard commutator bounds will in general scale with the system size, with lattice Hamiltonians generally leading to bounds on $W$ which are linear in the total number of sites \cite{childs_theory_2021}. It is therefore noteworthy that we observe the error coefficient to approach a constant as the system size is increased. This may be due to special structure in the time evolution of a local observable \cite{Heyl_local_observable_trotter_error}.

\cref{fig:vertex_degree_trotter_error} plots the same data, but now as a function of the degree of the vertex initially supporting $S_z(0)$. Here, we see a general linear scaling of $W$ with the vertex degree, which fits well even across system types. In particular for the final three kagome fragments, the size of the error is set mostly by the initial vertex degree, varying slightly as additional edges, or even sites, are added. This turns out to be a strong predictor of Trotter error even across different fragment types. Overall, since we observe only modest increases in Trotter error metrics between 1D and kagome lattices, and given the high accuracy observed in the 1D case, we expect that this Trotter step size is a reasonable choice to perform resource estimates for in the main case study.

\subsection{Sampling error analysis for spectral functions}
\label{app:sampling_error_analysis}

In this appendix, we perform analysis of the statistical error on a spectral function constructed with importance sampling, using the quench protocol of \cref{eq:quench_formula} and the circuits of \cref{fig:circuit_and_lattice}(a). We will show that the resulting estimator is unbiased in the absence of logical errors, that its error is shot-noise limited, $\mathcal{O}(1/\sqrt{n})$ in the number of shots $n$, and that the average sampling overhead due to quasi-probability rotation synthesis and error mitigation, \cref{eq:expected_sampling_overhead}, enters as a multiplicative factor on the number of shots.

We first write the spectral function in a form suitable for sampling. Combining \cref{eq:spectral_function_omega_approx,eq:quench_formula,eq:spectral_function_final_expression}, every spectral function considered in this work is a fixed linear combination of expectation values of Pauli operators in quenched states,
\begin{equation}
    \mathcal{A}(\omega) = \sum_{|k| \le \kmax} \sum_{r \in R} F_k \, c_r \, e^{\mathrm{i}k\omega\delta t} \, \langle \psi_r | Q_r(k\delta t) | \psi_r \rangle .
    \label{eq:I_target}
\end{equation}
Here the label $r = (\mu, a, b, \sigma, j, m)$ collects the sites $(\mu,a)$ and $(0,b)$ of \cref{eq:spectral_function_final_expression}, the quench label $(\sigma, j)$ of \cref{eq:quench_formula}, and the term $m$ of a Pauli decomposition $S^{\alpha}_{(\mu,a)} = \sum_m v_m Q_m$ of the measured operator. The quenched state is $|\psi_r\rangle = e^{\mathrm{i}(-1)^{\sigma}\frac{\pi}{4}P_j}|\psi\rangle$, where $P_j$ are the Pauli operators in the decomposition $S^{\beta}_{(0,b)} = \sum_j w_j P_j$, and $Q_r$ is the Pauli operator $Q_m$ acting on site $(\mu,a)$, including any Jordan--Wigner string. The coefficients
\begin{equation}
    c_r = -\frac{\mathrm{i}}{n_s} \, e^{-\mathrm{i}\mathbf{q}\cdot(\mathbf{R}_\mu + \mathbf{r}_a - \mathbf{r}_b)} \, (-1)^{\sigma} w_j \, v_m
    \label{eq:I_coeffs}
\end{equation}
are independent of $k$ and $\omega$. For the transverse spectral function, \cref{lemma:branch-cost} replaces the sum over quench states by the single quench $e^{\frac{\pi}{4}\gamma_{j\uparrow,1}\gamma_{j\downarrow,1}}$, which only changes the set $R$ and the coefficients $c_r$. We define the one-norms
\begin{equation}
    \lambda_F := \sum_{|k| \le \kmax} |F_k|, \qquad
    \lambda_c := \sum_{r \in R} |c_r|, \qquad
    \lambda_{\rm tot} := \lambda_F \lambda_c .
    \label{eq:I_norms}
\end{equation}
If, as discussed following \cref{eq:spectral_function_omega_approx_2}, the terms with $k<0$ are reconstructed from those with $k>0$, then the sum over $k$ runs over $0 \le k \le \kmax$ with coefficients $F_0$ and $2F_k$ for $k \geq 1$. The Fourier mode $e^{\mathrm{i}k\omega\delta t}$ below is then replaced by a function of modulus at most one, built from $\cos(k\omega\delta t)$ and $\sin(k\omega\delta t)$. This leaves $\lambda_F$ unchanged, and all bounds below continue to hold, with the equalities in \cref{eq:I_modulus,eq:I_mse,eq:I_pointwise} replaced by inequalities.

In the importance sampling procedure we sample $k \in \{ -k_{\max}, \ldots, k_{\max} \}$ from the probability distribution $p_k = |F_k|/\lambda_F$ (note, $|k|$ is the number of Trotter steps); combining $-k$ and $k$ gives the distribution of the main text, \cref{eq:importance_sample_probs}. We sample $r$ from $p_r = |c_r|/\lambda_c$. We denote by $k_j$ and $r_j$ the values obtained on the $j$'th sample. Once these are selected, we prepare $|\psi\rangle$, apply the quench to obtain $|\psi_{r_j}\rangle$, apply the Trotter step $U$ a total of $k_j$ times, and measure the Pauli operator $Q_{r_j}$. This returns a single outcome $s_j \in \{+1, -1\}$ such that
\begin{align}
    \mathbb{E}( s_j \,|\, k_j, r_j ) = \langle \psi_{r_j} | Q_{r_j}(k_j \delta t) | \psi_{r_j} \rangle,
\end{align}
where the time evolution is that of $U^{k_j}$. We consider this as returning a complex random variable
\begin{equation}
    z_j(\omega) := \lambda_{\rm tot} \, \mathrm{sgn}(F_{k_j}) \, \frac{c_{r_j}}{|c_{r_j}|} \, s_j \, e^{\mathrm{i}k_j\omega\delta t},
    \label{eq:I_sample}
\end{equation}
so that
\begin{equation}
    \mathbb{E}\big( z_j(\omega) \big)
    = \sum_{k} \sum_{r} \frac{|F_k|}{\lambda_F} \frac{|c_r|}{\lambda_c} \,
      \lambda_{\rm tot} \, \mathrm{sgn}(F_k) \frac{c_r}{|c_r|} \, \langle \psi_r | Q_r(k\delta t) | \psi_r \rangle \, e^{\mathrm{i}k\omega\delta t}
    = \mathcal{A}(\omega)
    \label{eq:I_unbiased}
\end{equation}
for every $\omega$. Hence $z_j$ is an unbiased estimator of the Trotterised spectral function. Note that this is the ``standard'' approach to analysis of statistical protocols like statistical phase estimation \cite{Lin2022}. Within this approach we consider $z_j(\omega)$ as a scalar random variable that is an unbiased estimator of the function $\mathcal{A}$ evaluated at the point $\omega$. However, we can also take the perspective that $z_j$ is a random \textit{function}. Every time we take a sample, $z_j$ takes the value given by \cref{eq:I_sample}. Each sample requires a single circuit with a single Pauli measurement, and the crucial fact is that its modulus is the same at every frequency,
\begin{equation}
    |z_j(\omega)| = \lambda_{\rm tot} \quad \text{for all } \omega .
    \label{eq:I_modulus}
\end{equation}
When $\mathcal{A}(\omega)$ is known to be real, one may instead use $\mathrm{Re}\, z_j(\omega)$, which is also unbiased and satisfies $|\mathrm{Re}\, z_j(\omega)| \le \lambda_{\rm tot}$; all bounds below continue to hold. Finally, because $|\langle\psi_r|Q_r(k\delta t)|\psi_r\rangle| \le 1$, we have $|\mathcal{A}(\omega)| \le \lambda_{\rm tot}$.

A single sample is a poor estimator and so we average over $n$ shots,
\begin{align}
    Z_n = \frac{1}{n} \sum_{j=1}^n z_j.
\end{align}
We will show that this is shot-noise limited, $\mathcal{O}(1/\sqrt{n})$, as expected, but first we establish a figure of merit with which to assess the randomly sampled spectral function.

When constructing a spectral function, $\mathcal{A}(\omega)$, we are interested in the function for the full range, rather than an estimate at a single value $\omega$. As such, we introduce the following inner product on functions
\begin{align}
    \langle f, h \rangle := \frac{1}{2c} \int_{-c}^{c} f^*(\omega) h(\omega) d\omega,
\end{align}
and this induces a natural 2-norm $\norm{f} = \sqrt{\langle f, f \rangle}$. We choose the limits of the integral ($\pm c$) to enforce an orthogonality relation on the sampled functions. In particular, we desire that for integer $k$ and $k'$
\begin{align}
    \langle e^{\mathrm{i}k\omega\delta t}, e^{\mathrm{i}k'\omega\delta t} \rangle = \delta_{k,k'}.
\end{align}
When $k = k'$ we have
\begin{align}
    \langle e^{\mathrm{i}k\omega\delta t}, e^{\mathrm{i}k\omega\delta t} \rangle = \frac{1}{2c} \int_{-c}^{c} |e^{\mathrm{i}k\omega\delta t}|^2 d\omega = \frac{1}{2c} \int_{-c}^{c} 1 \, d\omega = 1.
\end{align}
When $k' - k \neq 0$,
\begin{align}
    \langle e^{\mathrm{i}k\omega\delta t}, e^{\mathrm{i}k'\omega\delta t} \rangle & = \frac{1}{2c} \int_{-c}^{c} e^{\mathrm{i}\omega\delta t (k'-k)} d\omega \nonumber \\
    & = \frac{1}{2c\,\mathrm{i}\,\delta t (k'-k)} \left[ e^{\mathrm{i}c\delta t (k'-k)} - e^{-\mathrm{i}c\delta t (k'-k)} \right].
\end{align}
Setting $c$ such that $c\,\delta t = \pi$ means that $e^{\mathrm{i}c\delta t (k'-k)} = e^{-\mathrm{i}c\delta t (k'-k)} = (-1)^{k'-k}$, and so
\begin{align}
    \langle e^{\mathrm{i}k\omega\delta t}, e^{\mathrm{i}k'\omega\delta t} \rangle = 0.
\end{align}
Therefore, our inner product and norm are defined over a single period of the Fourier series,
\begin{equation}
    \langle f, h \rangle := \frac{\delta t}{2\pi} \int_{-\pi/\delta t}^{\pi/\delta t} f^*(\omega) h(\omega) d\omega,
    \label{eq:I_inner}
\end{equation}
which, by \cref{eq:upper_bound_time_step}, contains all transition energies of interest. From \cref{eq:I_modulus} and the normalisation of the inner product, $\norm{z_j}^2 = \lambda_{\rm tot}^2$ for every sample. Similarly, $\norm{\mathcal{A}} \le \lambda_{\rm tot}$.

We declare our figure of merit to be $\norm{Z_n - \mathcal{A}}$. Then,
\begin{align}
    \mathbb{E}( \norm{Z_n - \mathcal{A}}^2 ) = \mathbb{E}\left( \langle Z_n, Z_n \rangle \right) - \langle \mathcal{A}, \mathcal{A} \rangle.
\end{align}
Next, recall that $Z_n = (1/n) \sum_{j=1}^n z_j$ so
\begin{align}
    \mathbb{E}\left( \langle Z_n, Z_n \rangle \right) = \frac{1}{n^2} \sum_{j,j'} \mathbb{E}\left( \langle z_j, z_{j'} \rangle \right) = \frac{1}{n^2} \Big[ \sum_{j \neq j'} \mathbb{E}\left( \langle z_j, z_{j'} \rangle \right) + \sum_{j} \mathbb{E}\left( \langle z_j, z_j \rangle \right) \Big].
\end{align}
Since $\norm{z_j}^2 = \lambda_{\rm tot}^2$ we know $\langle z_j, z_j \rangle = \lambda_{\rm tot}^2$. And since for $j \neq j'$, $z_j$ and $z_{j'}$ are independent, we have $\mathbb{E}\left( \langle z_j, z_{j'} \rangle \right) = \langle \mathbb{E}(z_j), \mathbb{E}(z_{j'}) \rangle = \norm{\mathcal{A}}^2$. Putting this together,
\begin{align}
    \mathbb{E}\left( \langle Z_n, Z_n \rangle \right) = \frac{1}{n^2} \left[ n(n-1) \norm{\mathcal{A}}^2 + n \lambda_{\rm tot}^2 \right] = \left( 1 - \frac{1}{n} \right) \norm{\mathcal{A}}^2 + \frac{\lambda_{\rm tot}^2}{n}.
\end{align}
Therefore,
\begin{equation}
    \mathbb{E}( \norm{Z_n - \mathcal{A}}^2 ) = \frac{1}{n} \left( \lambda_{\rm tot}^2 - \norm{\mathcal{A}}^2 \right) \le \frac{\lambda_{\rm tot}^2}{n} = \mathcal{O}(1/n).
    \label{eq:I_mse}
\end{equation}
The same argument applied at a fixed frequency, using $|z_j(\omega)|^2 = \lambda_{\rm tot}^2$ from \cref{eq:I_modulus}, gives
\begin{equation}
    \mathbb{E}\left( |Z_n(\omega) - \mathcal{A}(\omega)|^2 \right) = \frac{1}{n} \left( \lambda_{\rm tot}^2 - |\mathcal{A}(\omega)|^2 \right) \le \frac{\lambda_{\rm tot}^2}{n} \quad \text{for all } \omega,
    \label{eq:I_pointwise}
\end{equation}
so the statistical noise is spread uniformly across the spectrum.

Next, we discuss confidence bounds. Roughly speaking, \cref{eq:I_mse} is the expected size of $\norm{Z_n - \mathcal{A}}^2$ and so we can reasonably expect that $\norm{Z_n - \mathcal{A}}$ is the square root of this result, and therefore $\mathcal{O}(1/\sqrt{n})$. More precisely, because $\norm{Z_n - \mathcal{A}}^2$ is a positive random variable we can use Markov's inequality, and therefore
\begin{align}
    \mathrm{Prob}( \norm{Z_n - \mathcal{A}} \geq \sqrt{a} ) = \mathrm{Prob}( \norm{Z_n - \mathcal{A}}^2 \geq a ) \leq \frac{ \mathbb{E}( \norm{Z_n - \mathcal{A}}^2 ) }{a}.
\end{align}
Setting $a = (1/p_f) \, \mathbb{E}( \norm{Z_n - \mathcal{A}}^2 )$ we obtain
\begin{align}
    \mathrm{Prob}\left( \norm{Z_n - \mathcal{A}} \geq \sqrt{ \mathbb{E}( \norm{Z_n - \mathcal{A}}^2 ) / p_f } \right) \leq p_f.
\end{align}
In other words, with probability at least $1 - p_f$ we have
\begin{equation}
    \norm{Z_n - \mathcal{A}} \le \lambda_{\rm tot} \sqrt{\frac{1}{n p_f}} = \mathcal{O}(1/\sqrt{n}).
    \label{eq:I_markov}
\end{equation}
The scaling with $p_f$ can be improved by using the Hoeffding inequality. Indeed, $Z_n$ is an average of independent random elements of a Hilbert space, and the centred samples are bounded by $\norm{z_j - \mathcal{A}} \le \norm{z_j} + \norm{\mathcal{A}} \le 2\lambda_{\rm tot}$. The vector-valued Hoeffding inequality~\cite{Pinelis1994} then gives $\mathrm{Prob}( \norm{Z_n - \mathcal{A}} \geq \epsilon ) \le 2 \exp\!\big( -n\epsilon^2 / 8\lambda_{\rm tot}^2 \big)$. For a failure probability $p_f$, it follows that with probability at least $1 - p_f$
\begin{equation}
    \norm{Z_n - \mathcal{A}} \le \lambda_{\rm tot} \sqrt{\frac{8 \ln(2/p_f)}{n}}.
    \label{eq:I_hoeffding}
\end{equation}
This provides an exponential improvement in the failure probability $p_f$ over the Markov-based bound in \cref{eq:I_markov}, replacing the $1/\sqrt{p_f}$ scaling with $\sqrt{\ln(1/p_f)}$.

So far this section has assumed noiseless circuits. Two changes are needed for the circuits that are actually run. First, with quasi-probability Clifford+T rotation synthesis, the circuit with $k$ Trotter steps is replaced by a quasi-probability mixture of Clifford+T circuits with one-norm $\lambda_{\mathrm{rs}}^{k}$; each shot samples a circuit from this mixture and records its sign $\varsigma_j = \pm 1$. Second, logical errors from the Clifford circuit components and faulty magic states are mitigated by rescaling, as described in \cref{sec:resource_estimates}. We assume that the noisy expectation values are damped by $\gamma^{k}$, where
\begin{equation}
\gamma = (1 - 1/Q)^{v}\,(1 - p_{\mathrm{cultivation}})^{m}
\end{equation}
is the probability that one Trotter step suffers no logical fault, and we divide each sample by $\gamma^{k_j}$. Writing $\lambda_{\mathrm{sample},k} = \lambda_{\mathrm{rs}}^{k}\,\gamma^{-k}$, so that
\begin{equation}
\lambda_{\mathrm{sample},k}^{2} = \lambda_{\mathrm{rs}}^{2k}\,(1 - 1/Q)^{-2vk}\,(1 - p_{\mathrm{cultivation}})^{-2mk}
\end{equation}
is the summand of \cref{eq:expected_sampling_overhead}, the sample of \cref{eq:I_sample} becomes
\begin{equation}
z_j(\omega) = \lambda_{\mathrm{tot}}\,\lambda_{\mathrm{sample},k_j}\,\varsigma_j\,\mathrm{sgn}(F_{k_j})\,\frac{c_{r_j}}{|c_{r_j}|}\,s_j\,e^{i k_j \omega \delta t}.
\end{equation}
Because $|s_j| = 1$ for every shot, we have that $\|z_j\|^{2} = \lambda_{\mathrm{tot}}^{2}\lambda_{\mathrm{sample},k_j}^{2}$. The mean of $z_j$, which we denote $\tilde{A}(\omega)$, is the Trotterised spectral function built from the noisy expectation values divided by $\gamma^{k}$. It equals the noiseless Trotterised spectral function $A(\omega)$ only if each noisy expectation value is exactly $\gamma^{k}$ times its ideal value. Otherwise, $\tilde{A} - A$ is a bias that does not decrease with more shots. Repeating the derivation of \cref{eq:I_mse} about the mean $\tilde{A}$ gives
\begin{equation}
\mathbb{E}\left(\|Z_n - A\|^{2}\right) = \mathbb{E}\left(\|Z_n - \tilde{A}\|^{2}\right) + \|\tilde{A} - A\|^{2} \le \frac{\lambda_{\mathrm{tot}}^{2}\,\lambda_{\mathrm{sample}}^{2}}{n} + \|\tilde{A} - A\|^{2},
\qquad
\lambda_{\mathrm{sample}}^{2} := \sum_{k} p_k\,\lambda_{\mathrm{sample},k}^{2},
\label{eq:I_mse_pec}
\end{equation}
where the cross term vanishes because $\mathbb{E}(Z_n) = \tilde{A}$, and similarly for \cref{eq:I_pointwise,eq:I_markov} with $A$ replaced by $\tilde{A}$. Here $\lambda_{\mathrm{sample}}^{2}$ is the expected sampling overhead per shot of \cref{eq:expected_sampling_overhead}, i.e. the averaged ``sampling overhead (total)'' reported in \cref{tab:costing_results}. It therefore acts as a multiplier on the number of shots required for a given statistical accuracy. In practice, the bias depends on how well the damping model describes the logical noise, and on how accurately $\gamma$ is calibrated, which is not analysed here.

We note that $p_k \propto |F_k|$ is not optimal when applying error mitigation. Choosing $p_k \propto |F_k| \lambda_{{\rm sample},k}$ instead gives $\mathbb{E}(\norm{z_j}^2) = \lambda_c^2 \big( \sum_k |F_k| \lambda_{{\rm sample},k} \big)^2 \le \lambda_{\rm tot}^2 \, \overline{\lambda_{\rm sample}^2}$ by the Cauchy--Schwarz inequality. This can reduce the number of shots further, at the cost of sampling longer circuits with higher probability.

Finally, we relate these bounds to the resolution of spectral features. For the Lorentzian broadening used in this work, $F_k = \frac{\delta t}{2\pi} e^{-\eta\delta t|k|} > 0$ (\cref{eq:periodic_broadening}), so that $\lambda_F = \sum_k F_k = \tilde{F}(0)$ is the peak height of a spectral line of unit weight in \cref{eq:explicit_spectral_function_with_broadening}. For $\eta\delta t \ll 1$ (\cref{eq:Lorentzian_approximation_condition}), summing the geometric series gives $\lambda_F \approx (1 - e^{-\eta\delta t\,\kmax})/(\pi\eta)$. For the parameters of \cref{app:broadening_function_and_explicit_representation}, where $\eta\delta t\,\kmax = \pi$, this is within $5\%$ of the peak height $1/(\pi\eta)$ of the Lorentzian of \cref{eq:lorentzian}. A spectral line with weight $C_{ij}$ therefore appears as a peak of height $\approx |C_{ij}|\lambda_F$. \Cref{eq:I_pointwise,eq:I_mse_pec} then give a statistical root-mean-square noise relative to this peak of at most
\begin{equation}
    \frac{\lambda_c}{|C_{ij}|} \sqrt{\frac{\overline{\lambda_{\rm sample}^2}}{n}} .
    \label{eq:I_relative}
\end{equation}
As an example, consider a site-resolved transverse correlator $C^{+-}_{ij}$ computed via \cref{lemma:branch-cost}. The measured operator $2\mathrm{i}S_i^+$ is a combination of four Pauli operators with one-norm $\lambda_c = 2$, \cref{eq:spin_pm_aop_majop}. Take the average total overhead $\overline{\lambda_{\rm sample}^2} = 15.7$ of the kagome circuits under SI1000 noise at $d = 15$ (from \cref{tab:costing_results}) and $n = 10^5$ shots. A line of unit weight is then resolved with a relative noise of about $2.5\%$. For the Optimistic noise model at $d = 13$ ($\overline{\lambda_{\rm sample}^2} = 3.10$) this falls to about $1.1\%$.

For the momentum-resolved function of \cref{eq:spectral_function_final_expression}, $\lambda_c$ grows with the number of site pairs summed over, if these are importance sampled. For example, for the longitudinal spectral function we have $\lambda_c = N_s/2$. In practice, however, the measured operators on different sites commute and can be measured simultaneously in every shot. In that case the bound of \cref{eq:I_mse} still holds but is typically very loose.

\end{document}